\documentclass[12pt,twoside,english]{extarticle}
\usepackage{lmodern}
\usepackage{helvet}
\usepackage[T1]{fontenc}
\usepackage[latin9]{inputenc}
\usepackage{geometry}
\usepackage{color}
\usepackage{babel}
\usepackage{float}
\usepackage{textcomp}
\usepackage{mathrsfs}
\usepackage{amsmath}
\usepackage{amsthm}
\usepackage{amssymb}
\usepackage{setspace}
\usepackage[unicode=true,pdfusetitle,
 bookmarks=true,bookmarksnumbered=false,bookmarksopen=false,
 breaklinks=true,pdfborder={0 0 0},pdfborderstyle={},backref=false,colorlinks=true]
 {hyperref}

\makeatletter

\newcommand{\lyxmathsym}[1]{\ifmmode\begingroup\def\b@ld{bold}
  \text{\ifx\math@version\b@ld\bfseries\fi#1}\endgroup\else#1\fi}

\providecommand{\tabularnewline}{\\}

\numberwithin{equation}{section}
\numberwithin{table}{section}
\numberwithin{figure}{section}
\usepackage[natbibapa]{apacite}
\theoremstyle{definition}
\newtheorem{defn}{\protect\definitionname}[section]
\theoremstyle{definition}
\newtheorem{example}{\protect\examplename}[section]
\theoremstyle{plain}
\newtheorem{assumption}{\protect\assumptionname}
\theoremstyle{plain}
\newtheorem{thm}{\protect\theoremname}[section]
\theoremstyle{plain}
\newtheorem{lem}{\protect\lemmaname}[section]
\theoremstyle{plain}
\newtheorem{prop}{\protect\propositionname}[section]
\theoremstyle{definition}
\newtheorem{condition}{\protect\conditionname}
\theoremstyle{plain}
\newtheorem{cor}{\protect\corollaryname}[section]

\usepackage{graphicx}
\usepackage{amsmath}
\usepackage{dcolumn}
\usepackage{listings}
\usepackage{color}
\usepackage{setspace}
\usepackage[normalem]{ulem}  
\definecolor{hellgelb}{rgb}{1,1,0.8}
\definecolor{colKeys}{rgb}{0,0,1}
\definecolor{colIdentifier}{rgb}{0,0,0}
\definecolor{colComments}{rgb}{1,0,0}
\definecolor{colString}{rgb}{0,0.5,0}

\usepackage{lmodern}
\usepackage[T1]{fontenc}
\usepackage{geometry}
\usepackage{fancyhdr}
\usepackage{amsthm}
\usepackage{thmtools}
\usepackage{amsmath}
\usepackage{amssymb}
\usepackage{esint}
\usepackage{multirow}
\usepackage{mathrsfs} 
\usepackage{textgreek}
\usepackage{amsfonts}
\usepackage{amssymb}
\usepackage{mathrsfs} 

\usepackage[USenglish]{isodate}
\usepackage{chngcntr}

\numberwithin{equation}{section}
\numberwithin{table}{section}
\numberwithin{assumption}{section}
\counterwithout{figure}{section}
\counterwithout{table}{section}

\makeatother

  \providecommand{\assumptionname}{Assumption}

  \providecommand{\corollaryname}{Corollary}
  
  \providecommand{\definitionname}{Definition}
  \providecommand{\examplename}{Example}

  \providecommand{\lemmaname}{Lemma}

  \providecommand{\propositionname}{Proposition}

  \providecommand{\theoremname}{Theorem}
 \providecommand{\corollaryname}{Corollary}
 \providecommand{\theoremname}{Theorem}
 
\usepackage{thmtools}

\newtheoremstyle{MyTheoremstyle}
  {\topsep} 
  {\topsep} 
  {} 
  {} 
  {\bfseries} 
  {.} 
  {.90em} 
  {} 
\theoremstyle{MyTheoremstyle} 
\theoremstyle{MyTheoremstyle} 
\theoremstyle{MyTheoremstyle} 
\theoremstyle{MyTheoremstyle} 
\theoremstyle{MyTheoremstyle}

\declaretheoremstyle[
    headfont=\bfseries,
    notefont=\normalfont,
    bodyfont=\itshape,
    headpunct=\newline,
    headformat={%
        \makebox{\NAME\ \NUMBER\ }{\NOTE}%
    },
]{theorem}

\newlength{\spacelength}
\declaretheoremstyle[
    headfont=\bfseries,
    notefont=\normalfont,
    bodyfont=\itshape,
    headpunct=\newline,
    headformat={%
        \makebox[0pt][l]{\NAME\ \NUMBER\ }\hskip-\spacelength{\NOTE}%
    },
]{theore}

 \usepackage{fancyhdr}
\usepackage{booktabs}
\usepackage{float}
\usepackage{graphicx}
\usepackage{epstopdf}
\usepackage{morefloats}
\usepackage[referable]{threeparttablex}
\usepackage{footnote}

\usepackage{setspace} 
\usepackage{graphicx,color}

\usepackage{listings}
\RequirePackage[T1]{fontenc}
\RequirePackage{ae,fancyvrb}
\DefineVerbatimEnvironment{Sinput}{Verbatim}{fontshape=sl}
\DefineVerbatimEnvironment{Soutput}{Verbatim}{}
\DefineVerbatimEnvironment{Scode}{Verbatim}{fontshape=sl}

\title{\bf Theory of Evolutionary Spectra for Heteroschesdasticity and Autocorrelation Robust Inference in possibly Misspecified and Nonstationary Models}
\author{
\textsc{\textcolor{MyBlue}{Alessandro Casini}}\thanks{Department of Economics and Finance, University of Rome Tor Vergata, Via Columbia 2, Rome 00133, IT. 
Email: 
\texttt{\textcolor{MyBlue}{{alessandro.casini@uniroma2.it}}}.} 
\
{/ University of Rome Tor Vergata}
}

\date{\small{\today} \\ \footnotesize{First Version: November 2018}}

 \makeatletter

\numberwithin{equation}{section}
\makeatother

\usepackage{babel}
\addto\captionsenglish{%
}

\usepackage{color}
\usepackage{xcolor}

\renewcommand*{\thesection}{\arabic{section}}

\definecolor{MyRed}{rgb}{0.8,0,0}
\definecolor{MyBlue}{rgb}{0,0,0.7}
\definecolor{Green}{rgb}{0,0.5,0}
\definecolor{hellgelb}{rgb}{1,1,0.8}
\definecolor{colKeys}{rgb}{0,0,1}
\definecolor{colIdentifier}{rgb}{0,0,0}
\definecolor{colComments}{rgb}{1,0,0}
\definecolor{colString}{rgb}{0,0.5,0}

\definecolor{MyLightRed}{rgb}{2.2,0.2,0.4} 
\definecolor{MyLightRed2}{rgb}{0.6,0.2,0.3} 
\definecolor{MyLightRed2temp}{rgb}{0.6,0.2,0.3}
\definecolor{MyLightRed3}{rgb}{0.8,0.1,0.1} 
\definecolor{MyRed}{rgb}{0.7,0.0,0}

\definecolor{MyLigthBlue13}{rgb}{0,0.2,0.7}
 \definecolor{MyLigthBlack}{rgb}{0.2,0.25,0.3} 

\hypersetup{%
  bookmarks=true
  pdftitle = {Title Here},
  pdfsubject = {},
  pdfkeywords = {Keywords},
  pdfauthor = {Alessandro Casini},}
\hypersetup{ 
  colorlinks = {true}, 
  citecolor = {MyBlue},
  linkcolor = {MyRed},
  filecolor= {Green},	
  urlcolor = {MyBlue},
  hyperindex = {true},
  linktocpage = {true},
  linkbordercolor ={1 0 0},
 citebordercolor ={0 1 1},
 urlbordercolor ={0 1 1},
hypertexnames=false
}

\usepackage{bookmark}

\usepackage[bottom]{footmisc}
\usepackage{multibib}
\newcites{ReferencesSupp}{References}
\newcites{ReferencesSupptwo}{References}

\makeatother

\providecommand{\assumptionname}{Assumption}
\providecommand{\conditionname}{Condition}
\providecommand{\corollaryname}{Corollary}
\providecommand{\definitionname}{Definition}
\providecommand{\examplename}{Example}
\providecommand{\lemmaname}{Lemma}
\providecommand{\propositionname}{Proposition}
\providecommand{\theoremname}{Theorem}

\begin{document}
\pagebreak{}

\setcounter{page}{0}

\raggedbottom
\title{\textbf{\Large{}Theory of Evolutionary Spectra for Heteroskedasticity
and Autocorrelation Robust Inference in Possibly Misspecified and
Nonstationary Models}\textbf{}\thanks{This paper is based on the first chapter of my doctoral dissertation
at Boston University. I am grateful to Pierre Perron for his continuous
support and advice. I thank Whitney Newey and Tim Vogelsang for helpful
discussions. I also thank Marine Carrasco, Giuseppe Cavaliere, Bin
Chen, Andrew Chesher,  Jean-Jacques Forneron, Massimo Franchi,
 Raffaella Giacomini, Liudas Giraitis, Jesus Gonzalo, Chris Hansen,
Jungbin Hwang, Hiroaki Kaido, Ivana Komunjer, Oliver Linton, Adam
McCloskey,   Zhongjun Qu, Myung Seo, Julius Vainora, Rasmus Varneskov,
Daniel Whilem, Dacheng Xiu, Yohei Yamamoto, and seminar participants
at   BC-BU Econometrics Workshop, Cambridge, Chicago Booth, Connecticut,
 KU Leuven, Georgetown and UCL for comments. }}
\maketitle
\begin{abstract}
{\footnotesize{}The literature on heteroskedasticity and autocorrelation
robust (HAR) inference is extensive but its usefulness relies on
 stationarity of the relevant process, say $V_{t}$, usually a function
of the data and estimated model residuals. Yet, a large body of work
shows widespread evidence of various forms of nonstationarity in the
latter. Also, many testing problems are such that $V_{t}$ is stationary
under the null hypothesis but nonstationary under the alternative.
In either case, the consequences are possible size distortions and,
especially, a reduction in power which can be substantial (e.g., non-monotonic
power), since all such estimates are based on weighted sums of the
sample autocovariances of $V_{t}$, which are inflated. We propose
HAR inference methods valid under a broad class of nonstationary processes,
labelled Segmented Local Stationary, which possess a spectrum that
varies both over frequencies and time. It is allowed to change either
slowly and continuously and/or abruptly at some time points, thereby
encompassing most nonstationary models used in applied work. We introduce
a double kernel estimator (DK-HAC) that applies a smoothing over both
lagged autocovariances and time. The optimal kernels and bandwidth
sequences are derived under a mean-squared error criterion. The data-dependent
bandwidths rely on the ``plug-in'' approach using approximating
parametric models having time-varying parameters estimated with standard
methods applied to local data. Our method yields tests with good size
and power under both stationary and nonstationary, thereby encompassing
previous methods. In particular, the power gains are achieved without
notable size distortions, the exact size being as good as those delivered
by the best fixed-$b$ approach, when the latter works well.}{\footnotesize\par}
\end{abstract}

\onehalfspacing
\thispagestyle{empty}

\pagebreak{}

\section{Introduction}

The literature on heteroskedasticity and autocorrelation robust (HAR)
inference is extensive and quite mature by now. For concreteness,
consider the linear model where $x_{t}$ is a vector of regressors
and $e_{t}$ is an unobservable disturbance, which can be serially
correlated. It is now common practice to  use OLS and correct the
standard errors. This entails the estimation of the covariance matrix
(referred to as the long-run variance, LRV) of $V_{t}=x_{t}e_{t}$
or ($2\pi$ times) the spectral density of $V_{t}$ at frequency zero
when the latter is stationary (of course, in general, the relevant
process $V_{t}$ can be generated from a more complex model; e.g.,
a moment condition in a GMM context). Early important contributions
in econometrics are \citeauthor{newey/west:87} (\citeyear{newey/west:87};
\citeyear{newey/west:94}) and \citet{andrews:91} who proposed heteroskedasticity
and autocorrelation consistent (HAC) estimators with some optimal
properties. This approach aims at devising good estimate of the LRV
of $V_{t}$. An alternative method foregoes that aim and concentrates
on having a test with a pivotal non-normal limit distribution that
is obtained through an inconsistent estimate of the LRV of $V_{t}$
that keeps the bandwidth at a fixed fraction of the sample size. This
is the so-called fixed-$b$ HAR inference initiated by \citet{Kiefer/vogelsang/bunzel:00}
and \citeauthor{Kiefer/vogelsang:02} (\citeyear{Kiefer/vogelsang:02};
\citeyear{kiefer/vogelsang:05}). The drawback of this approach is
that the limit distribution changes depending on the context and critical
values are to be obtained numerically on a case by case basis. The
literature since then has focused on providing various refinements,
mostly to have tests having exact sizes closer to the nominal level.\footnote{The fixed-$b$ or post-HAC literature is vast; see \citet{dou:18},
\citet{lazarus/lewis/stock:17}, Lazarus et al. \citeyearpar{lazarus/lewis/stock/watson:18},
\nocite{Goncalves/vogelsang:11} \citet{dejong/davidson:00}, \citet{ibragimov/muller:10},
\citet{jansson:04}, \citeauthor{muller:07} (\citeyear{muller:07};
\citeyear{mueller:14}), \citet{phillips:05}, \citet{politis:11},
\citet{preinerstorfer/potscher:16}, \citet{potscher/preinerstorfer:18},
\citet{rho/vogelsang:2020}, \citet{robinson:98}, \citeauthor{sun:13}
\citeyearpar{sun:13,sun:14,sun:14a}, \citet{sun/phillips/jin:08}
and \citet{zhang/shao:13}.} 

Most of this literature relies on stationarity with exception of the
consistency results in \citet{newey/west:87} and of some results
in \citet{andrews:91} which, however, do not provide accurate approximations.
Yet, another strand of the literature has argued convincingly that
the processes governing economic data $\{x_{t}\}$ and the errors
in the relevant regressions $\left\{ e_{t}\right\} $ are nonstationary.\footnote{By nonstationary we mean non-constant moments. As in the literature,
we consider processes whose sum of absolute autocovariances is finite.
That is, we rule out processes with unbounded second moments (e.g.,
unit root).} This can occur for several reasons: changes in the moments of $x_{t}$
induced by changes in the model parameters that govern the data {[}cf.
\citet{perron:89}, \citet{stock/watson:96} and the surveys of \citet{ng/wright:13}
and \citet{giacomini/rossi:15}{]}; changes in the moments of $e_{t}$
(think about the Great Moderation with the decline in variance for
many macroeconomic variables or the effects of the COVID-19 pandemic);
smooth changes in the distributions governing either processes that
arise from transitory dynamics; and so on. All these induces nonstationarity
in $\{V_{t}\}$, which then makes $\mathbb{E}(V_{t}V'_{t-k})$ depend
on both $k$ and $t$. Furthermore, even if the data and primitive
shocks $\{e_{t}\}$ are stationary, many HAR testing problems are
such that the relevant process $\{V_{t}\}$ is stationary under the
null hypothesis but is affected by changes in means (or other forms
of nonstationarity) under the alternative. This occurs, for instance,
when using tests involving structural breaks based on estimating the
model under the null hypothesis; e.g., popular tests for forecast
evaluation {[}e.g., \citet{diebold/mariano:95}{]}, tests for forecast
instability {[}cf. \citet{casini_CR_Test_Inst_Forecast}, \citet{giacomini/rossi:09}
and \citet{perron/yamamoto:18}{]}, tests for structural change {[}cf.
\citet{casini/perron_Oxford_Survey} and \citet{perron:06}{]}. When
standardized by classical HAC estimators such tests may suffer from
issues such as non-monotonic power, i.e., power that goes to zero
as the alternative gets farther away from the null value. Various
forms of misspecication and/or nonstationarity generate low frequency
contamination and make the series or residuals appear much more persistent.
As a consequence, HAC standard errors are too large and when used
as normalizing factors of test statistics, the tests lose power {[}see
\citet{casini/perron_Low_Frequency_Contam_Nonstat:2020} for formal
details{]}.\footnote{A partial list of works that present evidence of power issues with
HAR inference is \citet{altissimo/corradi:2003}, \citeauthor{casini/perron_Oxford_Survey}
(\citeyear{casini/perron_Oxford_Survey}, \citeyear{casini/perron_Lap_CR_Single_Inf},
\citeyear{casini/perron_SC_BP_Lap}), \citet{chan:2020},  \citet{crainiceanu/vogelsang:07},
 \citet{juhl/xiao:09}, \citet{kim/perron:09}, \citet{martins/perron:16},
\citet{perron/yamamoto:18}, \citet{shao/zhang:2010}, \citet{vogeslang:99},
\citet{xu:2013}, \citet{zhang/lavitas:2018}.} This applies even more forcefully to the fixed-$b$ type methods
and to the recent refinements by \citet{lazarus/lewis/stock:17} and
Lazarus et al. \citeyearpar{lazarus/lewis/stock/watson:18}, since
they involve more lagged autocovariances (or long bandwidths) and,
hence, larger contaminations. 

This points to the importance of extending the methods for HAR inference
so that they have the correct size and good power even under nonstationarity.
This is the aim of the paper. We first develop a theoretical framework
under which to analyze the statistical properties of our suggested
estimate. We introduce a class of nonstationary processes which possess
a spectrum that varies both over frequencies and time, thereby encompassing
the nonstationary models used in applied work. We work in an infill
asymptotic setting akin to the one used in nonparametric regression
{[}cf. \citet{robinson:89}{]}. For a process $V_{t}$, its spectrum
at frequency $\omega$ and time $u=t/T$, denoted by $f\left(u,\,\omega\right)$,
is allowed to change slowly yet continuously as well as to change
abruptly in $u$ at a finite number of time points; the latter allows
for structural breaks in the spectrum of $V_{t}$. We label this class
as Segmented Locally Stationary (SLS). It is related to the locally
stationary processes introduced by \citet{dahlhaus:96} that have
the characterizing property of behaving as a stationary process in
a small neighborhood of $u$. This is achieved via smoothness of
$f\left(u,\,\omega\right)$ in $u$. By allowing discontinuities across
some segments, we can deal with relevant features such as structural
change, regime switching-type and threshold models {[}cf. \citet{bai/perron:98},
\citeauthor{casini/perron_CR_Single_Break} (\citeyear{casini/perron_Oxford_Survey},
\citeyear{casini/perron_Lap_CR_Single_Inf}, \citeyear{casini/perron_SC_BP_Lap}
and \citeyear{casini/perron_CR_Single_Break}), \citet{hamilton:89}
and \citet{hansen:00ecma}{]}. The SLS class extends some of the analysis
of \citet{dahlhaus:96} to processes having a more general time-varying
spectrum.\footnote{A few authors used a notion of local stationarity that allows for
breaks {[}see, e.g., \citet{dahlhaus:2009} and \citet{last/shumway:08}{]}.
However, none of these works was concerned with HAR inference. \citet{dahlhaus:2009}
presented some results for local spectral density estimation and required
smoothness (see Example 4.2 there). \citet{last/shumway:08} considered
testing for change-points in a locally stationary series which under
the alternative hypothesis results in a piecewise locally stationary
series. Furthermore, our notion of SLS processes and related framework
are more general from a theoretical standpoint since we provide precise
definitions and establish theoretical results about the identification
of the local spectral density when there are discontinuities. } Our framework is of independent interest and can be useful in many
contexts in econometrics if one is interested in deriving the properties
of estimators or inference under nonstationarity. 

Under this framework, we introduce a double kernel HAC (DK-HAC) estimator
in order to flexibly account for nonstationarity and we show that
it is robust to low frequency contamination and other misspecifications.
This entails an extension of the classical HAC estimators since in
addition to the usual smoothing procedure over lagged autocovariances,
it applies a second smoothing over time for each lagged autocovariance,
involving a second kernel and bandwidth. If $\{V_{t}\}$ is Segmented
Locally Stationary, $\mathbb{E}(V_{t}V'_{t-k})$ changes smoothly
in $t$, as long as $t$ is away from the change-points in the spectrum
$f\left(t/T,\,\omega\right)$. Thus, the smoothing over time yields
good estimates for the time path of $\mathbb{E}(V_{t}V'_{t-k})$ for
all $k$. We determine the optimal kernels and optimal values for
both bandwidth sequences under a mean-squared error (MSE) criterion.
We establish new MSE bounds that show how nonstationarity affects
the bias-variance trade-off and are more informative than previously
established MSE bounds. We use them to construct data-dependent bandwidths
relying on the ``plug-in'' approach. Unlike \citet{andrews:91},
our candidate parametric models have time-varying parameters which
can be estimated by applying standard methods to local data, akin
to using rolling regressions. The procedure depends on three elements:
the bandwidths for the smoothing over autocovariances and over time,
and a block size to separate the regimes. In this paper, we consider
a sequential bandwidth selection procedure by first deriving the optimal
bandwidth for smoothing over time, then conditioning on this to obtain
the optimal bandwidth for smoothing over autocovariances.   

The DK-HAC estimators can result in HAR tests that are oversized when
there is high temporal dependence in the data, a well-known problem
for all methods, though for ours these distortions are relatively
minor compared to, e.g., the methods of \citet{newey/west:87} and
\citet{andrews:91}. Still, in order to improve the size control of
HAR tests, \citet{casini/perron_PrewhitedHAC}, using the theory of
this paper, propose a nonparametric nonlinear VAR prewhitened DK-HAC
estimators. This form of prewhitening differs from those discussed
previously {[}e.g., \citet{andrews/monahan:92} and \citet{rho/shao:13}{]}
in that it accounts explicitly for nonstationarity. HAR tests based
on prewhitened DK-HAC estimators have size control competitive to
fixed-$b$ HAR tests when the latter work well (i.e., under stationarity).
Notably, non-prewhitened and prewhitened DK-HAC have excellent power
properties even when existing HAR tests have serious issues with power.

\subsection*{Comparison to Existing Literature}

There are two main approaches to HAR inference differing on whether
the LRV estimator is consistent or not. The classical approach relies
on consistency, which results in HAC estimators {[}cf. \citeauthor{newey/west:87}
\citeyearpar{newey/west:87,newey/west:94}, \citet{andrews:91} and
\citet{hansen:92ecma}{]}, and on bandwidths chosen via MSE criterion.
Inference is standard because HAR tests follow asymptotically standard
distributions. The researcher then uses corrected standard errors
and asymptotic critical values. It was shown early that classical
HAC standard errors can result in oversized tests when there is substantial
temporal dependence. This stimulated a second approach based on inconsistent
LRV estimators that keep the bandwidth at some fixed fraction of the
sample size {[}cf. \citet{Kiefer/vogelsang/bunzel:00}{]}. Because
of the inconsistency, inference is nonstandard and HAR tests do not
asymptotically follow standard distributions. Critical values are
to be obtained numerically. Long bandwidths/fixed-$b$ methods require
stationarity and reduce the oversize problem of HAR tests. The bandwidth
choice is often based on testing-oriented criteria {[}e.g., \citet{sun/phillips/jin:08}{]}.
Our approach falls in the first category; we propose HAC estimators
and standard HAR inference. 

We now compare in detail our approach to the existing literature.
We believe that a fair comparison has to consider the following four
criteria: (1) applicability to general HAR inference tests; (2) size
of HAR tests; (3) power of HAR tests; (4) theoretical validity under
stationarity/nonstationarity. In terms of (1), it is clear that the
 HAC and DK-HAC estimators are generally and immediately applicable
to any HAR inference test and that they are simple to use in practice.
This also explains why the classical HAC estimators have become the
standard practice in econometrics. Methods that rely on long bandwidths/fixed-$b$
do not share the same property. They are not generally applicable
to HAR inference tests because a researcher would first need to derive
a new asymptotic non-standard fixed-$b$ distribution. This can be
unfeasible in non-standard testing problems {[}e.g., tests for parameter
instability, etc.{]}. Turning to (2), all existing HAR inference
tests are known to be oversized when there is strong serial dependence.
However, fixed-$b$ HAR tests (or versions thereof) are less oversized
than other tests based on the classical HAC estimators. The stronger
is the temporal dependence the larger is the difference in size between
the two approaches. Our prewhitened DK-HAC estimators are competitive
with fixed-$b$ HAR tests in controlling the size. Moving to (3),
prewhitened and non-prewhitened DK-HAC estimators have excellent power
under either stationarity or nonstationarity whereas existing methods
have serious problems with power under nonstationarity or under nonstationary
alternative hypotheses. These problems result in non-monotonic power
and little or no power in relevant circumstances especially in HAR
tests outside the stable linear regression model. Fixed-$b$ or long
bandwidths methods suffer most from these problems. Finally, turning
to (4), our method like the classical HAC approach is valid under
nonstationarity whereas methods using long bandwidths/fixed-$b$ are
only valid under stationarity {[}cf. \citet{casini_fixed_b_erp}{]}.
It should be mentioned that the fixed-$b$ approach is shown to achieve
(pointwise) higher-order refinements under stationarity while the
MSE-based optimality of the HAC or DK-HAC estimators pertains only
to the first-order but it holds under nonstationarity.\footnote{\citet{preinerstorfer/potscher:16} pointed out some limitations of
these approaches because optimality does not hold uniformly over DGPs.
They showed negative non-asymptotic results about size and power of
HAR tests when one allows for all correlation structures corresponding
to stationary Gaussian AR(1) processes. }

Recently, \citet{lazarus/lewis/stock:17} and Lazarus et al. \citeyearpar{lazarus/lewis/stock/watson:18}
made some progress on the applicability of fixed-$b$ methods. They
showed that the $t$-test using a LRV estimator based on equally-weighted
cosine (EWC) under fixed-$b$ asymptotics can achieve a $t$-distribution
with the degrees of freedom depending on the bandwidth choice.\footnote{\citet{hwang/sun:2017} proposed a modification to the trinity of
test statistics in the two-step GMM setting and showed that the modified
test statistics are asymptotically $F$ distributed under fixed-$b$
asymptotics and stationarity. However, following the logic of \citet{lazarus/lewis/stock:17},
procedures such as e.g., bootstrap-based autocorrelation robust tests,
modification of the test statistic, etc., are not as simple as HAC-based
inference and therefore may find less traction than HAC-based inference
in empirical work. } However, \citet{casini/perron_PrewhitedHAC} showed that EWC is oversized
relative to the original fixed-$b$ of \citet{Kiefer/vogelsang/bunzel:00}
and to the prewhitened DK-HAC when there is strong dependence. \citet{lazarus/lewis/stock:17}
relied on the Neyman-Pearson Lemma or simply ``apple-to-apple comparison''
to compare HAR tests. This is certainly a reasonable criterion. The
lemma suggests to compare the power of tests that have an empirical
size no greater than the significance level. However, the Neyman-Pearson
Lemma alone does not suffice to find the ``best'' test in this context
because all tests are oversized when there is strong dependence. Indeed,
it does not even apply in this context. We face a trade-off between
size and power. Our proposed method is competitive with the existing
methods which is least oversized {[}i.e., original fixed-$b$ of \citet{Kiefer/vogelsang/bunzel:00}{]}
and has excellent power even when existing HAR tests do not have any.
 We believe that our method strikes a good balance with respect to
criteria (1)-(4). 

Our approach is different from methods based on subsampling of $t$-statistics
{[}see, e.g., \citet{ibragimov/muller:10}{]}. The latter rely on
splitting the sample in subsamples and estimating the model within
each subsample. Under the assumption that the estimates from the subsamples
are asymptotically independent, the test statistic based on an average
of estimates across subsamples follows asymptotically a $t$-distribution.
In terms of point (1) above, this approach is not general enough compared
to the HAC/DK-HAC approach because this changes the test statistic
and its asymptotic distribution. Also, subsampling test statistics
which are not $t$-tests can be challenging/unfeasible and would require
at best extra work in general HAR inference contexts to derive the
new distribution. Finally, our simulation experience (not reported)
suggests that this method suffers from the same finite-sample issues
about size and power as the fixed-$b$ methods. 

\subsection*{Related Work}

This paper is part of a set of papers on HAR inference by the author
and collegues. The current paper provides the core theoretical and
empirical elements that are used in all other papers, which can be
viewed as providing extensions or refinements. \citet{casini/perron_PrewhitedHAC}
used our theoretical framework to derive minimax MSE bounds for LRV
estimation that are sharper than previously established and extended
some of our theoretical results to general nonstationarity. As a
finite-sample refinement, they also developed a new prewhitening procedure
robust to nonstationarity for DK-HAC estimators. Even though the latter
procedure is included in our simulations, we established the corresponding
theoretical results in a separate paper because of the extent of the
work needed in the analysis. \citet{casini/perron_Low_Frequency_Contam_Nonstat:2020}
showed analytically that the poor finite-sample performance of existing
LRV-based HAR tests under nonstationarity and misspecification is
induced by low frequency contamination. Belotti et al. \citeyearpar{belotti/casini/catania/grassi/perron_HAC_Sim_Bandws}
used our theoretical framework to propose alternative data-dependent
bandwidths for DK-HAC estimators that are optimal under a global MSE
criterion. \citet{casini:change-point-spectra} considered change-point
detection in time series with evolutionary spectra. Initially, it
was intended to be used in the method suggested in this paper to improve
the finite-sample size and power properties, given that we work with
segmented locally stationary processes. However, it turns out that
the current method to select the blocks (see Section \ref{subsec Choice-of nT})
is able to handle even abrupt structural change. 

\bigskip{}

The remainder of the paper is organized as follows. Section \ref{Section: Statistical Enviromnent}
introduces the statistical setting and the new HAC estimator. Section
\ref{Section HAC-Estimation-with Predetermined} presents consistency,
rates of convergence and the asymptotic MSE results for the DK-HAC
estimators. Asymptotically optimal kernels and bandwidths are derived
in Section \ref{Section Optimal-Kernels-and}. A data-dependent method
for choosing the bandwidths and its asymptotic properties are discussed
in Section \ref{Section Data-Dependent-Bandwidths}. Section \ref{Section Monte Carlo}
presents a Monte Carlo study. Section \ref{Section Conclusions}
concludes the paper. The supplemental materials {[}cf. \citet{casini_hac_supp}
and an additional supplement not for publication{]} contain some implementation
details and all mathematical proofs. The code to implement our methods
is provided in $\mathrm{\mathsf{Matlab}}$, $\mathrm{\mathrm{\mathsf{R}}}$
and $\mathrm{\mathrm{\mathsf{Stata}}}$ languages through a $\mathrm{\mathsf{Github}}$
repository.

\section{The Statistical Environment\label{Section: Statistical Enviromnent}}

To motivate our approach, consider the linear regression model estimated
by least-squares (LS): $y_{t}=x'_{t}\beta_{0}+e_{t}$ $(t=1,\ldots,\,T)$,
where $\beta_{0}\in\Theta\subset\mathbb{R}^{p}$, $y_{t}$ is an observation
on the dependent variable, $x_{t}$ is a $p$-vector of regressors
and $e_{t}$ is an unobserved disturbance. The LS estimator is given
by $\widehat{\beta}=(X'X)^{-1}X'Y$, where $Y=(y_{1},\ldots,\,y_{T})'$
and $X=(x_{1},\ldots,\,x_{T})'$. Classical inference about $\beta_{0}$
requires estimation of $\mathrm{Var}(\sqrt{T}(\widehat{\beta}-\beta_{0}))$
where
\begin{align*}
\mathrm{Var}(\sqrt{T}(\widehat{\beta}-\beta_{0})) & \triangleq\mathbb{E}\left[\left(T^{-1}\sum_{t=1}^{T}x_{t}x'_{t}\right)^{-1}T^{-1}\sum_{s=1}^{T}\sum_{t=1}^{T}e_{s}x_{s}(e_{t}x{}_{t})'\left(T^{-1}\sum_{t=1}^{T}x_{t}x'_{t}\right)^{-1}\right],
\end{align*}
where ``$\triangleq$'' is used for definitional equivalence. Consistent
estimation of $\mathrm{Var}(\sqrt{T}(\widehat{\beta}-\beta_{0}))$
relies on consistent estimation of $\mathrm{lim}_{T\rightarrow\infty}T^{-1}\sum_{s=1}^{T}\sum_{t=1}^{T}$
$\mathbb{E}(e_{s}x_{s}(e_{t}x{}_{t})')$. More generally, one needs
a consistent estimate of $J\triangleq\mathrm{lim}_{T\rightarrow\infty}J_{T}$
where $J_{T}=T^{-1}\sum_{s=1}^{T}\sum_{t=1}^{T}\mathbb{E}(V_{s}(\beta_{0})$
$V_{t}(\beta_{0})')$ with $V_{t}(\beta)$ being a random $p$-vector
for each $\beta\in\Theta$. For the linear regression model, $V_{t}(\beta)=(y_{t}-x'_{t}\beta)x_{t}$.
For HAR tests outside the regression model $V_{t}(\beta)$ takes different
forms.\footnote{If one suspects that $\beta_{0}$ may not be constant, one can use
appropriate tests for parameter instability. However, our discussions
and methods still apply because these tests are HAR inference tests
and one needs a LRV estimate of $J$ based on the appropriate $V_{t}.$ } Hence, our problem is to estimate $J$ when $\{V_{t}\}$ is a Segmented
Locally Stationary process, as defined in Section \ref{Subsec Segmented-Locally-Stationary}.\footnote{\citet{casini/perron_PrewhitedHAC} extended the results to the case
where $\left\{ V_{t}\right\} $ is generally nonstationary (i.e.,
$\left\{ V_{t}\right\} $ is a sequence of unconditionally heteroskedastic
random variables). } Such estimate can then be used to conduct HAR inference in the usual
way using the theory developed below. By a change of variables,
$J_{T}$ can be rewritten as 
\begin{align*}
J_{T} & =\sum_{k=-T+1}^{T-1}\Gamma_{T,k},\qquad\mathrm{where}\qquad\Gamma_{T,k}=\begin{cases}
T^{-1}\sum_{t=k+1}^{T}\mathbb{E}(V_{t}V'_{t-k}) & \mathrm{for\,}k\geq0\\
T^{-1}\sum_{t=-k+1}^{T}\mathbb{E}(V_{t+k}V'_{t}) & \mathrm{for\,}k<0
\end{cases},
\end{align*}
and $V_{t}=V_{t}\left(\beta_{0}\right)$. The rest of this section
is structured as follows. In Section \ref{Subsec Segmented-Locally-Stationary}
we introduce a new class of nonstationary time series  that we use
as the underlying framework for our theoretical analysis. Section
\ref{Subsection HAC-Estimation} presents the DK-HAC estimator. We
adopt the following notational conventions.  The $j$th element of
a vector $x$ is indicated by $x^{\left(j\right)}$ while the $\left(j,\,l\right)$th
element of a matrix $X$ is indicated by $X^{\left(j,\,l\right)}$.
$\mathrm{tr}(\cdot)$ denotes the trace and $\otimes$ denotes the
tensor product. The $p^{2}\times p^{2}$ matrix $C_{pp}$ is a commutation
matrix that transforms $\mathrm{vec}\left(A\right)$ into $\mathrm{vec}\left(A'\right)$,
i.e., $C_{pp}=\sum_{j=1}^{p}\sum_{l=1}^{p}\iota_{j}\iota_{l}'\otimes\iota_{l}\iota_{j}'$,
where $\iota_{j}$ is the $j$th elementary $p$-vector. $\lambda_{\max}\left(A\right)$
denotes the largest eigenvalue of $A$. $W$ and $\widetilde{W}$
are used for $p^{2}\times p^{2}$ weight matrices. $\mathbb{C}$
is used for the set of complex numbers. $\overline{A}$ is used for
the complex conjugate of $A\in\mathbb{C}$. Let $0=\lambda_{0}<\ldots<\lambda_{m+1}=1$.
A function $G\left(\cdot,\,\cdot\right):\,\left[0,\,1\right]\times\mathbb{R}\rightarrow\mathbb{C}$
is said to be piecewise (Lipschitz) continuous with $m+1$ segments
if it is (Lipschitz) continuous within each segment (e.g., it is piecewise
Lipschitz continuous if for each $j=1,\ldots,\,m+1$ it satisfies
$\sup_{u\neq v}\left|G\left(u,\,\omega\right)-G\left(v,\,\omega\right)\right|\leq K\left|u-v\right|$
for any $\omega\in\mathbb{R}$ with $\lambda_{j-1}<u,\,v\leq\lambda_{j}$
for some $K<\infty$). We define $G_{j}\left(u,\,\omega\right)=G\left(u,\,\omega\right)$
for $\lambda_{j-1}<u\leq\lambda_{j}$. If we say piecewise Lipschitz
continuous with index $\vartheta>0$, then the above inequality is
replaced by $\sup_{u\neq v}\left|G\left(u,\,\omega\right)-G\left(v,\,\omega\right)\right|\leq K\left|u-v\right|^{\vartheta}$.
A function $G\left(\cdot,\,\cdot\right):\,\left[0,\,1\right]\times\mathbb{R}\rightarrow\mathbb{C}$
is said to be left-differentiable at $u_{0}$ if $\partial G\left(u_{0},\omega\right)/\partial_{-}u\triangleq\lim_{u\rightarrow u_{0}^{-}}\left(G\left(u_{0},\,\omega\right)-G\left(u,\,\omega\right)\right)/\left(u_{0}-u\right)$
exists $\forall\omega\in\mathbb{R}$.   

\subsection{\label{Subsec Segmented-Locally-Stationary}Segmented Locally Stationary
Processes}

Suppose $\left\{ V_{t}\right\} _{t=1}^{T}$ is defined on an abstract
probability space $\left(\Omega,\,\mathscr{F},\,\mathbb{P}\right)$,
where $\Omega$ is the sample space, $\mathscr{F}$ is the $\sigma$-algebra
and $\mathbb{P}$ is a probability measure. In order to introduce
a framework to analyze time series models with a time-varying spectrum
it is necessary to introduce an infill asymptotic setting whereby
we rescale the original discrete time horizon $\left[1,\,T\right]$
by dividing each $t$ by $T.$ Letting $u=t/T$ and $T\rightarrow\infty,$
this defines a new time scale $u\in\left[0,\,1\right]$ which we interpret
as saying that as $T\rightarrow\infty$ we observe more and more realizations
of $V_{t}$ close to time $t$, i.e., we observe the rescaled process
$V_{Tu}$ on the interval $\left[u-\varepsilon,\,u+\varepsilon\right]$,
where $\varepsilon>0$ is a small number. 

In order to define a general class of nonstationary processes, we
shall start from processes that have a time-varying spectral representation
specified by:
\begin{align}
V_{t,T} & =\mu\left(t/T\right)+\int_{-\pi}^{\pi}\exp\left(i\omega t\right)A\left(t/T,\,\omega\right)d\xi\left(\omega\right),\label{Eq. 2.2 Definition V_t}
\end{align}
where $i\triangleq\sqrt{-1}$, $\mu\left(t/T\right)$ is the trend
function, $A\left(t/T,\,\omega\right)$ is the transfer function and
$\xi\left(\omega\right)$ is some stochastic process whose properties
are specified below. Observe that this representation is similar to
the spectral representation of stationary processes {[}see \citet{anderson:71},
\citet{brillinger:75}, \citet{hannan:70} and \citet{priestley:85}{]}.
We shall see that the main difference is that $A\left(t/T,\,\omega\right)$
and $\mu\left(t/T\right)$ are not constant in $t$.\footnote{In HAR inference, a minimal assumption on $V_{t}$ under the null
hypothesis is that it has zero mean (i.e., $\mu\left(t/T\right)=0$
for all $t$). However, in this subsection we allow for arbitrary
$\mu\left(t/T\right)$ so as to introduce a general framework, also
applicable under various alternative hypotheses in both within and
outside the regression model. } \citet{dahlhaus:96} used the time-varying spectral representation
to define the so-called locally stationary processes which are characterized,
broadly speaking, by smoothness conditions on $\mu\left(\cdot\right)$
and $A\left(\cdot,\,\cdot\right)$. Locally stationary processes have
been used widely in both statistics and economics, though in the latter
field they are best known as time-varying parameter processes {[}see,
e.g., \citet{cai:07} and \citet{chen/hong:12}{]}. The smoothness
restrictions exclude many prominent models that account for time variation
in the parameters. For example, structural change and regime switching-type
models do not belong to this class because parameter changes occur
suddenly at a particular point in time. We propose a class of nonstationarity
processes which allow both continuous and discontinuous changes in
the parameters. Stationarity and local stationarity are recovered
as special cases.
\begin{defn}
\label{Definition Segmented-Locally-Stationary}A sequence of stochastic
processes $\{V_{t,T}\}_{t=1}^{T}$ is called Segmented Locally Stationary
(SLS)\textbf{ }with $m_{0}+1$ regimes, transfer function $A^{0}$
 and trend $\mu$, if there exists a representation 
\begin{align}
V_{t,T} & =\mu_{j}\left(t/T\right)+\int_{-\pi}^{\pi}\exp\left(i\omega t\right)A_{j,t,T}^{0}\left(\omega\right)d\xi\left(\omega\right),\qquad\qquad\left(t=T_{j-1}^{0}+1,\ldots,\,T_{j}^{0}\right),\label{Eq. Spectral Rep of SLS}
\end{align}
for $j=1,\ldots,\,m_{0}+1$, where by convention $T_{0}^{0}=0$ and
$T_{m_{0}+1}^{0}=T$ and the following holds: 

(i) $\xi\left(\omega\right)$ is a stochastic process on $\left[-\pi,\,\pi\right]$
with $\overline{\xi\left(\omega\right)}=\xi\left(-\omega\right)$
and 
\begin{align*}
\mathrm{cum}\left\{ d\xi\left(\omega_{1}\right),\ldots,\,d\xi\left(\omega_{r}\right)\right\}  & =\varphi\left(\sum_{j=1}^{r}\omega_{j}\right)g_{r}\left(\omega_{1},\ldots,\,\omega_{r-1}\right)d\omega_{1}\ldots d\omega_{r},
\end{align*}
 where $\mathrm{cum}\left\{ \cdot\right\} $ is the cumulant  of
$r$th order, $g_{1}=0,\,g_{2}\left(\omega\right)=1$, $\left|g_{r}\left(\omega_{1},\ldots,\,\omega_{r-1}\right)\right|\leq M_{r}<\infty$
  and $\varphi\left(\omega\right)=\sum_{j=-\infty}^{\infty}\delta\left(\omega+2\pi j\right)$
is the period $2\pi$ extension of the Dirac delta function $\delta\left(\cdot\right)$.

(ii) There exists a constant $K>0$  and a piecewise continuous function
$A:\,\left[0,\,1\right]\times\mathbb{R}\rightarrow\mathbb{C}$ such
that, for each $j=1,\ldots,\,m_{0}+1$, there exists a $2\pi$-periodic
function $A_{j}:\,(\lambda_{j-1}^{0},\,\lambda_{j}^{0}]\times\mathbb{R}\rightarrow\mathbb{C}$
with $A_{j}\left(u,\,-\omega\right)=\overline{A_{j}\left(u,\,\omega\right)}$,
$\lambda_{j}^{0}\triangleq T_{j}^{0}/T$ and for all $T,$
\begin{align}
A\left(u,\,\omega\right)=A_{j}\left(u,\,\omega\right) & \,\mathrm{\,for\,}\,\lambda_{j-1}^{0}<u\leq\lambda_{j}^{0},\label{Eq A(u) =00003D Ai}\\
\sup_{1\leq j\leq m_{0}+1}\sup_{T_{j-1}^{0}<t\leq T_{j}^{0},\,\omega}\left|A_{j,t,T}^{0}\left(\omega\right)-A_{j}\left(t/T,\,\omega\right)\right| & \leq KT^{-1}.\label{Eq. 2.4 Smothenss Assumption on A}
\end{align}

(iii) $\mu_{j}\left(t/T\right)$ is piecewise continuous. 
\end{defn}
The smoothness properties of $A$ in $u$ guarantees that $V_{t,T}$
has a piecewise locally stationary behavior. Later we will require
additional smoothness properties for $A$. 
\begin{example}
(i) Suppose $X_{t}$ is a stationary process with spectral representation
$X_{t}=\int_{-\pi}^{\pi}\exp\left(i\omega t\right)$ $A\left(\omega\right)d\xi\left(\omega\right),$
and $\mu,\,\sigma:\,\left[0,\,1\right]\rightarrow\mathbb{R}$ are
piecewise continuous. Then, $V_{t,T}=\mu_{j}\left(t/T\right)+\sigma_{j}\left(t/T\right)X_{t}$,
with $T_{j-1}^{0}<t\leq T_{j}^{0}\,(1\leq j\leq m_{0}+1)$ is a SLS
process with $m_{0}+1$ regimes where $A_{j,t,T}^{0}\left(\omega\right)=A_{j}\left(t/T,\,\omega\right)=\sigma_{j}\left(t/T\right)A\left(\omega\right)$.
Within each segment, $V_{t,T}$ is locally stationary. When $t=Tu$
is away from the change-points, as $T\rightarrow\infty$ more and
more realizations of $V_{Tu,T}$ with $u\in\left[u-\varepsilon,\,u+\varepsilon\right]$
are observed, that is, realizations with amplitude close to $\sigma_{j}\left(u\right)$
for the appropriate $j$. 

(ii) Suppose $e_{t}$ is an $i.i.d.$ sequence and $V_{t,T}=\sum_{k=0}^{\infty}a_{j,k}\left(t/T\right)e_{t-k},\,T_{j-1}^{0}<t\leq T_{j}^{0}$
$(1\leq j\leq m_{0}+1)$. Then, $V_{t,T}$ is SLS with $A_{j,t,T}^{0}\left(\omega\right)=A_{j}\left(t/T,\,\omega\right)=\sum_{k=0}^{\infty}a_{j,k}\left(t/T\right)$
$\exp\left(-i\omega k\right).$ 

(iii) Autoregressive processes with time-varying coefficients, known
as TVAR, augmented with structural breaks are SLS. In this case,
we do not have the exact relationship $A_{j,t,T}^{0}\left(\omega\right)=A_{j}\left(t/T,\,\omega\right)$
but only the approximate relationship \eqref{Eq. 2.4 Smothenss Assumption on A}.
\end{example}
 If there is only a single regime (i.e., $m_{0}=0$) then $V_{t,T}$
is locally stationary {[}cf. \citet{dahlhaus:96}{]}. If $\mu$ and
$A^{0}$ do not depend on $t$, then $V_{t,T}$ is stationary and
the  spectral representation of stationary processes applies. However,
$m_{0}=0$ rules out structural change and regime switching models.
With $m_{0}>0$, we  propose a framework where parameter variation
can occur either smoothly or abruptly, both being relevant for economic
data.\footnote{Some authors have used alternative notions of local stationarity that
allow for discontinuities (i.e., piecewise locally stationary) and
have established some results in other contexts which are not related
to HAR inference {[}see, e.g., \citet{dahlhaus:2009}, \citet{last/shumway:08}
and \citet{zhou:2013}{]}. In particular, our framework is more general
because we also define (and work with) the covariance between observations
belonging to different regimes whereas previous works considered only
the covariance between observations belonging to the same regime thereby
using smoothness which restricts the framework substantially.} 

Let $\left\lfloor \cdot\right\rfloor $ denote the largest smaller
integer function and let $\mathcal{T}\triangleq\{T_{1}^{0},\,\ldots,\,T_{m_{0}}^{0}\}$.
We define the spectrum of $V_{t,T}$ in \eqref{Eq. 2.2 Definition V_t}
(for fixed $T$) as 
\begin{align*}
f_{j,T}\left(u,\,\omega\right) & \triangleq\begin{cases}
\left(2\pi\right)^{-1}\sum_{s=-\infty}^{\infty}\mathrm{Cov}\left(V_{\left\lfloor Tu-3\left|s\right|/2\right\rfloor ,T},\,V_{\left\lfloor Tu-\left|s\right|/2\right\rfloor ,T}\right)\exp\left(-i\omega s\right), & Tu\in\mathcal{T},\,u=\lambda_{j}^{0}\\
\left(2\pi\right)^{-1}\sum_{s=-\infty}^{\infty}\mathrm{Cov}\left(V_{\left\lfloor Tu-s/2\right\rfloor ,T},\,V_{\left\lfloor Tu+s/2\right\rfloor ,T}\right)\exp\left(-i\omega s\right), & Tu\notin\mathcal{T},\,u\in(\lambda_{j-1}^{0},\,\lambda_{j}^{0})
\end{cases}
\end{align*}
with $A_{1,t,T}^{0}\left(\omega\right)=A_{1}\left(0,\,\omega\right)$
for $t<1$ and $A_{m_{0}+1,t,T}^{0}\left(\omega\right)=A_{m_{0}+1}\left(1,\,\omega\right)$
for $t>T$. Our definition coincides with the Wigner-Ville spectrum
{[}cf. \citet{martin/flandrin:85}{]} when there are no change-points
(i.e., $m_{0}=0$). Below we show that $f_{j,T}\left(u,\,\omega\right)$
tends in mean-squared to $f_{j}\left(u,\,\omega\right)\triangleq\left|A_{j}\left(u,\,\omega\right)\right|^{2}$
for $T_{j-1}^{0}/T<u=t/T\leq T_{j}^{0}/T$ which is the spectrum that
corresponds to the spectral representation. Therefore, we call $f_{j}\left(u,\,\omega\right)$
the time-varying spectral density matrix of the process. 
\begin{assumption}
\label{Assumption Smothness of A}$A\left(u,\,\omega\right)$ is \textcolor{red}{
}piecewise Lipschitz continuous in the first component and uniformly
Lipschitz continuous in the second component, with index $\vartheta>1/2$
for both.
\end{assumption}
\begin{thm}
\label{Theorem 2.2 in Dal}Assume $V_{t,T}$ is Segmented Locally
Stationary with $m_{0}+1$ regimes and Assumption \ref{Eq. 2.2 Definition V_t}
holds. Then, for all $u\in\left(0,\,1\right)$, $\int_{-\pi}^{\pi}\sum_{j=1}^{m_{0}+1}\left|f_{j,T}\left(u,\,\omega\right)-f_{j}\left(u,\,\omega\right)\right|^{2}d\omega=o\left(1\right)$.
\end{thm}
Let $f\left(u,\,\omega\right)=f_{j}\left(u,\,\omega\right)$ if $Tu\in(T_{j-1}^{0},\,T_{j}^{0}]$
so as to suppress the subscript $j$ from $f$. It is well-known that,
even when $m_{0}=0,$ the spectral representation \eqref{Eq. Spectral Rep of SLS}
is not unique {[}cf. \citet{priestley:85}, Chapter 11.1{]}. A consequence
of Theorem \ref{Theorem 2.2 in Dal} is that $\{f_{j}\left(u,\,\omega\right)=\left|A_{j}\left(u,\,\omega\right)\right|^{2},\,j=1,\ldots,\,m_{0}+1\}$
is uniquely determined from the whole triangular array $\{V_{t,T}\}$.

For $Tu\notin\mathcal{T}$ with $T_{j-1}^{0}/T<u=t/T<T_{j}^{0}/T$,
only the realizations of $V_{t,T}$ in the time interval $u\in\left[u-n/T,\,u+n/T\right]$
with $n\rightarrow\infty$ contribute to $f_{j}\left(u,\,\omega\right)$.
Since this interval is fully contained in a segment $j$ where $A_{j}\left(u,\,\omega\right)$
is smooth, and given that the length of this interval tends to zero,
$V_{t,T}$ becomes ``asymptotically stationary'' on this interval.
 The length of the interval in which $V_{t,T}$ can be considered
stationary is given by $n\ln n/T^{\vartheta}\rightarrow0$ . For $Tu\in\mathcal{T}$,
the arguments are different. Suppose $Tu=T_{j}^{0}$. The spectrum
$f_{j,T}\left(u,\,\omega\right)$ is defined in such a way that only
observations prior to $T_{j}^{0}$ are used in order to construct
an approximation to $f_{j}\left(u,\,\omega\right)$. Since the length
of this interval tends to zero and $A_{j}\left(u,\,\omega\right)$
is left-Lipschitz continuous, then those observations become ``asymptotically
stationary'' and thus provide the same information about $f_{j}\left(u,\,\omega\right)$.

Given $f\left(u,\,\omega\right),$ we can define the local covariance
of $V_{t,T}$ at rescaled time $u$ with $Tu\notin\mathcal{T}$ and
lag $k\in\mathbb{Z}$ as 
\[
c\left(u,\,k\right)\triangleq\int_{-\pi}^{\pi}e^{i\omega k}f\left(u,\,\omega\right)d\omega.
\]
The same definition is also used when $Tu\in\mathcal{T}$ and $k\geq0$.
For $Tu\in\mathcal{T}$ and $k<0$ it is defined as $c\left(u,\,k\right)\triangleq\int_{-\pi}^{\pi}e^{i\omega k}A\left(u,\,\omega\right)A\left(u-k/T,\,-\omega\right)d\omega$.

\subsection{\label{Subsection HAC-Estimation}DK-HAC Estimation}

In model \eqref{Eq. Spectral Rep of SLS}, if $m_{0}=0$ and $A^{0}$
is constant in its first argument, then $\left\{ V_{t,T}\right\} $
is second-order stationary. Its spectral density matrix is then equal
to $f\left(\omega\right)\triangleq\left(2\pi\right)^{-1}\sum_{k=-\infty}^{\infty}\Gamma\left(k\right)e^{-i\omega k}$
where $\Gamma\left(k\right)\triangleq\mathbb{E}(V_{t,T}V'_{t-k,T})$.
When evaluated at frequency $\omega=0$ it plays a prominent role
because $\lim_{T\rightarrow\infty}J_{T}=2\pi f\left(0\right).$ Nonstationarity
implies that the spectral density  is time-varying since $\mathbb{E}(V_{t}V'_{t-k})$
now depends on $k$ as well as on $t$. The SLS processes introduced
above accommodate this property because they have a time-varying spectrum
$f\left(u,\,\omega\right)$.  Accordingly, we introduce the notation
$\Gamma_{u}\left(k\right)\triangleq\mathbb{E}(V_{Tu,T}V'_{Tu-k,T})$
where $u=t/T$. We show below that $\Gamma_{u}\left(k\right)=c\left(u,\,k\right)+O\left(T^{-1}\right)$
uniformly in $1\leq j\leq m+1$, $Tu\leq T_{j}^{0}$ and $k\in\mathbb{Z}.$
Under the rescaling $u=t/T,\,u\in\left[0,\,1\right]$, the limit of
$J_{T}$ for SLS processes is given by,
\begin{align*}
J\triangleq\lim_{T\rightarrow\infty}J_{T} & =\int_{0}^{1}c\left(u,\,0\right)du+\sum_{k=1}^{\infty}\int_{0}^{1}\left(c\left(u,\,k\right)+c\left(u,\,k\right)'\right)du.
\end{align*}
Using the definition of $f\left(u,\,\omega\right)$ it can be shown
that $J=2\pi\int_{0}^{1}f\left(u,\,0\right)du$. \citet{dahlhaus:2009}
discussed how to estimate $f\left(u,\,\omega\right)$ for the scalar
case under smoothness in both arguments using the smoothed local periodogram.
Our goals are to estimate $J$ using a time-domain method and to relax
the smoothness assumption in $u$. This is different from Dahlhaus'
work that considered local problems (i.e., estimation of $f\left(u,\,\omega\right)$
under smoothness) and not full-sample problems (i.e., estimation of
$J$).  The class of estimators of $J$ relies on double kernel smoothing
over lags and time,
\begin{align*}
\widehat{J}_{T} & =\widehat{J}_{T}\left(b_{1,T},\,b_{2,T}\right)\triangleq\frac{T}{T-p}\sum_{k=-T+1}^{T-1}K_{1}\left(b_{1,T}k\right)\widehat{\Gamma}\left(k\right),\,\,\mathrm{with}\,\,\\
\widehat{\Gamma}\left(k\right) & \triangleq\frac{n_{T}}{T-n_{T}}\sum_{r=0}^{\left\lfloor \left(T-n_{T}\right)/n_{T}\right\rfloor }\widehat{c}_{T}\left(rn_{T}/T,\,k\right),
\end{align*}
 where $K_{1}\left(\cdot\right)$ is a real-valued kernel in the class
$\boldsymbol{K}_{1}$ defined below, $b_{1,T}$ is a bandwidth sequence
discussed below, $n_{T}\rightarrow\infty$ satisfying the conditions
given below, and 
\begin{align}
\widehat{c}_{T}\left(rn_{T}/T,\,k\right) & \triangleq\begin{cases}
\left(Tb_{2,T}\right)^{-1}\sum_{s=k+1}^{T}K_{2}^{*}\left(\frac{\left(\left(r+1\right)n_{T}-\left(s-k/2\right)\right)/T}{b_{2,T}}\right)\widehat{V}_{s}\widehat{V}'_{s-k}, & k\geq0\\
\left(Tb_{2,T}\right)^{-1}\sum_{s=-k+1}^{T}K_{2}^{*}\left(\frac{\left(\left(r+1\right)n_{T}-\left(s+k/2\right)\right)/T}{b_{2,T}}\right)\widehat{V}_{s+k}\widehat{V}'_{s}, & k<0
\end{cases},\label{eq: Def. chat}
\end{align}
with $K_{2}^{*}$ being a real-valued kernel and $b_{2,T}$ is a
bandwidth sequence discussed below. $\widehat{c}_{T}\left(u,\,k\right)$
is an estimate of the local autocovariance $c\left(u,\,k\right)$
of lag $k$ at time $u=rn_{T}/T$. Estimation of $c\left(u,\,k\right)$
for locally stationary processes was considered by \citet{dahlhaus:12}.
 For positive semi-definiteness, it is necessary that $K_{2}^{*}$
takes the following form:
\[
K_{2}^{*}\left(\frac{\left(r+1\right)n_{T}-\left(s-k/2\right)}{Tb_{2,T}}\right)=\left(K_{2}\left(\frac{\left(r+1\right)n_{T}-s}{Tb_{2,T}}\right)K_{2}\left(\frac{\left(r+1\right)n_{T}-\left(s-k\right)}{Tb_{2,T}}\right)\right)^{1/2}\quad\mathrm{for}\,k\geq0,
\]
\[
K_{2}^{*}\left(\frac{\left(r+1\right)n_{T}-\left(s+k/2\right)}{Tb_{2,T}}\right)=\left(K_{2}\left(\frac{\left(r+1\right)n_{T}-s}{Tb_{2,T}}\right)K_{2}\left(\frac{\left(r+1\right)n_{T}-\left(s+k\right)}{Tb_{2,T}}\right)\right)^{1/2}\quad\mathrm{for}\,k<0.
\]
Setting $K_{2}\left(x\right)=(\int_{0}^{1}h\left(x\right)^{2}dx)^{-1}h\left(x+1/2\right)^{2}$
and $N_{T}=Tb_{2,T}$, we see that positive semi-definiteness requires
the use of a data taper $h\left(\cdot\right)$ with length $N_{T}$.
This follows because we need each $\widehat{V}_{t}$ $\left(t=1,\ldots,\,T\right)$
to be assigned the same weight across different $k$ for any given
$r$. Then, letting $\widehat{V}_{t}^{\lyxmathsym{\textdegree}}=\ensuremath{\left(K_{2}\left(\left(\left(r+1\right)n_{T}-t\right)/Tb_{2,T}\right)\right)^{1/2}\widehat{V}_{t}}$
we can use the same arguments as in \citet{andrews:91} applied now
to $\widehat{V}_{t}^{\lyxmathsym{\textdegree}}$ to show that $J_{T}$
is positive semi-definite for the appropriate choice of $K_{1}$. 

The estimator $\widehat{J}_{T}$ involves two kernels: $K_{1}$ smooths
the lagged sample autocovariances, akin to the classical HAC estimators,
while $K_{2}$ applies smoothing over time. The factor $T/\left(T-p\right)$
is an optional small-sample degrees of freedom adjustment. In Section
\ref{Section HAC-Estimation-with Predetermined}-\ref{Section Optimal-Kernels-and},
we consider estimators $\widehat{J}_{T}$ for which $b_{1,T}$ and
$b_{2,T}$ are given sequences. In Section \ref{Section Data-Dependent-Bandwidths},
we consider adaptive estimators $\widehat{J}_{T}$ for which $b_{1,T}$
and $b_{2,T}$ are data-dependent. Observe that the optimal $b_{2,T}$
actually depends on the properties of $\left\{ V_{t,T}\right\} $
in any given block. Since the order of $b_{2,T}\left(\cdot\right)$
is the same across blocks, we omit this notation for the developments
of the asymptotic results. However, when we determine the data-dependent
estimate of $b_{2,T}\left(\cdot\right)$, we will estimate $b_{2,T}\left(rn_{T}/T\right)$
for each $r$.  We consider the following class of kernels {[}cf.
\citet{andrews:91}{]}, 
\begin{align}
\boldsymbol{K}_{1} & =\{K_{1}\left(\cdot\right):\,\mathbb{R}\rightarrow\left[-1,\,1\right]:\,K_{1}\left(0\right)=1,\,K_{1}\left(x\right)=K_{1}\left(-x\right),\,\forall x\in\mathbb{R}\label{Eq. (2.6) K1 Kernel class}\\
 & \quad{\textstyle \int\nolimits _{-\infty}^{\infty}}K_{1}^{2}\left(x\right)dx<\infty,\,K_{1}\left(\cdot\right)\,\mathrm{is\,continuous\,at\,0\,and\,at\,all\,but\,finite\,numbers\,of\,points}\}.\nonumber 
\end{align}
Examples of kernels in $\boldsymbol{K}_{1}$ include the Truncated,
Bartlett, Parzen, Quadratic Spectral (QS) and Tukey-Hanning kernel.
We shall show below that the QS kernel has certain optimality properties:
\begin{align*}
K_{1}^{\mathrm{QS}}\left(x\right) & =\frac{25}{12\pi^{2}x^{2}}\left(\frac{\sin\left(6\pi x/5\right)}{6\pi x/5}-\cos\left(6\pi x/5\right)\right).
\end{align*}

\section{\label{Section HAC-Estimation-with Predetermined}HAC Estimation
with Predetermined Bandwidths}

In Section \ref{Subsection: Estimation-of-Local} we present some
asymptotic properties of  $\widehat{c}\left(\cdot,\,\cdot\right)$.
We use them in Section \ref{Subsec: Large-Sample-Results-on} in order
to establish consistency, rate of convergence and MSE properties of
predetermined bandwidths HAC estimators.  Let $\widetilde{J}_{T}$
denote the pseudo-estimator identical to $\widehat{J}_{T}$ but based
on  $\{V_{t,T}\}=\{V_{t,T}(\beta_{0})\}$ rather than on $\{\widehat{V}_{t,T}\}=\{V_{t,T}(\widehat{\beta})\}$.
We first require some smoothness of $A\left(u,\,\cdot\right)$ in
$u$. 
\begin{assumption}
\label{Assumption Smothness of A (for HAC)}(i) $\left\{ V_{t,T}\right\} $
is a mean-zero SLS process with $m_{0}+1$ regimes; (ii) $A\left(u,\,\omega\right)$
is \textcolor{red}{ }twice continuously differentiable in $u$ at
all $u\neq\lambda_{j}^{0}$ $(j=1,\ldots,\,m_{0}+1)$ with uniformly
bounded derivatives $\left(\partial/\partial u\right)A\left(u,\,\cdot\right)$
and $\left(\partial^{2}/\partial u^{2}\right)A\left(u,\,\cdot\right)$,
and Lipschitz continuous in the second component with index $\vartheta=1$;
(iii) $\left(\partial^{2}/\partial u^{2}\right)A\left(u,\,\cdot\right)$
is Lipschitz continuous at all $u\neq\lambda_{j}^{0}$ $(j=1,\ldots,\,m_{0}+1)$;
(iv) $A\left(u,\,\omega\right)$ is twice left-differentiable in $u$
at $u=\lambda_{j}^{0}$, $(j=1,\ldots,\,m_{0}+1)$ with uniformly
bounded derivatives $\left(\partial/\partial_{-}u\right)A\left(u,\,\cdot\right)$
and $\left(\partial^{2}/\partial_{-}u^{2}\right)A\left(u,\,\cdot\right),$
and has piecewise Lipschitz continuous derivative $\left(\partial^{2}/\partial_{-}u^{2}\right)A\left(u,\,\cdot\right)$.
\end{assumption}
We also need to impose conditions on the temporal dependence of $V_{t}=V_{t,T}$.
Let 
\begin{align*}
\kappa_{V,t}^{\left(a,b,c,d\right)}\left(u,\,v,\,w\right) & \triangleq\kappa^{\left(a,b,c,d\right)}\left(t,\,t+u,\,t+v,\,t+w\right)-\kappa_{\mathscr{N}}^{\left(a,b,c,d\right)}\left(t,\,t+u,\,t+v,\,t+w\right)\\
 & \triangleq\mathbb{E}(V_{t}^{\left(a\right)}V_{t+u}^{\left(b\right)}V_{t+v}^{\left(c\right)}V_{t+w}^{\left(d\right)})-\mathbb{E}(V_{\mathscr{N},t}^{\left(a\right)}V_{\mathscr{N},t+u}^{\left(b\right)}V_{\mathscr{N},t+v}^{\left(c\right)}V_{\mathscr{N},t+w}^{\left(d\right)}),
\end{align*}
where $\left\{ V_{\mathscr{N},t}\right\} $ is a Gaussian sequence
with the same mean and covariance structure as $\left\{ V_{t}\right\} $.
$\kappa_{V,t}^{\left(a,b,c,d\right)}\left(u,\,v,\,w\right)$ is the
time-$t$ fourth-order cumulant of $(V_{t}^{\left(a\right)},\,V_{t+u}^{\left(b\right)},\,V_{t+v}^{\left(c\right)},$
$\,V_{t+w}^{\left(d\right)})$ while $\kappa_{\mathscr{N}}^{\left(a,b,c,d\right)}(t,\,t+u,$
$\,t+v,\,t+w)$ is the time-$t$ centered fourth moment of $V_{t}$
if $V_{t}$ were Gaussian.
\begin{assumption}
\label{Assumption A - Dependence}(i) $\sum_{k=-\infty}^{\infty}\sup_{u\in\left[0,\,1\right]}$
$\left\Vert c\left(u,\,k\right)\right\Vert <\infty$, $\sum_{k=-\infty}^{\infty}\sup_{u\in\left[0,\,1\right]}\left\Vert \left(\partial^{2}/\partial u^{2}\right)c\left(u,\,k\right)\right\Vert <\infty$
and $\sum_{k=-\infty}^{\infty}\sum_{j=-\infty}^{\infty}\sum_{l=-\infty}^{\infty}\sup_{u\in\left[0,\,1\right]}|\kappa_{V,\left\lfloor Tu\right\rfloor }^{\left(a,b,c,d\right)}$
$\left(k,\,j,\,l\right)|<\infty$ for all $a,\,b,\,c,\,d\leq p$.
(ii) For all $a,\,b,\,c,\,d\leq p$ there exists a function $\widetilde{\kappa}_{a,b,c,d}:\,\left[0,\,1\right]\times\mathbb{Z}\times\mathbb{Z}\times\mathbb{Z}\rightarrow\mathbb{R}$
such that $\sup_{u\in\left(0,\,1\right)}|\kappa_{V,\left\lfloor Tu\right\rfloor }^{\left(a,b,c,d\right)}\left(k,\,s,\,l\right)$
$-\widetilde{\kappa}_{a,b,c,d}\left(u,\,k,\,s,\,l\right)|\leq KT^{-1}$
for some constant $K$; the function $\widetilde{\kappa}_{a,b,c,d}\left(u,\,k,\,s,\,l\right)$
is twice differentiable in $u$ at all $u\neq\lambda_{j}^{0}$, $(j=1,\ldots,\,m_{0}+1)$
with uniformly bounded derivatives $\left(\partial/\partial u\right)\widetilde{\kappa}_{a,b,c,d}\left(u,\cdot,\cdot,\cdot\right)$
and $\left(\partial^{2}/\partial u^{2}\right)\widetilde{\kappa}_{a,b,c,d}\left(u,\cdot,\cdot,\cdot\right)$,
and twice left-differentiable in $u$ at $u=\lambda_{j}^{0}$ $(j=1,\ldots,\,m_{0}+1)$
with uniformly bounded derivatives $\left(\partial/\partial_{-}u\right)\widetilde{\kappa}_{a,b,c,d}\left(u,\cdot,\cdot,\cdot\right)$
and $\left(\partial^{2}/\partial_{-}u^{2}\right)\widetilde{\kappa}_{a,b,c,d}$
$\left(u,\cdot,\cdot,\cdot\right),$ and piecewise Lipschitz continuous
derivative $\left(\partial^{2}/\partial_{-}u^{2}\right)\widetilde{\kappa}_{a,b,c,d}\left(u,\cdot,\cdot,\cdot\right)$.
\end{assumption}
If $\left\{ V_{t,T}\right\} $ is stationary then the cumulant condition
of Assumption \ref{Assumption A - Dependence}-(i) reduces to the
standard one used in the time series literature {[}see also Assumption
A in \citet{andrews:91}{]}. We do not require fourth-order stationarity
but only that the time-$t=Tu$ fourth order cumulant is locally constant
in a neighborhood of $u$. One can show that $\alpha$-mixing and
moment conditions imply that the cumulant condition of Assumption
\ref{Assumption A - Dependence} holds.

\subsection{\label{Subsection: Estimation-of-Local}Estimation of the Local Covariance}

Let $\widetilde{c}_{T}\left(u,\,k\right)$ denote the estimator that
uses $\left\{ V_{t,T}\right\} $.  We consider the following
class of kernels:
\begin{align}
\boldsymbol{K}_{2} & =\{K_{2}\left(\cdot\right):\,\mathbb{R}\rightarrow\left[0,\,\infty\right],\,K_{2}\left(x\right)=K_{2}\left(1-x\right),\,{\textstyle \int}K_{2}\left(x\right)dx=1,\label{Eq. K2 Kernel class}\\
 & \qquad\qquad K_{2}\left(x\right)=0,\,\mathrm{for\,}\,x\notin\left[0,\,1\right],\,K_{2}\left(\cdot\right)\,\mathrm{is\,continuous}\}.\nonumber 
\end{align}
 
\begin{lem}
\label{Lemma Rate of Convergence of ctilde - SLS}Suppose that Assumption
\ref{Assumption Smothness of A (for HAC)}-\ref{Assumption A - Dependence}
hold. If $b_{2,T}\rightarrow0$ and\textbf{ }$Tb_{2,T}^{5}\rightarrow\eta\in(0,\,\infty)$,
then $\widetilde{c}_{T}\left(u_{0},\,k\right)-c\left(u_{0},\,k\right)=O_{\mathbb{P}}(\sqrt{Tb_{2,T}})$
for all $u_{0}\in\left(0,\,1\right)$.
\end{lem}

\subsection{\label{Subsec: Large-Sample-Results-on}Results on DK-HAC Estimation
with Predetermined Bandwidths}

Following \citet{parzen:57}, we define $K_{1,q}\triangleq\lim_{x\downarrow0}\left(1-K_{1}\left(x\right)\right)/\left|x\right|^{q}$
for $q\in[0,\,\infty);$ $q$ increases with the smoothness of $K_{1}\left(\cdot\right)$
with the largest value being such that $K_{1,q}<\infty$. When $q$
is an even integer, $K_{1,q}=-\left(d^{q}K_{1}\left(x\right)/dx^{q}\right)|_{x=0}/q!$
and $K_{1,q}<\infty$ if and only if $K_{1}\left(x\right)$ is $q$
times differentiable at zero.  We define the index of smoothness
of $f\left(u,\,\omega\right)$ at $\omega=0$ by $f^{\left(q\right)}\left(u,\,0\right)\triangleq\left(2\pi\right)^{-1}\sum_{k=-\infty}^{\infty}\left|k\right|^{q}c\left(u,\,k\right)$,
for $q\in[0,\,\infty)$. If $q$ is even, then $f^{\left(q\right)}\left(u,\,0\right)=\left(-1\right)^{q/2}\left(d^{q}f\left(u,\,\omega\right)/d\omega^{q}\right)|_{\omega=0}$.
Further, $||f^{\left(q\right)}\left(u,\,0\right)||<\infty$ if and
only if $f\left(u,\,\omega\right)$ is $q$ times differentiable at
$\omega=0$. We define 
\begin{align}
\mathrm{MSE}\left(Tb_{1,T}b_{2,T},\,\widetilde{J}_{T},\,W\right) & =Tb_{1,T}b_{2,T}\mathbb{E}\left[\mathrm{vec}\left(\widetilde{J}_{T}-J_{T}\right)'W\mathrm{vec}\left(\widetilde{J}_{T}-J_{T}\right)\right].\label{Eq: 3.5 MSE}
\end{align}
\begin{thm}
\label{Theorem MSE J}Suppose $K_{1}\left(\cdot\right)\in\boldsymbol{K}_{1}$,
$K_{2}\left(\cdot\right)\in\boldsymbol{K}_{2}$, Assumption \ref{Assumption Smothness of A (for HAC)}-\ref{Assumption A - Dependence}
hold, $b_{1,T},\,b_{2,T}\rightarrow0$, $n_{T}\rightarrow\infty,\,n_{T}/T\rightarrow0$
and $1/Tb_{1,T}b_{2,T}\rightarrow0$. We have: (i)~
\begin{align*}
\lim_{T\rightarrow\infty} & Tb_{1,T}b_{2,T}\mathrm{Var}\left[\mathrm{vec}\left(\widetilde{J}_{T}\right)\right]\\
 & =4\pi^{2}\int K_{1}^{2}\left(y\right)dy\int_{0}^{1}K_{2}^{2}\left(x\right)dx\left(I+C_{pp}\right)\left(\int_{0}^{1}f\left(u,\,0\right)du\right)\otimes\left(\int_{0}^{1}f\left(v,\,0\right)dv\right).
\end{align*}

(ii) If $1/Tb_{1,T}^{q}b_{2,T}\rightarrow0$, $n_{T}/Tb_{1,T}^{q}\rightarrow0$
and $b_{2,T}^{2}/b_{1,T}^{q}\rightarrow0$ for some $q\in[0,\,\infty)$
for which $K_{1,q},$ $||\int_{0}^{1}f^{\left(q\right)}\left(u,\,0\right)du||\in[0,\,\infty)$,
then $\lim_{T\rightarrow\infty}b_{1,T}^{-q}\mathbb{E}(\widetilde{J}_{T}-J_{T})=-2\pi K_{1,q}\int_{0}^{1}f^{\left(q\right)}\left(u,\,0\right)du.$

(iii) If $n_{T}/Tb_{1,T}^{q}\rightarrow0$, $b_{2,T}^{2}/b_{1,T}^{q}\rightarrow0$
and $Tb_{1,T}^{2q+1}b_{2,T}\rightarrow\gamma\in\left(0,\,\infty\right)$
for some $q\in[0,\,\infty)$ for which $K_{1,q},\,||\int_{0}^{1}f^{\left(q\right)}\left(u,\,0\right)du||\in[0,\,\infty)$,
then
\begin{align*}
\lim_{T\rightarrow\infty} & \mathrm{MSE}\left(Tb_{1,T}b_{2,T},\,\widetilde{J}_{T},\,W\right)=4\pi^{2}\left[\gamma K_{1,q}^{2}\mathrm{vec}\left(\int_{0}^{1}f^{\left(q\right)}\left(u,\,0\right)du\right)'W\mathrm{vec}\left(\int_{0}^{1}f^{\left(q\right)}\left(u,\,0\right)du\right)\right.\\
 & \quad\left.+\int K_{1}^{2}\left(y\right)dy\int K_{2}^{2}\left(x\right)dx\,\mathrm{tr}\left(W\left(I_{p^{2}}+C_{pp}\right)\left(\int_{0}^{1}f\left(u,\,0\right)du\right)\otimes\left(\int_{0}^{1}f\left(v,\,0\right)dv\right)\right)\right].
\end{align*}
  
\end{thm}
If $b_{2,T}^{2}/b_{1,T}^{q}\rightarrow\nu<\infty$ replaces $b_{2,T}^{2}/b_{1,T}^{q}\rightarrow0$
in part (ii), then the asymptotic bias for the case of locally stationary
processes becomes
\begin{align}
\lim_{T\rightarrow\infty}b_{1,T}^{-q}\mathbb{E}(\widetilde{J}_{T}-J_{T}) & =-2\pi K_{1,q}\int_{0}^{1}f^{\left(q\right)}\left(u,\,0\right)du+\frac{\nu}{2}\int_{0}^{1}x^{2}K_{2}\left(x\right)\sum_{k=-\infty}^{\infty}\int_{0}^{1}\frac{\partial^{2}}{\partial u^{2}}c\left(u,\,k\right)du.\label{eq Asymptotic MSE full}
\end{align}
For the general case of SLS processes the term involving $\left(\partial^{2}/\partial^{2}u\right)c\left(u,\,k\right)$
is different. The second summand on the right-hand side of \eqref{eq Asymptotic MSE full}
cancels when $\int_{0}^{1}\left(\partial^{2}/\partial^{2}u\right)c\left(u,\,k\right)du=0$.
The latter occurs when the process is stationary. \citet{dahlhaus:12}
presented MSE results for a pointwise estimate of $f\left(u,\,\omega\right)$
under continuity in both components by applying smoothing over $u$
and $\omega.$ His results depends on the local behavior of $f\left(u,\,\omega\right)$
at time $u$ and frequency $\omega$ whereas in our problem the MSE
results depend on properties of the full time path of $f\left(u,\,0\right)$.
The theorem suggests that the optimal choice of $b_{1,T}$ hinges
on the degree of nonstationary in the data, a feature that does not
appear from the corresponding results in the literature. The results
 are derived as $n_{T}\rightarrow\infty$. It is possible and indeed
easier to keep $n_{T}$ fixed, in which case the  results are unchanged.
However, the case with $n_{T}$ fixed can have some disadvantages
when the spectrum is discontinuous because then the estimator would
be often dealing with observations from different regimes, which as
explained above might lead to low frequency contamination. We now
move to the results concerning $\widehat{J}_{T}$.
\begin{assumption}
\label{Assumption B}(i) $\sqrt{T}(\widehat{\beta}-\beta_{0})=O_{\mathbb{P}}\left(1\right)$;
(ii) $\sup_{u\in\left[0,\,1\right]}\mathbb{E}||V_{\left\lfloor Tu\right\rfloor }||^{2}<\infty$;
(iii) $\sup_{u\in\left[0,\,1\right]}\mathbb{E}\sup_{\beta\in\Theta}$
$||\left(\partial/\partial\beta'\right)V_{\left\lfloor Tu\right\rfloor }\left(\beta\right)||^{2}<\infty$;
(iv) $\int_{-\infty}^{\infty}\left|K_{1}\left(y\right)\right|dy,$
$\int_{0}^{1}\left|K_{2}\left(x\right)\right|dx<\infty.$
\end{assumption}
Assumption \ref{Assumption B}-(i,iii) is the same as Assumption B
in \citet{andrews:91}. As remarked above, we interpret $\beta_{0}$
as the pseudo-true parameter $\beta^{*}$ when the model is misspecified.
Part (iv) of the assumption is satisfied by most commonly used kernels.
 In order to obtain  rate of convergence results we replace Assumption
\ref{Assumption A - Dependence} with the following assumptions. 
\begin{assumption}
\label{Assumption C Andrews 91}(i) Assumption \ref{Assumption A - Dependence}
holds with $V_{t,T}$ replaced by 
\begin{align*}
\left(V'_{t},\,\mathrm{vec}\left(\left(\frac{\partial}{\partial\beta'}V_{t}\left(\beta_{0}\right)\right)-\mathbb{E}\left(\frac{\partial}{\partial\beta'}V_{t}\left(\beta_{0}\right)\right)\right)'\right)' & .
\end{align*}
(ii) $\sup_{u\in\left[0,\,1\right]}\mathbb{E}(\sup_{\beta\in\Theta}||\left(\partial^{2}/\partial\beta\partial\beta'\right)V_{\left\lfloor Tu\right\rfloor }^{\left(a\right)}\left(\beta\right)||^{2})<\infty$
for all $a=1,\ldots,\,p$.
\end{assumption}
\begin{assumption}
\label{Assumption W_T and unbounded kernel and Cumulant 8}Let $W_{T}$
denote a $p^{2}\times p^{2}$ weight matrix such that $W_{T}\overset{\mathbb{P}}{\rightarrow}W$.
\end{assumption}
\begin{thm}
\label{Theorem 1 -Consistency and Rate}Suppose $K_{1}\left(\cdot\right)\in\boldsymbol{K}_{1}$,
$K_{2}\left(\cdot\right)\in\boldsymbol{K}_{2}$, $b_{1,T},\,b_{2,T}\rightarrow0$,\textbf{
}$n_{T}\rightarrow\infty,\,n_{T}/Tb_{1,T}\rightarrow0,$ and $1/Tb_{1,T}b_{2,T}\rightarrow0$.
We have: 

(i) If Assumption \ref{Assumption Smothness of A (for HAC)}-\ref{Assumption B}
hold, $\sqrt{T}b_{1,T}\rightarrow\infty$, $b_{2,T}/b_{1,T}\rightarrow0$
then $\widehat{J}_{T}-J_{T}\overset{\mathbb{P}}{\rightarrow}0$ and
$\widehat{J}_{T}-\widetilde{J}_{T}\overset{\mathbb{P}}{\rightarrow}0$. 

(ii) If Assumption \ref{Assumption Smothness of A (for HAC)}, \ref{Assumption B}-\ref{Assumption C Andrews 91}
hold,  $n_{T}/Tb_{1,T}^{q}\rightarrow0$, $1/Tb_{1,T}^{q}b_{2,T}\rightarrow0$,
$b_{2,T}^{2}/b_{1,T}^{q}\rightarrow0$ and $Tb_{1,T}^{2q+1}b_{2,T}\rightarrow\gamma\in\left(0,\,\infty\right)$
for some $q\in[0,\,\infty)$ for which $K_{1,q},\,||\int_{0}^{1}f^{\left(q\right)}\left(u,\,0\right)du||\in[0,\,\infty)$,
then $\sqrt{Tb_{1,T}b_{2,T}}(\widehat{J}_{T}-J_{T})=O_{\mathbb{P}}\left(1\right)$
and $\sqrt{Tb_{1,T}}(\widehat{J}_{T}-\widetilde{J}_{T})=o_{\mathbb{P}}\left(1\right).$ 

(iii) Under the conditions of part (ii) and Assumption \ref{Assumption W_T and unbounded kernel and Cumulant 8},
\begin{align*}
\lim_{T\rightarrow\infty}\mathrm{MSE}\left(Tb_{1,T}b_{2,T},\,\widehat{J}_{T},\,W_{T}\right)=\lim_{T\rightarrow\infty}\mathrm{MSE}\left(Tb_{1,T}b_{2,T},\,\widetilde{J}_{T},\,W\right) & .
\end{align*}
\end{thm}
The consistency result of $\widehat{J}_{T}$ in part (i) applies to
kernels $K_{1}\left(\cdot\right)$ with unbounded support and to bandwidths
$b_{1,T}$ and $b_{2,T}$ such that $1/b_{1,T}b_{2,T}$ grows at rate
$o(\sqrt{T/b_{2,T}})$.  Part (ii) yields the consistency of $\widehat{J}_{T}$
with $b_{1,T}$ only required to be $o\left(Tb_{2,T}\right)$. This
rate is slower than the corresponding rate $o\left(T\right)$ of the
classical kernel HAC estimators as shown by \citet{andrews:91} in
his Theorem 1-(b). However, this property is of little practical import
because optimal growth rates typically are less than $T^{1/2}$; for
the QS kernel HAC estimator the optimal growth rate is $T^{1/5}$
while it is $T^{1/3}$ for the Newey-West HAC estimator. Part (ii)
of the theorem presents the rate of convergence of $\widehat{J}_{T}$
which is $\sqrt{Tb_{2,T}b_{1,T}}$. In Section \ref{Section Optimal-Kernels-and},
we compare the rate of convergence of $\widehat{J}_{T}$ with that
of the classical HAC estimators when the respective optimal bandwidths
are used. 

\section{Optimal Kernels, \label{Section Optimal-Kernels-and} Bandwidths
and Choice of $n_{T}$}

In this section, we show the optimality of quadratic-type kernels
under MSE criterion.\footnote{Besides \citet{andrews:91} and \citet{newey/west:87} in the context
of LRV estimation, the MSE-optimality criterion was also used more
recently by \citet{whilelm:2015} in a GMM context to determine the
optimal bandwidth of the nonparametric estimator of the optimal weighting
matrix. } For $K_{1},$ the result states that the QS kernel minimizes the
asymptotic MSE for any $K_{2}\left(\cdot\right)$. Let 
\begin{align*}
\mathrm{MSE} & (b_{2,T}^{-4},\,\widehat{c}_{T}\left(u_{0},\,k,\,\right),\,\widetilde{W}_{T})\\
 & \triangleq b_{2,T}^{-4}\mathbb{E}\left[\mathrm{vec}\left(\widehat{c}_{T}\left(u_{0},\,k\right)-c\left(u_{0},\,k\right)\right)\right]'\widetilde{W}_{T}\left[\mathrm{vec}\left(\widehat{c}_{T}\left(u_{0},\,k\right)-c\left(u_{0},\,k\right)\right)\right],
\end{align*}
 where $\widetilde{W}_{T}$ is some $p\times p$ positive semidefinite
matrix. The optimal bandwidths $b_{1,T}^{\mathrm{opt}}$ and $b_{2,T}^{\mathrm{opt}}$
satisfy the following sequential MSE criterion: 
\begin{align}
\mathrm{MSE} & \left(Tb_{1,T}^{\mathrm{opt}}\overline{b}_{2,T}^{\mathrm{opt}},\,\widehat{J}_{T}\left(b_{1,T}^{\mathrm{opt}},\,\overline{b}_{2,T}^{\mathrm{opt}}\right),\,W_{T}\right)\leq\mathrm{MSE}\left(Tb_{1,T}^{\mathrm{opt}}\overline{b}_{2,T}^{\mathrm{opt}},\,\widehat{J}_{T}\left(b_{1,T},\,\overline{b}_{2,T}^{\mathrm{opt}}\right),\,W_{T}\right)\label{eq (MSE criterio)}\\
 & \quad\mathrm{where}\quad\overline{b}_{2,T}^{\mathrm{opt}}=\int_{0}^{1}b_{2,T}^{\mathrm{opt}}\left(u\right)du\quad\nonumber \\
 & \quad\mathrm{and}\quad\quad b_{2,T}^{\mathrm{opt}}\left(u\right)=\underset{b_{2,T}}{\mathrm{argmin}}\,\,\,\mathrm{MSE}\left(b_{2,T}^{-4},\,\widehat{c}_{T}\left(u_{0},\,k\right),\,\widetilde{W}_{T}\right).\nonumber 
\end{align}
The first inequality above has to hold as $T\rightarrow\infty$.
The above criterion determines the globally optimal $b_{1,T}^{\mathrm{opt}}$
given the integrated locally optimal $b_{2,T}^{\mathrm{opt}}\left(u\right)$.
Thus, $b_{1,T}^{\mathrm{opt}}$ and $\overline{b}_{2,T}^{\mathrm{opt}}$
need not be the same as the bandwidths $(\widetilde{b}_{1,T}^{\mathrm{opt}},\,\widetilde{b}_{2,T}^{\mathrm{opt}})$
that jointly minimize the global asymptotic MSE, 
\begin{equation}
\lim_{T\rightarrow\infty}\mathrm{MSE}\left(Tb_{1,T}b_{2,T},\,\widehat{J}_{T}\left(b_{1,T},\,b_{2,T}\right),\,W_{T}\right).\label{Eq. (Global MSE)}
\end{equation}
 Theorem \ref{Theorem MSE J}-(ii) states that, under the condition
$b_{2,T}^{2}/b_{1,T}^{q}\rightarrow0$, the bias only depends on the
smoothing over lagged autocovariances but not on $b_{2,T}$. Then,
the global solution $\widetilde{b}_{2,T}^{\mathrm{opt}}$ would be
trivial: $b_{2,T}$ affects the MSE only through the variance term
and optimality requires to set the bandwidth as large as possible.
In contrast, the MSE criterion \eqref{eq (MSE criterio)} based on
the MSE given in Theorem \ref{Theorem MSE J}-(iii) leads to a unique
solution which can be obtained analytically.   Under the condition
$b_{2,T}^{2}/b_{1,T}^{q}\rightarrow\nu<\infty$, Belotti et al. \citeyearpar{belotti/casini/catania/grassi/perron_HAC_Sim_Bandws}\nocite{belotti/casini/catania/grassi/perron:2020}
determined the bandwidths $(\widetilde{b}_{1,T}^{\mathrm{opt}},\,\widetilde{b}_{2,T}^{\mathrm{opt}})$
that jointly minimize \eqref{Eq. (Global MSE)}. They showed that
$\widetilde{b}_{1,T}^{\mathrm{opt}},\,\widetilde{b}_{2,T}^{\mathrm{opt}}=O(T^{-1/6})$
while the optimal bandwidths $(b_{1,T}^{\mathrm{opt}},\,\overline{b}_{2,T}^{\mathrm{opt}})$
from \eqref{eq (MSE criterio)} satisfy $b_{1,T}^{\mathrm{opt}}=O(T^{-4/25})$
and $\overline{b}_{2,T}^{\mathrm{opt}}=O(T^{-1/5})$. Thus, the criterion
\eqref{eq (MSE criterio)} leads to a slightly shorter block length
relative to the global criterion \eqref{Eq. (Global MSE)} (i.e.,
$T\overline{b}_{2,T}^{\mathrm{opt}}<T\widetilde{b}_{2,T}^{\mathrm{opt}}$).
A shorter bock length is beneficial if there is substantial nonstationarity
and implies less sensitivity to low frequency contamination from not
properly accounting for nonstationarity {[}cf. Casini et al. \citeyearpar{casini/perron_Low_Frequency_Contam_Nonstat:2020}{]}.
For a throughout comparison between the two criteria see Belotti
et al. \citeyearpar{belotti/casini/catania/grassi/perron_HAC_Sim_Bandws}.\nocite{belotti/casini/catania/grassi/perron:2020} 

\subsection{Optimal $K_{2}\left(\cdot\right)$ and $b_{2,T}$}

Let $F\left(K_{2}\right)\triangleq\int_{0}^{1}K_{2}^{2}\left(x\right)dx$,
$H\left(K_{2}\right)=(\int_{0}^{1}x^{2}K_{2}\left(x\right)dx)^{2}$,
and for any $k\in\mathbb{Z}$,
\begin{align*}
D_{1}\left(u_{0}\right) & \triangleq\mathrm{\,vec}\left(\partial^{2}c\left(u_{0},\,k\right)/\partial u^{2}\right)'\widetilde{W}\mathrm{\,vec}\left(\partial^{2}c\left(u_{0},\,k\right)/\partial u^{2}\right),\quad\\
D_{2}\left(u_{0}\right) & \triangleq\mathrm{tr}\widetilde{W}\left(I_{p^{2}}+C_{pp}\right)\sum_{l=-\infty}^{\infty}c\left(u_{0},\,l\right)\otimes\left[c\left(u_{0},\,l\right)+c\left(u_{0},\,l+2k\right)\right].
\end{align*}

\begin{prop}
\label{Proposition: Optimal Local Covariance}Suppose Assumption \ref{Assumption Smothness of A (for HAC)},
\ref{Assumption B}-\ref{Assumption C Andrews 91} hold and $\widetilde{W}_{T}\overset{\mathbb{P}}{\rightarrow}\widetilde{W}$.
We have for all $a,\,b\leq p,$ 
\begin{align*}
\mathrm{MSE} & \left(1,\,\widehat{c}_{T}^{\left(a,b\right)}\left(u_{0},\,k\right),\,1\right)\\
 & =\frac{1}{4}b_{2,T}^{4}\left(\int_{0}^{1}xK_{2}\left(x\right)dx\right)^{2}\left(\frac{\partial^{2}}{\partial^{2}u}c^{\left(a,b\right)}\left(u_{0},\,k\right)\right)^{2}\\
 & \quad+\frac{1}{Tb_{2,T}}\int_{0}^{1}K_{2}^{2}\left(x\right)dx\sum_{l=-\infty}^{\infty}c^{\left(a,b\right)}\left(u_{0},\,l\right)\left[c^{\left(a,b\right)}\left(u_{0},\,l\right)+c^{\left(a,b\right)}\left(u_{0},\,l+2k\right)\right]\\
 & \quad+\frac{1}{Tb_{2,T}}\int_{0}^{1}K_{2}^{2}\left(x\right)dx\sum_{h_{1}=-\infty}^{\infty}\kappa_{V,\left\lfloor Tu_{0}\right\rfloor }^{\left(a,b,a,b\right)}\left(-k,\,h_{1},\,h_{1}-k\right)+o\left(b_{2,T}^{4}\right)+O\left(1/\left(b_{2,T}T\right)^{2}\right).
\end{align*}
 $\mathrm{MSE}(b_{2,T}^{-4},\,\widehat{c}_{T}\left(u_{0},\,k\right)-c\left(u_{0},\,k\right),\,\widetilde{W}_{T})$
is minimized with
\[
b_{2,T}^{\mathrm{opt}}\left(u_{0}\right)=[H\left(K_{2}^{\mathrm{opt}}\right)D_{1}\left(u_{0}\right)]^{-1/5}\left(F\left(K_{2}^{\mathrm{opt}}\right)\left(D_{2}\left(u_{0}\right)+D_{3}\left(u_{0}\right)\right)\right)^{1/5}T^{-1/5},
\]
where $D_{3}\left(u_{0}\right)$ depends on $\widetilde{\kappa}$
(for $p=1$, $D_{3}\left(u_{0}\right)=\sum_{h_{1}=-\infty}^{\infty}\kappa_{V,\left\lfloor Tu_{0}\right\rfloor }\left(-k,\,h_{1},\,h_{1}-k\right)$),
and $K_{2}^{\mathrm{opt}}\left(x\right)=6x\left(1-x\right),\,0\leq x\leq1$.
In addition if $V_{t}$ is Gaussian, then $D_{3}\left(u_{0}\right)=0$,
for $u_{0}\in\left(0,\,1\right)$. 
\end{prop}
The optimal kernel $K_{2}^{\mathrm{opt}}\left(x\right)$ is a transformation
of the Epanechnikov kernel. Optimality of quadratic kernels under
a MSE criterion has been shown in many contexts {[}cf. \citet{epanechnikov:69}
and \citet{priestley:85}{]}.  The optimal bandwidth sequence decreases
at rate $T^{-1/5}$ which is the same optimal rate derived in the
context of parameter estimation of locally stationary processes {[}see
e.g., \citet{Dahlhaus/Giraitis:98}{]}.  The term $D_{1}\left(u_{0}\right)$
is due to nonstationary, while the term $D_{2}\left(u_{0}\right)$
measures the variability of $\widehat{c}_{T}\left(u_{0},\,k\right)$.
The bandwidth $b_{2,T}^{\mathrm{opt}}$ converges to zero at a slower
rate as the process becomes closer to stationary (i.e., as the square
root of $D_{1}\left(u_{0}\right)$ decreases). 

\subsection{Optimal $K_{1}\left(\cdot\right)$ }

We next determine the optimal kernel $K_{1}$ and the optimal bandwidth
sequence $b_{1,T}$ given any $K_{2}$ and any $b_{2,T}$ of order
$O(T^{-1/5})$, i.e., the same order of $b_{2,T}^{\mathrm{opt}}\left(u\right)$
for any $u\in\left[0,\,1\right]$. Let $\widehat{J}_{T}^{\mathrm{QS}}$
denote $\widehat{J}_{T}$ when the latter is based on the QS kernel.
For some results below, we consider a subset of $\boldsymbol{K}_{1}$.
Let $\boldsymbol{\widetilde{K}}_{1}=\{K_{1}\left(\cdot\right)\in\boldsymbol{K}_{1}|\,\widetilde{K}\left(\omega\right)\geq0\,\forall\,\omega\in\mathbb{R}\}\bigr\}$
where $\widetilde{K}\left(\omega\right)=\left(2\pi\right)^{-1}\int_{-\infty}^{\infty}K_{1}\left(x\right)e^{-ix\omega}dx.$
The function $\widetilde{K}\left(\omega\right)$ is referred to as
the spectral window generator. The set $\widetilde{\boldsymbol{K}}_{1}$
contains all kernels $K_{1}$ that necessarily generate positive semidefinite
estimators in finite samples. 

We adopt the notation $\widehat{J}_{T}\left(b_{1,T}\right)=\widehat{J}_{T}\left(b_{1,T},\,b_{2,T},\,K_{2}\right)$
to denote the estimator $\widehat{J}_{T}$ that uses $b_{1,T},\,b_{2,T}=\overline{b}_{2,T}^{\mathrm{opt}}+o\left(T^{-1/5}\right)$
and $K_{2}^{\mathrm{}}\left(\cdot\right)$. We then compare two kernels
$K_{1}$ using comparable\textcolor{red}{{} }bandwidths $b_{1,T}$ which
are defined as follows. Given $K_{1}\left(\cdot\right)\in\widetilde{\boldsymbol{K}}_{1}$,
the QS kernel $K_{1}^{\mathrm{QS}}\left(\cdot\right)$, and a bandwidth
sequence $\left\{ b_{1,T}\right\} $ to be used with the QS kernel,
define a comparable bandwidth sequence $\left\{ b_{1,T,K_{1}}\right\} $
for use with $K_{1}\left(\cdot\right)$ such that both kernel/bandwidth
combinations have the same asymptotic variance when scaled by the
same factor $Tb_{1,T}b_{2,T}$. This means that
\begin{align*}
\lim_{T\rightarrow\infty}\mathrm{MSE} & (Tb_{1,T}b_{2,T},\,\widehat{J}_{T}^{\mathrm{QS}}(b_{1,T})-\mathbb{E}(\widetilde{J}_{T}^{\mathrm{QS}}(b_{1,T}))+J_{T},\,W_{T})\\
=\lim_{T\rightarrow\infty}\mathrm{MSE} & (Tb_{1,T}b_{2,T},\,\widehat{J}_{T}^{\mathrm{}}(b_{1,T,K_{1}})-\mathbb{E}(\widetilde{J}_{T}(b_{1,T,K_{1}}))+J_{T},\,W_{T}).
\end{align*}
This definition yields $b_{1,T,K_{1}}=b_{1,T}/(\int K_{1}^{2}\left(x\right)dx)$
and  $b_{1,T,\mathrm{QS}}=b_{1,T}$ since $\int\left(K_{1}^{\mathrm{QS}}\right)^{2}\left(x\right)dx=1$.
\begin{thm}
\label{Theorem Optimal Kernels}Suppose Assumption \ref{Assumption Smothness of A (for HAC)},
\ref{Assumption B}-\ref{Assumption W_T and unbounded kernel and Cumulant 8}
hold, $\int_{0}^{1}||f^{\left(2\right)}\left(u,\,0\right)||du<\infty$,
$b_{2,T}\rightarrow0$, $b_{2,T}^{5}T\rightarrow\eta\in\left(0,\,\infty\right)$,
$(\mathrm{vec}(\int_{0}^{1}f^{\left(q\right)}\left(u,\,0\right)du))'W\mathrm{vec}(\int_{0}^{1}f^{\left(q\right)}\left(u,\,0\right)du)>0$
and $W$ is positive semidefinite. For any bandwidth sequence $\left\{ b_{1,T}\right\} $
such that $b_{2,T}/b_{1,T}\rightarrow0$, $n_{T}/Tb_{1,T}^{2}\rightarrow0$
and $Tb_{1,T}^{5}b_{2,T}\rightarrow\gamma\in\left(0,\,\infty\right)$,
and for any kernel $K_{1}\left(\cdot\right)\in\boldsymbol{\widetilde{K}}_{1}$
used to construct $\widehat{J}_{T}^{\mathrm{}}$, the QS kernel is
preferred to $K_{1}\left(\cdot\right)$ in the sense that
\begin{align*}
\lim_{T\rightarrow\infty} & \left(\mathrm{MSE}\left(Tb_{1,T}b_{2,T},\,\widehat{J}_{T}^{\mathrm{}}\left(b_{1,T,K_{1}}\right),\,W_{T}\right)-\mathrm{MSE}\left(Tb_{1,T}b_{2,T},\,\widehat{J}_{T}^{\mathrm{QS}}\left(b_{1,T}\right),\,W_{T}\right)\right)\\
 & =4\gamma\pi^{2}\left(\mathrm{vec}\left(\int_{0}^{1}f^{\left(2\right)}\left(u,\,0\right)du\right)\right)'W\mathrm{vec}\left(\int_{0}^{1}f^{\left(2\right)}\left(u,\,0\right)du\right)\int_{0}^{1}\left(K_{2}^{\mathrm{opt}}\left(x\right)\right)^{2}dx\\
 & \quad\times\left[K_{1,2}^{2}\left(\int K_{1}^{2}\left(y\right)dy\right)^{4}-\left(K_{1,2}^{\mathrm{QS}}\right)^{2}\right]\geq0.
\end{align*}
 The inequality is strict if $K_{1}\left(x\right)\neq K_{1}^{\mathrm{QS}}\left(x\right)$
with positive Lebesgue measure.
\end{thm}
The requirement $\int_{0}^{1}||f^{\left(2\right)}\left(u,\,0\right)||du<\infty$
is not stringent and it reduces to the one used by \citet{andrews:91}
when $\left\{ V_{t,T}\right\} $ is stationary. If $\int_{0}^{1}||f^{\left(q\right)}\left(u,\,0\right)||du<\infty$
only for some $1\leq q<2$, one can show that any kernel with $K_{1,q}=0$
has smaller asymptotic MSE than a kernel with $K_{1,q}>0$. The QS,
Parzen, and Tukey-Hanning kernels have $K_{1,q}=0$ for $1\leq q<2$,
whereas the Bartlett has $K_{1,q}>0$ for $1\leq q<2$. Thus, the
asymptotic superiority of the former kernels over the Bartlett kernel
holds even if $\int_{0}^{1}||f^{\left(q\right)}\left(u,\,0\right)||du<\infty$
only for $1\leq q<2$. 

\subsection{Optimal Predetermined Bandwidth Sequence $b_{1,T}$ }

We now present the predetermined bandwidth sequence that minimizes
the asymptotic MSE given $b_{2,T}=O(b_{2,T}^{\mathrm{opt}})$ and
$K_{2}=K_{2}^{\mathrm{opt}}.$ This optimality result applies to
each kernel $K_{1}\left(\cdot\right)\in\boldsymbol{K}_{1}$ for which
$K_{1,q}\in\left(0,\,\infty\right)$ for some $q\in\left(0,\,\infty\right)$.
Thus, most commonly used kernels are allowed with the exception of
the truncated kernel. Let 
\begin{align*}
\phi\left(q\right) & =\frac{\mathrm{vec}\left(\int_{0}^{1}f^{\left(q\right)}\left(u,\,0\right)du\right)'W\mathrm{vec}\left(\int_{0}^{1}f^{\left(q\right)}\left(u,\,0\right)du\right)}{\mathrm{tr}W\left(I_{p^{2}}+C_{pp}\right)\left(\int_{0}^{1}f\left(u,\,0\right)du\right)\otimes\left(\int_{0}^{1}f\left(v,\,0\right)dv\right)}.
\end{align*}
The optimal bandwidth is $b_{1,T}^{\mathrm{opt}}=(2qK_{1,q}^{2}\phi\left(q\right)Tb_{2,T}^{\mathrm{\mathrm{opt}}}/(\int K_{1}^{2}\left(y\right)dy\int_{0}^{1}K_{2}^{2}\left(x\right)dx))^{-1/\left(2q+1\right)}$,
where $\phi\left(q\right)$ is a function of the\textcolor{red}{{} }unknown
spectral density $f\left(\cdot,\,\cdot\right)$. Hence, the optimal
bandwidth $b_{1,T}^{\mathrm{opt}}$ is unknown in practice, and we
consider data-dependent estimates of $\phi\left(q\right)$ in Section
\ref{Section Data-Dependent-Bandwidths}. 
\begin{condition}
\label{Condition b_T}$b_{1,T},\,b_{2,T}\rightarrow0$ with $b_{2,T}/b_{1,T}\rightarrow0$,
 and $Tb_{1,T}^{2q+1}b_{2,T}\rightarrow\gamma\in\left(0,\,\infty\right)$
for some $q\in[0,\,\infty)$ for which $K_{1,q},\,||\int_{0}^{1}f^{\left(q\right)}\left(u,\,0\right)du||\in[0,\,\infty)$,
where $b_{2,T}^{\mathrm{}}=O(T^{-1/5})$. 
\end{condition}
\begin{cor}
\label{Corollary 1 -Optimal b1 }Suppose Assumption \ref{Assumption Smothness of A (for HAC)},
\ref{Assumption B}-\ref{Assumption W_T and unbounded kernel and Cumulant 8}
hold, $||\int_{0}^{1}f^{\left(q\right)}\left(u,\,\omega\right)du||<\infty$,
$\phi\left(q\right)\in\left(0,\,\infty\right)$, and $W$ is positive
definite. Consider $K_{1}\left(\cdot\right)\in\boldsymbol{K}_{1}$
for which $K_{1,q}\in\left(0,\,\infty\right)$ for some $q\in\left(0,\,\infty\right)$.
Then, $\{b_{1,T}^{\mathrm{opt}}\}$ is optimal among the sequences
$\{b_{1,T}\}$ that satisfy Condition \ref{Condition b_T} in the
sense that, 
\begin{align*}
\lim_{T\rightarrow\infty} & \Biggl(\mathrm{MSE}\left(\left(Tb_{2,T}^{\mathrm{}}\right)^{2q/\left(2q+1\right)},\,\widehat{J}_{T}^{\mathrm{}}\left(b_{1,T},\,b_{2,T}\right),\,W_{T}\right)\\
 & -\mathrm{MSE}\left(\left(Tb_{2,T}^{\mathrm{}}\right)^{2q/\left(2q+1\right)},\,\widehat{J}_{T}^{\mathrm{}}\left(b_{1,T}^{\mathrm{opt}},\,b_{2,T}\right),\,W_{T}\right)\Biggr)\geq0.
\end{align*}
The inequality is strict unless $b_{1,T}=b_{1,T}^{\mathrm{opt}}+o((Tb_{2,T})^{-1/\left(2q+1\right)})$. 
\end{cor}
In Corollary \ref{Corollary 1 -Optimal b1 }, $q=2$ for the QS kernel
and so $b_{1,T}^{\mathrm{opt}}=0.6584(\phi\left(2\right)Tb_{2,T}^{\mathrm{\mathrm{opt}}})^{-1/5}(\int_{0}^{1}K_{2}^{2}\left(y\right)dy)^{1/5}.$
For $K_{2}\left(y\right)=K_{2}^{\mathrm{opt}}\left(y\right),$  the
latter reduces to,
\begin{align}
b_{1,T}^{\mathrm{opt}} & =0.6828(\phi\left(2\right)Tb_{2,T}^{\mathrm{\mathrm{opt}}})^{-1/5}.\label{eq: b1 (opt)}
\end{align}
 The optimal bandwidth is of order $T^{-4/25}.$ Thus, the optimal
bandwidth sequence decreases to zero at a slower rate than the optimal
bandwidth sequence for the QS kernel-based HAC estimator of \citet{andrews:91},
for which the rate is of order $T^{-1/5}$. The slower rate is due
to the fact that our estimator  smooths the spectrum over time through
$K_{2}\left(\cdot\right)$ and this restricts the smoothing of $K_{1}\left(\cdot\right)$.
In particular, the optimal choice of $b_{1,T}$ hinges on the degree
of nonstationary through $b_{2,T}^{\mathrm{\mathrm{opt}}}$. The
more nonstationary are the data, the smaller is $b_{2,T}^{\mathrm{\mathrm{opt}}}$
and the large is $b_{1,T}^{\mathrm{\mathrm{opt}}}$ which means that
less weight is given to $\widehat{\Gamma}\left(k\right)$ for $k\neq0$.
In contrast, the optimal choice of $b_{1,T}$ for the methods proposed
in the literature is independent of the degree of nonstationarity.
 When $b_{1,T}$ and $b_{2,T}$ are chosen optimally, the convergence
rate from Theorem \ref{Theorem 1 -Consistency and Rate} reduces to
$T^{8/25}$.  Thus, the rate is slower than the corresponding one
for the QS kernel HAC estimator considered in \citet{andrews:91}.
However, it is misleading to compare our DK-HAC estimator with the
classical HAC estimators only on the basis of the rate of convergence.
In fact, the DK-HAC estimators account flexibly for nonstationarity
and are robust to low frequency contamination induced by nonstationarity/misspecification
whereas the classical HAC estimators are not in general {[}cf. Casini
et al. \citeyearpar{casini/perron_Low_Frequency_Contam_Nonstat:2020}{]}.

\subsection{Choice of $n_{T}$\label{subsec Choice-of nT}}

Our MSE analysis does not indicate an optimal value for $n_{T}$.
It only suggests growth rate bounds. When $K_{1}^{\mathrm{QS}}$
is used, $n_{T}$ cannot grow faster than $T^{2/3}$. We set $n_{T}=T^{0.66}$
for the $\mathrm{QS}$ kernel. That is, we choose  $n_{T}$ to be
the largest possible value allowed by the condition. Our sensitivity
analysis (not reported) suggests that choosing a smaller $n_{T}$
might result in excessive overlapping of regimes when the process
is SLS (i.e., $m_{0}>0$). See Belotti et al. \citeyearpar{belotti/casini/catania/grassi/perron_HAC_Sim_Bandws}
for more details.

\section{Data-Dependent Bandwidths\label{Section Data-Dependent-Bandwidths}}

In this section we consider estimators $\widehat{J}_{T}$ that use
bandwidths $b_{1,T}$ and $b_{2,T}$ whose values are determined via
data-dependent methods.  We use the ``plug-in'' method which
is characterized by plugging-in estimates of unknown quantities into
an asymptotic formula for an optimal bandwidth parameter (i.e., the
expressions for $b_{1,T}^{\mathrm{opt}}$  and $b_{2,T}^{\mathrm{opt}}$
from Section \ref{Section Optimal-Kernels-and}).  Section \ref{subsec:Implementation}
explains how to construct the automatic bandwidths while Section \ref{subsec:Theoretical-Results}
presents the corresponding theoretical results. 

\subsection{\label{subsec:Implementation}Implementation}

Let us begin with $b_{1,T}^{\mathrm{opt}}$ and then move to $b_{2,T}^{\mathrm{opt}}$.
The first step for the construction of data-dependent bandwidth parameters
is to specify $p$ univariate  parametric models for the elements
of $V_{t}=(V_{t}^{\left(1\right)},\ldots,\,V_{t}^{\left(p\right)})'$.
The second step involves the estimation of the parameters.  In our
context, the logical estimation methods to use are local (weighted)
least-squares (LS) (i.e., LS method applied to rolling windows) and
nonparametric kernel methods. In a third step, we replace the unknown
parameters in $\phi\left(q\right)$ with corresponding estimates.
Such estimate $\widehat{\phi}\left(q\right)$  is then substituted
into the expression for $b_{1,T}^{\mathrm{opt}}$ to yield the data-dependent
bandwidth $\widehat{b}_{1,T}$: 
\begin{align}
\widehat{b}_{1,T} & =\left(2qK_{1,q}^{2}\widehat{\phi}\left(q\right)T\widehat{\overline{b}}_{2,T}/\left(\int K_{1}^{2}\left(y\right)dy\int_{0}^{1}K_{2}^{2}\left(x\right)dx\right)\right)^{-1/\left(2q+1\right)},\label{Eq. (6.1) Andrews 91}
\end{align}
where $\widehat{\overline{b}}_{2,T}=\left(n_{T}/T\right)\sum_{r=1}^{\left\lfloor T/n_{T}\right\rfloor -1}\widehat{b}_{2,T}\left(rn_{T}/T\right)$.
$\widehat{\overline{b}}_{2,T}$ is an average of the estimates $\widehat{b}_{2,T}\left(\cdot\right)$.
Since $b_{2,T}$ depends on $u$, it is more efficient to estimate
it for each block as its optimal value can change over time. In practice,
a reasonable candidate for an approximating parametric model is the
class of first order autoregressive {[}AR(1){]} models for $\{V_{t}^{\left(r\right)}\},\,r=1,\ldots,\,p$
(with different parameters for each $r$) or a first order vector
autoregressive {[}VAR(l){]} model for $\{V_{t}\}$. These classes
were also used by \citet{andrews:91}. However, in our context it
is reasonable to allow the parameters of the AR(1) model to be time-varying.
For parsimony, we consider a time-varying AR(1) with no breaks in
$f\left(u,\,\omega\right)$, i.e., $V_{t}^{\left(r\right)}=a_{1}\left(t/T\right)V_{t-1}^{\left(r\right)}+u_{t}^{\left(r\right)},$
where the $u_{t}^{\left(r\right)}$ need not be independent across
$r$.

The use of $p$ univariate parametric models requires a simple form
for the weight matrix $W$. In particular, $W$ has to be a diagonal
matrix which in turn implies that $\phi\left(q\right)$ reduces to
\begin{align*}
\phi\left(q\right) & =2^{-1}\sum_{r=1}^{p}W^{\left(r,r\right)}\left(\int_{0}^{1}f^{\left(q\right)\left(r,r\right)}\left(u,\,0\right)du\right)^{2}/\sum_{r=1}^{p}W^{\left(r,r\right)}\left(\int_{0}^{1}f^{\left(r,r\right)}\left(u,\,0\right)du\right)^{2}.
\end{align*}
The usual choice is $W^{\left(r,r\right)}=1$ for all $r$ except
that which corresponds to an intercept for which it is set to zero.
 An estimate of $f^{\left(r,r\right)}\left(u,\,0\right)$ is $\widehat{f}^{\left(r,r\right)}\left(u,\,0\right)=\left(2\pi\right)^{-1}(\widehat{\sigma}^{\left(r\right)}\left(u\right))^{2}(1-\widehat{a}_{1}^{\left(r\right)}\left(u\right))^{-2}$
while $f^{\left(2\right)\left(r,r\right)}\left(u,\,0\right)$ can
be estimated by $\widehat{f}^{\left(2\right)\left(r,r\right)}\left(u,\,0\right)=3\pi^{-1}$
$((\widehat{\sigma}^{\left(r\right)}\left(u\right))^{2}\widehat{a}_{1}^{\left(r\right)}\left(u\right))(1-\widehat{a}_{1}^{\left(r\right)}\left(u\right))^{-4}$
where $\widehat{a}_{1}^{\left(r\right)}\left(u\right)$ and $\widehat{\sigma}^{\left(r\right)}\left(u\right)$
are the LS estimates computed using local data to the left of $u=t/T$:
\begin{align}
\widehat{a}_{1}^{\left(r\right)}\left(u\right) & =\frac{\sum_{j=\left\lfloor Tu\right\rfloor -n_{2,T}+1}^{\left\lfloor Tu\right\rfloor }\widehat{V}_{j}^{\left(r\right)}\widehat{V}_{j-1}^{\left(r\right)}}{\sum_{j=\left\lfloor Tu\right\rfloor -n_{2,T}+1}^{\left\lfloor Tu\right\rfloor }\left(\widehat{V}_{j-1}^{\left(r\right)}\right)^{2}},\qquad\label{eq: alpha and sigma}\\
\widehat{\sigma}^{\left(r\right)}\left(u\right) & =\left(\sum_{j=\left\lfloor Tu\right\rfloor -n_{2,T}+1}^{\left\lfloor Tu\right\rfloor }\left(\widehat{V}_{j}^{\left(r\right)}-\widehat{a}_{1}^{\left(r\right)}\left(u\right)\widehat{V}_{j-1}^{\left(r\right)}\right)^{2}\right)^{1/2},\nonumber 
\end{align}
where $n_{2,T}\rightarrow\infty$. Then,  for the QS kernel $K_{1}$,
\begin{align*}
\widehat{\phi}\left(2\right) & =\sum_{r=1}^{p}W^{\left(r,r\right)}\left(18\left(\frac{n_{3,T}}{T}\sum_{j=0}^{\left\lfloor T/n_{3,T}\right\rfloor -1}\frac{\left(\widehat{\sigma}^{\left(r\right)}\left(\left(jn_{3,T}+1\right)/T\right)\widehat{a}_{1}^{\left(r\right)}\left(\left(jn_{3,T}+1\right)/T\right)\right)^{2}}{\left(1-\widehat{a}_{1}^{\left(r\right)}\left(\left(jn_{3,T}+1\right)/T\right)\right)^{4}}\right)^{2}\right)/\\
 & \quad\sum_{r=1}^{p}W^{\left(r,r\right)}\left(\frac{n_{3,T}}{T}\sum_{j=0}^{\left\lfloor T/n_{3,T}\right\rfloor -1}\frac{\left(\widehat{\sigma}^{\left(r\right)}\left(\left(jn_{3,T}+1\right)/T\right)\right)^{2}}{\left(1-\widehat{a}_{1}^{\left(r\right)}\left(\left(jn_{3,T}+1\right)/T\right)\right)^{2}}\right)^{2}.
\end{align*}
For most of the results below we can take $n_{3,T}=n_{2,T}=n_{T}.$
After plugging-in $\widehat{\phi}\left(2\right)$ into the formula
\eqref{eq: b1 (opt)}, we have $\widehat{b}_{1,T}=0.6828(\widehat{\phi}\left(2\right)T\widehat{\overline{b}}_{2,T})^{-1/5}.$

We now propose a data-dependent procedure for  $b_{2,T}\left(u_{r}\right),$
where $u_{r}=rn_{T}/T$ for $r=1,\ldots,\,\left\lfloor \left(T-n_{T}\right)/n_{T}\right\rfloor $.
We assume  that the parameters of the approximating time-varying
AR(1) models change slowly such that the smoothness of $f\left(\cdot,\,\omega\right)$
and thus of $c\left(\cdot,\,k\right)$ is the same as the one that
would arise if $a_{1}\left(u\right)=0.8\left(\cos1.5+\cos4\pi u\right)$
and $\sigma\left(u\right)=\sigma=1$ for all $u\in\left[0,\,1\right]$
{[}cf. \citet{dahlhaus:12}{]}. The reason for imposing this condition
is that it is otherwise difficult to estimate $\left(\partial^{2}/\partial u^{2}\right)c\left(u,\,k\right)$,
which enters $D_{1}\left(u\right)$, from the data. Under the above
specification, the exact expression of $D_{1}\left(u\right)$ can
be computed analytically: 
\begin{align*}
D_{1}\left(u\right) & \triangleq(\int_{-\pi}^{\pi}\left[\frac{3}{\pi}\left(1+0.8\left(\cos1.5+\cos4\pi u\right)\exp\left(-i\omega\right)\right)^{-4}\left(0.8\left(-4\pi\sin\left(4\pi u\right)\right)\right)\exp\left(-i\omega\right)\right.\\
 & \quad\left.-\frac{1}{\pi}\left|1+0.8\left(\cos1.5+\cos4\pi u\right)\exp\left(-i\omega\right)\right|^{-3}\left(0.8\left(-16\pi^{2}\cos\left(4\pi u\right)\right)\right)\exp\left(-i\omega\right)\right]d\omega)^{2}.
\end{align*}
An estimate of $D_{1}\left(u\right)$ is given by
\begin{align*}
\widehat{D}_{1}\left(u\right) & \triangleq(\left[S_{\omega}\right]^{-1}\sum_{s\in S_{\omega}}\left[\frac{3}{\pi}\left(1+0.8\left(\cos1.5+\cos4\pi u\right)\exp\left(-i\omega_{s}\right)\right)^{-4}\left(0.8\left(-4\pi\sin\left(4\pi u\right)\right)\right)\exp\left(-i\omega_{s}\right)\right.\\
 & \quad\left.-\frac{1}{\pi}\left|1+0.8\left(\cos1.5+\cos4\pi u\right)\exp\left(-i\omega_{s}\right)\right|^{-3}\left(0.8\left(-16\pi^{2}\cos\left(4\pi u\right)\right)\right)\exp\left(-i\omega_{s}\right)\right])^{2},
\end{align*}
 where $\left[S_{\omega}\right]$ is the cardinality of $S_{\omega}$
and $\omega_{s+1}>\omega_{s}$ with $\omega_{1}=-\pi,\,\omega_{\left[S_{\omega}\right]}=\pi.$
In our simulations we use $S_{\omega}=\left\{ -\pi,\,-3,\,-2,\,-1,\,0,\,1,\,2,\,3,\,\pi\right\} $.
 Note that we have computed $\widehat{D}_{1}\left(u\right)$ for
$k=0$ because it makes the computation simpler. Further, this is
consistent with our sequential MSE criterion because $k=0$ is the
only lag for which $K_{1}(0)=1$ for all $K_{1}$ so that the choice
of $K_{1}$ does not influence $b_{2}^{\mathrm{opt}}\left(\cdot\right)$.
It remains to derive an estimate of $D_{2}\left(u\right)$ since $F\left(K_{2}\right)$
and $H\left(K_{2}\right)$ can be computed for a given $K_{2}\left(\cdot\right)$.
We assume that the innovations of the approximating time-varying
AR(1) model satisfy $\mathbb{E}(u_{t}^{\left(r\right)})=0$, $\mathbb{E}((u_{t}^{\left(r\right)})^{2})=\sigma^{2}$
and $\mathbb{E}((u_{t}^{\left(r\right)})^{4})=3\sigma^{4}$ so that
$D_{3}\left(u\right)=0$ for all $u\in\left(0,\,1\right)$. That is,
the term involving the cumulant drops from $b_{2}^{\mathrm{opt}}\left(u\right)$.
In practice this is convenient because it is complex to deal with
consistent estimation of cumulant terms. Note also that $D_{3}\left(u\right)=0$
if $u_{t}$ is Gaussian. Since $c\left(u,\,k\right)$ can be consistently
estimated by $\widehat{c}_{T}\left(u,\,k\right)$, an estimate of
$D_{2}\left(u\right)$ is given by
\[
\widehat{D}_{2}\left(u_{0}\right)\triangleq p^{-1}\sum_{r=1}^{p}\sum_{l=-\left\lfloor T^{4/25}\right\rfloor }^{\left\lfloor T^{4/25}\right\rfloor }\widehat{c}_{T}^{\left(r,r\right)}\left(u_{0},\,l\right)\left[2\widehat{c}_{T}^{\left(r,r\right)}\left(u_{0},\,l\right)\right],
\]
where the number of summands  grows at the same rate as $(b_{1,T}^{\mathrm{opt}})^{-1}$;
a different choice is allowed as long as it grows at a slower rate
than $T^{2/5}$.  Hence, the estimate of the optimal $b_{2,T}$ is
given by 
\begin{align*}
\widehat{b}_{2,T}\left(u_{r}\right) & =1.6786\left(\widehat{D}_{1}\left(u_{r}\right)\right){}^{-1/5}(\widehat{D}_{2}\left(u_{r}\right))^{1/5}T^{-1/5},\qquad\mathrm{where}\qquad u_{r}=rn_{T}/T.
\end{align*}

\subsection{\label{subsec:Theoretical-Results}Theoretical Results}

Next, we establish consistency, rate of convergence and asymptotic
MSE results for the estimator $\widehat{J}_{T}(\widehat{b}_{1,T},\,\widehat{\overline{b}}_{2,T})$
that uses the data-dependent bandwidths $\widehat{b}_{1,T}$ and $\widehat{\overline{b}}_{2,T}$.
As in \citet{andrews:91}, we need to restrict the class of admissible
kernels to the following class: 
\begin{align}
\boldsymbol{K}_{3} & =\{K_{3}\left(\cdot\right)\in\boldsymbol{K}_{1}:\,\left(i\right)\,\left|K_{1}\left(x\right)\right|\leq C_{1}\left|x\right|^{-b}\,\mathrm{with\,}b>\max\left(1+1/q,\,3\right)\label{Eq. (2.6) K3 Kernel class}\\
 & \quad\,\,\,\mathrm{for}\,\left|x\right|\in\left[\overline{x}_{L},\,D_{T}h_{T}\overline{x}_{U}\right],\,b_{1,T}^{2}h_{T}\rightarrow\infty,\,D_{T}>0,\,\overline{x}_{L},\,\overline{x}_{U}\in\mathbb{R},\,1\leq\overline{x}_{L}<\overline{x}_{U},\,\mathrm{and}\,\nonumber \\
 & \quad\,\,\,\mathrm{with\,}b>1+1/q\,\mathrm{\,for}\,\left|x\right|\notin\left[\overline{x}_{L},\,D_{T}h_{T}\overline{x}_{U}\right],\,\mathrm{and\,some\,}C_{1}<\infty,\,\mathrm{where}\,q\in\left(0,\,\infty\right)\nonumber \\
 & \quad\,\,\,\mathrm{is\,such\,that\,}K_{1,q}\in\left(0,\,\infty\right),\,\left(ii\right)\,\left|K_{1}\left(x\right)-K_{1}\left(y\right)\right|\leq C_{2}\left|x-y\right|\,\forall x,\,y\in\mathbb{R}\,\mathrm{for\,some}\nonumber \\
 & \quad\,\,\,\mathrm{costant}\,C_{2}<\infty,\,\mathrm{and}\,(iii)\,q<34/4\}.\nonumber 
\end{align}
Let $\widehat{\theta}$ denote the estimator of the parameter of the
approximate (time-varying) parametric model(s) introduced above. For
example, with univariate AR(1) approximating parametric models, $\widehat{\theta}=(\int_{0}^{1}\widehat{a}_{1}^{\left(1\right)}\left(u\right)du,\,\int_{0}^{1}(\widehat{\sigma}^{\left(1\right)}\left(u\right))^{2}du,\ldots,\,\int_{0}^{1}\widehat{a}_{1}^{\left(p\right)}\left(u\right)du,\,\int_{0}^{1}(\widehat{\sigma}^{\left(p\right)}\left(u\right))^{2}du)'$.
Let $\theta^{*}$ denote the probability limit of $\widehat{\theta}$.
$\widehat{\phi}\left(q\right)$ is the value of $\phi\left(q\right)$
with $\widehat{\theta}$ instead of $\theta$. Its probability limit
is denoted by $\phi_{\theta^{*}}$.
\begin{assumption}
\label{Assumption E-F-G}(i) $\widehat{\phi}\left(q\right)=O\mathbb{_{P}}\left(1\right)$
and $1/\widehat{\phi}\left(q\right)=O\mathbb{_{P}}\left(1\right)$;
(ii) $\inf\{T/n_{3,T},\,\sqrt{n_{2,T}}\}(\widehat{\phi}\left(q\right)-\phi_{\theta^{*}})=O_{\mathbb{P}}\left(1\right)$
for some $\phi_{\theta^{*}}\in\left(0,\,\infty\right)$ where $n_{2,T}/T+n_{3,T}/T\rightarrow0,$
$n_{2,T}^{10/6}/T\rightarrow[c_{2},\,\infty),$ $n_{3,T}^{10/6}/T\rightarrow[c_{3},\,\infty)$
with $0<c_{2},\,c_{3}<\infty$; (iii) $\sup_{u\in\left[0,\,1\right]}\lambda_{\max}(\Gamma_{u}\left(k\right))\leq C_{3}k^{-l}$
for all $k\geq0$ for some $C_{3}<\infty$ and some $l>\max\left\{ 2,\,1+48q/\left(46+20q\right),\,1+q/\left(3/4+q/2\right)\right\} $,
where $q$ is as in $\boldsymbol{K}_{3}$; (iv) uniformly in $u\in\left[0,\,1\right]$,
$\widehat{D}_{1}\left(u\right),\,\widehat{D}_{2}\left(u\right)=O\mathbb{_{P}}\left(1\right)$
and $1/\widehat{D}_{1}\left(u\right),\,1/\widehat{D}_{2}\left(u\right)=O\mathbb{_{P}}\left(1\right)$;
(v) $|\omega_{s+1}-\omega_{s}|=O\left(T^{-1}\right)$ and $\left[S_{\omega}\right]=O\left(T\right)$;
(vi) $\sqrt{Tb_{2,T}\left(u\right)}(\widehat{D}_{2}\left(u\right)-D_{2}\left(u\right))=O_{\mathbb{P}}\left(1\right)$
for all $u\in\left[0,\,1\right]$; (vii) $\boldsymbol{K}_{2}$ includes
kernels that satisfy $|K_{2}\left(x\right)-K_{2}\left(y\right)|\leq C_{4}\left|x-y\right|$
for all $x,\,y\in\mathbb{R}$ and some constant $C_{4}<\infty$.
\end{assumption}
Parts (i)-(ii) of Assumption \ref{Assumption E-F-G} are the nonparametric
analogue to Assumption E and F, respectively, in \citet{andrews:91}.
Part (iii) is satisfied if $\left\{ V_{t}\right\} $ is strong mixing
with mixing numbers that are less stringent than those sufficient
for the cumulant condition in Assumption \ref{Assumption A - Dependence}-(i).
Part (iv) and (vi) extend (i)-(ii) to $\widehat{D}_{1}$ and $\widehat{D}_{2}$.
Part (v) is needed to apply the convergence of Riemann sums. Part
(vi) follows from the asymptotic results about $\widehat{c}_{T}\left(u,\,k\right)$.
Part (vii) requires $K_{2}$ to satisfy Lipschitz continuity.  Note
that $\phi_{\theta^{*}}$ coincides with the optimal value $\phi\left(q\right)$
only when the approximate parametric model indexed by $\theta^{*}$
corresponds to the true data-generating mechanism.  

Let $b_{\theta_{1},T}=(2qK_{1,q}^{2}\phi_{\theta^{*}}T\overline{b}_{\theta_{2},T}/\int K_{1}^{2}\left(y\right)dy\int_{0}^{1}K_{2}^{2}\left(x\right)dx)^{-1/\left(2q+1\right)},$
where $\overline{b}_{\theta_{2},T}\triangleq\int_{0}^{1}b_{2,T}^{\mathrm{opt}}\left(u\right)du$.
The asymptotic properties of $\widehat{J}_{T}(\widehat{b}_{1,T},\,\widehat{\overline{b}}_{2,T})$
are shown to be equivalent to those of $\widehat{J}_{T}(b_{\theta_{1},T},\,b_{\theta_{2},T})$.
\begin{thm}
\label{Theorem 3 Andrews 91}Suppose $K_{1}\left(\cdot\right)\in\boldsymbol{K}_{3}$,
$q$ is as in $\boldsymbol{K}_{3}$, $K_{2}\left(\cdot\right)\in\boldsymbol{K}_{2}$,\textbf{
}$n_{T}\rightarrow\infty,$ $n_{T}/Tb_{\theta_{1},T}\rightarrow0,$
and $||\int_{0}^{1}f^{\left(q\right)}\left(u,\,0\right)du||<\infty$.
Then, 

(i) If Assumption \ref{Assumption Smothness of A (for HAC)}-\ref{Assumption B}
and \ref{Assumption E-F-G}-(i,iv,vii) hold, $n_{3,T}=n_{2,T}=n_{T},$
and $q>1/2$, then $\widehat{J}_{T}(\widehat{b}_{1,T},\,\widehat{\overline{b}}_{2,T})-J_{T}\overset{\mathbb{P}}{\rightarrow}0$.

(ii) If Assumption \ref{Assumption Smothness of A (for HAC)}, \ref{Assumption B}-\ref{Assumption C Andrews 91}
and \ref{Assumption E-F-G}-(ii,iii,v,vi,vii) hold and $n_{T}/Tb_{\theta_{1},T}^{2}\rightarrow0,$
then $\sqrt{Tb_{\theta_{1},T}b_{\theta_{2},T}}$ $(\widehat{J}_{T}(\widehat{b}_{1,T},\,\widehat{\overline{b}}_{2,T})-J_{T})=O_{\mathbb{P}}\left(1\right)$
and $\sqrt{Tb_{\theta_{1},T}b_{\theta_{2},T}}(\widehat{J}_{T}(\widehat{b}_{1,T},\,\widehat{\overline{b}}_{2,T})-\widehat{J}_{T}(b_{\theta_{1},T},\,b_{\theta_{2},T}))=o_{\mathbb{P}}\left(1\right)$.

(iii) Let $\gamma_{\theta}=2qK_{1,q}^{2}\phi_{\theta}/(\int K_{1}^{2}\left(y\right)dy\int_{0}^{1}K_{2}^{2}\left(x\right)dx)$.
If Assumption \ref{Assumption Smothness of A (for HAC)}, \ref{Assumption B}-\ref{Assumption W_T and unbounded kernel and Cumulant 8}
and \ref{Assumption E-F-G}-(ii,iii,v,vi,vii) hold, then
\begin{align*}
\lim_{T\rightarrow\infty} & \mathrm{MSE}\left(T^{4q/10\left(2q+1\right)},\,\widehat{J}_{T}\left(\widehat{b}_{1,T},\,\widehat{\overline{b}}_{2,T}\right),\,W_{T}\right)\\
 & =\lim_{T\rightarrow\infty}\mathrm{MSE}\left(Tb_{\theta_{1},T}b_{\theta_{2},T},\,\widehat{J}_{T}\left(b_{\theta_{1},T},\,b_{\theta_{2},T}\right),\,W_{T}\right)\\
 & =4\pi^{2}\left[\gamma_{\theta}K_{1,q}^{2}\mathrm{vec}\left(\int_{0}^{1}f^{\left(q\right)}\left(u,\,0\right)du\right)'W\mathrm{vec}\left(\int_{0}^{1}f^{\left(q\right)}\left(u,\,0\right)du\right)\right]\\
 & \quad+\int K_{1}^{2}\left(y\right)dy\int K_{2}^{2}\left(x\right)dx\,\mathrm{tr}\left(W\left(I_{p^{2}}-C_{pp}\right)\left(\int_{0}^{1}f\left(u,\,0\right)du\right)\otimes\left(\int_{0}^{1}f\left(v,\,0\right)dv\right)\right).
\end{align*}
\end{thm}
When the chosen parametric model indexed by $\theta$ is correct,
it follows that $\phi_{\theta^{*}}=\phi\left(q\right)$ and $\widehat{\phi}\left(q\right)\overset{\mathbb{P}}{\rightarrow}\phi\left(q\right)$.
The theorem then implies that $\widehat{J}_{T}(\widehat{b}_{1,T},\,\widehat{\overline{b}}_{2,T})$
exhibits the same optimality properties presented in Theorem \ref{Theorem Optimal Kernels}
and Corollary \ref{Corollary 1 -Optimal b1 }. We omit the details.

\section{\label{Section Monte Carlo}Small-Sample Evaluations}

We conduct a Monte Carlo analysis to evaluate the properties of HAR
inference based on the HAC estimator $\widehat{J}_{T}$. We consider
HAR tests in the linear regression model as well as HAR tests used
in the forecast evaluation literature, namely the \citeauthor{diebold/mariano:95}'s
\citeyearpar{diebold/mariano:95} test\nocite{diebold/mariano:95}
and the forecast breakdown test of \citet{giacomini/rossi:09}. The
linear regression models have an intercept and a stochastic regressor.
 We focus on the $t$-statistics $t_{r}=\sqrt{T}(\widehat{\beta}^{\left(r\right)}-\beta_{0}^{\left(r\right)})/\sqrt{\widehat{J}_{X,T}^{\left(r,r\right)}}$
where
\begin{align*}
\widehat{J}_{X,T}= & \left(T^{-1}\sum_{t=1}^{T}x_{t}x'_{t}\right)^{-1}\widehat{J}_{T}\left(T^{-1}\sum_{t=1}^{T}x_{t}x'_{t}\right)^{-1},
\end{align*}
 is a consistent estimate of the limit of $\mathrm{Var}(\sqrt{T}(\widehat{\beta}-\beta_{0}))$
and $r=1,\,2$. $t_{1}$ is the $t$-statistic for the parameter associated
with the intercept while $t_{2}$ is associated with the stochastic
regressor $x_{t}$. Results for the $F$-test are qualitatively similar
{[}see \citet{casini_diss}{]}. Six basic regression models are considered.
We run a $t$-test on the intercept in model M1 and M5 whereas a $t$-test
on the coefficient of $x_{t}$ is run in model M2-M4 and M6.  The
models are based on, 
\begin{align}
y_{t} & =\beta_{0}^{\left(1\right)}+\delta+\beta_{0}^{\left(2\right)}x_{t}+e_{t},\qquad\qquad t=1,\ldots,\,T,\label{eq: Model P1}
\end{align}
for the $t$-test on the intercept (i.e., $t_{1}$) and
\begin{align}
y_{t} & =\beta_{0}^{\left(1\right)}+\left(\beta_{0}^{\left(2\right)}+\delta\right)x_{t}+e_{t},\qquad\qquad t=1,\ldots,\,T,\label{eq Model P1 beta2}
\end{align}
for the $t$-test on $\beta_{0}^{\left(2\right)}$ (i.e., $t_{2}$)
where $\delta=0$ under the null hypotheses. In Model M1 $e_{t}=0.5e_{t-1}+u_{t},\,u_{t}\sim\mathrm{i.i.d.\,}\mathscr{N}\left(0,\,0.5\right),$
$x_{t}\sim\mathrm{i.i.d.\,}\mathscr{N}\left(1,\,1\right)$, $\beta_{0}^{\left(1\right)}=0$
and $\beta_{0}^{\left(2\right)}=1.$\footnote{For the results with AR coefficient 0.9 see Table 1 in \citet{casini/perron_PrewhitedHAC}
and footnote \ref{fn:10} below.} Model M2 involves $e_{t}=0.8e_{t-1}+u_{t},\,u_{t}\sim\mathrm{i.i.d.\,}\mathscr{N}\left(0,\,1\right),$
$x_{t}\sim\mathrm{i.i.d.\,}\mathscr{N}\left(1,\,1\right)$, and $\beta_{0}^{\left(1\right)}=\beta_{0}^{\left(2\right)}=0.$
 In Model M3 we have segmented locally stationary errors $e_{t}=\rho_{t}e_{t-1}+u_{t},\,u_{t}\sim\mathrm{i.i.d.\,}\mathscr{N}\left(0,\,1\right),\,\rho_{t}=\max\left\{ 0,\,-1\left(\cos\left(1.5-\cos\left(5t/T\right)\right)\right)\right\} $
for $t<4T/5$ and $e_{t}=0.9e_{t-1}+u_{t},\,u_{t}\sim\mathrm{\mathrm{i.i.d.\,}}\mathscr{N}\left(0,\,1\right)$
for $t\geq4T/5,$ and $x_{t}=0.4x_{t-1}+u_{X,t},\,u_{X,t}\sim\mathrm{i.i.d.\,}\mathscr{N}\left(0,\,1\right)$.
Note that $\rho_{t}$ varies smoothly between 0 and 0.8071. Model
M4  involves some misspecification that induces nonstationarity in
the errors,
\begin{align*}
y_{t} & =\beta_{0}^{\left(1\right)}+\left(\beta_{0}^{\left(2\right)}+\delta\right)x_{t}+w_{t}\mathbf{1}\left\{ t\geq4T/5\right\} +e_{t},\qquad\qquad t=1,\ldots,\,T,
\end{align*}
where $e_{t}=\rho_{t}e_{t-1}+u_{t},\,u_{t}\sim\mathrm{i.i.d.\,}\mathscr{N}\left(0,\,1\right),\,\rho_{t}$
as in M3, $x_{t}\sim\mathrm{i.i.d.\,}\mathscr{N}\left(1,\,1\right)$,
and $w_{t}\sim\mathrm{i.i.d.\,}\mathscr{N}\left(2,\,1\right)$ independent
from $x_{t}$. Model M5  involves misspecification under $H_{1}$
via a smooth change in the coefficient $\beta_{0}^{\left(2\right)}$
toward the end of the sample. This situation is very common in practice
and it is motivated by the model for the variable ``cay'' from \citet{bianchi/lettau/ludvigson:18}
(cf. Figure 3 in their paper). The model is given by
\begin{align*}
y_{t} & =\beta_{0}^{\left(1\right)}+\delta+\left(\beta_{0}^{\left(2\right)}+d_{t}\mathbf{1}\left\{ t\geq4.5T/5\right\} \right)x_{t}+e_{t},\qquad\qquad t=1,\ldots,\,T,
\end{align*}
 where $d_{t}=1.5\delta\left(t-4.5T/5\right)/T$, $e_{t}=\rho_{t}e_{t-1}+u_{t},\,u_{t}\sim\mathrm{i.i.d.\,}\mathscr{N}\left(0,\,1\right),\,\rho_{t}=0.8(\cos(1.5-\cos(t/$
$2T)))$ for $t\in\left\{ 1,\ldots,\,T/2-1\right\} \cup\left\{ T/2+T/4+1,\ldots,\,T\right\} $
and $e_{t}=0.2e_{t-1}+2u_{t},\,u_{t}\sim\mathrm{\mathrm{i.i.d.\,}}\mathscr{N}\left(0,\,1\right)$
for $T/2\leq t\leq T/2+T/4$, and $x_{t}=2+0.5x_{t-1}+u_{X,t},\,u_{X,t}\sim\mathrm{i.i.d.}\,\mathscr{N}\left(0,\,1\right)$.
That is, $\rho_{t}$ varies smoothly between 0 and 0.7021. Model M6
 is given by \eqref{eq Model P1 beta2} where $e_{t}=\rho_{t}e_{t-1}+u_{t},\,u_{t}\sim\mathrm{i.i.d.\,}\mathscr{N}\left(0,\,1\right),\,\rho_{t}=\max\left\{ 0,\,0.3\left(\cos\left(1.5-\cos\left(t/5T\right)\right)\right)\right\} $
for $t\in\left\{ 1,\ldots,\,T/2-1\right\} \cup\{T/2+4,\ldots,$ $T-16\}$
and $e_{t}=0.99e_{t-1}+2u_{t},\,u_{t}\sim\mathrm{\mathrm{i.i.d.\,}}\mathscr{N}\left(0,\,1\right)$
for $T/2\leq t\leq T/2+3$ and $e_{t}=0.9e_{t-1}+2u_{t},\,u_{t}\sim\mathrm{\mathrm{i.i.d.\,}}\mathscr{N}\left(0,\,1\right)$
for $T-15\leq t\leq T$, and $x_{t}\sim\mathrm{i.i.d.}\,\mathscr{N}\left(1,\,1\right)$.
Note that $\rho_{t}\in\left[0,\,0.2633\right]$.

Next, we move to the forecast evaluation tests. The Diebold-Mariano
test statistic is defined as $t_{\mathrm{DM}}\triangleq\sqrt{T_{n}}\,\overline{d}_{L}/\sqrt{\widehat{J}_{d_{L},T}}$,
where $\overline{d}_{L}$ is the average of the loss differentials
between two competing forecast models, $\widehat{J}_{d_{L},T}$ is
an estimate of the asymptotic variance of the the loss differential
series and $T_{n}$ is the number of observations in the out-of-sample.
Throughout we use the quadratic loss. In model M7, we consider an
out-of-sample forecasting exercise with a fixed  scheme where, given
a sample of $T$ observations, $0.5T$ observations are used for the
in-sample and the remaining half is used for prediction. The true
model for the target variable is given by $y_{t}=\beta_{0}^{\left(1\right)}+\beta_{0}^{\left(2\right)}x_{t-1}^{(0)}+e_{t}$
where $x_{t-1}^{(0)}\sim\mathrm{i.i.d.}\,\mathscr{N}\left(1,\,1\right)$,
$e_{t}=0.3e_{t-1}+u_{t}$ with $u_{t}\sim\mathrm{i.i.d.\,}\mathscr{N}\left(0,\,1\right)$
and we set $\beta_{0}^{\left(1\right)}=\beta_{0}^{\left(2\right)}=1.$
The two competing models both involve an intercept but differ on the
predictor used in place of $x_{t}^{(0)}$. The first forecast model
uses $x_{t}^{(1)}$ while the second uses $x_{t}^{(2)}$ where $x_{t}^{(1)}$
and $x_{t}^{(2)}$ are independent $\mathrm{i.i.d.}\,\mathscr{N}\left(1,\,1\right)$
sequences, both independent from $x_{t}^{(0)}$. Each forecast model
generates a sequence of $\tau\left(=1\right)$-step ahead out-of-sample
losses $L_{t}^{(i)}$ $\left(i=1,\,2\right)$ for $t=T/2+1,\ldots,\,T-\tau.$
Then $d_{t}\triangleq L_{t}^{(2)}-L_{t}^{(1)}$ denotes the loss differential
at time $t$. The Diebold-Mariano test rejects the null of equal predictive
ability when (after normalization) $\overline{d}$ is sufficiently
far from zero. 

Finally, we consider model M8 which we use to investigate the performance
of a $t$-test for forecast breakdown {[}cf. \citet{giacomini/rossi:09}{]}.
Suppose we want to forecast a variable $y_{t}$ following the equation
$y_{t}=\beta_{0}^{\left(1\right)}+\beta_{0}^{\left(2\right)}x_{t-1}+e_{t}$
where $x_{t}\sim\mathrm{i.i.d.\,}\mathscr{N}\left(1,\,1.5\right)$
and $e_{t}=0.3e_{t-1}+u_{t}$ with $u_{t}\sim\mathrm{i.i.d.\,}\mathscr{N}\left(0,\,1\right)$.
For a given forecast model and forecasting scheme, the test of \citet{giacomini/rossi:09}
(GR) detects a forecast breakdown when the average of the out-of-sample
losses differs significantly from the average of the in-sample losses.
The in-sample is used to obtain estimates of $\beta_{0}^{\left(1\right)}$
and $\beta_{0}^{\left(2\right)}$ which are in turn used to construct
out-of-sample forecasts $\widehat{y}_{t}=\widehat{\beta}_{0}^{\left(1\right)}+\widehat{\beta}_{0}^{\left(2\right)}x_{t-1}$.
We set $\beta_{0}^{\left(1\right)}=\beta_{0}^{\left(2\right)}=1.$
We consider a fixed forecasting scheme and one-step ahead forecasts.
The GR's (2009) test statistic is defined as $t^{\mathrm{GR}}\triangleq\sqrt{T_{n}}\overline{SL}/\sqrt{\widehat{J}_{SL}}$
where $\overline{SL}\triangleq T_{n}^{-1}\sum_{t=T_{m}}^{T-\tau}SL_{t+\tau}$,
$SL_{t+\tau}$ is the surprise loss at time $t+\tau$, i.e., the difference
between the time $t+\tau$ out-of-sample loss and in-sample loss,
$SL_{t+\tau}=L_{t+\tau}-\overline{L}_{t+\tau}$, $T_{n}$ is the sample
size in the out-of-sample, $T_{m}$ is the sample size in the in-sample
and $\widehat{J}_{SL}$ is an HAC estimator. We restrict attention
to $\tau=1.$

Throughout our study we consider the following LRV estimators: $\widehat{J}_{T}$
with automatic bandwidths; $\widehat{J}_{T}$ with automatic bandwidths
and the prewhitening of \citet{casini/perron_PrewhitedHAC}; Andrews's
(1991) HAC estimator with automatic bandwidth; Andrews's (1991) HAC
estimator with automatic bandwidth and the prewhitening procedure
of \citet{andrews/monahan:92}; Newey and West's (1987) HAC estimator
with the automatic bandwidth as proposed in \citet{newey/west:94};
Newey and West's (1987) HAC estimator with the automatic bandwidth
as proposed in \citet{newey/west:94} and the prewhitening procedure;
Newey-West with the fixed-$b$ method of \citet{Kiefer/vogelsang/bunzel:00}.\footnote{\label{fn:10}To save space, we do not report results for the Empirical
Weighted Periodogram (EWP) or Empirical Weighted Cosine (EWC) of Lazarus
et al. \citeyearpar{lazarus/lewis/stock:17} and Lazarus et al. \citeyearpar{lazarus/lewis/stock/watson:18},
respectively. Their performance is similar to the method of \citet{Kiefer/vogelsang/bunzel:00}.
The LRV estimator of \citet{Kiefer/vogelsang/bunzel:00} leads to
HAR tests that have better size control. \citet{casini/perron_PrewhitedHAC}
showed that EWC leads to oversized tests when there is strong dependence
in the data relative to the fixed-$b$ method of \citet{Kiefer/vogelsang/bunzel:00}
and to the prewhitened DK-HAC. The power properties of tests normalized
by the EWP and EWC are similar to those using the method of \citet{Kiefer/vogelsang/bunzel:00}.} \citet{casini/perron_PrewhitedHAC} proposed three forms of prewhitening:
(1) $\widehat{J}_{T,\mathrm{pw},1}$ uses a stationary model to whiten
the data; (2) $\widehat{J}_{T,\mathrm{pw},\mathrm{SLS}}$ uses a nonstationary
model to whiten the data; (3) $\widehat{J}_{T,\mathrm{pw},\mathrm{SLS},\mu}$
is the same as $\widehat{J}_{T,\mathrm{pw},\mathrm{SLS}}$ but it
adds a time-varying intercept in the VAR to whiten the data. For model
M7 we also report results using $\widehat{J}_{T}$ and $\widehat{J}_{T,\mathrm{pw},\mathrm{SLS}}$
with the pre-test for breaks in the spectrum as developed in \citet{casini:change-point-spectra}.
We do not report the results for the pre-test for model M1-M6 and
M8 because they are equivalent to those without the pre-test.

For all versions of $\widehat{J}_{T}$ we use $K_{1}^{\mathrm{opt}}$
and $K_{2}^{\mathrm{opt}}$. We set $n_{T}=T^{0.66}$ as explained
in Section \ref{subsec Choice-of nT} and $n_{2,T}=n_{3,T}=n_{T}.$
We consider the following sample sizes for M1-M6: $T=200,\,400$.
Simulation results for additional data-generating processes involving
ARMA, ARCH and heteroskedastic errors are not discussed here because
the results are qualitatively equivalent {[}see, e.g., \citet{casini_diss}
and \citet{casini/perron_PrewhitedHAC}{]}. The significance level
is $\alpha=0.05$ throughout the study.

\subsection{Empirical Sizes of HAR Inference Tests}

Table \ref{Table S1-S2}-\ref{Table Size Forecasting DM-GR} report
the rejection rates for model M1-M8. We begin with the $t$-test in
the linear regression models. As a general pattern, we confirm previous
evidence that the Newey-West's (1987) and Andrews' (1991) HAC estimators
lead to $t$-tests that are oversized when the data are stationary
{[}cf. model M1-M2{]}. The same problem occurs for the Newey-West
(1987) HAC estimator using the usual ``rule'' to determine the number
of lags (not reported). For extreme temporal dependence, simulations
in \citet{casini/perron_PrewhitedHAC} showed that the size distortions
can be even larger especially for the $t$-test on the intercept.
Prewhitening is often effective in helping the HAC estimators to better
control the size under stationarity. However, the simulation results
in \citet{casini/perron_PrewhitedHAC} and in the literature show
that the prewhitened HAC estimators can lead to oversized tests when
there is high serial dependence. The rejection rates of tests normalized
by the Newey-West estimator with fixed-$b$ are the most accurate
in model M1-M2 for $T=200$. Overall, the results in the literature
along with those in \citet{casini_diss} and \citet{casini/perron_PrewhitedHAC}
showed that under stationarity the original fixed-$b$ method of KVB
is in general the least oversized across different degrees of dependence
among all existing methods. Table \ref{Table S1-S2} shows that for
the $t$-test on the intercept the non-prewhitened DK-HAC leads to
HAR tests that are oversized while they are accurate for the $t$-test
on the coefficient on the stochastic regressor. The table also shows
that the prewhitened DK-HAC estimators are competitive with the KVB's
fixed-$b$ in controlling the size. $\widehat{J}_{T,\mathrm{pw},1}$
is the most accurate among the DK-HAC estimators. Since $\widehat{J}_{T,\mathrm{pw},1}$
uses a stationarity VAR model to whiten the data, it works better
than $\widehat{J}_{T,\mathrm{pw},\mathrm{SLS}}$ and $\widehat{J}_{T,\mathrm{pw},\mathrm{SLS},\mu}$
when stationarity actually holds which is consistent with the results
of Table \ref{Table S1-S2}.

Turning to nonstationarity, Table \ref{Table S3-S4} casts concerns
about the  performance of existing methods in this context. For both
model M3 and M4, existing LRV estimators lead to HAR tests that have
either size equal or close to zero. The methods that use long bandwidths
(i.e., many lags) such as KVB's fixed-$b$ suffer most from this problem
relative to the classical HAC estimators. This is consistent with
the argument in \citet{casini/perron_Low_Frequency_Contam_Nonstat:2020}
who showed analytically that nonstationarity induces a positive bias
for each sample autocovariance. That bias is constant across different
lags. Since existing LRV estimators are weighted sum of sample autocovariances,
the larger the bandwidth (i.e., the more lagged autocovariances are
included) the larger the positive bias. Thus, LRV estimators are inflated
and HAR tests have rejection rates lower than the significance level.
This mechanism has consequences for power as well, as we show below
that traditional HAR tests have low power. In model M3-M4 the non-prewhitened
DK-HAC and the prewhitened DK-HAC (except $\widehat{J}_{T,\mathrm{pw},1}$)
perform well. $\widehat{J}_{T,\mathrm{pw},1}$ suffers from the same
problem as the existing estimators because it uses stationarity and
when this is violated its performance is affected. In model M5, the
classical HAC estimators yield HAR tests that are oversized. Also
the non-prewhitened DK-HAC is oversized. In contrast, the KVB's fixed-$b$
and the prewhitened DK-HAC have rejection rates close to the significance
level. In model M6, the KVB's fixed-$b$ HAR tests tend to be undersized
whereas the HAC and DK-HAC estimators lead to tests that control the
size more accurately. 

Turning to the HAR tests for forecast evaluations, Table \ref{Table Size Forecasting DM-GR}
shows that for model M7 the KVB's fixed-$b$ HAR test has size essentially
equal to zero while the classical HAC estimators yield HAR tests that
are somewhat oversized. In contrast, the tests normalized by the prewhitened
DK-HAC estimators have most accurate rejection rates. In model M8,
the KVB's fixed-$b$ HAR tests are well-sized whereas the classical
HAC estimators lead to tests that are severely undersized. The DK-HAC
estimators control the size reasonably well. 

In summary, the prewhitened DK-HAC estimators yield $t$-tests in
regression models with rejection rates that are relatively close to
the nominal size. The non-prewhitened DK-HAC can lead to oversized
tests for the $t$-tests on the intercept if there is  high dependence.
Our results confirm the oversize problem of the HAR tests normalized
by the classical HAC estimators documented in the literature under
stationarity. The Fixed-$b$ HAR tests control the size well when
the data are stationary but can show severe undersized issues under
nonstationarity, a problem also affecting the tests normalized by
the classical HAC estimators. Thus, with regards to size control,
the prewhitened DK-HAC estimators are competitive with fixed-$b$
methods under stationarity and they also perform well when the data
are nonstationary. 

\subsection{Empirical Power of HAR Inference Tests}

For model M1-M6 we report the values of the power in Table \ref{Table M1 Power}-\ref{Table Power M6}.
The sample size is $T=200$.  Power functions for the Diebold-Maraino
and for the forecast breakdown test are presented next. For model
M1, the non-prewhitened HAC and DK-HAC lead to tests that have the
highest power but they were more oversized than the other methods.
The KVB's fixed-$b$ LRV leads to $t$-tests that sacrify some power
relative to the prewhitened HAC and DK-HAC estimators. In model M2,
a similar picture arises. HAR tests normalized by either classical
HAC or DK-HAC estimators have similarly good power while HAR tests
based on KVB's fixed-$b$ have relatively less power. In model M3,
the prewhitening HAC estimators and $\widehat{J}_{T,\mathrm{pw},1}$
(which uses a stationary model to whiten the data) have low power.
The best power is achieved by tests normalized by Andrews' (1991)
HAC estimator and $\widehat{J}_{T}$, followed by $\widehat{J}_{T,\mathrm{pw},\mathrm{SLS}}$
and $\widehat{J}_{T,\mathrm{pw},\mathrm{SLS},\mu}$. The KVB's fixed-$b$
leads to relatively less power than the latter methods. The Newey-West's
(1987) estimator leads to tests that have good power but they were
shown to be oversized. Similar comments apply to model M4. Here Andrews'
(1991) HAC estimator leads to tests that have better power for small
to medium breaks while tests based on $\widehat{J}_{T}$ have better
power for large breaks. In model M5, prewhitening HAC estimators and
KVB's fixed-$b$ lead to HAR tests that have non-monotonic power and
reach zero as $\delta$ increases. This does not occur for the classical
HAC estimators which, however, were oversized. HAR tests based on
$\widehat{J}_{T,\mathrm{QS}}$, $\widehat{J}_{T,\mathrm{pw},\mathrm{SLS}}$
and $\widehat{J}_{T,\mathrm{pw},\mathrm{SLS},\mu}$ perform best for
this model. $\widehat{J}_{T,\mathrm{pw},1}$ results in HAR tests
that have lower power relative to the tests based on the other DK-HAC
because stationarity is violated. In model M6, all HAR tests enjoy
monotonic power with small differences.

Next, let us move to the evaluation of the power properties of the
$t$-tests used in the forecasting literature. We begin with the Diebold-Mariano
test. For this test, the separation between the null and alternative
hypotheses does not depend on the value of a single parameter. Thus,
the data-generating mechanism is different from the one under the
null. The two competing forecast models are as follows: the first
model uses the actual true data-generating process while the second
model differs in that in place of $x_{t-1}^{(0)}$ it uses $x_{t-1}^{(2)}=x_{t-1}^{(0)}+u_{X_{2},t}$
for $t\leq3T/4$ and $x_{t-1}^{(2)}=\delta+x_{t-1}^{(0)}+u_{X_{2},t}$
for $t>3T/4$ with $u_{X_{2},t}\sim\mathrm{i.i.d.\,}\mathscr{N}\left(0,\,1\right)$.
Evidently, the null hypothesis of equal predictive ability should
be rejected by the Diebold-Mariano test whenever $\delta>0$. Table
\ref{Table Power DM Test} reports the power for several values of
$\delta.$ The HAR tests based on existing estimators have lower power
relative to the $\widehat{J}_{T}$ DK-HAC estimators for small values
of $\delta$. When we raise $\delta$ the tests based on the HAC estimators
of \citet{andrews:91} and Newey and West (1987), and KVB's fixed-$b$
method display non-monotonic power gradually converging to zero. In
contrast, the DK-HAC estimators lead to tests that have monotonic
power that reach and maintain unit power. The only exception is the
test based on $\widehat{J}_{T,\mathrm{pw},1}$ that has lower power
because stationarity is violated. The table also reports $\widehat{J}_{T}$
and $\widehat{J}_{T,\mathrm{pw},\mathrm{SLS}}$ with the pre-test
for breaks in the spectrum {[}cf. \citet{casini:change-point-spectra}{]}
that is used for choosing more efficiently how to split the sample
in blocks to compute $\widehat{\Gamma}\left(k\right)$. The pre-test
yields HAR tests with higher power while having the same size as the
corresponding HAR tests with no pre-test. We have not reported the
results with the pre-test for model M1-M6 and M8 because they are
the same as with no pre-test.

Finally, we move to the $t$-test of \citet{giacomini/rossi:09}.
The data-generating process under $H_{1}:\,\mathbb{E}\left(\overline{SL}\right)\neq0$
is given by $y_{t}=1+x_{t-1}+\delta x_{t-1}\mathbf{1}\left\{ t>T_{1}^{0}\right\} +e_{t}$,
where $x_{t-1}\sim\mathrm{i.i.d.}\,\mathscr{N}\left(1.5,\,1\right)$,
$e_{t}=0.3e_{t-1}+u_{t}$, $u_{t}\sim\mathrm{i.i.d.}\mathscr{N}\left(0,\,1\right)$
and $T_{1}^{0}=T\lambda_{1}^{0}$ with $\lambda_{1}^{0}=0.8$. Under
this specification there is a break in the coefficient associated
to the predictor $x_{t-1}$. Thus, there is a forecast failure and
the test of \citet{giacomini/rossi:09} should reject $H_{0}$. From
Table \ref{Table Power GR Test} it appears that all versions of the
classical HAC estimators of \citet{andrews:91} and \citet{newey/west:87},
and KVB's fixed-$b$ lead to $t$-tests that have, essentially, zero
power for all $\delta$. The only exception is Andrews' (1991) HAC
estimator with prewhitening that shows some power but it is not monotonic.
In contrast, the $t$-test based on the DK-HAC estimators have good
power. The failure of existing LRV estimators cannot be attributed
to the sample size because as we raise the sample size to 400 or 800,
the tests still display no power {[}see \citet{casini_diss}{]}.

The failure of the HAR tests based on the existing LRV estimators
occurring in some of the data-generating mechanisms reported here
can be simply reconciled with the fact that in such models the spectrum
of $V_{t}$ is not constant. In other words, the autocovariance of
$V_{t}$ depends not only on the lag order but also on $t$. Existing
LRV estimators estimate an average of a time-varying spectrum. Because
of this instability in the spectrum, they overestimate the extent
of the dependence or variation in $V_{t}$. This is explained analytically
in Casini et al. \citeyearpar{casini/perron_Low_Frequency_Contam_Nonstat:2020}
who showed in a general setting that nonstationarity/misspecification
alters the low frequency components of a time series making the latter
appear as more persistent. Since traditional LRV estimators are a
weighted sum of a large number of low frequency periodogram ordinates,
these estimates turn to be inflated. Similarly, LRV estimators using
long bandwidths (i.e., fixed-$b$) are weighted sum of a large number
of sample autocovariances. Each sample autocovariance is biased upward
so that the latter estimates are even more inflated than the classical
HAC estimators that use a smaller number of sample autocovariances.
This explains why KVB's fixed-$b$ HAR tests are subject to more power
problems, even though the classical HAC estimators are also largely
affected. 

The introduction of the smoothing over time in the DK-HAC estimators
avoids the low frequency contamination because observations belonging
to different regimes are not mixed up when computing sample autocovariances.
This guarantees good power properties also under nonstationarity/misspecification
or under nonstationary alternative hypotheses (e.g., HAR tests for
forecast evaluation discussed above). Casini et al. \citeyearpar{casini/perron_Low_Frequency_Contam_Nonstat:2020}
reconciled this issue with some results in the unit root and long
memory literature. Tests for a unit root are known to struggle to
reject the unit root hypotheses if a process is second-order stationary
(i.e., no unit root) but it is contaminated by breaks in the mean
or trend {[}cf. \citeauthor{perron:89} (\citeyear{perron:89}, \citeyear{perron:90}){]}.
Similarly, a short memory sequence contaminated by structural breaks
can approximate a long memory series in that the autocorrelation function
has the same properties as that of a long memory series {[}cf. \citet{diebold/inoue:01},
\citet{hillebrand:05}, \citet{mccloskey/hill:2017} and \citet{mikosh/starica:04}{]}.

\section{\label{Section Conclusions}Conclusions}

Economic time series are highly nonstationary. Methods constructed
under the assumption of stationarity might then have undesirable properties.
This paper developed a theoretical framework for inference in settings
where the data may be nonstationary. A new class of double kernel
heteroskedasticity and autocorrelation consistent (DK-HAC) estimators
was presented. In addition to the usual smoothing procedure over lagged
autocovariances, the estimator applies smoothing over time. This
is important in order to account flexibly for the variation over time
of the structural properties of the economic time series. Optimality
results under MSE criterion concerning bandwidths and kernels have
been established. A data-dependent method based on the ``plug-in''
approach has been proposed. There are empirical relevant circumstances
where HAR tests, either in linear regression models or other contexts,
standardized by existing LRV estimators perform poorly. These may
result in size distortions as well as significant power losses, even
when the sample size is large. In contrast, when the proposed DK-HAC
estimator is used the same HAR tests do not suffer from those issues.
DK-HAC estimators lead to HAR tests that have competitive size control
relative to fixed-$b$ HAR tests, when the latter work well, and have
good power, irrespective of whether there is weak or strong dependence
in the data. 

\newpage{}

\bibliographystyle{elsarticle-harv}
\bibliography{References_JoE}
\addcontentsline{toc}{section}{References}

\newpage{}

\newpage{}

\clearpage 
\pagenumbering{arabic}
\renewcommand*{\thepage}{A-\arabic{page}}
\appendix

\section{Appendix}

{\footnotesize{}}
\begin{table}[H]
{\footnotesize{}\caption{\label{Table S1-S2}Empirical small-sample size for model M1-M2}
}{\footnotesize\par}
\begin{centering}
{\footnotesize{}}%
\begin{tabular}{lcccc}
\hline 
 & \multicolumn{2}{c}{{\footnotesize{}Model M1, $t_{1}$}} & \multicolumn{2}{c}{{\footnotesize{}Model M2, $t_{2}$}}\tabularnewline
{\footnotesize{}5\% nominal size} & {\footnotesize{}$T=200$} & {\footnotesize{}$T=400$ } & {\footnotesize{}$T=200$} & {\footnotesize{}$T=400$ }\tabularnewline
\hline 
\hline 
{\footnotesize{}$\widehat{J}_{T}$} & {\footnotesize{}0.086} & {\footnotesize{}0.067} & {\footnotesize{}0.054} & {\footnotesize{}0.038}\tabularnewline
{\footnotesize{}$\widehat{J}_{T,\mathrm{pw},1}$ } & {\footnotesize{}0.052} & {\footnotesize{}0.047} & {\footnotesize{}0.060} & {\footnotesize{}0.043 }\tabularnewline
{\footnotesize{}$\widehat{J}_{T,\mathrm{pw},\mathrm{SLS}}$ } & {\footnotesize{}0.053} & {\footnotesize{}0.041} & {\footnotesize{}0.069} & {\footnotesize{}0.037 }\tabularnewline
{\footnotesize{}$\widehat{J}_{T,\mathrm{pw},\mathrm{SLS},\mu}$ } & {\footnotesize{}0.048} & {\footnotesize{}0.044} & {\footnotesize{}0.065} & {\footnotesize{}0.039 }\tabularnewline
{\footnotesize{}Andrews (1991)} & {\footnotesize{}0.081} & {\footnotesize{}0.060} & {\footnotesize{}0.082} & {\footnotesize{}0.072}\tabularnewline
{\footnotesize{}Andrews (1991), prewhite} & {\footnotesize{}0.059} & {\footnotesize{}0.048} & {\footnotesize{}0.062 } & {\footnotesize{}0.047}\tabularnewline
{\footnotesize{}Newey-West (1987)} & {\footnotesize{}0.091} & {\footnotesize{}0.068} & {\footnotesize{}0.058} & {\footnotesize{}0.052}\tabularnewline
{\footnotesize{}Newey-West (1987), prewhite} & {\footnotesize{}0.073} & {\footnotesize{}0.054} & {\footnotesize{}0.071} & {\footnotesize{}0.064}\tabularnewline
{\footnotesize{}Newey-West (1987), fixed-$b$ (KVB)} & {\footnotesize{}0.057} & {\footnotesize{}0.054} & {\footnotesize{}0.059} & {\footnotesize{}0.059}\tabularnewline
\hline 
\end{tabular}{\footnotesize\par}
\par\end{centering}
{\footnotesize{}}{\footnotesize\par}
\end{table}

{\footnotesize{}}
\begin{table}[H]
{\footnotesize{}\caption{\label{Table S3-S4}Empirical small-sample size for model M3-M4}
}{\footnotesize\par}
\begin{centering}
{\footnotesize{}}%
\begin{tabular}{lcccc}
\hline 
 & \multicolumn{2}{c}{{\footnotesize{}Model M3, $t_{2}$}} & \multicolumn{2}{c}{{\footnotesize{}Model M4, $t_{2}$}}\tabularnewline
{\footnotesize{}5\% nominal size} & {\footnotesize{}$T=200$} & {\footnotesize{}$T=400$ } & {\footnotesize{}$T=200$} & {\footnotesize{}$T=400$}\tabularnewline
\hline 
\hline 
{\footnotesize{}$\widehat{J}_{T}$} & {\footnotesize{}0.063} & {\footnotesize{}0.056} & {\footnotesize{}0.064} & {\footnotesize{}0.054}\tabularnewline
{\footnotesize{}$\widehat{J}_{T,\mathrm{pw},1}$ } & {\footnotesize{}0.013} & {\footnotesize{}0.011} & {\footnotesize{}0.008} & {\footnotesize{}0.000}\tabularnewline
{\footnotesize{}$\widehat{J}_{T,\mathrm{pw},\mathrm{SLS}}$ } & {\footnotesize{}0.062} & {\footnotesize{}0.061} & {\footnotesize{}0.063} & {\footnotesize{}0.043}\tabularnewline
{\footnotesize{}$\widehat{J}_{T,\mathrm{pw},\mathrm{SLS},\mu}$ } & {\footnotesize{}0.056} & {\footnotesize{}0.054} & {\footnotesize{}0.034} & {\footnotesize{}0.042}\tabularnewline
{\footnotesize{}Andrews (1991)} & {\footnotesize{}0.047} & {\footnotesize{}0.025} & {\footnotesize{}0.043} & {\footnotesize{}0.016}\tabularnewline
{\footnotesize{}Andrews (1991), prewhite} & {\footnotesize{}0.013} & {\footnotesize{}0.019} & {\footnotesize{}0.000} & {\footnotesize{}0.000}\tabularnewline
{\footnotesize{}Newey-West (1987)} & {\footnotesize{}0.072} & {\footnotesize{}0.064} & {\footnotesize{}0.023} & {\footnotesize{}0.032}\tabularnewline
{\footnotesize{}Newey-West (1987), prewhite} & {\footnotesize{}0.014} & {\footnotesize{}0.016} & {\footnotesize{}0.000} & {\footnotesize{}0.000}\tabularnewline
{\footnotesize{}Newey-West (1987), fixed-$b$ (KVB)} & {\footnotesize{}0.003} & {\footnotesize{}0.001} & {\footnotesize{}0.000} & {\footnotesize{}0.000}\tabularnewline
\hline 
\end{tabular}{\footnotesize\par}
\par\end{centering}
{\footnotesize{}}{\footnotesize\par}
\end{table}
{\footnotesize{} }{\footnotesize\par}

{\footnotesize{}}
\begin{table}[H]
{\footnotesize{}\caption{\label{Table S5-S6}Empirical small-sample size for model M5-M6}
}{\footnotesize\par}
\begin{centering}
{\footnotesize{}}%
\begin{tabular}{lcccc}
\hline 
 & \multicolumn{2}{c}{{\footnotesize{}M5, $t_{1}$ }} & \multicolumn{2}{c}{{\footnotesize{}M6, $t_{2}$  }}\tabularnewline
{\footnotesize{}5\% nominal size} & {\footnotesize{}$T=200$} & {\footnotesize{}$T=400$} & {\footnotesize{}$T=200$} & {\footnotesize{}$T=400$}\tabularnewline
\hline 
\hline 
{\footnotesize{}$\widehat{J}_{T}$} & {\footnotesize{}0.095} & {\footnotesize{}0.093} & {\footnotesize{}0.065 } & {\footnotesize{}0.060}\tabularnewline
{\footnotesize{}$\widehat{J}_{T,\mathrm{pw},1}$ } & {\footnotesize{}0.057} & {\footnotesize{}0.056} & {\footnotesize{}0.044} & {\footnotesize{}0.049}\tabularnewline
{\footnotesize{}$\widehat{J}_{T,\mathrm{pw},\mathrm{SLS}}$ } & {\footnotesize{}0.064} & {\footnotesize{}0.059} & {\footnotesize{}0.056} & {\footnotesize{}0.052}\tabularnewline
{\footnotesize{}$\widehat{J}_{T,\mathrm{pw},\mathrm{SLS},\mu}$ } & {\footnotesize{}0.067} & {\footnotesize{}0.065} & {\footnotesize{}0.059} & {\footnotesize{}0.056}\tabularnewline
{\footnotesize{}Andrews (1991)} & {\footnotesize{}0.081} & {\footnotesize{}0.058} & {\footnotesize{}0.049} & {\footnotesize{}0.048}\tabularnewline
{\footnotesize{}Andrews (1991), prewhite} & {\footnotesize{}0.069} & {\footnotesize{}0.051} & {\footnotesize{}0.040} & {\footnotesize{}0.044}\tabularnewline
{\footnotesize{}Newey-West (1987)} & {\footnotesize{}0.111} & {\footnotesize{}0.084} & {\footnotesize{}0.052} & {\footnotesize{}0.048}\tabularnewline
{\footnotesize{}Newey-West (1987), prewhite} & {\footnotesize{}0.078} & {\footnotesize{}0.057} & {\footnotesize{}0.048} & {\footnotesize{}0.045}\tabularnewline
{\footnotesize{}Newey-West (1987), fixed-$b$ (KVB)} & {\footnotesize{}0.063} & {\footnotesize{}0.054} & {\footnotesize{}0.034} & {\footnotesize{}0.035}\tabularnewline
\hline 
\end{tabular}{\footnotesize\par}
\par\end{centering}
{\footnotesize{}}{\footnotesize\par}
\end{table}

{\footnotesize{}}
\begin{table}[H]
{\footnotesize{}\caption{\label{Table Size Forecasting DM-GR}Empirical small-sample size for
model M7-M8}
}{\footnotesize\par}
\begin{centering}
{\footnotesize{}}%
\begin{tabular}{lcccc}
\hline 
 & \multicolumn{2}{c}{{\footnotesize{}DM test}} & \multicolumn{2}{c}{{\footnotesize{}GR test}}\tabularnewline
{\footnotesize{}5\% nominal size} & {\footnotesize{}$T_{n}=200$} & {\footnotesize{}$T_{n}=400$} & {\footnotesize{}$T_{n}=240$} & {\footnotesize{}$T_{n}=380$}\tabularnewline
\hline 
\hline 
{\footnotesize{}$\widehat{J}_{T}$} & {\footnotesize{}0.035} & {\footnotesize{}0.063} & {\footnotesize{}0.029} & {\footnotesize{}0.043}\tabularnewline
{\footnotesize{}$\widehat{J}_{T,\mathrm{pw},1}$ } & {\footnotesize{}0.026} & {\footnotesize{}0.031} & {\footnotesize{}0.028} & {\footnotesize{}0.033}\tabularnewline
{\footnotesize{}$\widehat{J}_{T,\mathrm{pw},\mathrm{SLS}}$ } & {\footnotesize{}0.045} & {\footnotesize{}0.042} & {\footnotesize{}0.036} & {\footnotesize{}0.039}\tabularnewline
{\footnotesize{}$\widehat{J}_{T,\mathrm{pw},\mathrm{SLS},\mu}$ } & {\footnotesize{}0.043} & {\footnotesize{}0.046} & {\footnotesize{}0.047} & {\footnotesize{}0.045}\tabularnewline
{\footnotesize{}Andrews (1991)} & {\footnotesize{}0.083} & {\footnotesize{}0.085} & {\footnotesize{}0.000} & {\footnotesize{}0.000}\tabularnewline
{\footnotesize{}Andrews (1991), prewhite} & {\footnotesize{}0.082} & {\footnotesize{}0.085} & {\footnotesize{}0.000} & {\footnotesize{}0.003}\tabularnewline
{\footnotesize{}Newey-West (1987)} & {\footnotesize{}0.080} & {\footnotesize{}0.083} & {\footnotesize{}0.000} & {\footnotesize{}0.000}\tabularnewline
{\footnotesize{}Newey-West (1987), prewhite} & {\footnotesize{}0.079} & {\footnotesize{}0.083} & {\footnotesize{}0.000} & {\footnotesize{}0.000}\tabularnewline
{\footnotesize{}Newey-West (1987), fixed-$b$ (KVB)} & {\footnotesize{}0.002} & {\footnotesize{}0.002} & {\footnotesize{}0.068} & {\footnotesize{}0.049}\tabularnewline
\hline 
\end{tabular}{\footnotesize\par}
\par\end{centering}
{\footnotesize{}}{\footnotesize\par}
\end{table}

{\footnotesize{}}
\begin{table}[H]
{\footnotesize{}\caption{\label{Table M1 Power}Empirical small-sample power for model M1}
}{\footnotesize\par}
\begin{centering}
{\footnotesize{}}%
\begin{tabular}{lcccc}
\hline 
 & \multicolumn{3}{c}{} & \tabularnewline
{\footnotesize{}5\% nominal size, $T=200$} & {\footnotesize{}$\delta=0.2$} & {\footnotesize{}$\delta=0.4$} & {\footnotesize{}$\delta=0.8$} & {\footnotesize{}$\delta=1.6$}\tabularnewline
\hline 
\hline 
{\footnotesize{}$\widehat{J}_{T}$} & {\footnotesize{}0.481} & {\footnotesize{}0.924} & {\footnotesize{}1.000} & {\footnotesize{}1.000}\tabularnewline
{\footnotesize{}$\widehat{J}_{T,\mathrm{pw},1}$ } & {\footnotesize{}0.394} & {\footnotesize{}0.887} & {\footnotesize{}1.000} & {\footnotesize{}1.000}\tabularnewline
{\footnotesize{}$\widehat{J}_{T,\mathrm{pw},\mathrm{SLS}}$ } & {\footnotesize{}0.381} & {\footnotesize{}0.907} & {\footnotesize{}1.000} & {\footnotesize{}1.000}\tabularnewline
{\footnotesize{}$\widehat{J}_{T,\mathrm{pw},\mathrm{SLS},\mu}$ } & {\footnotesize{}0.370} & {\footnotesize{}0.907} & {\footnotesize{}1.000} & {\footnotesize{}1.000}\tabularnewline
{\footnotesize{}Andrews (1991)} & {\footnotesize{}0.479} & {\footnotesize{}0.943} & {\footnotesize{}1.000} & {\footnotesize{}1.000}\tabularnewline
{\footnotesize{}Andrews (1991), prewhite} & {\footnotesize{}0.436} & {\footnotesize{}0.899} & {\footnotesize{}1.000} & {\footnotesize{}1.000}\tabularnewline
{\footnotesize{}Newey-West (1987)} & {\footnotesize{}0.549} & {\footnotesize{}0.961} & {\footnotesize{}1.000} & {\footnotesize{}1.000}\tabularnewline
{\footnotesize{}Newey-West (1987), prewhite} & {\footnotesize{}0.454} & {\footnotesize{}0.934} & {\footnotesize{}1.000} & {\footnotesize{}1.000}\tabularnewline
{\footnotesize{}Newey-West (1987), fixed-$b$ (KVB)} & {\footnotesize{}0.323} & {\footnotesize{}0.769} & {\footnotesize{}0.998} & {\footnotesize{}1.000}\tabularnewline
\hline 
\end{tabular}{\footnotesize\par}
\par\end{centering}
{\footnotesize{}}{\footnotesize\par}
\end{table}

{\footnotesize{}}
\begin{table}[H]
{\footnotesize{}\caption{\label{Table Power M2}Empirical small-sample power for model M2}
}{\footnotesize\par}
\begin{centering}
{\footnotesize{}}%
\begin{tabular}{lccccc}
\hline 
 & \multicolumn{5}{c}{}\tabularnewline
{\footnotesize{}5\% nominal size, $T=200$} & {\footnotesize{}$\delta=0.1$ } & {\footnotesize{}$\delta=0.2$ } & {\footnotesize{}$\delta=0.4$ } & {\footnotesize{}$\delta=0.6$ } & {\footnotesize{}$\delta=0.8$ }\tabularnewline
\hline 
\hline 
{\footnotesize{}$\widehat{J}_{T}$} & {\footnotesize{}0.153} & {\footnotesize{}0.403} & {\footnotesize{}0.906} & {\footnotesize{}0.996} & {\footnotesize{}1.000}\tabularnewline
{\footnotesize{}$\widehat{J}_{T,\mathrm{pw},1}$ } & {\footnotesize{}0.150} & {\footnotesize{}0.366} & {\footnotesize{}0.858} & {\footnotesize{}0.987} & {\footnotesize{}1.000}\tabularnewline
{\footnotesize{}$\widehat{J}_{T,\mathrm{pw},\mathrm{SLS}}$ } & {\footnotesize{}0.177} & {\footnotesize{}0.390} & {\footnotesize{}0.878} & {\footnotesize{}0.992} & {\footnotesize{}1.000}\tabularnewline
{\footnotesize{}$\widehat{J}_{T,\mathrm{pw},\mathrm{SLS},\mu}$ } & {\footnotesize{}0.175} & {\footnotesize{}0.388} & {\footnotesize{}0.876} & {\footnotesize{}0.990} & {\footnotesize{}1.000}\tabularnewline
{\footnotesize{}Andrews (1991)} & {\footnotesize{}0.203} & {\footnotesize{}0.503} & {\footnotesize{}0.930} & {\footnotesize{}0.997} & {\footnotesize{}0.999}\tabularnewline
{\footnotesize{}Andrews (1991), prewhite} & {\footnotesize{}0.148} & {\footnotesize{}0.416} & {\footnotesize{}0.914} & {\footnotesize{}0.997} & {\footnotesize{}1.000}\tabularnewline
{\footnotesize{}Newey-West (1987)} & {\footnotesize{}0.163} & {\footnotesize{}0.448} & {\footnotesize{}0.925} & {\footnotesize{}0.998} & {\footnotesize{}1.000}\tabularnewline
{\footnotesize{}Newey-West (1987), prewhite} & {\footnotesize{}0.178} & {\footnotesize{}0.463} & {\footnotesize{}0.924} & {\footnotesize{}0.997} & {\footnotesize{}1.000}\tabularnewline
{\footnotesize{}Newey-West (1987), fixed-$b$ (KVB)} & {\footnotesize{}0.133} & {\footnotesize{}0.332} & {\footnotesize{}0.781} & {\footnotesize{}0.957} & {\footnotesize{}0.995}\tabularnewline
\hline 
\end{tabular}{\footnotesize\par}
\par\end{centering}
{\footnotesize{}}{\footnotesize\par}
\end{table}
{\footnotesize{} }{\footnotesize\par}

{\footnotesize{}}
\begin{table}[H]
{\footnotesize{}\caption{\label{Table Power M3}Empirical small-sample power for model M3}
}{\footnotesize\par}
\begin{centering}
{\footnotesize{}}%
\begin{tabular}{lcccccc}
\hline 
 & \multicolumn{6}{c}{}\tabularnewline
{\footnotesize{}5\% nominal size, $T=200$} & {\footnotesize{}$\delta=0.1$ } & {\footnotesize{}$\delta=0.2$} & {\footnotesize{}$\delta=0.4$ } & {\footnotesize{}$\delta=0.8$ } & {\footnotesize{}$\delta=1.6$} & {\footnotesize{}$\delta=2.5$ }\tabularnewline
\hline 
\hline 
{\footnotesize{}$\widehat{J}_{T}$} & {\footnotesize{}0.165} & {\footnotesize{}0.230} & {\footnotesize{}0.488} & {\footnotesize{}0.811} & {\footnotesize{}0.975} & {\footnotesize{}1.000}\tabularnewline
{\footnotesize{}$\widehat{J}_{T,\mathrm{pw},1}$ } & {\footnotesize{}0.020} & {\footnotesize{}0.047} & {\footnotesize{}0.189} & {\footnotesize{}0.545} & {\footnotesize{}0.913} & {\footnotesize{}1.000}\tabularnewline
{\footnotesize{}$\widehat{J}_{T,\mathrm{pw},\mathrm{SLS}}$ } & {\footnotesize{}0.080} & {\footnotesize{}0.131} & {\footnotesize{}0.303} & {\footnotesize{}0.661} & {\footnotesize{}0.954} & {\footnotesize{}1.000}\tabularnewline
{\footnotesize{}$\widehat{J}_{T,\mathrm{pw},\mathrm{SLS},\mu}$ } & {\footnotesize{}0.068} & {\footnotesize{}0.105} & {\footnotesize{}0.275} & {\footnotesize{}0.651} & {\footnotesize{}0.931} & {\footnotesize{}1.000}\tabularnewline
{\footnotesize{}Andrews (1991)} & {\footnotesize{}0.097} & {\footnotesize{}0.242} & {\footnotesize{}0.570} & {\footnotesize{}0.836} & {\footnotesize{}0.967} & {\footnotesize{}1.000}\tabularnewline
{\footnotesize{}Andrews (1991), prewhite} & {\footnotesize{}0.026} & {\footnotesize{}0.074} & {\footnotesize{}0.254} & {\footnotesize{}0.599} & {\footnotesize{}0.874} & {\footnotesize{}1.000}\tabularnewline
{\footnotesize{}Newey-West (1987)} & {\footnotesize{}0.108} & {\footnotesize{}0.195} & {\footnotesize{}0.448} & {\footnotesize{}0.793} & {\footnotesize{}0.976} & {\footnotesize{}1.000}\tabularnewline
{\footnotesize{}Newey-West (1987), prewhite} & {\footnotesize{}0.035} & {\footnotesize{}0.094} & {\footnotesize{}0.298} & {\footnotesize{}0.627} & {\footnotesize{}0.874} & {\footnotesize{}1.000}\tabularnewline
{\footnotesize{}Newey-West (1987), fixed-$b$ (KVB)} & {\footnotesize{}0.012} & {\footnotesize{}0.061} & {\footnotesize{}0.254} & {\footnotesize{}0.605} & {\footnotesize{}0.882} & {\footnotesize{}0.996}\tabularnewline
\hline 
\end{tabular}{\footnotesize\par}
\par\end{centering}
{\footnotesize{}}{\footnotesize\par}
\end{table}

{\footnotesize{}}
\begin{table}[H]
{\footnotesize{}\caption{\label{Table Power M4}Empirical small-sample power for model M4}
}{\footnotesize\par}
\begin{centering}
{\footnotesize{}}%
\begin{tabular}{lcccccc}
\hline 
 & \multicolumn{6}{c}{}\tabularnewline
{\footnotesize{}5\% nominal size, $T=200$} & {\footnotesize{}$\delta=0.1$ } & {\footnotesize{}$\delta=0.2$} & {\footnotesize{}$\delta=0.4$ } & {\footnotesize{}$\delta=0.8$} & {\footnotesize{}$\delta=1.6$} & {\footnotesize{}$\delta=3$}\tabularnewline
\hline 
\hline 
{\footnotesize{}$\widehat{J}_{T}$} & {\footnotesize{}0.112} & {\footnotesize{}0.135} & {\footnotesize{}0.310} & {\footnotesize{}0.645} & {\footnotesize{}0.969} & {\footnotesize{}1.000}\tabularnewline
{\footnotesize{}$\widehat{J}_{T,\mathrm{pw},1}$ } & {\footnotesize{}0.010} & {\footnotesize{}0.021} & {\footnotesize{}0.073} & {\footnotesize{}0.339} & {\footnotesize{}0.856} & {\footnotesize{}1.000}\tabularnewline
{\footnotesize{}$\widehat{J}_{T,\mathrm{pw},\mathrm{SLS}}$ } & {\footnotesize{}0.064} & {\footnotesize{}0.089} & {\footnotesize{}0.166} & {\footnotesize{}0.431} & {\footnotesize{}0.874} & {\footnotesize{}1.000}\tabularnewline
{\footnotesize{}$\widehat{J}_{T,\mathrm{pw},\mathrm{SLS},\mu}$ } & {\footnotesize{}0.039} & {\footnotesize{}0.051} & {\footnotesize{}0.104} & {\footnotesize{}0.332} & {\footnotesize{}0.832} & {\footnotesize{}1.000}\tabularnewline
{\footnotesize{}Andrews (1991)} & {\footnotesize{}0.108} & {\footnotesize{}0.218} & {\footnotesize{}0.484} & {\footnotesize{}0.749} & {\footnotesize{}0.915} & {\footnotesize{}0.995}\tabularnewline
{\footnotesize{}Andrews (1991), prewhite} & {\footnotesize{}0.000} & {\footnotesize{}0.000} & {\footnotesize{}0.007} & {\footnotesize{}0.186} & {\footnotesize{}0.708} & {\footnotesize{}0.956}\tabularnewline
{\footnotesize{}Newey-West (1987)} & {\footnotesize{}0.031} & {\footnotesize{}0.071} & {\footnotesize{}0.200} & {\footnotesize{}0.538} & {\footnotesize{}0.931} & {\footnotesize{}1.000}\tabularnewline
{\footnotesize{}Newey-West (1987), prewhite} & {\footnotesize{}0.000} & {\footnotesize{}0.000} & {\footnotesize{}0.033} & {\footnotesize{}0.280} & {\footnotesize{}0.740} & {\footnotesize{}0.965}\tabularnewline
{\footnotesize{}Newey-West (1987), fixed-$b$ (KVB)} & {\footnotesize{}0.000} & {\footnotesize{}0.009} & {\footnotesize{}0.096} & {\footnotesize{}0.398} & {\footnotesize{}0.753} & {\footnotesize{}0.952}\tabularnewline
\hline 
\end{tabular}{\footnotesize\par}
\par\end{centering}
{\footnotesize{}}{\footnotesize\par}
\end{table}

{\footnotesize{}}
\begin{table}[H]
{\footnotesize{}\caption{\label{Table Power M5}Empirical small-sample power for model M5}
}{\footnotesize\par}
\begin{centering}
{\footnotesize{}}%
\begin{tabular}{lccccc}
\hline 
 & \multicolumn{3}{c}{} &  & \tabularnewline
{\footnotesize{}5\% nominal size, $T=200$} & {\footnotesize{}$\delta=0.2$} & {\footnotesize{}$\delta=0.4$} & {\footnotesize{}$\delta=0.8$} & {\footnotesize{}$\delta=1.6$} & {\footnotesize{}$\delta=2.5$}\tabularnewline
\hline 
\hline 
{\footnotesize{}$\widehat{J}_{T}$} & {\footnotesize{}0.365} & {\footnotesize{}0.705} & {\footnotesize{}0.935} & {\footnotesize{}0.977} & {\footnotesize{}1.000}\tabularnewline
{\footnotesize{}$\widehat{J}_{T,\mathrm{pw},1}$ } & {\footnotesize{}0.213} & {\footnotesize{}0.446} & {\footnotesize{}0.717} & {\footnotesize{}0.795} & {\footnotesize{}0.890}\tabularnewline
{\footnotesize{}$\widehat{J}_{T,\mathrm{pw},\mathrm{SLS}}$ } & {\footnotesize{}0.232} & {\footnotesize{}0.511} & {\footnotesize{}0.792} & {\footnotesize{}0.908} & {\footnotesize{}1.000}\tabularnewline
{\footnotesize{}$\widehat{J}_{T,\mathrm{pw},\mathrm{SLS},\mu}$ } & {\footnotesize{}0.242} & {\footnotesize{}0.542} & {\footnotesize{}0.804} & {\footnotesize{}0.902} & {\footnotesize{}1.000}\tabularnewline
{\footnotesize{}Andrews (1991)} & {\footnotesize{}0.249} & {\footnotesize{}0.427} & {\footnotesize{}0.532} & {\footnotesize{}0.816} & {\footnotesize{}0.718}\tabularnewline
{\footnotesize{}Andrews (1991), prewhite} & {\footnotesize{}0.214} & {\footnotesize{}0.320} & {\footnotesize{}0.122} & {\footnotesize{}0.035} & {\footnotesize{}0.340}\tabularnewline
{\footnotesize{}Newey-West (1987)} & {\footnotesize{}0.319} & {\footnotesize{}0.737} & {\footnotesize{}0.849} & {\footnotesize{}0.918} & {\footnotesize{}0.937}\tabularnewline
{\footnotesize{}Newey-West (1987), prewhite} & {\footnotesize{}0.212} & {\footnotesize{}0.563} & {\footnotesize{}0.146} & {\footnotesize{}0.062} & {\footnotesize{}0.403}\tabularnewline
{\footnotesize{}Newey-West (1987), fixed-$b$ (KVB)} & {\footnotesize{}0.095} & {\footnotesize{}0.108} & {\footnotesize{}0.127} & {\footnotesize{}0.132} & {\footnotesize{}0.143}\tabularnewline
\hline 
\end{tabular}{\footnotesize\par}
\par\end{centering}
{\footnotesize{}}{\footnotesize\par}
\end{table}

{\footnotesize{}}
\begin{table}[H]
{\footnotesize{}\caption{\label{Table Power M6}Empirical small-sample power for model M6}
}{\footnotesize\par}
\begin{centering}
{\footnotesize{}}%
\begin{tabular}{lccccc}
\hline 
 & \multicolumn{5}{c}{{\footnotesize{}M6, $t_{2}$}}\tabularnewline
{\footnotesize{}5\% nominal size, $T=200$} & {\footnotesize{}$\delta=0.1$ } & {\footnotesize{}$\delta=0.2$} & {\footnotesize{}$\delta=0.4$} & {\footnotesize{}$\delta=0.8$} & {\footnotesize{}$\delta=1.6$}\tabularnewline
\hline 
\hline 
{\footnotesize{}$\widehat{J}_{T}$} & {\footnotesize{}0.186} & {\footnotesize{}0.504 } & {\footnotesize{}0.945} & {\footnotesize{}1.000} & {\footnotesize{}1.000}\tabularnewline
{\footnotesize{}$\widehat{J}_{T,\mathrm{pw},1}$ } & {\footnotesize{}0.112} & {\footnotesize{}0.360} & {\footnotesize{}0.888} & {\footnotesize{}0.995} & {\footnotesize{}1.000}\tabularnewline
{\footnotesize{}$\widehat{J}_{T,\mathrm{pw},\mathrm{SLS}}$ } & {\footnotesize{}0.103} & {\footnotesize{}0.339} & {\footnotesize{}0.884} & {\footnotesize{}0.996} & {\footnotesize{}1.000}\tabularnewline
{\footnotesize{}$\widehat{J}_{T,\mathrm{pw},\mathrm{SLS},\mu}$ } & {\footnotesize{}0.102} & {\footnotesize{}0.334} & {\footnotesize{}0.888} & {\footnotesize{}0.996} & {\footnotesize{}1.000}\tabularnewline
{\footnotesize{}Andrews (1991)} & {\footnotesize{}0.201} & {\footnotesize{}0.564} & {\footnotesize{}0.936} & {\footnotesize{}0.998} & {\footnotesize{}1.000}\tabularnewline
{\footnotesize{}Andrews (1991), prewhite} & {\footnotesize{}0.169} & {\footnotesize{}0.499} & {\footnotesize{}0.916} & {\footnotesize{}0.996} & {\footnotesize{}1.000}\tabularnewline
{\footnotesize{}Newey-West (1987)} & {\footnotesize{}0.223} & {\footnotesize{}0.543} & {\footnotesize{}0.935} & {\footnotesize{}0.990} & {\footnotesize{}1.000}\tabularnewline
{\footnotesize{}Newey-West (1987), prewhite} & {\footnotesize{}0.215} & {\footnotesize{}0.530} & {\footnotesize{}0.925} & {\footnotesize{}0.996} & {\footnotesize{}1.000}\tabularnewline
{\footnotesize{}Newey-West (1987), fixed-$b$ (KVB)} & {\footnotesize{}0.131} & {\footnotesize{}0.368} & {\footnotesize{}0.776} & {\footnotesize{}0.974} & {\footnotesize{}1.000}\tabularnewline
\hline 
\end{tabular}{\footnotesize\par}
\par\end{centering}
{\footnotesize{}}{\footnotesize\par}
\end{table}

{\footnotesize{}}
\begin{table}[H]
{\footnotesize{}\caption{\label{Table Power DM Test}Empirical small-sample power of the DM
(1995) test}
}{\footnotesize\par}
\begin{centering}
{\footnotesize{}}%
\begin{tabular}{lccccc}
\hline 
 & \multicolumn{5}{c}{{\footnotesize{}Model M7}}\tabularnewline
{\footnotesize{}5\% nominal size, $T=400$} & {\footnotesize{}$\delta=0.2$} & {\footnotesize{}$\delta=0.5$} & {\footnotesize{}$\delta=2$} & {\footnotesize{}$\delta=5$} & {\footnotesize{}$\delta=10$}\tabularnewline
\hline 
\hline 
{\footnotesize{}$\widehat{J}_{T}$} & {\footnotesize{}0.323} & {\footnotesize{}0.451} & {\footnotesize{}0.925} & {\footnotesize{}0.970} & {\footnotesize{}1.000}\tabularnewline
{\footnotesize{}$\widehat{J}_{T,\mathrm{pw},1}$ } & {\footnotesize{}0.245} & {\footnotesize{}0.365} & {\footnotesize{}0.914} & {\footnotesize{}0.964} & {\footnotesize{}0.972}\tabularnewline
{\footnotesize{}$\widehat{J}_{T,\mathrm{pw},\mathrm{SLS}}$ } & {\footnotesize{}0.351} & {\footnotesize{}0.505} & {\footnotesize{}0.922} & {\footnotesize{}0.962} & {\footnotesize{}1.000}\tabularnewline
{\footnotesize{}$\widehat{J}_{T,\mathrm{pw},\mathrm{SLS},\mu}$ } & {\footnotesize{}0.341} & {\footnotesize{}0.499} & {\footnotesize{}0.934} & {\footnotesize{}1.000} & {\footnotesize{}1.000}\tabularnewline
{\footnotesize{}$\widehat{J}_{T}$, auto, pretest} & {\footnotesize{}0.329} & {\footnotesize{}0.457} & {\footnotesize{}0.932} & {\footnotesize{}1.000} & {\footnotesize{}1.000}\tabularnewline
{\footnotesize{}$\widehat{J}_{T,\mathrm{pw},\mathrm{SLS}}$, pretest} & {\footnotesize{}0.372} & {\footnotesize{}0.516} & {\footnotesize{}0.942} & {\footnotesize{}1.000} & {\footnotesize{}1.000}\tabularnewline
{\footnotesize{}Andrews (1991)} & {\footnotesize{}0.300} & {\footnotesize{}0.350} & {\footnotesize{}0.151} & {\footnotesize{}0.000} & {\footnotesize{}0.000}\tabularnewline
{\footnotesize{}Andrews (1991), prewhite} & {\footnotesize{}0.293} & {\footnotesize{}0.345} & {\footnotesize{}0.371} & {\footnotesize{}0.080} & {\footnotesize{}0.000}\tabularnewline
{\footnotesize{}Newey-West (1987)} & {\footnotesize{}0.297} & {\footnotesize{}0.350} & {\footnotesize{}0.598} & {\footnotesize{}0.817} & {\footnotesize{}0.782}\tabularnewline
{\footnotesize{}Newey-West (1987), prewhite} & {\footnotesize{}0.288} & {\footnotesize{}0.314} & {\footnotesize{}0.191} & {\footnotesize{}0.000} & {\footnotesize{}0.000}\tabularnewline
{\footnotesize{}Newey-West (1987), fixed-$b$ (KVB)} & {\footnotesize{}0.231} & {\footnotesize{}0.201} & {\footnotesize{}0.000} & {\footnotesize{}0.000} & {\footnotesize{}0.000}\tabularnewline
\hline 
\end{tabular}{\footnotesize\par}
\par\end{centering}
{\footnotesize{}}{\footnotesize\par}
\end{table}

{\footnotesize{}}
\begin{table}[H]
{\footnotesize{}\caption{\label{Table Power GR Test}Empirical small-sample power of the GR
(2009) test}
}{\footnotesize\par}
\begin{centering}
{\footnotesize{}}%
\begin{tabular}{lccccc}
\hline 
 & \multicolumn{5}{c}{{\footnotesize{}Model M8}}\tabularnewline
{\footnotesize{}5\% nominal size, $T=800$} & {\footnotesize{}$\delta=0.2$} & {\footnotesize{}$\delta=0.4$} & {\footnotesize{}$\delta=0.8$} & {\footnotesize{}$\delta=1.6$} & {\footnotesize{}$\delta=2.5$}\tabularnewline
\hline 
\hline 
{\footnotesize{}$\widehat{J}_{T}$} & {\footnotesize{}0.066} & {\footnotesize{}0.496} & {\footnotesize{}0.999} & {\footnotesize{}1.000} & {\footnotesize{}1.000}\tabularnewline
{\footnotesize{}$\widehat{J}_{T,\mathrm{pw},1}$ } & {\footnotesize{}0.059} & {\footnotesize{}0.491} & {\footnotesize{}0.997} & {\footnotesize{}1.000} & {\footnotesize{}1.000}\tabularnewline
{\footnotesize{}$\widehat{J}_{T,\mathrm{pw},\mathrm{SLS}}$ } & {\footnotesize{}0.082} & {\footnotesize{}0.406} & {\footnotesize{}0.995} & {\footnotesize{}1.000} & {\footnotesize{}1.000}\tabularnewline
{\footnotesize{}$\widehat{J}_{T,\mathrm{pw},\mathrm{SLS},\mu}$ } & {\footnotesize{}0.104} & {\footnotesize{}0.560} & {\footnotesize{}0.996} & {\footnotesize{}1.000} & {\footnotesize{}1.000}\tabularnewline
{\footnotesize{}Andrews (1991)} & {\footnotesize{}0.000} & {\footnotesize{}0.350} & {\footnotesize{}0.000} & {\footnotesize{}0.000} & {\footnotesize{}0.000}\tabularnewline
{\footnotesize{}Andrews (1991), prewhite} & {\footnotesize{}0.000} & {\footnotesize{}0.345} & {\footnotesize{}0.133} & {\footnotesize{}0.591} & {\footnotesize{}0.742}\tabularnewline
{\footnotesize{}Newey-West (1987)} & {\footnotesize{}0.000} & {\footnotesize{}0.350} & {\footnotesize{}0.598} & {\footnotesize{}0.000} & {\footnotesize{}0.000}\tabularnewline
{\footnotesize{}Newey-West (1987), prewhite} & {\footnotesize{}0.000} & {\footnotesize{}0.314} & {\footnotesize{}0.191} & {\footnotesize{}0.000} & {\footnotesize{}0.000}\tabularnewline
{\footnotesize{}Newey-West (1987), fixed-$b$ (KVB)} & {\footnotesize{}0.026} & {\footnotesize{}0.201} & {\footnotesize{}0.000} & {\footnotesize{}0.000} & {\footnotesize{}0.000}\tabularnewline
\hline 
\end{tabular}{\footnotesize\par}
\par\end{centering}
{\footnotesize{}}{\footnotesize\par}
\end{table}

\section{Supplemental Materials}

The supplement for online publication {[}cf. \citet{casini_hac_supp}{]}
reviews how to apply the proposed DK-HAC estimator in GMM and IV contexts
and contains the proofs of the results of Section \ref{Section HAC-Estimation-with Predetermined}.
An additional supplement, not for publication, includes the proofs
of the results of Section \ref{Section: Statistical Enviromnent}
and \ref{Section Optimal-Kernels-and}-\ref{Section Data-Dependent-Bandwidths}.

\newpage{}

\pagebreak{}

\section*{}
\addcontentsline{toc}{part}{Supplemental Material}

\begin{center}
\title{\textbf{\Large{Supplement to ``Theory of Evolutionary Spectra for Heteroskedasticity and Autocorrelation Robust Inference in Possibly Misspecified and Nonstationary Models"}}} 
\maketitle
\end{center}
\medskip{} 
\medskip{} 
\medskip{} 
\thispagestyle{empty}

\begin{center}
\author{\textsc{\textcolor{MyBlue}{Alessandro Casini}}}\\ 
\medskip{}
\medskip{} 
\medskip{} 

\small{{Department of Economics and Finance}}\\
\small{{University of Rome Tor Vergata}}\\
\medskip{}
\medskip{} 
\medskip{} 
\medskip{} 
\date{\small{\today}} 
\medskip{} 
\medskip{} 
\medskip{} 
\end{center}
\begin{abstract}
{\footnotesize{}This supplemental material is for online publication
only. It contains the proofs of the results of Section \ref{Section HAC-Estimation-with Predetermined}
in the paper.}{\footnotesize\par}
\end{abstract}
\setcounter{page}{0}
\setcounter{section}{0}
\renewcommand*{\theHsection}{\the\value{section}}

\newpage{}

\begin{singlespace} 
\noindent 
\small

\allowdisplaybreaks


\renewcommand{\thepage}{S-\arabic{page}}   
\renewcommand{\thesection}{S.\Alph{section}}   
\renewcommand{\theequation}{S.\arabic{equation}}




\section{\label{Section Mathematical-Appendix}Appendix: Proofs of the Results
of Section \ref{Section HAC-Estimation-with Predetermined}}

In the proofs below, we discard the degrees of freedom adjustment
$T/\left(T-p\right)$ from the derivations since asymptotically it
does not play any role. Similarly, we use $T/n_{T}$ in place of $\left(T-n_{T}\right)/n_{T}$
in the expression for $\widehat{\Gamma}\left(k\right)$.  In some
of the proofs below we first consider the locally stationary case
under Assumption \ref{Assumption Smothness of A (for HAC)- Locally Stationary - Supp}-\ref{Assumption A - Dependence Locally Stationary - Supp}
and then extend the results to the SLS case. Note that Assumption
\ref{Assumption Smothness of A (for HAC)- Locally Stationary - Supp}-\ref{Assumption A - Dependence Locally Stationary - Supp}
are implied by Assumption \ref{Assumption Smothness of A (for HAC)}-\ref{Assumption A - Dependence}
since the former are weaker because local stationarity does not allow
for break points in the spectrum. A function $G\left(\cdot,\,\cdot\right):\,\left[0,\,1\right]\times\mathbb{R}\rightarrow\mathbb{C}$
is said to be right-differentiable at $u_{0}$ if $\partial G\left(u_{0},\omega\right)/\partial_{+}u\triangleq\lim_{u\rightarrow u_{0}^{_{+}}}\left(G\left(u_{0},\,\omega\right)-G\left(u,\,\omega\right)\right)/\left(u-u_{0}\right)$
exists for any $\omega\in\mathbb{R}$.  We sometimes use $\sum_{t}$
omitting the limits of the summation for the sum in $\widetilde{c}_{T}\left(u,\,k\right)$. 
\begin{assumption}
\label{Assumption Smothness of A (for HAC)- Locally Stationary - Supp}$\{V_{t,T}\}$
is a mean-zero  locally stationary process, $A\left(u,\,\omega\right)$
is \textcolor{red}{ }twice differentiable in $u$ with uniformly
bounded and Lipschitz continuous derivatives $\left(\partial/\partial u\right)A\left(u,\,\cdot\right)$
and $\left(\partial^{2}/\partial u^{2}\right)A\left(u,\,\cdot\right)$,
and Lipschitz continuous in the second component with index $\vartheta=1$. 
\end{assumption}
\begin{assumption}
\label{Assumption A - Dependence Locally Stationary - Supp}(i) 
$\sum_{k=-\infty}^{\infty}\sup_{u\in\left[0,\,1\right]}$ $\left\Vert c\left(u,\,k\right)\right\Vert <\infty$,
$\sum_{k=-\infty}^{\infty}\sup_{u\in\left[0,\,1\right]}\left\Vert \left(\partial^{2}/\partial u^{2}\right)c\left(u,\,k\right)\right\Vert <\infty$
and $\sum_{k=-\infty}^{\infty}\sum_{j=-\infty}^{\infty}\sum_{l=-\infty}^{\infty}\sup_{u\in\left[0,\,1\right]}\kappa_{V,\left\lfloor Tu\right\rfloor }^{\left(a,b,c,d\right)}$
$\left(k,\,j,\,l\right)<\infty$ for all $a,\,b,\,c,\,d\leq p$; (ii)
For all $a,\,b,\,c,\,d\leq p$ there exists a function $\widetilde{\kappa}_{a,b,c,d}:\,\left[0,\,1\right]\times\mathbb{Z}\times\mathbb{Z}\times\mathbb{Z}\rightarrow\mathbb{R}$
such that $\sup_{u\in\left[0,\,1\right]}|\kappa_{V,\left\lfloor Tu\right\rfloor }^{\left(a,b,c,d\right)}\left(k,\,s,\,l\right)-\widetilde{\kappa}_{a,b,c,d}\left(u,\,k,\,s,\,l\right)$
$|\leq KT^{-1}$ for some constant $K$; the function $\widetilde{\kappa}_{a,b,c,d}\left(u,\,k,\,s,\,l\right)$
is twice differentiable in $u$ with uniformly bounded and Lipschitz
continuous derivative $\left(\partial^{2}/\partial u^{2}\right)\widetilde{\kappa}_{a,b,c,d}\left(u,\cdot,\cdot,\cdot\right)$. 
\end{assumption}

\subsection{Preliminary Lemmas}
\begin{lem}
\label{Lemma eq. (73) in Dalhaus 2012 - SLS}Under Assumption \ref{Assumption Smothness of A (for HAC)}-\ref{Assumption A - Dependence},
\begin{align*}
\sup_{u\in\{\{\left(0,\,1\right)\}/\{\lambda_{j}^{0},\,j=1,\ldots,\,m_{0}\}\}}\sup_{k\in\mathbb{Z}}\left\Vert \mathrm{Cov}\left(V_{\left\lfloor Tu\right\rfloor ,T},\,V_{\left\lfloor Tu\right\rfloor -k,T}\right)-c\left(u,\,k\right)\right\Vert  & =O(T^{-1}),\\
\sup_{u\in\left(0,\,1\right)}\sup_{k\geq0}\left\Vert \mathrm{Cov}\left(V_{\left\lfloor Tu\right\rfloor ,T},\,V_{\left\lfloor Tu\right\rfloor -k,T}\right)-c\left(u,\,k\right)\right\Vert  & =O(T^{-1}),\\
\max_{u=\lambda_{j}^{0},\,j=1,\ldots,\,m_{0}}\sup_{k<0}\left\Vert \mathrm{Cov}\left(V_{\left\lfloor Tu\right\rfloor ,T},\,V_{\left\lfloor Tu\right\rfloor -k,T}\right)-c\left(u,\,-k\right)\right\Vert  & =O(T^{-1}).
\end{align*}
\end{lem}
\noindent\textit{Proof of Lemma} \ref{Lemma eq. (73) in Dalhaus 2012 - SLS}.
It is sufficient to consider the scalar case $p=1$. Consider first
$Tu\notin\mathcal{T},\,\lambda_{j-1}^{0}<u<\lambda_{j}^{0}.$ Using
the spectral representation \eqref{Eq. Spectral Rep of SLS}, \eqref{Eq. 2.4 Smothenss Assumption on A}
and Assumption \ref{Assumption Smothness of A} leads to 
\begin{align}
\mathrm{Cov}\left(V_{\left\lfloor Tu\right\rfloor ,T},\,V_{\left\lfloor Tu\right\rfloor -k,T}\right) & =\int_{-\pi}^{\pi}\exp\left(i\omega k\right)A_{j,\left\lfloor Tu\right\rfloor ,T}^{0}\left(\omega\right)A_{j,\left\lfloor Tu\right\rfloor -k,T}^{0}\left(-\omega\right)d\omega\nonumber \\
 & =\int_{-\pi}^{\pi}\exp\left(i\omega k\right)A_{j}\left(u,\,\omega\right)A_{j}\left(u-k/T,\,-\omega\right)d\omega+O(T^{-1})\nonumber \\
 & =c\left(u,\,k\right)+O(T^{-1}),\label{eq. (1) Lemma cov=00003Dc}
\end{align}
where the $O(T^{-1})$ term is uniform in $u\in\{\{\left(0,\,1\right)\}/\{\lambda_{j}^{0},\,j=1,\ldots,\,m_{0}\}\}$
and $k.$ Now consider the case $Tu\in\mathcal{T},\,u=T_{j}^{0}/T$
and $k\geq0$. Using \eqref{Eq. Spectral Rep of SLS} and \eqref{Eq. 2.4 Smothenss Assumption on A}
yields 
\begin{align}
\mathrm{Cov}\left(V_{\left\lfloor Tu\right\rfloor ,T},\,V_{\left\lfloor Tu\right\rfloor -k,T}\right) & =\int_{-\pi}^{\pi}\exp\left(i\omega k\right)A_{j,\left\lfloor Tu\right\rfloor ,T}^{0}\left(\omega\right)A_{j+1,\left\lfloor Tu\right\rfloor -k,T}^{0}\left(-\omega\right)d\omega\nonumber \\
 & =\int_{-\pi}^{\pi}\exp\left(i\omega k\right)A_{j}\left(u,\,\omega\right)A_{j+1}\left(u-k/T,\,-\omega\right)d\omega+O(T^{-1})\nonumber \\
 & =c\left(u,\,-k\right)+O(T^{-1}),\label{eq. (2) Lemma cov=00003Dc}
\end{align}
 where the $O\left(T^{-1}\right)$ term is uniform in $u$ and $k\geq0$.
The argument for the case $Tu\in\mathcal{T}$ and $k<0$ is the same
as for the case $Tu\notin\mathcal{T}.$ $\square$
\begin{lem}
\label{Lemma GammaTu =00003D GammaTu+k - Locally Stationary}Under
Assumption \ref{Assumption Smothness of A (for HAC)- Locally Stationary - Supp}-\ref{Assumption A - Dependence Locally Stationary - Supp},
$\sup_{u\in\left(0,\,1\right)}\sup_{v,\,k\in\mathbb{Z}}||\Gamma_{u}\left(v\right)-\Gamma_{u+k/T}\left(v\right)||=O(T^{-1})$.
\end{lem}
\noindent\textit{Proof of Lemma} \ref{Lemma GammaTu =00003D GammaTu+k - Locally Stationary}.
We know that $\Gamma_{u}\left(v\right)=c\left(u,\,v\right)+O(T^{-1})$
uniformly in $u$ and $v$ by Lemma \ref{Lemma eq. (73) in Dalhaus 2012 - SLS}
where $c\left(u,\,v\right)=\int_{-\pi}^{\pi}e^{i\omega v}f\left(u,\,\omega\right)d\omega.$
Using Assumption \ref{Assumption Smothness of A (for HAC)- Locally Stationary - Supp},
\begin{align*}
c\left(u,\,v\right) & =\int_{-\pi}^{\pi}e^{i\omega v}f\left(u+k/T,\,\omega\right)d\omega+O\left(k/T\right)\\
 & =c\left(u+k/T,\,v\right)+O\left(k/T\right)\\
 & =\Gamma_{u+k/T}\left(v\right)+O\left(k/T\right)+O(T^{-1}),
\end{align*}
uniformly in $u\in\left(0,\,1\right)$ and $v,\,k\in\mathbb{Z}$.
$\square$

\medskip{}
 Let 
\begin{align*}
\mathrm{MSE}\left(\widetilde{c}_{T}\left(u_{0},\,k\right)\right) & =Tb_{2,T}\mathbb{E}\left[\mathrm{vec}\left(\widetilde{c}_{T}\left(u_{0},\,k\right)-c\left(u_{0},\,k\right)\right)'W\mathrm{vec}\left(\widetilde{c}_{T}\left(u_{0},\,k\right)-c\left(u_{0},\,k\right)\right)\right],
\end{align*}
where $W$ is some $p^{2}\times p^{2}$ weight matrix. 
\begin{lem}
\label{Lemma Bias, MSE of c(u,k), scalar case - Locally Stationary}Suppose
Assumption \ref{Assumption Smothness of A (for HAC)- Locally Stationary - Supp}-\ref{Assumption A - Dependence Locally Stationary - Supp}
hold and $b_{2,T}\rightarrow0$ as $T\rightarrow\infty$. Then, for
all $u_{0}\in\left(0,\,1\right)$,
\begin{align}
\mathbb{E}\left[\widetilde{c}_{T}\left(u_{0},\,k\right)\right] & =c\left(u_{0},\,k\right)+\frac{1}{2}b_{2,T}^{2}\int_{0}^{1}x^{2}K_{2}\left(x\right)dx\left[\frac{\partial^{2}}{\partial^{2}u}c\left(u_{0},\,k\right)\right]+o\left(b_{2,T}^{2}\right)+O\left(1/\left(Tb_{2,T}\right)\right),\label{eq. Bias Locally Stationary}
\end{align}
and for all $j,\,l,\,r,\,w\leq p$,
\begin{align}
\mathrm{Cov} & \left[\widetilde{c}_{T}^{\left(j,l\right)}\left(u_{0},\,k\right),\,\widetilde{c}_{T}^{\left(r,w\right)}\left(u_{0},\,k\right)\right]\label{eq: Covariance jlrw Locally Stationary}\\
 & =\frac{1}{Tb_{2,T}}\int_{0}^{1}K_{2}^{2}\left(x\right)dx\sum_{l=-\infty}^{\infty}\left[c^{\left(j,r\right)}\left(u_{0},\,l\right)c^{\left(l,w\right)}\left(u_{0},\,l\right)+c^{\left(j,w\right)}\left(u_{0},\,l\right)c^{\left(l,r\right)}\left(u_{0},\,l+2k\right)\right]\nonumber \\
 & \quad+\frac{1}{Tb_{2,T}}\int_{0}^{1}K_{2}^{2}\left(x\right)dx\sum_{h_{1}=-\infty}^{\infty}\widetilde{\kappa}_{j,l,r,w}\left(u_{0},\,-k,\,h_{1},\,h_{1}-k\right)+o\left(\frac{1}{Tb_{2,T}}\right).\nonumber 
\end{align}
If $Tb_{2,T}^{5}\rightarrow\eta\in(0,\,\infty)$, then, for all $u_{0}\in\left(0,\,1\right)$,
$\widetilde{c}_{T}\left(u_{0},\,k\right)-c\left(u_{0},\,k\right)=O_{\mathbb{P}}\left(\sqrt{Tb_{2,T}}\right)$. 

If in addition $V_{t,T}$ is Gaussian, then for all $u_{0}\in\left(0,\,1\right)$,
\begin{align}
\mathrm{Cov} & \left[\widetilde{c}_{T}^{\left(j,l\right)}\left(u_{0},\,k\right),\,\widetilde{c}_{T}^{\left(r,w\right)}\left(u_{0},\,k\right)\right]\nonumber \\
 & =\frac{1}{Tb_{2,T}}\int_{0}^{1}K_{2}^{2}\left(x\right)dx\sum_{l=-\infty}^{\infty}\left[c^{\left(j,r\right)}\left(u_{0},\,l\right)c^{\left(l,w\right)}\left(u_{0},\,l\right)+c^{\left(j,w\right)}\left(u_{0},\,l\right)c^{\left(l,r\right)}\left(u_{0},\,l+2k\right)\right]\label{eq (Cov Gaussian)}\\
 & \quad+o\left(1/\left(Tb_{2,T}\right)\right),\nonumber 
\end{align}
and if $Tb_{2,T}^{5}\rightarrow\eta\in(0,\,\infty)$, then
\begin{align*}
\lim_{T\rightarrow\infty}\mathrm{MSE}\left(\widetilde{c}_{T}\left(u_{0},\,k\right)\right) & =\frac{\eta}{4}\left(\int_{0}^{1}x^{2}K_{2}\left(x\right)dx\right)^{2}\left[\frac{\partial^{2}}{\partial^{2}u}\mathrm{vec}\left(c\left(u_{0},\,k\right)\right)\right]'W\left[\frac{\partial^{2}}{\partial^{2}u}\mathrm{vec}\left(c\left(u_{0},\,k\right)\right)\right]\\
 & \quad+\int_{0}^{1}K_{2}^{2}\left(x\right)dx\,\mathrm{tr}W\sum_{l=-\infty}^{\infty}\mathrm{vec}\left(c\left(u_{0},\,l\right)\right)\left[\mathrm{vec}\left(c\left(u_{0},\,l\right)\right)'+\mathrm{vec}\left(c\left(u_{0},\,l+2k\right)\right)'\right].
\end{align*}
\end{lem}
\noindent\textit{Proof of Lemma} \ref{Lemma Bias, MSE of c(u,k), scalar case - Locally Stationary}.
The bias expression follows from \citeReferencesSupp{dahlhaus:96}.
For the second moment and MSE of $\widetilde{c}_{T}\left(u_{0},\,k\right)$,
we first present the proof for the case where $V_{t,T}$ is Gaussian
and $p=1$. Evaluating the expectation, we have for $k<0$,
\begin{align*}
\mathrm{Var} & \left[\widetilde{c}_{T}\left(u_{0},\,k\right)\right]\\
 & =\frac{1}{\left(Tb_{2,T}\right)^{2}}\sum_{t}\sum_{s}K_{2}^{*}\left(\frac{u_{0}-\left(t+k/2\right)/T}{b_{T}}\right)K_{2}^{*}\left(\frac{u_{0}-\left(s+k/2\right)/T}{b_{T}}\right)\mathbb{E}\left(V_{t,T}V_{s,T}\right)\mathbb{E}\left(V_{t+k,T}V_{s+k,T}\right)\\
 & \quad+\frac{1}{\left(Tb_{2,T}\right)^{2}}\sum_{t}\sum_{s}K_{2}^{*}\left(\frac{u_{0}-\left(t+k/2\right)/T}{b_{T}}\right)K_{2}^{*}\left(\frac{u_{0}-\left(s+k/2\right)/T}{b_{T}}\right)\mathbb{E}\left(V_{t,T}V_{t+k,T}\right)\mathbb{E}\left(V_{s,T}V_{s+k,T}\right)\\
 & \quad+\frac{1}{\left(Tb_{2,T}\right)^{2}}\sum_{t}\sum_{s}K_{2}^{*}\left(\frac{u_{0}-\left(t+k/2\right)/T}{b_{T}}\right)K_{2}^{*}\left(\frac{u_{0}-\left(s+k/2\right)/T}{b_{T}}\right)\mathbb{E}\left(V_{t,T}V_{s+k,T}\right)\mathbb{E}\left(V_{s,T}V_{t+k,T}\right)\\
 & \quad-\left[\mathbb{E}\left(\widetilde{c}_{T}\left(u_{0},\,k\right)\right)\right]^{2}.
\end{align*}
Using the continuity of $K_{2}$, $\left(s-t\right)/T\rightarrow0$
for fixed $s$ and $t$, the smoothness of $\Gamma_{u}\left(\cdot\right)$
and Lemma \ref{Lemma eq. (73) in Dalhaus 2012 - SLS}, implies that
the first term on the right-hand side is equal to 
\begin{align*}
\frac{1}{Tb_{2,T}}\int_{0}^{1}x^{2}K_{2}\left(x\right)dx\sum_{l=-\infty}^{\infty}c\left(u_{0},\,l\right)^{2} & .
\end{align*}
 For the second and third terms we use a similar argument with in
addition Lemma 6.2.1 in \citeReferencesSupp{fuller:1995} so that
\begin{align}
\mathrm{Var} & \left[\widetilde{c}_{T}\left(u_{0},\,k\right)\right]\nonumber \\
 & =\frac{1}{Tb_{2,T}}\int_{0}^{1}x^{2}K_{2}\left(x\right)dx\sum_{l=-\infty}^{\infty}c\left(u_{0},\,l\right)^{2}+\frac{1}{Tb_{2,T}}\int_{0}^{1}x^{2}K_{2}\left(x\right)dx\sum_{l=-\infty}^{\infty}c\left(u_{0},\,l\right)c\left(u_{0},\,l+2k\right)+o\left(\frac{1}{Tb_{2,T}}\right).\label{eq. (Var ctilde Locally Stationary)}
\end{align}
Next, \eqref{eq (Cov Gaussian)} follows similarly. We have 
\begin{align*}
\mathrm{Cov} & \left[\widetilde{c}_{T}^{\left(j,l\right)}\left(u_{0},\,k\right),\,\widetilde{c}_{T}^{\left(r,w\right)}\left(u_{0},\,k\right)\right]\\
 & =\frac{1}{\left(Tb_{2,T}\right)^{2}}\sum_{t}\sum_{s}K_{2}^{*}\left(\frac{u_{0}-\left(t+k/2\right)/T}{b_{T}}\right)K_{2}^{*}\left(\frac{u_{0}-\left(s+k/2\right)/T}{b_{T}}\right)\mathbb{E}\left(V_{t,T}^{\left(j\right)}V_{s,T}^{\left(r\right)}\right)\mathbb{E}\left(V_{t+k,T}^{\left(l\right)}V_{s+k,T}^{\left(w\right)}\right)\\
 & \quad+\frac{1}{\left(Tb_{2,T}\right)^{2}}\sum_{t}\sum_{s}K_{2}^{*}\left(\frac{u_{0}-\left(t+k/2\right)/T}{b_{T}}\right)K_{2}^{*}\left(\frac{u_{0}-\left(s+k/2\right)/T}{b_{T}}\right)\mathbb{E}\left(V_{t,T}^{\left(j\right)}V_{s+k,T}^{\left(w\right)}\right)\mathbb{E}\left(V_{t+k,T}^{\left(l\right)}V_{s,T}^{\left(r\right)}\right)\\
 & =\frac{1}{Tb_{2,T}}\int_{0}^{1}K_{2}^{2}\left(x\right)dx\sum_{l=-\infty}^{\infty}\left[c^{\left(j,r\right)}\left(u_{0},\,l\right)c^{\left(l,w\right)}\left(u_{0},\,l\right)+c^{\left(j,w\right)}\left(u_{0},\,l\right)c^{\left(l,r\right)}\left(u_{0},\,l+2k\right)\right]+o\left(1/\left(Tb_{2,T}\right)\right).
\end{align*}
Using a standard bias-variance argument, we have $\widetilde{c}_{T}\left(u_{0},\,k\right)-c\left(u_{0},\,k\right)=o_{\mathbb{P}}\left(1\right)$.
If $Tb_{2,T}^{5}\rightarrow\eta\in(0,\,\infty)$, the asymptotic MSE
of $\widetilde{c}_{T}\left(u_{0},\,k\right)$ is given by 
\begin{align}
\lim_{T\rightarrow\infty}\mathrm{MSE}\left(\widetilde{c}_{T}\left(u_{0},\,k\right)\right) & =\frac{\eta}{4}\left(\int_{0}^{1}x^{2}K_{2}\left(x\right)dx\right)^{2}\left[\frac{\partial^{2}}{\partial^{2}u}c\left(u_{0},\,k\right)\right]^{2}\nonumber \\
 & \quad+\int_{0}^{1}K_{2}^{2}\left(x\right)dx\sum_{l=-\infty}^{\infty}c\left(u_{0},\,l\right)\left[c\left(u_{0},\,l\right)+c\left(u_{0},\,l+2k\right)\right].\label{eq: (MSE) Scalar case}
\end{align}
The latter suggests that if $Tb_{2,T}^{5}\rightarrow\eta\in(0,\,\infty)$,
then $\widetilde{c}_{T}\left(u_{0},\,k\right)-c\left(u_{0},\,k\right)=O_{\mathbb{P}}\left(\sqrt{Tb_{2,T}}\right)$
for all $u_{0}\in\left(0,\,1\right)$. The MSE expression for the
multivariate case follows from \eqref{eq: (MSE) Scalar case}.

Consider now the second moment of $\widetilde{c}_{T}\left(u_{0},\,k\right)$
for the general case. When $V_{t,T}$ is non-Gaussian, there is an
extra term in $\mathrm{Cov}[\widetilde{c}_{T}^{\left(j,l\right)}\left(u_{0},\,k\right),\,\widetilde{c}_{T}^{\left(r,w\right)}\left(u_{0},\,k\right)]$,
namely 
\begin{align*}
\frac{1}{\left(Tb_{2,T}\right)^{2}}\sum_{t}\sum_{s}K_{2}^{*}\left(\frac{u_{0}-\left(t+k/2\right)/T}{b_{T}}\right)K_{2}^{*}\left(\frac{u_{0}-\left(s+k/2\right)/T}{b_{T}}\right)\kappa_{V,t}^{\left(j,l,r,w\right)}\left(-k,\,s-t,\,s-t-k\right) & .
\end{align*}
 By Assumption \ref{Assumption A - Dependence Locally Stationary - Supp}
with $u=t/T,$ 
\begin{align*}
\sup_{u\in\left(0,\,1\right)}\left|\kappa_{V,Tu}^{\left(j,l,r,w\right)}\left(-k,\,s-Tu,\,s-Tu-k\right)-\widetilde{\kappa}_{j,l,r,w}\left(u,\,-k,\,s-Tu,\,s-Tu-k\right)\right| & =O(T^{-1}).
\end{align*}
Taking a second-order Taylor's expansion of $\kappa_{V,Tu}^{\left(j,l,r,w\right)}$
around $u_{0}$ we have
\begin{align*}
\frac{1}{\left(Tb_{2,T}\right)^{2}} & \sum_{t}\sum_{s}K_{2}^{*}\left(\frac{u_{0}-\left(t+k/2\right)/T}{b_{T}}\right)K_{2}^{*}\left(\frac{u_{0}-\left(s+k/2\right)/T}{b_{T}}\right)\kappa_{V,Tu}^{\left(j,l,r,w\right)}\left(-k,\,s-Tu,\,s-Tu-k\right)\\
 & =\frac{1}{\left(Tb_{2,T}\right)^{2}}\sum_{t}\sum_{s}K_{2}^{*}\left(\frac{u_{0}-\left(t+k/2\right)/T}{b_{T}}\right)K_{2}^{*}\left(\frac{u_{0}-\left(s+k/2\right)/T}{b_{T}}\right)\\
 & \quad\times\widetilde{\kappa}_{j,l,r,w}\left(u_{0},\,-k,\,s-Tu_{0},\,s-Tu_{0}-k\right)\\
 & \quad+\frac{1}{\left(Tb_{2,T}\right)^{2}}\sum_{t}\sum_{s}K_{2}^{*}\left(\frac{u_{0}-\left(t+k/2\right)/T}{b_{T}}\right)K_{2}^{*}\left(\frac{u_{0}-\left(s+k/2\right)/T}{b_{T}}\right)\\
 & \quad\times\frac{\partial\widetilde{\kappa}_{j,l,r,w}}{\partial u}\left(u_{0},\,-k,\,s-Tu_{0},\,s-Tu_{0}-k\right)\left(u_{0}-u\right)\\
 & \quad+\frac{1}{\left(Tb_{2,T}\right)^{2}}\sum_{t}\sum_{s}K_{2}^{*}\left(\frac{u_{0}-\left(t+k/2\right)/T}{b_{T}}\right)K_{2}^{*}\left(\frac{u_{0}-\left(s+k/2\right)/T}{b_{T}}\right)\\
 & \quad\times\frac{\partial^{2}\widetilde{\kappa}_{j,l,r,w}}{\partial u^{2}}\left(u_{0},\,-k,\,s-Tu_{0},\,s-Tu_{0}-k\right)\left(u_{0}-u\right)^{2}\\
 & =\frac{1}{Tb_{2,T}}\int_{0}^{1}K_{2}^{2}\left(x\right)dx\sum_{h_{1}=-\infty}^{\infty}\widetilde{\kappa}_{j,l,r,w}\left(u_{0},\,-k,\,h_{1},\,h_{1}-k\right)+o\left(\frac{1}{Tb_{2,T}}\right).\,\square
\end{align*}

\begin{lem}
\label{Lemma Bias, MSE of c(u,k) Breaks SLS}Suppose Assumption \ref{Assumption Smothness of A (for HAC)}-\ref{Assumption A - Dependence}
hold and $b_{2,T}\rightarrow0$ as $T\rightarrow\infty$. For each
$T\lambda_{j}^{0}=Tu_{0}\in\mathcal{T}$ $\left(j=1,\ldots,\,m_{0}\right)$
and $\left|k\right|/Tb_{2,T}\rightarrow\eta_{2}\in(0,\,\lambda_{j+1}^{0}-\lambda_{j}^{0})$,
\begin{align*}
\mathbb{E}\left[\widetilde{c}_{T}\left(u_{0},\,k\right)\right] & =c\left(u_{0},\,k\right)+\frac{1}{2}b_{2,T}^{2}\int_{0}^{1}x^{2}K_{2}\left(x\right)dx\\
 & \quad\times\int_{-\pi}^{\pi}\exp\left(i\omega k\right)\left(C_{1}\left(u_{0},\,\omega\right)+C_{2}\left(u_{0},\,\omega\right)+C_{3}\left(u_{0},\,\omega\right)\right)d\omega+O\left(\frac{1}{Tb_{2,T}}\right)+o\left(b_{2,T}^{2}\right),
\end{align*}
 where 
\begin{align*}
C_{1}\left(u_{0},\,\omega\right)=2\frac{\partial A_{j}\left(u_{0},\,-\omega\right)}{\partial_{-}u}\frac{\partial A_{j+1}\left(v_{0},\,\omega\right)}{\partial_{+}v}, & \qquad C_{2}\left(u_{0},\,\omega\right)=\frac{\partial^{2}A_{j+1}\left(v_{0},\,\omega\right)}{\partial_{+}v^{2}}A_{j}\left(u_{0},\,-\omega\right)\\
C_{3}\left(u_{0},\,\omega\right)= & \frac{\partial^{2}A_{j}\left(u_{0},\,\omega\right)}{\partial_{-}u^{2}}A_{j+1}\left(v_{0},\,\omega\right),
\end{align*}
and $v_{0}=u_{0}-k/2T$. For $Tu_{0}\notin\mathcal{T}$ or for $Tu_{0}\in\mathcal{T}$
and $\left|k\right|/Tb_{2,T}\rightarrow0$, \eqref{eq. Bias Locally Stationary}
and \eqref{eq: Covariance jlrw Locally Stationary} hold. For all
$u_{0}\in\left(0,\,1\right),$ $\lim_{T\rightarrow\infty}b_{2,T}^{-2}$
$\mathbb{E}\left[\widetilde{c}_{T}\left(u_{0},\,k\right)-c\left(u_{0},\,k\right)\right]<\infty$,
and if further it holds that $Tb_{2,T}^{5}\rightarrow\eta\in\left(0,\,\infty\right),$
then $\lim_{T\rightarrow\infty}Tb_{2,T}\mathrm{Var}\left[\widetilde{c}_{T}\left(u_{0},\,k\right)\right]<\infty.$
Furthermore, we have $\widehat{c}_{T}\left(u_{0},\,k\right)-c\left(u_{0},\,k\right)=O_{\mathbb{P}}\left(\sqrt{Tb_{2,T}}\right)$
for all $u_{0}\in\left(0,\,1\right)$.
\end{lem}
\noindent\textit{Proof of Lemma} \ref{Lemma Bias, MSE of c(u,k) Breaks SLS}.
If $Tu_{0}\notin\mathcal{T}$ then the result follows from Lemma \ref{Lemma Bias, MSE of c(u,k), scalar case - Locally Stationary}.
Suppose $Tu_{0}\in\mathcal{T}$ and $k/Tb_{2,T}\rightarrow0$ (the
case $k<0$ is similar and omitted). We omit the subscript $j$ from
$A_{j,s-k,T}^{0}\left(\omega\right)$ and from $A_{j}\left(\left(s-k\right)/T,\,\omega\right)$
since the value $j$ is determined by $s$ and $s-k$, respectively,
and can thus be omitted. Using \eqref{Eq. Spectral Rep of SLS} we
have, 
\begin{align*}
\mathbb{E}\left[\widetilde{c}_{T}\left(u_{0},\,k\right)\right] & =\frac{1}{Tb_{2,T}}\sum_{s=k+1}^{T}K_{2}^{*}\left(\frac{u_{0}-\left(s-k/2\right)/T}{b_{2,T}}\right)\int_{-\pi}^{\pi}\exp\left(i\omega k\right)A_{s-k,T}^{0}\left(\omega\right)A_{s,T}^{0}\left(-\omega\right)d\omega.
\end{align*}
Since $K_{2}\left(x\right)=0$ for $x<0$, the above sum runs up to
$s=Tu_{0}+k/2T$. Hence, the behavior of $A_{s,T}^{0}\left(\omega\right)$
only matters on a left neighborhood of $u_{0}$. Using \eqref{Eq. 2.4 Smothenss Assumption on A}
we have,
\begin{align*}
\mathbb{E}\left[\widetilde{c}_{T}\left(u_{0},\,k\right)\right] & =\frac{1}{Tb_{2,T}}\sum_{s=k+1}^{T}K_{2}^{*}\left(\frac{u_{0}-\left(s-k/2\right)/T}{b_{2,T}}\right)\int_{-\pi}^{\pi}\exp\left(i\omega k\right)A\left(\frac{s-k}{T},\,\omega\right)A\left(\frac{s}{T},\,-\omega\right)d\omega+O\left(T^{-1}\right).
\end{align*}
By the definition of $f\left(\cdot,\,\cdot\right)$, it follows that,
\begin{align*}
\mathbb{E}\left[\widetilde{c}_{T}\left(u_{0},\,k\right)\right] & =\frac{1}{Tb_{2,T}}\sum_{s=k+1}^{T}K_{2}^{*}\left(\frac{u_{0}-\left(s-k/2\right)/T}{b_{2,T}}\right)\int_{-\pi}^{\pi}\exp\left(i\omega k\right)f\left(\frac{s-k/2}{T},\,\omega\right)d\omega+O\left(T^{-1}\right).
\end{align*}
Let $u_{\epsilon,T}=u_{0}-\epsilon_{T},$ where $\epsilon_{T}>0.$
Since $f\left(u,\,\omega\right)$ is twice differentiable in $u$
at $u\neq\lambda_{j}^{0}$ (cf. Assumption \ref{Assumption Smothness of A (for HAC)}),
by taking a second-order Taylor's expansion of $f$ around $u_{\epsilon,T}$
we have 
\begin{align*}
\mathbb{E}\left[\widetilde{c}_{T}\left(u_{0},\,k\right)\right] & =\frac{1}{Tb_{2,T}}\sum_{s=k+1}^{T}K_{2}^{*}\left(\frac{u_{0}-\left(s-k/2\right)/T}{b_{2,T}}\right)\int_{-\pi}^{\pi}\exp\left(i\omega k\right)f\left(u_{\epsilon,T},\,\omega\right)d\omega\\
 & \quad+\frac{1}{Tb_{2,T}}\sum_{s=k+1}^{T}K_{2}^{*}\left(\frac{u_{0}-\left(s-k/2\right)/T}{b_{2,T}}\right)\int_{-\pi}^{\pi}\exp\left(i\omega k\right)\frac{\partial f\left(u_{\epsilon,T},\,\omega\right)}{\partial u}\left(\frac{s-k/2}{T}-u_{\epsilon,T}\right)d\omega\\
 & \quad+\frac{1}{2}\frac{1}{Tb_{2,T}}\sum_{s=k+1}^{T}K_{2}^{*}\left(\frac{u_{0}-\left(s-k/2\right)/T}{b_{2,T}}\right)\int_{-\pi}^{\pi}\exp\left(i\omega k\right)\frac{\partial^{2}f\left(u_{\epsilon,T},\,\omega\right)}{\partial u^{2}}\left(\frac{s-k/2}{T}-u_{\epsilon,T}\right)^{2}d\omega\\
 & \quad+o\left(b_{2,T}^{2}\right)+O\left(\frac{1}{Tb_{2,T}}\right).
\end{align*}
 Choose  $\epsilon_{T}=o_{\mathbb{P}}(\max\{b_{2,T}^{2},\,1/(Tb_{2,T})\})$.
Using $\int_{0}^{1}K_{2}\left(x\right)dx=1,$ $K_{2}\left(x\right)=K_{2}\left(1-x\right)$
and the definition of $c\left(u_{\epsilon,T},\,k\right)$, the right-hand
side above is equal to
\begin{align*}
c\left(u_{\epsilon_{T}},\,k\right) & +\frac{1}{2}b_{2,T}^{2}\int_{0}^{1}x^{2}K_{2}\left(x\right)dx\int_{-\pi}^{\pi}\exp\left(i\omega k\right)\frac{\partial^{2}f\left(u_{\epsilon_{T}},\,\omega\right)}{\partial u^{2}}d\omega+O\left(\frac{1}{Tb_{2,T}}\right)+o\left(b_{2,T}^{2}\right).
\end{align*}
 Since $c\left(u_{0},\,k\right)$ and $\partial^{2}f\left(u_{0},\,\omega\right)/\partial u^{2}$
are left-Lipschitz continuous by Assumption \ref{Assumption Smothness of A (for HAC)}-(iii),
\begin{align*}
c\left(u_{\epsilon_{T}},\,k\right)-c\left(u_{0},\,k\right) & =O_{\mathbb{P}}\left(\epsilon_{T}\right),\qquad\frac{\partial^{2}f\left(u_{\epsilon_{T}},\,\omega\right)}{\partial u^{2}}-\frac{\partial^{2}f\left(u_{0},\,\omega\right)}{\partial_{-}u^{2}}=O_{\mathbb{P}}\left(\epsilon_{T}\right).
\end{align*}
Then,
\begin{align*}
\mathbb{E} & \left[\widetilde{c}_{T}\left(u_{0},\,k\right)-c\left(u_{0},\,k\right)\right]=\frac{1}{2}b_{2,T}^{2}\int_{0}^{1}x^{2}K_{2}\left(x\right)dx\int_{-\pi}^{\pi}\exp\left(i\omega k\right)\frac{\partial^{2}f\left(u_{0},\,\omega\right)}{\partial_{-}u^{2}}d\omega+O\left(\frac{1}{Tb_{2,T}}\right)+o\left(b_{2,T}^{2}\right).
\end{align*}
It remains to consider the case $Tu_{0}=T\lambda_{j}^{0}\in\mathcal{T}$
and $\left|k\right|/T\rightarrow\eta_{2}\in(0,\,\lambda_{j+1}^{0}-\lambda_{j}^{0})$.
Suppose $k<0$ (the case $k>0$ is similar and omitted). The derivations
for the bias expression are different. Again, using \eqref{Eq. Spectral Rep of SLS}
we have, 
\begin{align*}
\mathbb{E}\left[\widetilde{c}_{T}\left(u_{0},\,k\right)\right] & =\frac{1}{Tb_{2,T}}\sum_{s=k+1}^{T}K_{2}^{*}\left(\frac{u_{0}-\left(s+k/2\right)/T}{b_{2,T}}\right)\int_{-\pi}^{\pi}\exp\left(i\omega k\right)A_{s+k,T}^{0}\left(\omega\right)A_{s,T}^{0}\left(-\omega\right)d\omega.
\end{align*}
Using \eqref{Eq. 2.4 Smothenss Assumption on A}, we have
\begin{align*}
\mathbb{E}\left[\widetilde{c}_{T}\left(u_{0},\,k\right)\right] & =\frac{1}{Tb_{2,T}}\sum_{s=k+1}^{T}K_{2}^{*}\left(\frac{u_{0}-\left(s+k/2\right)/T}{b_{2,T}}\right)\int_{-\pi}^{\pi}\exp\left(i\omega k\right)A\left(\frac{s+k}{T},\,\omega\right)A\left(\frac{s}{T},\,-\omega\right)d\omega+O\left(T^{-1}\right).
\end{align*}
We cannot use the property $f_{j}\left(u,\,\omega\right)=\left|A_{j}\left(u,\,\omega\right)\right|^{2}$
for $T_{j-1}^{0}/T<u=t/T\leq T_{j}^{0}/T$ because now $u_{0}=s+k/2$
implies $s=Tu_{0}-k/2>Tu_{0}$. That is, $A_{j}\left(\left(s+k\right)/T,\,\omega\right)A_{j+1}\left(s/T,\,-\omega\right)$
cannot be approximated by $f_{j}\left(s-k/2,\,\omega\right)$ for
those $s$ such that $s>T_{j}^{0}$. However, by taking a second-order
Taylor's expansion of $A_{j}$ about $u_{0}-\epsilon_{1,T}$ and of
$A_{j+1}$ about $v_{0}+\epsilon_{2,T}$ where $v_{0}=u_{0}-k/2T$
and $\epsilon_{1,T},\,\epsilon_{2,T}>0$, we have
\begin{align}
\mathbb{E} & \left[\widetilde{c}_{T}\left(u_{0},\,k\right)\right]\nonumber \\
 & =\frac{1}{Tb_{2,T}}\sum_{s=k+1}^{T}K_{2}^{*}\left(\frac{u_{0}-\left(s+k/2\right)/T}{b_{2,T}}\right)\int_{-\pi}^{\pi}\exp\left(i\omega k\right)A_{j+1}\left(v_{0}+\epsilon_{2,T},\,\omega\right)A_{j}\left(u_{0}-\epsilon_{1,T},\,-\omega\right)d\omega\nonumber \\
 & \quad+\frac{1}{Tb_{2,T}}\sum_{s=k+1}^{T}K_{2}^{*}\left(\frac{u_{0}-\left(s+k/2\right)/T}{b_{2,T}}\right)\int_{-\pi}^{\pi}\exp\left(i\omega k\right)\left[\frac{\partial A_{j+1}\left(v_{0}+\epsilon_{2,T},\,\omega\right)}{\partial v}\right.\nonumber \\
 & \quad\times A_{j}\left(u_{0}-\epsilon_{1,T},\,-\omega\right)\left(\frac{s}{T}-v_{0}-\epsilon_{2,T}\right)\nonumber \\
 & \quad\left.+\frac{\partial A_{j}\left(u_{0}-\epsilon_{1,T},\,-\omega\right)}{\partial u}A_{j+1}\left(v_{0}+\epsilon_{2,T},\,\omega\right)\left(\frac{s+k/2}{T}-u_{0}+\epsilon_{1,T}\right)\right]d\omega\nonumber \\
 & \quad+\frac{1}{2}\frac{1}{Tb_{2,T}}\sum_{s=k+1}^{T}K_{2}^{*}\left(\frac{u_{0}-\left(s+k/2\right)/T}{b_{2,T}}\right)\int_{-\pi}^{\pi}\exp\left(i\omega k\right)\left[\frac{\partial^{2}A_{j+1}\left(v_{0}+\epsilon_{2,T},\,\omega\right)}{\partial v^{2}}\right.\nonumber \\
 & \quad\times A_{j}\left(u_{0}-\epsilon_{1,T},\,-\omega\right)\left(\frac{s}{T}-v_{0}-\epsilon_{2,T}\right)^{2}\nonumber \\
 & \quad\left.+\frac{\partial^{2}A_{j}\left(u_{0}-\epsilon_{1,T},\,-\omega\right)}{\partial u^{2}}A_{j+1}\left(v_{0}+\epsilon_{2,T},\,\omega\right)\left(\frac{s+k/2}{T}-u_{0}+\epsilon_{1,T}\right)^{2}\right]d\omega\nonumber \\
 & \quad+\frac{1}{Tb_{2,T}}\sum_{s=k+1}^{T}K_{2}^{*}\left(\frac{u_{0}-\left(s+k/2\right)/T}{b_{2,T}}\right)\nonumber \\
 & \quad\times\int_{-\pi}^{\pi}\exp\left(i\omega k\right)\left[\frac{\partial A_{j+1}\left(v_{0}+\epsilon_{2,T},\,\omega\right)}{\partial v}\frac{\partial A_{j}\left(u_{0}-\epsilon_{1,T},\,-\omega\right)}{\partial u}\left(\frac{s}{T}-v_{0}-\epsilon_{2,T}\right)\left(\frac{s+k/2}{T}-u_{0}+\epsilon_{1,T}\right)\right]d\omega\label{Eq(E ctilde)}\\
 & \quad+o\left(b_{2,T}^{2}\right).\nonumber 
\end{align}
By Assumption \ref{Assumption Smothness of A (for HAC)}, $A_{j}\left(u,\,-\omega\right)$,
$\partial A_{j}\left(u,\,-\omega\right)/\partial u$ and $\partial^{2}A_{j}\left(u,\,-\omega\right)/\partial u^{2}$
are left-continuous at $u=u_{0}$, and $A_{j+1}\left(u,\,\omega\right)$,
$\partial A_{j+1}\left(u,\,\omega\right)/\partial u$ and $\partial^{2}A_{j+1}\left(u,\,\omega\right)/\partial u^{2}$
are right-continuous at $u=v_{0}$, thus we have,
\begin{align*}
A_{j}\left(u_{0}-\epsilon_{1,T},\,-\omega\right)-A_{j}\left(u_{0},\,-\omega\right)=O_{\mathbb{P}}\left(\epsilon_{1,T}\right) & ,\qquad\frac{\partial A_{j}\left(u_{0}-\epsilon_{1,T},\,-\omega\right)}{\partial u}-\frac{\partial A_{j}\left(u_{0},\,-\omega\right)}{\partial_{-}u}=O_{\mathbb{P}}\left(\epsilon_{1,T}\right),\\
\frac{\partial^{2}A_{j}\left(u_{0}-\epsilon_{1,T},\,-\omega\right)}{\partial u^{2}} & -\frac{\partial^{2}A_{j}\left(u_{0},\,-\omega\right)}{\partial_{-}u^{2}}=O_{\mathbb{P}}\left(\epsilon_{1,T}\right)\\
A_{j+1}\left(v_{0}+\epsilon_{2,T},\,\omega\right)-A_{j+1}\left(v_{0},\,\omega\right)=O_{\mathbb{P}}\left(\epsilon_{2,T}\right) & ,\qquad\frac{\partial A_{j+1}\left(v_{0}+\epsilon_{2,T},\,\omega\right)}{\partial v}-\frac{\partial A_{j+1}\left(v_{0},\,\omega\right)}{\partial_{+}v}=O_{\mathbb{P}}\left(\epsilon_{2,T}\right),\\
\frac{\partial^{2}A_{j+1}\left(v_{0}+\epsilon_{2,T},\,\omega\right)}{\partial v^{2}} & -\frac{\partial^{2}A_{j+1}\left(v_{0},\,\omega\right)}{\partial_{+}v^{2}}=O_{\mathbb{P}}\left(\epsilon_{2,T}\right).
\end{align*}
 Choose $\epsilon_{1,T}=o_{\mathbb{P}}(\max\{b_{2,T}^{2},\,1/\left(Tb_{2,T}\right)\})$
and $\epsilon_{2,T}=o_{\mathbb{P}}(\max\{b_{2,T}^{2},\,1/\left(Tb_{2,T}\right)\})$.
Using the definition of $c\left(u_{0},\,k\right)$ for $k<0$, \eqref{Eq(E ctilde)}
is equal to, 
\begin{align*}
c\left(u_{0},\,k\right) & +b_{2,T}^{2}\int_{0}^{1}x^{2}K_{2}\left(x\right)dx\int_{-\pi}^{\pi}\exp\left(i\omega k\right)\left(C_{1}\left(u_{0},\,\omega\right)+C_{2}\left(u_{0},\,\omega\right)+C_{3}\left(u_{0},\,\omega\right)\right)d\omega\\
 & \quad+O\left(\frac{1}{Tb_{2,T}}\right)+o\left(b_{2,T}^{2}\right).
\end{align*}
 For $Tu_{0}\notin\mathcal{T},$ \eqref{eq. Bias Locally Stationary}
and \eqref{eq: Covariance jlrw Locally Stationary} follow by a similar
proof as for Lemma \ref{Lemma Bias, MSE of c(u,k), scalar case - Locally Stationary}.
Next, let us consider $\mathrm{Var}\left[\widetilde{c}_{T}\left(u_{0},\,k\right)\right]$
for $p=1$ and $V_{t,T}$ Gaussian. Assume $u_{0}=\lambda_{j}^{0}$
and $\left|k\right|/Tb_{2,T}\rightarrow\eta_{2}\in(0,\,\lambda_{j+1}^{0}-\lambda_{j}^{0})$,
we have for $k<0$,
\begin{align*}
\mathrm{Var} & \left[\widetilde{c}_{T}\left(u_{0},\,k\right)\right]\\
 & =\frac{1}{\left(Tb_{2,T}\right)^{2}}\sum_{t}\sum_{s}K_{2}^{*}\left(\frac{u_{0}-\left(t+k/2\right)/T}{b_{2,T}}\right)K_{2}^{*}\left(\frac{u_{0}-\left(s+k/2\right)/T}{b_{2,T}}\right)\mathbb{E}\left(V_{t,T}V_{s,T}\right)\mathbb{E}\left(V_{t+k,T}V_{s+k,T}\right)\\
 & \quad+\frac{1}{\left(Tb_{2,T}\right)^{2}}\sum_{t}\sum_{s}K_{2}^{*}\left(\frac{u_{0}-\left(t+k/2\right)/T}{b_{2,T}}\right)K_{2}^{*}\left(\frac{u_{0}-\left(s+k/2\right)/T}{b_{2,T}}\right)\mathbb{E}\left(V_{t,T}V_{s+k,T}\right)\mathbb{E}\left(V_{s,T}V_{t+k,T}\right).
\end{align*}
By \eqref{Eq. 2.4 Smothenss Assumption on A}, $A_{s+k,T}^{0}\left(\omega\right)A_{t,T}^{0}\left(-\omega\right)=A_{j}\left(\left(s+k\right)/T,\,\omega\right)A_{j+1}\left(t/T,\,-\omega\right)+O(T^{-1})$
and $A_{t,T}^{0}\left(\omega\right)A_{s,T}^{0}\left(-\omega\right)=A_{j+1}\left(t/T,\,\omega\right)A_{j+1}\left(s/T,\,-\omega\right)+O(T^{-1})$
for $s,\,t=Tu_{0}-k/2$. Now take a second order Taylor's expansion
of $A_{j+1}$ around $v_{0}=u_{0}-k/2T+\epsilon_{2,T}$ and of $A_{j}$
around $u_{\epsilon,T}=u_{0}-\epsilon_{T},$ where $\epsilon_{2,T},\,\epsilon_{T}>0.$
Applying the manipulations in \eqref{Eq(E ctilde)} involving $A_{j}$
and $A_{j+1}$ combined with the same derivations that led to \eqref{eq. (Var ctilde Locally Stationary)}
we obtain,
\begin{align}
\mathrm{Var}\left[\widetilde{c}_{T}\left(u_{0},\,k\right)\right] & =\int_{0}^{1}K_{2}\left(x\right)^{2}dx\left\{ \sum_{l=-\infty}^{\infty}\left[c\left(v_{0},\,l\right)c\left(u_{0},\,l+2k\right)\right]\right.\nonumber \\
 & \quad\left.+\sum_{l=-\infty}^{0}\left[c\left(u_{0},\,l\right)c\left(u_{0},\,l\right)\right]+\sum_{l=1}^{\infty}\left[c\left(v_{0},\,l\right)c\left(v_{0},\,l\right)\right]\right\} ,\label{eq: (Var ctilde Breaks)}
\end{align}
 where $c\left(u_{0},\,\cdot\right)$ in the second line above takes
the form {[}cf. the definition of $c\left(u_{0},\,l\right)$ for $l<0$
at the end of Section \ref{Subsec Segmented-Locally-Stationary}{]},
\begin{align*}
c\left(u_{0},\,l\right) & =\int_{-\pi}^{\pi}\exp\left(i\omega l\right)A_{j}\left(u_{0},\,\omega\right)A_{j+1}\left(u_{0}-l/T,\,\omega\right)d\omega.
\end{align*}
   When $V_{t,T}$ is non-Gaussian, there is an extra term
in $\mathrm{Cov}[\widetilde{c}_{T}^{\left(j,l\right)}\left(u_{0},\,k\right),\,\widetilde{c}_{T}^{\left(r,w\right)}\left(u_{0},\,k\right)]$,
namely
\begin{align}
\frac{1}{\left(Tb_{2,T}\right)^{2}}\sum_{t}\sum_{s}K_{2}^{*}\left(\frac{u_{0}-\left(t+k/2\right)/T}{b_{T}}\right)K_{2}^{*}\left(\frac{u_{0}-\left(s+k/2\right)/T}{b_{T}}\right)\kappa_{V,t}^{\left(j,l,r,w\right)}\left(-k,\,s-t,\,s-t-k\right) & .\label{eq(Cov, Multi)}
\end{align}
 By Assumption \ref{Assumption A - Dependence} with $u=t/T,$ 
\[
\sup_{1\leq j\leq m_{0}+1}\sup_{\lambda_{j-1}^{0}<u\leq\lambda_{j}^{0}}\left|\kappa_{V,Tu}^{\left(j,l,r,w\right)}\left(-k,\,s-Tu,\,s-Tu-k\right)-\widetilde{\kappa}_{j,l,r,w}\left(u,\,-k,\,s-Tu,\,s-Tu-k\right)\right|=O\left(T^{-1}\right).
\]
Taking a second-order Taylor's expansion of $\kappa_{V,Tu}^{\left(j,l,r,w\right)}$
with respect to the first argument around $v_{0}=u_{0}-k/2T+\epsilon_{2,T}$
with $\epsilon_{2,T}>0$, we have
\begin{align}
\frac{1}{\left(Tb_{2,T}\right)^{2}} & \sum_{t}\sum_{s}K_{2}^{*}\left(\frac{u_{0}-\left(t+k/2\right)/T}{b_{2,T}}\right)K_{2}^{*}\left(\frac{u_{0}-\left(s+k/2\right)/T}{b_{2,T}}\right)\kappa_{V,t}^{\left(j,l,r,w\right)}\left(-k,\,s-t,\,s-t-k\right)\nonumber \\
 & =\frac{1}{\left(Tb_{2,T}\right)^{2}}\sum_{t}\sum_{s}K_{2}^{*}\left(\frac{u_{0}-\left(t+k/2\right)/T}{b_{2,T}}\right)K_{2}^{*}\left(\frac{u_{0}-\left(s+k/2\right)/T}{b_{2,T}}\right)\label{eq. (var c kumulant)}\\
 & \quad\times\widetilde{\kappa}_{j,l,r,w}\left(v_{0},\,-k,\,s-Tv_{0},\,s-Tv_{0}-k\right)\nonumber \\
 & \quad+\frac{1}{\left(Tb_{2,T}\right)^{2}}\sum_{t}\sum_{s}K_{2}^{*}\left(\frac{u_{0}-\left(t+k/2\right)/T}{b_{2,T}}\right)K_{2}^{*}\left(\frac{u_{0}-\left(s+k/2\right)/T}{b_{2,T}}\right)\nonumber \\
 & \quad\times\frac{\partial\widetilde{\kappa}_{j,l,r,w}}{\partial v}\left(v_{0},\,-k,\,s-Tv_{0},\,s-Tv_{0}-k\right)\left(v_{0}-t/T\right)\nonumber \\
 & \quad+\frac{1}{\left(Tb_{2,T}\right)^{2}}\sum_{t}\sum_{s}K_{2}^{*}\left(\frac{u_{0}-\left(t+k/2\right)/T}{b_{2,T}}\right)K_{2}^{*}\left(\frac{u_{0}-\left(s+k/2\right)/T}{b_{2,T}}\right)\nonumber \\
 & \quad\times\frac{\partial^{2}\widetilde{\kappa}_{j,l,r,w}}{\partial v^{2}}\left(v_{0},\,-k,\,s-Tv_{0},\,s-Tv_{0}-k\right)\left(v_{0}-t/T\right)^{2}+O\left(T^{-1}\right).\nonumber 
\end{align}
Let $\epsilon_{2,T}=o_{\mathbb{P}}(\max\{b_{2,T}^{2},\,1/\left(Tb_{2,T}\right)\})$.
Since $\widetilde{\kappa}_{j,l,r,w}\left(v_{0},\,\cdot,\,\cdot,\,\cdot\right)$
is uniformly piecewise Lipschitz continuous by Assumption \ref{Assumption A - Dependence}-(ii),
the first term on the right-hand side above is equal to 
\begin{align*}
\frac{1}{\left(Tb_{2,T}\right)^{2}} & \sum_{t}\sum_{s}K_{2}^{*}\left(\frac{u_{0}-\left(t+k/2\right)/T}{b_{T}}\right)K_{2}^{*}\left(\frac{u_{0}-\left(s+k/2\right)/T}{b_{T}}\right)\\
 & \quad\times\left(\widetilde{\kappa}_{j,l,r,w}\left(v_{0},\,-k,\,s-Tv_{0},\,s-Tv_{0}-k\right)+O\left(T^{-1}\right)\right).
\end{align*}
The second and third term of \eqref{eq. (var c kumulant)} are of
smaller order $o\left(1/Tb_{2,T}\right).$ Thus, \eqref{eq(Cov, Multi)}
is equal to 
\begin{align*}
\frac{1}{Tb_{2,T}}\int_{0}^{1}K_{2}^{2}\left(x\right)dx\sum_{h_{1}=-\infty}^{\infty}\widetilde{\kappa}_{j,l,r,w}\left(v_{0},\,-k,\,h_{1},\,h_{1}-k\right)+o\left(\frac{1}{Tb_{2,T}}\right) & .
\end{align*}
It remains to derive the expressions for $\mathrm{Var}\left[\widetilde{c}_{T}\left(u_{0},\,k\right)\right]$
and $\mathrm{Cov}[\widetilde{c}_{T}^{\left(j,l\right)}\left(u_{0},\,k\right),\,\widetilde{c}_{T}^{\left(r,w\right)}\left(u_{0},\,k\right)]$
for the case $\left|k\right|/Tb_{2,T}\rightarrow0$. As seen when
studying the bias, the behavior of $A_{\cdot,T}^{0}\left(\cdot\right)$
only matters on a left neighborhood of $u_{0}$ and thus the result
remains the same as in the locally stationary case. The argument involves
using first a Taylor's expansion around $u_{0}-\epsilon_{1,T}$ with
$\epsilon_{1,T}>0$ and then exploiting left-Lipschitz continuity.
As in the proof of Lemma \ref{Lemma Bias, MSE of c(u,k), scalar case - Locally Stationary},
basic manipulations lead to the bound for the MSE. Then, consistency
and the rate of convergence follow from the same arguments used there.
$\square$
\begin{lem}
\label{Lemma: Tb2*Var(Gamma_k)}Cosnider $p=1.$ Under Assumption
\ref{Assumption Smothness of A (for HAC)}-\ref{Assumption A - Dependence},
$\sup_{k\geq1}Tb_{2,T}\mathrm{Var}(\widetilde{\Gamma}(k))=O\left(1\right).$ 
\end{lem}
\noindent\textit{Proof of Lemma} \ref{Lemma: Tb2*Var(Gamma_k)}.
We have for $k\geq0$,
\begin{align*}
\mathrm{Var}\left(\widetilde{\Gamma}\left(k\right)\right) & =\left(\frac{n_{T}}{T}\right)^{2}\sum_{r=0}^{\left\lfloor T/n_{T}\right\rfloor }\sum_{w=0}^{\left\lfloor T/n_{T}\right\rfloor }\mathrm{Cov}\left(\widetilde{c}_{T}\left(rn_{T}/T,\,k\right),\,\widetilde{c}_{T}\left(wn_{T}/T,\,k\right)\right),
\end{align*}
with
\begin{align*}
\mathrm{Cov} & \left(\widetilde{c}_{T}\left(rn_{T}/T,\,k\right),\,\widetilde{c}_{T}\left(wn_{T}/T,\,k\right)\right)\\
 & =\frac{1}{\left(Tb_{2,T}\right)^{2}}\sum_{t}\sum_{s}K_{2}^{*}\left(\frac{rn_{T}/T-\left(t+k/2\right)/T}{b_{T}}\right)K_{2}^{*}\left(\frac{wn_{T}/T-\left(s+k/2\right)/T}{b_{T}}\right)\\
 & \quad\times\mathbb{E}\left(V_{t,T}V_{t+k,T}V_{s,T}V_{s+k,T}\right)\\
 & \quad+\frac{1}{\left(Tb_{2,T}\right)^{2}}\sum_{t}\sum_{s}K_{2}^{*}\left(\frac{rn_{T}/T-\left(t+k/2\right)/T}{b_{T}}\right)K_{2}^{*}\left(\frac{wn_{T}/T-\left(s+k/2\right)/T}{b_{T}}\right)\\
 & \quad\times\mathbb{E}\left(V_{t,T}V_{s,T}\right)\mathbb{E}\left(V_{t+k,T}V_{s+k,T}\right)\\
 & \quad+\frac{1}{\left(Tb_{2,T}\right)^{2}}\sum_{t}\sum_{s}K_{2}^{*}\left(\frac{rn_{T}/T-\left(t+k/2\right)/T}{b_{T}}\right)K_{2}^{*}\left(\frac{wn_{T}/T-\left(s+k/2\right)/T}{b_{T}}\right)\\
 & \quad\times\mathbb{E}\left(V_{t,T}V_{t+k,T}\right)\mathbb{E}\left(V_{s,T}V_{s+k,T}\right)\\
 & \quad+\frac{1}{\left(Tb_{2,T}\right)^{2}}\sum_{t}\sum_{s}K_{2}^{*}\left(\frac{rn_{T}/T-\left(t+k/2\right)/T}{b_{T}}\right)K_{2}^{*}\left(\frac{wn_{T}/T-\left(s+k/2\right)/T}{b_{T}}\right)\\
 & \quad\times\mathbb{E}\left(V_{t,T}V_{s+k,T}\right)\mathbb{E}\left(V_{s,T}V_{t+k,T}\right)-\mathbb{E}\left(\widetilde{c}_{T}\left(rn_{T}/T,\,k\right)\right)\mathbb{E}\left(\widetilde{c}_{T}\left(wn_{T}/T,\,k\right)\right).
\end{align*}
Proceeding as in the proof of Lemma \ref{Lemma Bias, MSE of c(u,k) Breaks SLS},
we have 
\begin{align*}
\mathrm{Cov} & \left(\widetilde{c}_{T}\left(rn_{T}/T,\,k\right),\,\widetilde{c}_{T}\left(wn_{T}/T,\,k\right)\right)\\
 & =\frac{1}{Tb_{2,T}}\int_{0}^{1}K_{2}\left(x\right)dx\sum_{h_{1}=-\infty}^{\infty}\widetilde{\kappa}\left(rn_{T}/T,-k,\,h_{1},\,h_{1}-k\right)\\
 & \quad+\frac{1}{Tb_{2,T}}\int_{0}^{1}x^{2}K_{2}\left(x\right)dx\sum_{l=-\infty}^{\infty}c\left(rn_{T}/T,\,l\right)c\left(wn_{T}/T,\,l\right)\\
 & \quad+\frac{1}{Tb_{2,T}}\int_{0}^{1}x^{2}K_{2}\left(x\right)dx\sum_{l=-\infty}^{\infty}c\left(rn_{T}/T,\,l\right)c\left(wn_{T}/T,\,l+2k\right)+o\left(\frac{1}{Tb_{2,T}}\right),
\end{align*}
where $\widetilde{\kappa}=\widetilde{\kappa}_{1,1,1,1}$ is the cumulant
for the univariate case. Note that
\begin{align*}
\sum_{l=-\infty}^{\infty}c\left(rn_{T}/T,\,l\right)c\left(wn_{T}/T,\,l+2k\right) & \leq\sum_{l=-\infty}^{\infty}\left|c\left(rn_{T}/T,\,l\right)\right|\sum_{s=-\infty}^{\infty}\left|c\left(wn_{T}/T,\,s+2k\right)\right|\\
 & \leq\sum_{l=-\infty}^{\infty}\left|c\left(rn_{T}/T,\,l\right)\right|\sum_{s=-\infty}^{\infty}\left|c\left(wn_{T}/T,\,s\right)\right|.
\end{align*}
The desired result then follows by Assumption \ref{Assumption A - Dependence}-(i)
and the convergence of approximation to Riemann sums. $\square$

\subsection{Proofs of the Results of Section \ref{Section HAC-Estimation-with Predetermined}}

\subsubsection{Proof of Lemma \ref{Lemma Rate of Convergence of ctilde - SLS}}

It follows by Lemma \ref{Lemma Bias, MSE of c(u,k) Breaks SLS}. $\square$

\subsubsection{Proof of Theorem \ref{Theorem MSE J}}

We first prove the result for the locally stationary case (i.e.,
$m=0$) and then extend it to the general case $m>0.$ We begin with
the result for the scalar case $\left(p=1\right)$ and then extend
it to the vector case. 
\begin{lem}
\label{Lemma Theorem 1 -Locally Stationary - Scalar case}Suppose
$p=1,$ $K_{1}\left(\cdot\right)\in\boldsymbol{K}_{1}$, $K_{2}\left(\cdot\right)\in\boldsymbol{K}_{2}$,
Assumption \ref{Assumption Smothness of A (for HAC)- Locally Stationary - Supp}-\ref{Assumption A - Dependence Locally Stationary - Supp}
hold, $b_{1,T},\,b_{2,T}\rightarrow0$, $n_{T}\rightarrow\infty,\,n_{T}/T\rightarrow0$
and $1/Tb_{1,T}b_{2,T}\rightarrow0$. We have:

(i)  $\lim_{T\rightarrow\infty}Tb_{1,T}b_{2,T}\mathrm{Var}(\widetilde{J}_{T})=4\pi^{2}\int K_{1}^{2}\left(y\right)dy\int_{0}^{1}K_{2}^{2}\left(x\right)dx(\int_{0}^{1}f\left(u,\,0\right)du)^{2}$. 

(ii) If $1/Tb_{1,T}^{q}b_{2,T}\rightarrow0$, $n_{T}/Tb_{1,T}^{q}\rightarrow0$
and $b_{2,T}^{2}/b_{1,T}^{q}\rightarrow0$ for some $q\in[0,\,\infty)$
for which $K_{1,q},\,|\int_{0}^{1}$ $f^{\left(q\right)}\left(u,\,0\right)du|\in[0,\,\infty)$,
then $\lim_{T\rightarrow\infty}b_{1,T}^{-q}[\mathbb{E}(\widetilde{J}_{T}-J_{T})]=-2\pi K_{1,q}\int_{0}^{1}f^{\left(q\right)}\left(u,\,0\right)du.$

(iii) If $n_{T}/Tb_{1,T}^{q}\rightarrow0$, $b_{2,T}^{2}/b_{1,T}^{q}\rightarrow0$
and $Tb_{1,T}^{2q}b_{2,T}\rightarrow\gamma\in\left(0,\,\infty\right)$
for some $q\in[0,\,\infty)$ for which $K_{1,q},\,|\int_{0}^{1}f^{\left(q\right)}\left(u,\,0\right)du|\in[0,\,\infty)$,
then 
\begin{align*}
\lim_{T\rightarrow\infty} & \mathrm{MSE}\left(Tb_{1,T}b_{2,T},\,\widehat{J}_{T},\,1\right)\\
 & =4\pi^{2}\left[\gamma K_{1,q}^{2}\left(\int_{0}^{1}f^{\left(q\right)}\left(u,\,0\right)du\right)^{2}+\int K_{1}^{2}\left(y\right)dy\int K_{2}^{2}\left(x\right)dx\,\left(\int_{0}^{1}f\left(u,\,0\right)du\right)^{2}\right].
\end{align*}
 
\end{lem}
\noindent\textit{Proof of Lemma }\ref{Lemma Theorem 1 -Locally Stationary - Scalar case}.
We begin with part (i). Note that for any fixed non-negative $\tau_{1},\,\tau_{2}\in\mathbb{R}$,
\begin{align*}
\mathrm{Cov} & \left(V_{s}V_{s-\tau_{1}},\,V_{l}V_{l-\tau_{2}}\right)\\
 & =\mathbb{E}\left[\left(V_{s}V_{s-\tau_{1}}-\mathbb{E}\left(V_{s}V_{s-\tau_{1}}\right)\right)\left(V_{l}V_{l-\tau_{2}}-\mathbb{E}\left(V_{l}V_{l-\tau_{2}}\right)\right)\right]\\
 & =\mathbb{E}\left(V_{s}V_{s-\tau_{1}}V_{l}V_{l-\tau_{2}}\right)-\Gamma_{s/T}\left(\tau_{1}\right)\Gamma_{l/T}\left(\tau_{2}\right)-\Gamma_{s/T}\left(\tau_{1}\right)\Gamma_{l/T}\left(\tau_{2}\right)-\Gamma_{l/T}\left(l-s\right)\Gamma_{\left(l-\tau_{2}\right)/T}\left(l-s-\tau_{2}+\tau_{1}\right)\\
 & \quad-\Gamma_{\left(l-\tau_{2}\right)/T}\left(l-s-\tau_{2}\right)\Gamma_{l/T}\left(l-s+\tau_{1}\right)+\Gamma_{s/T}\left(\tau_{1}\right)\Gamma_{l/T}\left(\tau_{2}\right)+\Gamma_{l/T}\left(l-s\right)\Gamma_{\left(l-\tau_{2}\right)/T}\left(l-s-\tau_{2}+\tau_{1}\right)\\
 & \quad+\Gamma_{\left(l-\tau_{2}\right)/T}\left(l-s-\tau_{2}\right)\Gamma_{l/T}\left(l-s+\tau_{1}\right)\\
 & =\kappa_{V,s}\left(-\tau_{1},\,l-s,\,l-s-\tau_{2}\right)+\Gamma_{s/T}\left(s-l\right)\Gamma_{\left(l-\tau_{2}\right)/T}\left(l-s-\tau_{2}+\tau_{1}\right)\\
 & \quad+\Gamma_{s/T}\left(s-l+\tau_{2}\right)\Gamma_{l/T}\left(l-s+\tau_{1}\right)\\
 & =\kappa_{V,s}\left(-\tau_{1},\,l-s,\,l-s-\tau_{2}\right)+\Gamma_{l/T}\left(l-s\right)\Gamma_{\left(l-\tau_{2}\right)/T}\left(l-s-\tau_{2}+\tau_{1}\right)\\
 & \quad+\Gamma_{\left(l-\tau_{2}\right)/T}\left(l-s-\tau_{2}\right)\Gamma_{l/T}\left(l-s+\tau_{1}\right).
\end{align*}
 For large $T$, we have by Lemma \ref{Lemma GammaTu =00003D GammaTu+k - Locally Stationary}:
$\Gamma_{\left(l-\tau_{2}\right)/T}\left(k\right)-\Gamma_{l/T}\left(k\right)=O\left(\tau_{2}/T\right)$,
and $\Gamma_{\left(s-\tau_{1}\right)/T}\left(k\right)=\Gamma_{s/T}\left(k\right)+O\left(\tau_{1}/T\right)$
uniformly in $k,\,l$ and $s$.  Apply the changes in variables $w=s-l$
and $v=l$,  then
\begin{align}
\sum_{s=\tau_{1}+1}^{T} & \sum_{l=\tau_{2}+1}^{T}\mathrm{Cov}\left(V_{s}V_{s-\tau_{1}},\,V_{l}V_{l-\tau_{2}}\right)\nonumber \\
 & =\sum_{s=\tau_{1}+1}^{T}\sum_{l=\tau_{2}+1}^{T}\kappa_{V,s}\left(-\tau_{1},\,l-s,\,l-s-\tau_{2}\right)\nonumber \\
 & \quad+\sum_{v=\tau_{2}+1}^{T}\sum_{w=\tau_{1}+1-v}^{T-v}\left[\Gamma_{v/T}\left(-w\right)\Gamma_{v/T}\left(-w+\tau_{2}-\tau_{1}\right)+\Gamma_{v/T}\left(-w-\tau_{2}\right)\Gamma_{v/T}\left(-w+\tau_{1}\right)\right]\nonumber \\
 & \quad+\sum_{v=\tau_{2}+1}^{T}\sum_{w=\tau_{1}+1-v}^{T-v}\left[\Gamma_{v/T}\left(-w\right)O\left(\tau_{2}/T\right)+O\left(\tau_{2}/T\right)\Gamma_{v/T}\left(-w+\tau_{1}\right)\right].\label{eq (3.10 in Parzen)}
\end{align}
  A bound for the term involving $\Gamma_{v/T}\left(-w\right)O\left(\tau_{2}/T\right)$
in \eqref{eq (3.10 in Parzen)} is
\begin{align}
\sum_{v=\tau_{2}+1}^{T}\sum_{w=\tau_{1}+1-v}^{T-v}\left|\Gamma_{v/T}\left(-w\right)\right|O\left(\tau_{2}/T\right) & \leq O\left(\tau_{2}/T\right)\sum_{v=\tau_{2}+1}^{T}\sum_{w=\tau_{1}+1-v}^{T-v}\sup_{\left(v/T\right)\in\left[0,\,1\right]}\left|\Gamma_{v/T}\left(w\right)\right|\nonumber \\
 & \leq O\left(T^{-1}\right),\label{eq (AT)}
\end{align}
 where we have used Assumption \ref{Assumption A - Dependence Locally Stationary - Supp}-(i).
The argument for the term involving $O\left(\tau_{2}/T\right)\Gamma_{v/T}\left(-w+\tau_{1}\right)$
is analogous. We next evaluate the covariance of $\widetilde{c}_{T}\left(t/T,\,k\right)$.
For any $1\leq t_{1},\,t_{2}\leq T$ and (without loss of generality)
non-negative integers $\tau_{1},\,\tau_{2}\in\mathbb{R},$ apply the
following changes in variables $w=s-l$ and $v=l$,  so that
\begin{align*}
Tb_{2,T} & \mathrm{Cov}\left[\widetilde{c}_{T}\left(t_{1}/T,\,\tau_{1}\right),\,\widetilde{c}_{T}\left(t_{2}/T,\,\tau_{2}\right)\right]\\
 & =Tb_{2,T}\left(\frac{1}{Tb_{2,T}}\right)^{2}\sum_{s=\tau_{1}+1}^{T}\sum_{l=\tau_{2}+1}^{T}\\
 & \quad\times K_{2}^{*}\left(\frac{\left(t_{1}-\left(s-\tau_{1}/2\right)\right)/T}{b_{2,T}}\right)K_{2}^{*}\left(\frac{\left(t_{2}-\left(l-\tau_{2}/2\right)\right)/T}{b_{2,T}}\right)\mathrm{Cov}\left(V_{s}V_{s-\tau_{1}},\,V_{l}V_{l-\tau_{2}}\right)\\
 & =\frac{1}{Tb_{2,T}}\sum_{v=\tau_{2}+1}^{T}\sum_{w=\tau_{1}+1-v}^{T-v}K_{2}^{*}\left(\frac{\left(t_{1}-\left(v+w-\tau_{1}/2\right)\right)/T}{b_{2,T}}\right)K_{2}^{*}\left(\frac{\left(t_{2}-\left(v-\tau_{2}/2\right)\right)/T}{b_{2,T}}\right)\\
 & \quad\times\left\{ \left[\Gamma_{v/T}\left(-w\right)\Gamma_{v/T}\left(-w+\tau_{2}-\tau_{1}\right)+\Gamma_{v/T}\left(-w-\tau_{2}\right)\Gamma_{v/T}\left(-w+\tau_{1}\right)\right]\right\} \\
 & \quad+\frac{1}{Tb_{2,T}}\sum_{s=\tau_{1}+1}^{T}\sum_{l=\tau_{2}+1}^{T}K_{2}^{*}\left(\frac{\left(t_{1}-\left(s-\tau_{1}/2\right)\right)/T}{b_{2,T}}\right)\\
 & \quad\times K_{2}^{*}\left(\frac{\left(t_{2}-\left(v-\tau_{2}/2\right)\right)/T}{b_{2,T}}\right)\kappa_{V,s}\left(-\tau_{1},\,l-s,\,l-s-\tau_{2}\right)+A_{T},
\end{align*}
 where
\begin{align*}
A_{T} & \triangleq\frac{1}{Tb_{2,T}}\sum_{v=\tau_{2}+1}^{T}\sum_{w=\tau_{1}+1-v}^{T-v}K_{2}^{*}\left(\frac{\left(t_{1}-\left(v+w-\tau_{1}/2\right)\right)/T}{b_{2,T}}\right)K_{2}^{*}\left(\frac{\left(t_{2}-\left(v-\tau_{2}/2\right)\right)/T}{b_{2,T}}\right)\\
 & \quad\times\left\{ \left[\Gamma_{v/T}\left(w\right)O\left(\tau_{2}/T\right)+O\left(\tau_{2}/T\right)\Gamma_{v/T}\left(w+\tau_{1}\right)\right]\right\} .
\end{align*}
 Using \eqref{eq (AT)}, we have $A_{T}=o\left(T^{-1}\right)$. Then,
using the change of variable $z=v/Tb_{2,T}$, 
\begin{align}
Tb_{2,T} & \mathrm{Cov}\left[\widetilde{c}_{T}\left(t_{1}/T,\,\tau_{1}\right),\,\widetilde{c}_{T}\left(t_{2}/T,\,\tau_{2}\right)\right]\nonumber \\
 & =\frac{1}{Tb_{2,T}}\sum_{v=\tau_{2}+1}^{T}\sum_{w=\tau_{1}+1-v}^{T-v}K_{2}^{*}\left(\frac{\left(t_{1}-v-w+\tau_{1}/2+v-v\right)/T}{b_{2,T}}\right)K_{2}^{*}\left(\frac{\left(t_{2}-v+\tau_{2}/2\right)/T}{b_{2,T}}\right)\nonumber \\
 & \quad\times\left\{ \left[\Gamma_{v}\left(-w\right)\Gamma_{v}\left(-w+\tau_{2}-\tau_{1}\right)+\Gamma_{v}\left(-w-\tau_{2}\right)\Gamma_{v}\left(-w+\tau_{1}\right)\right]\right\} \nonumber \\
 & \quad+\frac{1}{Tb_{2,T}}\sum_{s=\tau_{1}+1}^{T}\sum_{l=\tau_{2}+1}^{T}K_{2}^{*}\left(\frac{\left(t_{1}-\left(s-\tau_{1}/2\right)\right)/T}{b_{2,T}}\right)K_{2}^{*}\left(\frac{\left(t_{2}-\left(v-\tau_{2}/2\right)\right)/T}{b_{2,T}}\right)\nonumber \\
 & \quad\times\kappa_{V,s}\left(-\tau_{1},\,l-s,\,l-s-\tau_{2}\right)+A_{T}\nonumber \\
 & =\frac{1}{Tb_{2,T}}\sum_{z=\left(\tau_{2}+1\right)/Tb_{2,T}}^{1/b_{2,T}}\sum_{w=\tau_{1}+1-Tb_{2,T}z}^{T-Tb_{2,T}z}K_{2}^{*}\left(\frac{\left(t_{1}+w+\tau_{1}/2\right)/T}{b_{2,T}}-z\right)K_{2}^{*}\left(\frac{\left(t_{2}+\tau_{2}/2\right)/T}{b_{2,T}}-z\right)\label{Eq. 5.16}\\
 & \quad\times\left\{ \left[\Gamma_{zTb_{2,T}}\left(-w\right)\Gamma_{zTb_{2,T}}\left(-w+\tau_{2}-\tau_{1}\right)+\Gamma_{zTb_{2,T}}\left(-w-\tau_{2}\right)\Gamma_{zTb_{2,T}}\left(-w+\tau_{1}\right)\right]\right\} \nonumber \\
 & \quad+\frac{1}{Tb_{2,T}}\sum_{s=\tau_{1}+1}^{T}\sum_{l=\tau_{2}+1}^{T}K_{2}^{*}\left(\frac{\left(t_{1}-\left(s+\tau_{1}/2\right)\right)/T}{b_{2,T}}\right)K_{2}^{*}\left(\frac{\left(t_{2}-\left(v+\tau_{2}/2\right)\right)/T}{b_{2,T}}\right)\nonumber \\
 & \quad\times\kappa_{V,s}\left(-\tau_{1},\,l-s,\,l-s-\tau_{2}\right)+A_{T}.\nonumber 
\end{align}
Thus, with $u=t_{1}/T$ and $v=t_{2}/T$, the limit of the first
term of \eqref{Eq. 5.16} is equal to 
\begin{align*}
\int_{0}^{1} & K_{2}^{2}\left(x\right)dx\left\{ \sum_{w=-\infty}^{\infty}\left[\Gamma_{u}\left(w\right)\Gamma_{v}\left(-w+\tau_{2}-\tau_{1}\right)+\Gamma_{u}\left(w+\tau_{2}\right)\Gamma_{v}\left(-w+\tau_{1}\right)\right]\right\} .
\end{align*}
When $\tau_{1}=\tau_{2}=k$ and $t=t_{1}=t_{2}$, we have
\begin{align*}
Tb_{2,T}\mathrm{Var}\left(\widetilde{c}_{T}\left(t/T,\,k\right)\right) & =\int_{0}^{1}K_{2}\left(x\right)^{2}dx\left\{ \sum_{w=-\infty}^{\infty}\left[\Gamma_{u}\left(w\right)\Gamma_{u}\left(w\right)+\Gamma_{u}\left(w+k\right)\Gamma_{u}\left(w-k\right)\right]\right\} \\
 & =\int_{0}^{1}K_{2}\left(x\right)^{2}dx\left\{ \sum_{h=-\infty}^{\infty}\left[\Gamma_{u}\left(h\right)\Gamma_{u}\left(h\right)+\Gamma_{u}\left(h+2k\right)\Gamma_{u}\left(h\right)\right]\right\} ,
\end{align*}
 where $u=t/T$ and we have used the change in variable $h=w-k$.
Next, we consider $\mathrm{Cov}[\widetilde{\Gamma}\left(\tau_{1}\right),\,\widetilde{\Gamma}\left(\tau_{2}\right)].$
Note that,
\begin{align*}
Tb_{2,T} & \mathrm{Cov}\left[\widetilde{\Gamma}\left(\tau_{1}\right),\,\widetilde{\Gamma}\left(\tau_{2}\right)\right]\\
 & \rightarrow\int_{0}^{1}K_{2}^{2}\left(x\right)dx\int_{0}^{1}\int_{0}^{1}\left\{ \sum_{h=-\infty}^{\infty}\left[\Gamma_{u}\left(h\right)\Gamma_{u}\left(h-\tau_{2}+\tau_{1}\right)+\Gamma_{v}\left(-h-\tau_{2}\right)\Gamma_{v}\left(-h-\tau_{1}\right)\right]\right\} dvdu.
\end{align*}
The latter can be used to evaluate $\mathrm{Var}[\sum_{k=-T+1}^{T-1}K_{1}\left(b_{1,T}k\right)\widetilde{\Gamma}\left(k\right)]$
  as follows,
\begin{align}
Tb_{1,T}b_{2,T} & \mathrm{Var}\left[\sum_{k=-T+1}^{T-1}K_{1}\left(b_{1,T}k\right)\widetilde{\Gamma}\left(k\right)\right]\nonumber \\
 & =2b_{1,T}\sum_{k=-T+1}^{T-1}\sum_{j=0}^{T-1}K_{1}\left(b_{1,T}k\right)K_{1}\left(b_{1,T}j\right)\label{eq (Var) 5.16 EP}\\
 & \quad\times\left(\frac{n_{T}}{T}\right)^{2}\sum_{r=0}^{T/n_{T}}\sum_{b=0}^{T/n_{T}}\frac{1}{Tb_{2,T}}\sum_{s=k+1}^{T}\sum_{l=j+1}^{T}K_{2}^{*}\left(\frac{\left(rn_{T}+1\right)-\left(s+k/2\right)}{Tb_{2,T}}\right)K_{2}^{*}\left(\frac{\left(bn_{T}+1\right)-\left(l+j/2\right)}{Tb_{2,T}}\right)\nonumber \\
 & \quad\times\biggl(\left[\Gamma_{l/T}\left(l-s\right)\Gamma_{l/T}\left(l-s-j+k\right)+\Gamma_{l/T}\left(-s+l-\tau_{2}\right)\Gamma_{l/T}\left(-s+l+k\right)\right]\nonumber \\
 & \quad+\kappa_{V,s}\left(-k,\,l-s,\,l-s-j\right)\biggr)+o\left(1\right),\nonumber 
\end{align}
 where the $o\left(1\right)$ term follows from $A_{T}=o\left(b_{1,T}/T\right)$.
The term involving $\kappa_{V,s}\left(-k,\,l-s,\,l-s-j\right)$ is
dominated by
\begin{align*}
Cb_{1,T}\left|\sum_{k=-\infty}^{\infty}\sum_{j=0}^{\infty}\sum_{w=-\infty}^{\infty}\sup_{s}\kappa_{V,s}\left(-k,\,-w,\,-w-j\right)\right| & =O\left(b_{1,T}\right),
\end{align*}
where $C<\infty$ and we have used  Assumption \ref{Assumption A - Dependence Locally Stationary - Supp}-(i).
Now let $w=s-l$ and $v=l$ and rewrite \eqref{eq (Var) 5.16 EP}
as 
\begin{align*}
Tb_{1,T}b_{2,T} & \mathrm{Var}\left[\sum_{k=-T+1}^{T-1}K_{1}\left(b_{1,T}k\right)\widetilde{\Gamma}\left(k\right)\right]\\
 & =2b_{1,T}\sum_{k=-T+1}^{T-1}\sum_{j=0}^{T-1}K_{1}\left(b_{1,T}k\right)K_{1}\left(b_{1,T}j\right)\\
 & \quad\times\left(\frac{n_{T}}{T}\right)^{2}\sum_{r=0}^{T/n_{T}}\sum_{b=0}^{T/n_{T}}\frac{1}{Tb_{2,T}}\sum_{v=j+1}^{T}\sum_{w=k+1-v}^{T-v}\\
 & \quad\times K_{2}^{*}\left(\frac{\left(rn_{T}+1\right)-\left(w+v-k/2\right)}{Tb_{2,T}}\right)K_{2}^{*}\left(\frac{\left(bn_{T}+1\right)-\left(v-j/2\right)}{Tb_{2,T}}\right)\\
 & \quad\times\left[\Gamma_{v/T}\left(-w\right)\Gamma_{v/T}\left(-w+j-k\right)+\Gamma_{v/T}\left(-w-j\right)\Gamma_{v/T}\left(-w+k\right)\right]+o\left(1\right)+O\left(b_{1,T}\right).
\end{align*}
We next show that the term involving $\Gamma_{v/T}\left(-w-j\right)\Gamma_{v/T}\left(-w+k\right)$
vanishes in the limit. Using a change in variables $z_{1}=j+k$ and
$z=w+j$, the latter is bounded by  
\begin{align}
4b_{1,T} & \sum_{j=0}^{T-1}\sum_{z_{1}=j}^{T-1+j}K_{1}\left(b_{1,T}\left(z_{1}-j\right)\right)K_{1}\left(b_{1,T}j\right)\left(\frac{n_{T}}{T}\right)^{2}\sum_{r=0}^{T/n_{T}}\sum_{b=0}^{T/n_{T}}\nonumber \\
 & \quad\times\frac{1}{Tb_{2,T}}\sum_{v=j+1}^{T}\sum_{z=\left(z_{1}-j\right)+1-v+j}^{T-v+j}K_{2}^{*}\left(\frac{\left(rn_{T}+1\right)-\left(z-j+v-\left(z_{1}-j\right)/2\right)}{Tb_{2,T}}\right)\nonumber \\
 & \quad\times K_{2}^{*}\left(\frac{\left(\left(bn_{T}+1\right)-\left(v-j/2\right)\right)/T}{b_{2,T}}\right)\left[\Gamma_{v/T}\left(-z\right)\Gamma_{v/T}\left(-z+z_{1}\right)\right].\label{Eq (2b1)}
\end{align}
Making the change in variable $z_{2}=jb_{1,T}$, \eqref{Eq (2b1)}
can be expressed as, 
\begin{align*}
4b_{1,T} & \sum_{z_{2}=0}^{\left(T-1\right)/b_{1,T}}\sum_{z_{1}=z_{2}/b_{1,T}}^{T-1+z_{2}/b_{1,T}}K_{1}\left(b_{1,T}\left(z_{1}-z_{2}/b_{1,T}\right)\right)K_{1}\left(z_{2}\right)\left(\frac{n_{T}}{T}\right)^{2}\sum_{r=0}^{T/n_{T}}\sum_{b=0}^{T/n_{T}}\\
 & \quad\times\frac{1}{Tb_{2,T}}\sum_{v=z_{2}/b_{1,T}+1}^{T}\sum_{z=z_{1}+1-v}^{T-v+z_{2}/b_{1,T}}K_{2}^{*}\left(\frac{\left(rn_{T}+1\right)-\left(z-z_{2}/b_{1,T}+v-\left(z_{1}-z_{2}/b_{1,T}\right)/2\right)}{Tb_{2,T}}\right)\\
 & \quad\times K_{2}^{*}\left(\frac{\left(\left(bn_{T}+1\right)-\left(v-z_{2}/2b_{1,T}\right)\right)/T}{b_{2,T}}\right)\left[\Gamma_{v/T}\left(-z\right)\Gamma_{v/T}\left(-z+z_{1}\right)\right],
\end{align*}
which converges to zero because the range of summation over $z_{1}$
tends to infinity. 

Next, let us consider the term of \eqref{eq (Var) 5.16 EP} involving
$\Gamma_{v/T}\left(-w\right)\Gamma_{v/T}\left(-w+j-k\right)$. With
the changes in variables $u_{1}=k-j$ and $u_{2}=j$, this term becomes
\begin{align}
b_{1,T} & \sum_{u_{2}=-T+1}^{T-1}\sum_{u_{1}=-u_{2}-T+1}^{T-1-u_{2}}K_{1}\left(b_{1,T}\left(u_{2}+u_{1}\right)\right)K_{1}\left(b_{1,T}u_{2}\right)\left(\frac{n_{T}}{T}\right)^{2}\sum_{r=0}^{T/n_{T}}\sum_{b=0}^{T/n_{T}}\frac{1}{Tb_{2,T}}\sum_{v=u_{2}+1}^{T}\sum_{w=u_{2}+u_{1}+1-v}^{T-v}\label{eq (Of the form)}\\
 & \quad\times K_{2}^{*}\left(\frac{\left(rn_{T}+1\right)-\left(w+v-\left(u_{1}+u_{2}\right)/2\right)}{Tb_{2,T}}\right)K_{2}^{*}\left(\frac{\left(bn_{T}+1\right)-\left(v-u_{2}/2\right)}{Tb_{2,T}}\right)\nonumber \\
 & \quad\times\left[\Gamma_{v/T}\left(w\right)\Gamma_{v/T}\left(-w-u_{1}\right)\right].\nonumber 
\end{align}
Apply the change in variable $z=b_{1,T}u_{2}$ and consider the lattice
points $z_{n}=nb_{1,T}$, where $n=-T,\ldots,\,T$. As $T\rightarrow\infty$,
the distance between the lattice points $z_{n}=nb_{1,T}$ converges
to zero and the highest lattice point converges to infinity. Hence,
\eqref{eq (Of the form)} can be expressed as,
\begin{align}
\sum_{z_{n}=-\left(T-1\right)b_{1,T}}^{\left(T-1\right)b_{1,T}} & \sum_{u_{1}=-z_{n}/b_{1,T}-T+1}^{T-1-z_{n}/b_{1,T}}K_{1}\left(b_{1,T}u_{1}+z_{n}\right)K_{1}\left(z_{n}\right)\label{eq. (5.19)}\\
 & \quad\times\left(\frac{n_{T}}{T}\right)^{2}\sum_{r=0}^{T/n_{T}}\sum_{b=0}^{T/n_{T}}\frac{1}{Tb_{2,T}}\nonumber \\
 & \quad\times\sum_{v=z_{n}/b_{1,T}+1}^{T}\sum_{w=z_{n}/b_{1,T}+u_{1}+1-v}^{T-v}K_{2}\left(\frac{\left(\left(rn_{T}+1\right)-\left(w+v-\left(z_{n}/b_{1,T}+u_{1}\right)/2\right)\right)/T}{b_{2,T}}\right)\nonumber \\
 & \quad\times K_{2}\left(\frac{\left(\left(bn_{T}+1\right)-\left(v-z/2b_{1,T}\right)\right)/T}{b_{2,T}}\right)\left[\Gamma_{v/T}\left(w\right)\Gamma_{v/T}\left(-w-u_{1}\right)\right].\nonumber 
\end{align}
By Lemma \ref{Lemma eq. (73) in Dalhaus 2012 - SLS}, $\Gamma_{v/T}\left(w\right)\Gamma_{v/T}\left(-w-u_{1}\right)=c\left(v/T,\,-w\right)c\left(v/T,\,w+u_{1}\right)+O\left(T^{-1}\right)$.
By taking a second order Taylor's expansion of $c\left(v/T,\,-w\right)$
around $rn_{T}/T$ and of $c\left(\left(v-u_{1}/1\right)/T,\,w+u_{1}/2\right)$
around $bn_{T}/T$, we have
\begin{align*}
\sum_{v=z_{n}/b_{1,T}+1}^{T} & \sum_{w=z_{n}/b_{1,T}+u_{1}+1-v}^{T-v}K_{2}\left(\frac{\left(\left(rn_{T}+1\right)-\left(w+v-\left(z_{n}/b_{1,T}+u_{1}\right)/2\right)\right)/T}{b_{2,T}}\right)\\
 & \quad\times K_{2}\left(\frac{\left(\left(bn_{T}+1\right)-\left(v-z/2b_{1,T}\right)\right)/T}{b_{2,T}}\right)\left[c\left(v/T,\,-w\right)c\left(v/T,\,w+u_{1}\right)\right]\\
 & =\int_{0}^{1}K_{2}\left(x\right)^{2}dx\,c\left(rn_{T}/T,\,-w\right)c\left(bn_{T}/T,\,w+u_{1}\right)\\
 & \quad+b_{2,T}^{2}\int_{0}^{1}x^{2}K_{2}\left(x\right)^{2}dx\frac{\partial}{\partial v}c\left(v,\,-w\right)|_{v=rn_{T}/T}\frac{\partial}{\partial v}c\left(v,\,w+u_{1}\right)|_{v=bn_{T}/T}\\
 & \quad+2^{-1}b_{2,T}^{2}\int_{0}^{1}x^{2}K_{2}\left(x\right)^{2}dx\frac{\partial^{2}}{\partial v^{2}}c\left(v,\,-w\right)|_{v=rn_{T}/T}c\left(bn_{T}/T,\,w+u_{1}\right)\\
 & \quad+2^{-1}b_{2,T}^{2}\int_{0}^{1}x^{2}K_{2}\left(x\right)^{2}dx\,c\left(rn_{T}/T,\,-w\right)\frac{\partial^{2}}{\partial v^{2}}c\left(v,\,w+u_{1}\right)|_{v=bn_{T}/T}\\
 & \quad+o\left(b_{2,T}^{2}\right)+O\left(\frac{1}{Tb_{2,T}}\right).
\end{align*}
We can now use Lemma \ref{Lemma eq. (73) in Dalhaus 2012 - SLS} backward
to show that the limit of \eqref{eq. (5.19)} is equal to 
\begin{align*}
\int K_{1}\left(y\right)^{2}dy & \int_{0}^{1}K_{2}\left(x\right)^{2}dx\int_{0}^{1}\int_{0}^{1}\sum_{u_{1}=\infty}^{\infty}\sum_{w=-\infty}^{\infty}\left[\Gamma_{u}\left(w\right)\Gamma_{a}\left(w+u_{1}\right)\right]duda\\
= & 4\pi^{2}\int K_{1}\left(y\right)^{2}dy\int_{0}^{1}K_{2}\left(x\right)^{2}dx\left(\int_{0}^{1}f\left(u,\,0\right)du\right)\left(\int_{0}^{1}f\left(a,\,0\right)da\right).
\end{align*}
 This proves the result of part (i). We now move to part (ii). Let
\begin{align*}
J_{c,T} & \triangleq\int_{0}^{1}c\left(u,\,0\right)+2\sum_{k=1}^{T-1}\int_{0}^{1}c\left(u,\,k\right)du.
\end{align*}
We begin with the following relationship,
\begin{align*}
\mathbb{E}\left(\widetilde{J}_{T}-J_{T}\right) & =\sum_{k=-T+1}^{T-1}K_{1}\left(b_{1,T}k\right)\mathbb{E}\left(\widetilde{\Gamma}\left(k\right)\right)-J_{c,T}+\left(J_{c,T}-J_{T}\right).
\end{align*}
Using Lemma \ref{Lemma Bias, MSE of c(u,k), scalar case - Locally Stationary},
we have for any $-T+1\leq k\leq T-1$, 
\begin{align*}
\mathbb{E} & \left(\frac{n_{T}}{T}\sum_{r=0}^{T/n_{T}}\widetilde{c}_{T}\left(rn_{T}/T,\,k\right)-\int_{0}^{1}c\left(u,\,k\right)du\right)\\
 & =\frac{n_{T}}{T}\sum_{r=0}^{T/n_{T}}\left(c\left(rn_{T}/T,\,k\right)+\frac{1}{2}b_{2,T}^{2}\int_{0}^{1}x^{2}K_{2}\left(x\right)dx\frac{\partial^{2}}{\partial^{2}u}c\left(u,\,k\right)|_{u=rn_{T}/T}+o\left(b_{2,T}^{2}\right)+O\left(\frac{1}{b_{2,T}T}\right)\right)\\
 & \quad-\int_{0}^{1}c\left(u,\,k\right)du\\
 & =O\left(\frac{n_{T}}{T}\right)+\frac{1}{2}b_{2,T}^{2}\int_{0}^{1}x^{2}K_{2}\left(x\right)dx\int_{0}^{1}\frac{\partial^{2}}{\partial^{2}u}c\left(u,\,k\right)du+o\left(b_{2,T}^{2}\right)+O\left(\frac{1}{Tb_{2,T}}\right),
\end{align*}
where the last equality follows from the convergence of approximations
to Riemann sums. This leads to,
\begin{align*}
b_{1,T}^{-q} & \mathbb{E}\left(\widetilde{J}_{T}-J_{c,T}\right)\\
 & =-b_{1,T}^{-q}\sum_{k=-T+1}^{T}\left(1-K_{1}\left(b_{1,T}k\right)\right)\int_{0}^{1}c\left(u,\,k\right)du\\
 & \quad+\frac{1}{2}\frac{b_{2,T}^{2}}{b_{1,T}^{q}}\int_{0}^{1}x^{2}K_{2}\left(x\right)dx\sum_{k=-T+1}^{T}K_{1}\left(b_{1,T}k\right)\int_{0}^{1}\frac{\partial^{2}}{\partial^{2}u}c\left(u,\,k\right)du+O\left(\frac{1}{Tb_{1,T}^{q}b_{2,T}}\right)+O\left(\frac{n_{T}}{Tb_{1,T}^{q}}\right)\\
 & =-b_{1,T}^{-q}\sum_{k=-T+1}^{T}\left(1-K_{1}\left(b_{1,T}k\right)\right)\int_{0}^{1}c\left(u,\,k\right)du\\
 & \quad-\frac{1}{2}b_{2,T}^{2}\int_{0}^{1}x^{2}K_{2}\left(x\right)dxO\left(1\right)+\frac{1}{2}\frac{b_{2,T}^{2}}{b_{1,T}^{q}}\int_{0}^{1}x^{2}K_{2}\left(x\right)dxO\left(1\right)+O\left(\frac{1}{Tb_{1,T}^{q}b_{2,T}}\right)+O\left(\frac{n_{T}}{Tb_{1,T}^{q}}\right),
\end{align*}
 since $|\sum_{k=-\infty}^{\infty}\left|k\right|^{q}\int_{0}^{1}\left(\partial^{2}/\partial^{2}u\right)c\left(u,\,k\right)du|<\infty$
by Assumption \ref{Assumption A - Dependence Locally Stationary - Supp}-(i).
Since $J_{c,T}-J_{T}=O\left(T^{-1}\right)$, we conclude that
\begin{align*}
\lim_{T\rightarrow\infty}b_{1,T}^{-q}\mathbb{E}\left(\widetilde{J}_{T}-J_{T}\right) & =-2\pi K_{1,q}\int_{0}^{1}f^{\left(q\right)}\left(u,\,0\right)du,
\end{align*}
because $b_{2,T}^{2}/b_{1,T}^{q}\rightarrow0.$ It remains to show
part (iii). Note that $Tb_{1,T}b_{2,T}=Tb_{1,T}b_{2,T}b_{1,T}^{2q}/b_{1,T}^{2q}=b_{1,T}^{-2q}/(1/Tb_{1,T}^{2q+1}b_{2,T})=b_{1,T}^{-2q}/\left(1/\left(\gamma+o\left(1\right)\right)\right).$
Hence, using part (i)-(ii), we deduce the desired result, namely,
\begin{align*}
\lim_{T\rightarrow\infty} & \mathrm{MSE}\left(Tb_{1,T}b_{2,T},\,\widetilde{J}_{T},\,1\right)\\
 & =\lim_{T\rightarrow\infty}b_{1,T}^{-2q}\mathbb{E}\left[\left(\widetilde{J}_{T}-J_{T}\right)^{2}\right]\left(\gamma+o\left(1\right)\right)+\lim_{T\rightarrow\infty}Tb_{1,T}b_{2,T}\mathrm{Var}\left(\widetilde{J}_{T}\right)\\
 & =4\pi^{2}\left[\gamma K_{1,q}^{2}\left(\int_{0}^{1}f^{\left(q\right)}\left(u,\,0\right)du\right)^{2}+\int K_{1}^{2}\left(y\right)dy\int_{0}^{1}K_{2}^{2}\left(x\right)dx\left(\int_{0}^{1}f\left(u,\,0\right)du\right)^{2}\right].\qquad\square
\end{align*}

\begin{lem}
\label{Lemma: Theorem MSE J, Locally Stationary}Suppose $K_{1}\left(\cdot\right)\in\boldsymbol{K}_{1},\,K_{2}\left(\cdot\right)\in\boldsymbol{K}_{2}$,
Assumption \ref{Assumption Smothness of A (for HAC)- Locally Stationary - Supp}-\ref{Assumption A - Dependence Locally Stationary - Supp}
hold, $b_{1,T},\,b_{2,T}\rightarrow0$, $n_{T}\rightarrow\infty,\,n_{T}/T\rightarrow0$
and $1/Tb_{1,T}b_{2,T}\rightarrow0$. Then, part (i)-(iii) of Theorem
\ref{Theorem MSE J} hold. 
\end{lem}
\noindent\textit{Proof of Lemma }\ref{Lemma: Theorem MSE J, Locally Stationary}.
We begin with part (i). We provide the expression for the asymptotic
covariance between the $\left(a,\,l\right)$ and $\left(m,\,n\right)$
elements of $\widetilde{J}_{T}$:   

\begin{align}
T & b_{1,T}b_{2,T}\mathrm{Cov}\left[\sum_{k=-T+1}^{T-1}K_{1}\left(b_{1,T}k\right)\widetilde{\Gamma}^{\left(a,l\right)}\left(k\right),\,\sum_{j=-T+1}^{T-1}K_{1}\left(b_{1,T}j\right)\widetilde{\Gamma}^{\left(m,n\right)}\left(j\right)\right]\nonumber \\
 & =4b_{1,T}\sum_{k=0}^{T-1}\sum_{j=0}^{T-1}K_{1}\left(b_{1,T}k\right)K_{1}\left(b_{1,T}j\right)\left(\frac{n_{T}}{T}\right)^{2}\sum_{r=0}^{T/n_{T}}\sum_{b=0}^{T/n_{T}}\frac{1}{Tb_{2,T}}\sum_{s=k+1}^{T}\sum_{h=j+1}^{T}\label{eq (Var) 5.16 EP-1}\\
 & \quad\times K_{2}^{*}\left(\frac{\left(\left(rn_{T}+1\right)-\left(s-k/2\right)\right)/T}{b_{2,T}}\right)K_{2}^{*}\left(\frac{\left(\left(bn_{T}+1\right)-\left(h-j/2\right)\right)/T}{b_{2,T}}\right)\nonumber \\
 & \quad\times\left\{ \kappa_{V,s}^{\left(a,l,m,n\right)}\left(-k,\,h-s,\,h-s-j\right)\right.\nonumber \\
 & \quad\left.+\left[\Gamma_{h/T}^{\left(a,m\right)}\left(h-s\right)\Gamma_{h/T}^{\left(l,n\right)}\left(h-s-j+k\right)+\Gamma_{h/T}^{\left(a,n\right)}\left(h-s-j\right)\Gamma_{h/T}^{\left(l,m\right)}\left(h-s+k\right)\right]\right\} +o\left(1\right),\nonumber 
\end{align}
 where the $o\left(1\right)$ term follows from using \eqref{eq (AT)}.
The term involving $\kappa_{V,s}^{\left(a,l,m,n\right)}\left(-k,\,h-s,\,h-s-j\right)$
is negligible as for the scalar case. The limit of the term involving
$\Gamma_{h/T}^{\left(a,m\right)}\left(h-s\right)\Gamma_{h/T}^{\left(l,n\right)}\left(h-s+j-k\right)$
is, according to the derivations to prove part (i) of Lemma \ref{Lemma Theorem 1 -Locally Stationary - Scalar case},
\begin{align}
4\pi^{2}\int K_{1}\left(y\right)^{2}dy\int_{0}^{1}K_{2}\left(x\right)^{2}dx\left(\int_{0}^{1}f^{\left(a,m\right)}\left(u,\,0\right)du\right)\left(\int_{0}^{1}f^{\left(l,n\right)}\left(v,\,0\right)dv\right) & .\label{Eq. (3) p. 314}
\end{align}
 Similarly, the limit of the term involving $\Gamma_{h/T}^{\left(a,n\right)}\left(s-h-j\right)\Gamma_{h/T}^{\left(l,m\right)}\left(s-h+k\right)$
is the same as \eqref{Eq. (3) p. 314} but with $m$ and $n$ interchanged.
The commutation-tensor product formula arises from the fact that the
asymptotic covariances between $\widetilde{J}_{T}^{\left(a,l\right)}$
and $\widetilde{J}_{T}^{\left(m,n\right)}$ for $a,\,l,\,m,\,n\leq p$
are of the same form as the covariances between $X_{a}X_{l}$ and
$X_{m}X_{n}$, where $X=\left(X_{1},\ldots,\,X_{p}\right)'\sim\mathscr{N}\left(0,\,\Sigma\right)$.
The formula then follows from $\mathrm{Var}\left(\mathrm{vec}\left(XX'\right)\right)=\mathrm{Var}\left(X\otimes X\right)=\left(I+C_{pp}\right)\Sigma\otimes\Sigma$.
The proof of part (ii) of the lemma follows that of the scalar case
with minor changes. Since part (iii) simply uses part (i)-(ii), it
follows that 
\begin{align*}
\lim_{T\rightarrow\infty} & \mathrm{MSE}\left(Tb_{1,T}b_{2,T},\,\widetilde{J}_{T},\,W\right)\\
 & =\lim_{T\rightarrow\infty}\gamma b_{1,T}^{-2q}\mathbb{E}\left(\widetilde{J}_{T}-J_{T}\right)'W\mathbb{E}\left(\widetilde{J}_{T}-J_{T}\right)+\lim_{T\rightarrow\infty}Tb_{1,T}b_{2,T}\,\mathrm{tr}W\,\mathrm{Var}\left(\mathrm{vec}\left(\widetilde{J}_{T}\right)\right),
\end{align*}
 converges to the desired limit. $\square$

\begin{lem}
\label{Lemma Theorem 1 -Segmented Locally Stationary - Scalar case}Suppose
$p=1,$ $K_{1}\left(\cdot\right)\in\boldsymbol{K}_{1}$, $K_{2}\left(\cdot\right)\in\boldsymbol{K}_{2}$,
Assumption \ref{Assumption Smothness of A (for HAC)}-\ref{Assumption A - Dependence}
hold, $b_{1,T},\,b_{2,T}\rightarrow0$, $n_{T}\rightarrow\infty,\,n_{T}/T\rightarrow0$
and $1/Tb_{1,T}b_{2,T}\rightarrow0$. Then, (i)-(iii) of Lemma \ref{Lemma Theorem 1 -Locally Stationary - Scalar case}
continue to hold. 
\end{lem}
\noindent\textit{Proof of Lemma }\ref{Lemma Theorem 1 -Segmented Locally Stationary - Scalar case}.
We assume without loss of generality that $m_{0}=1$ and provide
the proof only for the single break case. Hence, the break date is
$T_{2}^{0}$ (i.e., $T_{1}^{0}=0$ and $T_{3}^{0}=T$). Note that
by standard properties of approximations to Riemann sums,  $\overline{\Gamma}\left(k\right)\overset{}{\rightarrow}\int_{0}^{1}\left(c\left(u,\,k\right)\right)du$
even when $c\left(\cdot,\,k\right)$ has a finite number of discontinuities
in $u$, where 
\begin{align*}
\overline{\Gamma}\left(k\right) & \triangleq\frac{n_{T}}{T-n_{T}}\sum_{r=0}^{\left\lfloor \left(T-n_{T}\right)/n_{T}\right\rfloor }c\left(rn_{T}/T,\,k\right).
\end{align*}
 Since the results in Lemma \ref{Lemma Bias, MSE of c(u,k) Breaks SLS}
about the order of the bias and variance of $\widetilde{c}_{T}\left(u_{0},\,k\right)$
are the same to their counterpart results in Lemma \ref{Lemma Bias, MSE of c(u,k), scalar case - Locally Stationary},
the proof of Lemma \ref{Lemma Theorem 1 -Locally Stationary - Scalar case}
can be repeated with the following changes. We begin with part (i).
For any fixed non-negative $\tau_{1},\,\tau_{2}\in\mathbb{R}$, 
\begin{align*}
\mathrm{Cov} & \left(V_{s}V_{s-\tau_{1}},\,V_{l}V_{l-\tau_{2}}\right)\\
 & =\kappa_{V,s}\left(-\tau_{1},\,l-s,\,l-s-\tau_{2}\right)+\Gamma_{l/T}\left(l-s\right)\Gamma_{\left(l-\tau_{2}\right)/T}\left(l-s-\tau_{2}+\tau_{1}\right)\\
 & \quad+\Gamma_{\left(l-\tau_{2}\right)/T}\left(l-s-\tau_{2}\right)\Gamma_{l/T}\left(l-s+\tau_{1}\right).
\end{align*}
 When $l=T_{2}^{0}$ and $\tau_{2}<0$, Lemma \ref{Lemma GammaTu =00003D GammaTu+k - Locally Stationary}
cannot be applied because of the discontinuity in the spectrum of
$\left\{ V_{t,T}\right\} $ at time $t=T_{2}^{0}$. Thus, the relation
$\Gamma_{\left(l-\tau_{2}\right)/T}\left(k\right)-\Gamma_{l/T}\left(k\right)=\left(\tau_{2}/T\right)$
for $l=T_{2}^{0}$ and $\tau_{2}<0$ does not hold.  One has to
carry $\Gamma_{\left(l-\tau_{2}\right)/T}\left(k\right)$ through
the proof. Applying the changes in variables $w=s-l$ and $v=l$,
 we have
\begin{align}
\sum_{s=\tau_{1}+1}^{T} & \sum_{l=\tau_{2}+1}^{T}\mathrm{Cov}\left(V_{s/T}V_{\left(s-\tau_{1}\right)/T},\,V_{l/T}V_{\left(l-\tau_{2}\right)/T}\right)\nonumber \\
 & =\sum_{s=\tau_{1}+1}^{T}\sum_{l=\tau_{2}+1}^{T}\kappa_{V,s}\left(-\tau_{1},\,l-s,\,l-s-\tau_{2}\right)\nonumber \\
 & \quad+\sum_{v=\tau_{2}+1}^{T}\sum_{w=\tau_{1}+1-v}^{T-\tau_{2}-v}\left[\Gamma_{v/T}\left(-w\right)\Gamma_{\left(v-\tau_{2}\right)/T}\left(-w+\tau_{2}-\tau_{1}\right)+\Gamma_{\left(v-\tau_{2}\right)/T}\left(-w-\tau_{2}\right)\Gamma_{v/T}\left(-w+\tau_{1}\right)\right].\label{eq (3.10 in Parzen)-1}
\end{align}
 We next evaluate the covariance of $\widetilde{c}_{T}\left(t/T,\,k\right)$.
For any $1\leq t_{1},\,t_{2}\leq T$ and (without loss of generality)
non-negative integers $\tau_{1},\,\tau_{2}\in\mathbb{R},$ 
\begin{align*}
Tb_{2,T} & \mathrm{Cov}\left[\widetilde{c}_{T}\left(t_{1}/T,\,\tau_{1}\right),\,\widetilde{c}_{T}\left(t_{2}/T,\,\tau_{2}\right)\right]\\
 & =Tb_{2,T}\left(\frac{1}{Tb_{2,T}}\right)^{2}\sum_{s=\tau_{1}+1}^{T}\sum_{v=\tau_{2}+1}^{T}\\
 & \quad\times K_{2}^{*}\left(\frac{\left(t_{1}-\left(s-\tau_{1}/2\right)\right)/T}{b_{2,T}}\right)K_{2}^{*}\left(\frac{\left(t_{2}-\left(v-\tau_{2}/2\right)\right)/T}{b_{2,T}}\right)\mathrm{Cov}\left(V_{s}V_{s-\tau_{1}},\,V_{l}V_{l-\tau_{2}}\right)\\
 & =\frac{1}{Tb_{2,T}}\sum_{v=\tau_{2}+1}^{T}\sum_{w=\tau_{1}+1-v}^{T-v}K_{2}^{*}\left(\frac{\left(t_{1}-\left(v+w-\tau_{1}/2\right)\right)/T}{b_{2,T}}\right)K_{2}^{*}\left(\frac{\left(t_{2}-\left(v-\tau_{2}/2\right)\right)/T}{b_{2,T}}\right)\\
 & \quad\times\left\{ \left[\Gamma_{v/T}\left(-w\right)\Gamma_{\left(v-\tau_{2}\right)/T}\left(-w+\tau_{2}-\tau_{1}\right)+\Gamma_{\left(v-\tau_{2}\right)/T}\left(-w-\tau_{2}\right)\Gamma_{v/T}\left(-w+\tau_{1}\right)\right]\right\} \\
 & \quad+\frac{1}{Tb_{2,T}}\sum_{s=\tau_{1}+1}^{T}\sum_{l=\tau_{2}+1}^{T}K_{2}^{*}\left(\frac{\left(t_{1}-\left(s-\tau_{1}/2\right)\right)/T}{b_{2,T}}\right)\\
 & \quad\times K_{2}^{*}\left(\frac{\left(t_{2}-\left(v-\tau_{2}/2\right)\right)/T}{b_{2,T}}\right)\kappa_{V,s}\left(-\tau_{1},\,l-s,\,l-s-\tau_{2}\right).
\end{align*}
Then, using the change of variable $z=v/Tb_{2,T}$, 
\begin{align}
Tb_{2,T} & \mathrm{Cov}\left[\widetilde{c}_{T}\left(t_{1}/T,\,\tau_{1}\right),\,\widetilde{c}_{T}\left(t_{2}/T,\,\tau_{2}\right)\right]\nonumber \\
 & =\frac{1}{Tb_{2,T}}\sum_{v=\tau_{2}+1}^{T}K_{2}^{*}\left(\frac{\left(t_{1}-v-w-\tau_{1}/2+v-v\right)/T}{b_{2,T}}\right)K_{2}^{*}\left(\frac{\left(t_{2}-zTb_{2,T}-\tau_{2}/2\right)/T}{b_{2,T}}\right)\nonumber \\
 & \quad\times\left\{ \left[\Gamma_{zb_{2,T}}\left(-w\right)\Gamma_{zb_{2,T}-\tau_{2}/T}\left(-w+\tau_{2}-\tau_{1}\right)+\Gamma_{zb_{2,T}-\tau_{2}/T}\left(-w-\tau_{2}\right)\Gamma_{zb_{2,T}}\left(-w+\tau_{1}\right)\right]\right\} \nonumber \\
 & \quad+\frac{1}{Tb_{2,T}}\sum_{s=\tau_{1}+1}^{T}\sum_{l=\tau_{2}+1}^{T}K_{2}^{*}\left(\frac{\left(t_{1}-\left(s-\tau_{1}/2\right)\right)/T}{b_{2,T}}\right)\nonumber \\
 & \quad\times K_{2}^{*}\left(\frac{\left(t_{2}-\left(v+\tau_{2}/2\right)\right)/T}{b_{2,T}}\right)\kappa_{V,s}\left(-\tau_{1},\,l-s,\,l-s-\tau_{2}\right)\nonumber \\
 & =\frac{1}{Tb_{2,T}}\sum_{z=\left(\tau_{2}+1\right)/Tb_{2,T}}^{1/b_{2,T}}\sum_{w=\tau_{1}+1-zTb_{2,T}}^{T-z/Tb_{2,T}}K_{2}^{*}\left(\frac{\left(t_{1}+w-\tau_{1}/2\right)/T}{b_{2,T}}-z\right)K_{2}^{*}\left(\frac{\left(t_{2}-\tau_{2}/2\right)/T}{b_{2,T}}-z\right)\label{Eq. 5.16-1}\\
 & \quad\times\left\{ \left[\Gamma_{zb_{2,T}}\left(-w\right)\Gamma_{zb_{2,T}-\tau_{2}/T}\left(-w+\tau_{2}-\tau_{1}\right)+\Gamma_{zb_{2,T}-\tau_{2}/T}\left(-w-\tau_{2}\right)\Gamma_{zb_{2,T}}\left(-w+\tau_{1}\right)\right]\right\} \nonumber \\
 & \quad+\frac{1}{Tb_{2,T}}\sum_{s=\tau_{1}+1}^{T}\sum_{l=\tau_{2}+1}^{T}K_{2}^{*}\left(\frac{\left(t_{1}-\left(s-\tau_{1}/2\right)\right)/T}{b_{2,T}}\right)\nonumber \\
 & \quad\times K_{2}^{*}\left(\frac{\left(t_{2}-\left(v+\tau_{2}/2\right)\right)/T}{b_{2,T}}\right)\kappa_{V,s}\left(-\tau_{1},\,l-s,\,l-s-\tau_{2}\right).\nonumber 
\end{align}
By Lemma \ref{Lemma eq. (73) in Dalhaus 2012 - SLS}, $\Gamma_{zb_{2,T}}\left(-w\right)\Gamma_{zb_{2,T}-\tau_{2}/T}\left(-w+\tau_{2}-\tau_{1}\right)=c\left(zb_{2,T}/T,\,w\right)c\left(zb_{2,T}-\tau_{2}/T,\,w-\tau_{2}+\tau_{1}\right)+O\left(T^{-1}\right)$.
We need to distinguish two cases. The first case involves both $t_{1}$
and $t_{2}$ being continuity points (i.e., $t_{1},\,t_{2}\neq T_{2}^{0}$).
The second case involves either $t_{1}$ or $t_{2}$ (or both) being
discontinuity points (i.e., $t_{1}=T_{2}^{0}$ or $t_{2}=T_{2}^{0}$,
or $t_{1}=t_{2}=T_{2}^{0}$). The first case is the one considered
in Lemma \ref{Lemma Theorem 1 -Locally Stationary - Scalar case}
and thus we omit the details. For the second case, we cannot apply
the same argument as in Lemma \ref{Lemma Theorem 1 -Locally Stationary - Scalar case}.
Suppose $t_{1}=T_{2}^{0}$ whereas $t_{2}\neq T_{2}^{0}$. Let $u_{1,\epsilon,T}=t_{1}/T-\epsilon_{1,T},\,\epsilon_{1,T}>0.$
We proceed as in \eqref{Eq(E ctilde)} by taking a second order Taylor's
expansion of $c\left(zb_{2,T}/T,\,w\right)$ around $u_{1,\epsilon,T}$
and then use the left-Lipschitz continuity at $t_{1}/T$. Repeat this
argument for $c\left(zb_{2,T}-\tau_{2}/T,\,w+\tau_{2}\right).$ For
$c\left(zb_{2,T}-\tau_{2}/T,\,w-\tau_{2}+\tau_{1}\right)$ and $c\left(zb_{2,T}/T,\,w-\tau_{1}\right)$,
take a Taylor's expansion around $t_{2}/T$. Finally, use Lemma \ref{Lemma eq. (73) in Dalhaus 2012 - SLS}
backward to obtain
\begin{align*}
c\left(t_{1}/T,\,w\right)c\left(t_{2},\,w-\tau_{2}+\tau_{1}\right) & =\Gamma_{t_{1}/T}\left(-w\right)\Gamma_{t_{2}/T}\left(-w+\tau_{2}-\tau_{1}\right)+O\left(T^{-1}\right).
\end{align*}
Thus, with $u=t_{1}/T$ and $v=t_{2}/T$, the limit of the first term
of \eqref{Eq. 5.16-1} is equal to 
\begin{align}
\int_{0}^{1} & K_{2}^{2}\left(x\right)dx\left\{ \sum_{w=-\infty}^{\infty}\left[\Gamma_{u}\left(w\right)\Gamma_{v}\left(-w+\tau_{2}-\tau_{1}\right)+\Gamma_{u}\left(w+\tau_{2}\right)\Gamma_{v}\left(-w+\tau_{1}\right)\right]\right\} .\label{Eq. 1st term SLS}
\end{align}
For the sub-case where only $t_{2}$ is a discontinuity point, use
a Taylor's expansion of $c\left(zb_{2,T}/T,\,w\right)$ and $c\left(zb_{2,T}-\tau_{2}/T,\,w+\tau_{2}\right)$
around $t_{1}/T$, and proceed as in \eqref{Eq(E ctilde)} by taking
a second order Taylor's expansion of $c\left(zb_{2,T}-\tau_{2}/T,\,w-\tau_{2}+\tau_{1}\right)$
and $c\left(zb_{2,T}/T,\,w-\tau_{1}\right)$ around $u_{2,\epsilon,T}=t_{2}/T-\epsilon_{2,T},\,\epsilon_{2,T}>0$
and then use the left-Lipschitz continuity at $t_{2}/T$. Again using
Lemma \ref{Lemma eq. (73) in Dalhaus 2012 - SLS} backward leads to
\eqref{Eq. 1st term SLS}. For the final case where $t_{1}=t_{2}=T_{2}^{0}$
we need to proceed as in the previous two sub-cases with $t_{1}=T_{2}^{0}$
and $t_{2}=T_{2}^{0}$ being discontinuity points. This would lead
to \eqref{Eq. 1st term SLS}. We can use \eqref{Eq. 1st term SLS}
to obtain,
\begin{align*}
Tb_{2,T} & \mathrm{Cov}\left[\widetilde{\Gamma}\left(\tau_{1}\right),\,\widetilde{\Gamma}\left(\tau_{2}\right)\right]\\
 & \rightarrow\int_{0}^{1}K_{2}^{2}\left(x\right)dx\int_{0}^{1}\int_{0}^{1}\left\{ \sum_{h=-\infty}^{\infty}\left[\Gamma_{u}\left(h\right)\Gamma_{u}\left(h-\tau_{2}+\tau_{1}\right)+\Gamma_{v}\left(-h-\tau_{2}\right)\Gamma_{v}\left(-h-\tau_{1}\right)\right]\right\} dvdu.
\end{align*}
In \eqref{Eq. 5.16-1} the term involving $\kappa_{V,s}\left(-\tau_{1},\,l-s,\,l-s-\tau_{2}\right)$
is negligible as in Lemma \ref{Lemma Theorem 1 -Locally Stationary - Scalar case}
while the term involving $\Gamma_{\left(l-\tau_{2}\right)/T}$ $\left(-w-j\right)\Gamma_{l/T}\left(-w+k\right)$
vanishes in the limit  using the same argument as in the proof of
Lemma \ref{Lemma Theorem 1 -Locally Stationary - Scalar case}. 
This proves the result of part (i).

We move to part (ii). Let 
\[
J_{c,T}=\int_{0}^{1}c\left(u,\,0\right)du+2\sum_{k=1}^{T-1}\int_{0}^{1}c\left(u,\,k\right)du,
\]
and $\mathcal{T}_{C}\triangleq\{\left\{ 0,\,n_{T},\ldots,\,T-n_{T},\,T\right\} /\mathcal{T}\}$.
We begin with the following relationship,
\begin{align*}
\mathbb{E}\left(\widetilde{J}_{T}-J_{T}\right) & =\sum_{k=-T+1}^{T-1}K_{1}\left(b_{1,T}k\right)\mathbb{E}\left(\widetilde{\Gamma}\left(k\right)\right)-J_{c,T}+\left(J_{c,T}-J_{T}\right).
\end{align*}
Using Lemma \ref{Lemma Bias, MSE of c(u,k) Breaks SLS}, we have for
any $-T+1\leq k\leq T-1$, 
\begin{align*}
\mathbb{E} & \left(\frac{n_{T}}{T}\sum_{r=0}^{T/n_{T}}\widetilde{c}_{T}\left(rn_{T}/T,\,k\right)-\int_{0}^{1}c\left(u,\,k\right)du\right)\\
 & =\frac{n_{T}}{T}\sum_{r=0}^{T/n_{T}}c\left(rn_{T}/T,\,k\right)-\int_{0}^{1}c\left(u,\,k\right)du\\
 & \quad+\frac{1}{2}b_{2,T}^{2}\int_{0}^{1}x^{2}K_{2}\left(x\right)dx\int_{0}^{1}\frac{\partial^{2}}{\partial^{2}u}c\left(u,\,k\right)du+o\left(b_{2,T}^{2}\right)+O\left(\frac{1}{Tb_{2,T}}\right)\\
 & \quad+\frac{1}{2}b_{2,T}^{2}\int_{0}^{1}x^{2}K_{2}\left(x\right)dx\\
 & \quad\times\int_{0}^{1}\left(\int_{-\pi}^{\pi}\exp\left(i\omega k\right)\left(C_{1}\left(u,\,\omega\right)+C_{2}\left(u,\,\omega\right)+C_{3}\left(u,\,\omega\right)\right)d\omega\mathbf{1}\left\{ Tu\in\mathcal{T}\right\} \right)du\\
 & \quad+o\left(b_{2,T}^{2}\right)+O\left(\frac{1}{Tb_{2,T}}\right)\\
 & =O\left(\frac{n_{T}}{T}\right)+\frac{1}{2}b_{2,T}^{2}\int_{0}^{1}x^{2}K_{2}\left(x\right)dx\int_{0}^{1}\frac{\partial^{2}}{\partial^{2}u}c\left(u,\,k\right)du+o\left(b_{2,T}^{2}\right)+O\left(\frac{1}{Tb_{2,T}}\right),
\end{align*}
where the last equality follows from the convergence of approximations
to Riemann sums and from the fact that $\mathbf{1}\left\{ Tu\in\mathcal{T}\right\} $
has zero Lebesgue measure. Thus, $b_{1,T}^{-q}\mathbb{E}(\widetilde{J}_{T}-J_{c,T})$
has the same form as in the locally stationary case. The relation
$J_{c,T}-J_{T}=O(T^{-1})$ continues to hold for SLS processes in
virtue of Lemma \ref{Lemma eq. (73) in Dalhaus 2012 - SLS}. Hence,
$\lim_{T\rightarrow\infty}b_{1,T}^{-q}\mathbb{E}(\widetilde{J}_{T}-J_{T})=-2\pi K_{1,q}\int_{0}^{1}f^{\left(q\right)}\left(u,\,0\right)du.$
Part (iii) follows from part (i)-(ii). $\square$\medskip{}

\noindent\textit{Proof of Theorem }\ref{Theorem MSE J}. We can
now complete the proof of Theorem \ref{Theorem MSE J}. We begin with
part (i). We provide the expression for the asymptotic covariance
between the $\left(a,\,l\right)$ and $\left(m,\,n\right)$ elements
of $\widetilde{J}_{T}$:   

\begin{align}
T & b_{1,T}b_{2,T}\mathrm{Cov}\left[\sum_{k=-T+1}^{T-1}K_{1}\left(b_{1,T}k\right)\widetilde{\Gamma}^{\left(a,l\right)}\left(k\right),\,\sum_{j=-T+1}^{T-1}K_{1}\left(b_{1,T}j\right)\widetilde{\Gamma}^{\left(m,n\right)}\left(j\right)\right]\nonumber \\
 & =b_{1,T}\sum_{k=-T+1}^{T-1}\sum_{j=-T+1}^{T-1}K_{1}\left(b_{1,T}k\right)K_{1}\left(b_{1,T}j\right)\left(\frac{n_{T}}{T}\right)^{2}\sum_{r=0}^{T/n_{T}}\sum_{b=0}^{T/n_{T}}\frac{1}{Tb_{2,T}}\sum_{s=k+1}^{T}\sum_{h=j+1}^{T}\label{eq (Var) 5.16 EP-1-1}\\
 & \quad\times K_{2}^{*}\left(\frac{\left(\left(rn_{T}+1\right)-\left(s-k/2\right)\right)/T}{b_{2,T}}\right)K_{2}^{*}\left(\frac{\left(\left(bn_{T}+1\right)-\left(h-j/2\right)\right)/T}{b_{2,T}}\right)\nonumber \\
 & \quad\times\left\{ \kappa_{V,s}^{\left(a,l,m,n\right)}\left(-k,\,h-s,\,h-s-j\right)\right.\nonumber \\
 & \quad\left.+\left[\Gamma_{h/T}^{\left(a,m\right)}\left(h-s\right)\Gamma_{\left(h-j\right)/T}^{\left(l,n\right)}\left(h-s-j+k\right)+\Gamma_{\left(h-j\right)/T}^{\left(a,n\right)}\left(h-s-j\right)\Gamma_{h/T}^{\left(l,m\right)}\left(h-s+k\right)\right]\right\} .\nonumber 
\end{align}
 As for the scalar case, the term involving $\kappa_{V,s}^{\left(a,l,m,n\right)}\left(-k,\,h-s,\,h-s-j\right)$
is negligible. The limit of the term involving $\Gamma_{h/T}^{\left(a,m\right)}\left(h-s\right)\Gamma_{\left(h-j\right)/T}^{\left(l,n\right)}\left(h-s-j+k\right)$
is, according to the derivations for the proof of part (i) of Lemma
\ref{Lemma Theorem 1 -Segmented Locally Stationary - Scalar case},
\begin{align}
4\pi^{2}\int K_{1}\left(y\right)^{2}dy\int_{0}^{1}K_{2}\left(x\right)^{2}dx\left(\int_{0}^{1}f^{\left(a,m\right)}\left(u,\,0\right)du\right)\left(\int_{0}^{1}f^{\left(l,n\right)}\left(v,\,0\right)dv\right) & .\label{Eq: (3) p. 314-1}
\end{align}
 Similarly, the limit of the term involving $\Gamma_{\left(h-j\right)/T}^{\left(a,n\right)}\left(s-h-j\right)\Gamma_{h/T}^{\left(l,m\right)}\left(s-h+k\right)$
is the same as \eqref{Eq: (3) p. 314-1} but with $m$ and $n$ interchanged.
The commutation-tensor product formula follows from the same argument
as in Lemma \ref{Lemma: Theorem MSE J, Locally Stationary}. The
proof of part (ii) of the theorem follows from that of the scalar
case with minor changes. Since part (iii) simply uses part (i)-(ii),
it follows that 
\begin{align*}
\lim_{T\rightarrow\infty} & \mathrm{MSE}\left(Tb_{1,T}b_{2,T},\,\widetilde{J}_{T},\,W\right)\\
 & =\lim_{T\rightarrow\infty}\gamma b_{1,T}^{-2q}\mathbb{E}\left(\widetilde{J}_{T}-J_{T}\right)'W\mathbb{E}\left(\widetilde{J}_{T}-J_{T}\right)+\lim_{T\rightarrow\infty}Tb_{1,T}b_{2,T}\,\mathrm{tr}W\,\mathrm{Var}\left(\mathrm{vec}\left(\widetilde{J}_{T}\right)\right),
\end{align*}
 converges to the desired limit. $\square$

\subsubsection{Proof of Theorem \ref{Theorem 1 -Consistency and Rate}}

Under Assumption \ref{Assumption A - Dependence}, $||\int_{0}^{1}f^{\left(0\right)}\left(u,\,0\right)||<\infty$.
In view of $K_{1,0}=0$, Theorem \ref{Theorem MSE J}-(i,ii) {[}with
$q=0$ in part (ii){]} implies $\widetilde{J}_{T}-J_{T}=o_{\mathbb{P}}\left(1\right)$.
Noting that $\widehat{J}_{T}-\widetilde{J}_{T}=o_{\mathbb{P}}\left(1\right)$
if and only if $b'\widehat{J}_{T}b-b'\widetilde{J}_{T}b=o_{\mathbb{P}}\left(1\right)$
for arbitrary $b\in\mathbb{R}^{p}$ we shall provide the proof only
for the scalar case. We first show that $\sqrt{T}b_{1,T}(\widehat{J}_{T}-\widetilde{J}_{T})=O_{\mathbb{P}}\left(1\right)$
under Assumption \ref{Assumption B}. Let $\widetilde{J}_{T}\left(\beta\right)$
denote the estimator that uses $\left\{ V_{t,T}\left(\beta\right)\right\} $.
A mean-value expansion of $\widetilde{J}_{T}(\widehat{\beta})(=\widehat{J}_{T})$
about $\beta_{0}$ yields 
\begin{align}
\sqrt{T}b_{1,T}(\widehat{J}_{T}-\widetilde{J}_{T}) & =b_{1,T}\frac{\partial}{\partial\beta'}\widetilde{J}_{T}(\bar{\beta})\sqrt{T}(\widehat{\beta}-\beta_{0})\nonumber \\
 & =b_{1,T}\sum_{k=-T+1}^{T-1}K_{1}\left(b_{1,T}k\right)\frac{\partial}{\partial\beta'}\widehat{\Gamma}\left(k\right)|_{\beta=\bar{\beta}}\sqrt{T}(\widehat{\beta}-\beta_{0}),\label{eq (A.9) Andrews}
\end{align}
for some $\bar{\beta}$ on the line segment joining $\widehat{\beta}$
and $\beta_{0}$. Note also that $\widehat{c}\left(rn_{T}/T,\,k\right)$
depends on $\beta$ although we omit it. We have for $k\geq0$ (the
case $k<0$ is similar and omitted), 

\begin{align}
\Bigl\Vert & \frac{\partial}{\partial\beta'}\widehat{c}\left(rn_{T}/T,\,k\right)\Bigr\Vert|_{\beta=\bar{\beta}}\label{Eq. A.10 Andrews 91}\\
 & =\Biggl\Vert\left(Tb_{2,T}\right)^{-1}\sum_{s=k+1}^{T}K_{2}^{*}\left(\frac{\left(r+1\right)n_{T}-\left(s-k/2\right)}{Tb_{2,T}}\right)\nonumber \\
 & \quad\times\left(V_{s}\left(\beta\right)\frac{\partial}{\partial\beta'}V{}_{s-k}\left(\beta\right)+\frac{\partial}{\partial\beta'}V_{s}\left(\beta\right)V{}_{s-k}\left(\beta\right)\right)\Biggr\Vert|_{\beta=\bar{\beta}}\nonumber \\
 & \leq2\left(\left(Tb_{2,T}\right)^{-1}\sum_{s=1}^{T}K_{2}^{*}\left(\frac{\left(r+1\right)n_{T}-\left(s-k/2\right)}{Tb_{2,T}}\right)^{2}\sup_{s\geq1}\sup_{\beta\in\Theta}\left(V_{s}\left(\beta\right)\right)^{2}\right)^{1/2}\nonumber \\
 & \quad\times\left(\left(Tb_{2,T}\right)^{-1}\sum_{s=1}^{T}K_{2}^{*}\left(\frac{\left(r+1\right)n_{T}-\left(s-k/2\right)}{Tb_{2,T}}\right)^{2}\sup_{s\geq1}\sup_{\beta\in\Theta}\left\Vert \frac{\partial}{\partial\beta'}V_{s}\left(\beta\right)\right\Vert ^{2}\right)^{1/2}\nonumber \\
 & =O_{\mathbb{P}}\left(1\right),\nonumber 
\end{align}
where we have used the boundedness of the kernel $K_{2}$ (and thus
of $K_{2}^{*}$), Assumption \ref{Assumption B}-(ii,iii) and Markov's
inequality to each term in parentheses; also $\sup_{s\geq1}\mathbb{E}\sup_{\beta\in\Theta}\left\Vert V_{s}\left(\beta\right)\right\Vert ^{2}<\infty$
under Assumption \ref{Assumption B}-(ii,iii) by a mean-value expansion
and, 
\begin{align*}
\left(Tb_{2,T}\right)^{-1}\sum_{s=k+1}^{T}K_{2}^{*}\left(\left(\left(r+1\right)n_{T}-\left(s+k/2\right)\right)/Tb_{2,T}\right)^{2} & \rightarrow\int_{0}^{1}K_{2}^{2}\left(x\right)dx<\infty.
\end{align*}
 Then, \eqref{eq (A.9) Andrews} becomes
\begin{align*}
b_{1,T} & \sum_{k=T+1}^{T-1}K_{1}\left(b_{1,T}k\right)\frac{\partial}{\partial\beta'}\widehat{\Gamma}\left(k\right)|_{\beta=\bar{\beta}}\sqrt{T}\left(\widehat{\beta}-\beta_{0}\right)\\
 & \leq b_{1,T}\sum_{k=-T+1}^{T-1}\left|K_{1}\left(b_{1,T}k\right)\right|\frac{n_{T}}{T}\sum_{r=0}^{T/n_{T}}O_{\mathbb{P}}\left(1\right)O_{\mathbb{P}}\left(1\right)\\
 & =O_{\mathbb{P}}\left(1\right),
\end{align*}
where the last equality uses $b_{1,T}\sum_{k=-T+1}^{T-1}\left|K_{1}\left(b_{1,T}k\right)\right|\rightarrow\int\left|K_{1}\left(x\right)\right|dx<\infty.$
This concludes the proof of part (i) of Theorem \ref{Theorem 1 -Consistency and Rate}
because $\sqrt{T}b_{1,T}\rightarrow\infty$ by assumption.

The next step is to show that $\sqrt{Tb_{1,T}}(\widehat{J}_{T}-\widetilde{J}_{T})=o_{\mathbb{P}}\left(1\right)$
under the assumptions of Theorem \ref{Theorem 1 -Consistency and Rate}-(ii).
A second-order Taylor's expansion gives
\begin{align*}
\sqrt{Tb_{1,T}}\left(\widehat{J}_{T}-\widetilde{J}_{T}\right) & =\left[\sqrt{b_{1,T}}\frac{\partial}{\partial\beta'}\widetilde{J}_{T}\left(\beta_{0}\right)\right]\sqrt{T}\left(\widehat{\beta}-\beta_{0}\right)\\
 & \quad+\frac{1}{2}\sqrt{T}\left(\widehat{\beta}-\beta_{0}\right)'\left[\sqrt{b_{1,T}}\frac{\partial^{2}}{\partial\beta\partial\beta'}\widetilde{J}_{T}\left(\overline{\beta}\right)/\sqrt{T}\right]\sqrt{T}\left(\widehat{\beta}-\beta_{0}\right)\\
 & \triangleq G_{T}'\sqrt{T}\left(\widehat{\beta}-\beta_{0}\right)+\frac{1}{2}\sqrt{T}\left(\widehat{\beta}-\beta_{0}\right)'H_{T}\sqrt{T}\left(\widehat{\beta}-\beta_{0}\right).
\end{align*}
Proceeding as in \eqref{Eq. A.10 Andrews 91} but now using Assumption
\ref{Assumption C Andrews 91}-(ii), 
\begin{align*}
\biggl\Vert & \frac{\partial^{2}}{\partial\beta\partial\beta'}\widehat{c}\left(rn_{T}/T,\,k\right)\biggr\Vert\biggl|_{\beta=\bar{\beta}}\\
 & =\left\Vert \left(Tb_{2,T}\right)^{-1}\sum_{s=k+1}^{T}K_{2}^{*}\left(\frac{\left(\left(r+1\right)n_{T}-\left(s+k/2\right)\right)/T}{b_{2,T}}\right)\left(\frac{\partial^{2}}{\partial\beta\partial\beta'}V_{s}\left(\beta\right)V{}_{s-k}\left(\beta\right)\right)\right\Vert \biggl|_{\beta=\bar{\beta}}\\
 & =O_{\mathbb{P}}\left(1\right),
\end{align*}
 and thus, 
\begin{align*}
\left\Vert H_{T}\right\Vert  & \leq\left(\frac{b_{1,T}}{T}\right)^{1/2}\sum_{k=-T+1}^{T-1}\left|K_{1}\left(b_{1,T}k\right)\right|\sup_{\beta\in\Theta}\left\Vert \frac{\partial^{2}}{\partial\beta\partial\beta'}\widehat{\Gamma}\left(k\right)\right\Vert \\
 & \leq\left(\frac{b_{1,T}}{T}\right)^{1/2}\sum_{k=-T+1}^{T-1}\left|K_{1}\left(b_{1,T}k\right)\right|O_{\mathbb{P}}\left(1\right)\\
 & \leq\left(\frac{1}{Tb_{1,T}}\right)^{1/2}b_{1,T}\sum_{k=-T+1}^{T-1}\left|K_{1}\left(b_{1,T}k\right)\right|O_{\mathbb{P}}\left(1\right)=o_{\mathbb{P}}\left(1\right),
\end{align*}
since $Tb_{1,T}\rightarrow\infty$. Next, we want to show that $G_{T}=o_{\mathbb{P}}\left(1\right)$.
We apply the results of Theorem \ref{Theorem MSE J}-(i,ii) to $\widetilde{J}_{T}$
where the latter is constructed using $\left(V'_{t},\,\partial V_{t}/\partial\beta'-\mathbb{E}\left(\partial V_{t}/\partial\beta'\right)\right)'$
rather than just with $V_{t}$. The first row and column of the off-diagonal
elements of $\widetilde{J}_{T}$ are now (written as column vectors)
\begin{align*}
A_{1} & \triangleq\sum_{k=-T+1}^{T-1}K_{1}\left(b_{1,T}k\right)\frac{n_{T}}{T}\sum_{r=0}^{T/n_{T}}\frac{1}{Tb_{2,T}}\\
 & \quad\times\sum_{s=k+1}^{T}K_{2}^{*}\left(\frac{\left(\left(r+1\right)n_{T}-\left(s+k/2\right)\right)/T}{b_{2,T}}\right)V_{s}\left(\frac{\partial}{\partial\beta}V{}_{s-k}-\mathbb{E}\left(\frac{\partial}{\partial\beta}V{}_{s}\right)\right)\\
A_{2} & \triangleq\sum_{k=-T+1}^{T-1}K_{1}\left(b_{1,T}k\right)\frac{n_{T}}{T}\sum_{r=0}^{T/n_{T}}\frac{1}{Tb_{2,T}}\\
 & \quad\times\sum_{s=k+1}^{T}K_{2}^{*}\left(\frac{\left(\left(r+1\right)n_{T}-\left(s+k/2\right)\right)/T}{b_{2,T}}\right)\left(\frac{\partial}{\partial\beta}V{}_{s}-\mathbb{E}\left(\frac{\partial}{\partial\beta}V{}_{s}\right)\right)V_{s-k}.
\end{align*}
By Theorem \ref{Theorem MSE J}-(i,ii), each expression above is $O_{\mathbb{P}}\left(1\right)$.
Since 
\begin{align*}
G_{T} & =\sqrt{b_{1,T}}\left(A_{1}+A_{2}\right)+\sqrt{b_{1,T}}\sum_{k=-T+1}^{T-1}K_{1}\left(b_{1,T}k\right)\frac{n_{T}}{T}\sum_{r=0}^{T/n_{T}}\frac{1}{Tb_{2,T}}\\
 & \quad\times\sum_{s=k+1}^{T}K_{2}^{*}\left(\frac{\left(\left(r+1\right)n_{T}-\left(s+k/2\right)\right)/T}{b_{2,T}}\right)\left(V_{s}+V_{s-k}\right)\mathbb{E}\left(\frac{\partial}{\partial\beta}V{}_{s}\right)\\
 & \leq\sqrt{b_{1,T}}\left(A_{1}+A_{2}\right)+A_{3}\sup_{s\leq T}\left|\mathbb{E}\left(\frac{\partial}{\partial\beta}V{}_{s}\right)\right|,
\end{align*}
where 
\begin{align*}
A_{3} & =\sqrt{b_{1,T}}\sum_{k=-T+1}^{T-1}\left|K_{1}\left(b_{1,T}k\right)\right|\frac{n_{T}}{T}\sum_{r=0}^{T/n_{T}}\frac{1}{Tb_{2,T}}\\
 & \quad\times\sum_{s=k+1}^{T}\left|K_{2}^{*}\left(\frac{\left(\left(r+1\right)n_{T}-\left(s-k/2\right)\right)/T}{b_{2,T}}\right)\right|\left|\left(V_{s}+V_{s-k}\right)\right|,
\end{align*}
 it remains to show that $A_{3}$ is $o_{\mathbb{P}}\left(1\right).$
Note that 
\begin{align*}
\mathbb{E}\left(A_{3}^{2}\right) & \leq b_{1,T}\sum_{k=-T+1}^{T-1}\sum_{j=-T+1}^{T-1}\left|K_{1}\left(b_{1,T}k\right)K_{1}\left(b_{1,T}j\right)\right|4\left(\frac{n_{T}}{T}\right)^{2}\sum_{r=0}^{T/n_{T}}\sum_{b=0}^{T/n_{T}}\\
 & \quad\times\frac{1}{Tb_{2,T}}\frac{1}{Tb_{2,T}}\sum_{s=1}^{T}\sum_{l=1}^{T}K_{2}^{*}\left(\frac{\left(\left(r+1\right)n_{T}-\left(s-k/2\right)\right)/T}{b_{2,T}}\right)\\
 & \quad\times K_{2}^{*}\left(\frac{\left(\left(b+1\right)n_{T}-\left(l-j/2\right)\right)/T}{b_{2,T}}\right)\left|\mathbb{E}\left(V_{s}V_{l}\right)\right|,
\end{align*}
 and that $\mathbb{E}\left(V_{s}V_{l}\right)=c\left(u,\,h\right)+O\left(T^{-1}\right)$
where $h=s-l$ and $u=s/T$ by Lemma \ref{Lemma eq. (73) in Dalhaus 2012 - SLS}.
Since $\sum_{h=-\infty}^{\infty}\sup_{u\in\left[0,\,1\right]}$ $\left|c\left(u,\,h\right)\right|<\infty$,
we have  
\begin{align*}
\mathbb{E}\left(A_{3}^{2}\right) & \leq\frac{1}{Tb_{1,T}b_{2,T}}\left(b_{1,T}\sum_{k=-T+1}^{T-1}\left|K_{1}\left(b_{1,T}k\right)\right|\right)^{2}\int_{0}^{1}K_{2}^{2}\left(x\right)dx\int_{0}^{1}\sum_{h=-\infty}^{\infty}\left|c\left(u,\,h\right)\right|du=o\left(1\right).
\end{align*}
This implies $G_{T}=o_{\mathbb{P}}\left(1\right)$. It follows that
$\sqrt{Tb_{1,T}}(\widehat{J}_{T}-\widetilde{J}_{T})=o_{\mathbb{P}}\left(1\right)$
which concludes the proof of part (ii) because $\sqrt{Tb_{1,T}b_{2,T}}(\widetilde{J}_{T}-J_{T})=O_{\mathbb{P}}\left(1\right)$
by Theorem \ref{Theorem MSE J}-(iii).

Finally, we need to consider part (iii). Let 
\[
\xi_{T}\triangleq Tb_{1,T}\left(\mathrm{vec}\left(\widehat{J}_{T}-J_{T}\right)'W\mathrm{vec}\left(\widehat{J}_{T}-J_{T}\right)-\mathrm{vec}\left(\widetilde{J}_{T}-J_{T}\right)'W\mathrm{vec}\left(\widetilde{J}_{T},-J_{T}\right)\right).
\]
 By part (ii), we know that $\sqrt{Tb_{1,T}}(\widehat{J}_{T}-J_{T})=O_{\mathbb{P}}\left(1\right)$
and $\sqrt{Tb_{1,T}}(\widehat{J}_{T}-\widetilde{J}_{T})=o_{\mathbb{P}}\left(1\right)$.
This implies 
\[
Tb_{1,T}\left(\mathrm{vec}\left(\widehat{J}_{T}-J_{T}\right)'W_{T}\mathrm{vec}\left(\widehat{J}_{T}-J_{T}\right)-\mathrm{vec}\left(\widetilde{J}_{T}-J_{T}\right)'W_{T}\mathrm{vec}\left(\widetilde{J}_{T},-J_{T}\right)\right)\overset{\mathbb{P}}{\rightarrow}0.
\]
Then, using Assumption \ref{Assumption W_T and unbounded kernel and Cumulant 8},
$\xi_{T}=o_{\mathbb{P}}\left(1\right)$ and since $\left|\xi_{T}\right|$
is bounded we have $\mathbb{E}\left(\xi_{T}\right)\rightarrow0$ by
Lemma A1 in \citeReferencesSupp{andrews:91}. $\square$

\bibliographystyleReferencesSupp{elsarticle-harv}  
\bibliographyReferencesSupp{References_Supp}

\clearpage{}

\end{singlespace}

\newpage{}

\pagebreak{}

\section*{}
\addcontentsline{toc}{part}{Supplemental Material not for Publication}
\begin{center}
\Large{{Supplemental Material} not for Publication to} 
\end{center}

\begin{center}
\title{\textbf{\Large{Theory of Evolutionary Spectra for Heteroskedasticity and Autocorrelation Robust Inference in Possibly Misspecified and Nonstationary Models}}} 
\maketitle
\end{center}
\medskip{} 
\medskip{} 
\medskip{} 
\thispagestyle{empty}

\begin{center}
\author{\textsc{\textcolor{MyBlue}{Alessandro Casini}}}\\ 
\medskip{}
\medskip{} 
\medskip{} 

\small{{Department of Economics and Finance}}\\
\small{{University of Rome Tor Vergata}}\\
\medskip{}
\medskip{} 
\medskip{} 
\medskip{} 
\date{\small{\today}} 
\medskip{} 
\medskip{} 
\medskip{} 
\end{center}
\begin{abstract}
{\footnotesize{}This supplemental material is not for publication
and is structured as follows. Section \ref{sec:Implementations--HAC}
reviews how to apply the proposed DK-HAC estimator in GMM and IV contexts.
Section \ref{Section: Proofs of the Results of Section 2, 4-5} contains
the proofs of the results of Section \ref{Section: Statistical Enviromnent}
and \ref{Section Optimal-Kernels-and}-\ref{Section Data-Dependent-Bandwidths}. }{\footnotesize\par}
\end{abstract}
\setcounter{page}{0}
\setcounter{section}{0}
\renewcommand*{\theHsection}{\the\value{section}}

\newpage{}

\begin{singlespace} 
\noindent 
\small

\allowdisplaybreaks


\renewcommand{\thepage}{N-\arabic{page}}   
\renewcommand{\thesection}{N.\Alph{section}}   
\renewcommand{\theequation}{N.\arabic{equation}}




\section{\label{sec:Implementations--HAC}Implementation of DK-HAC in GMM
and IV Models}

Section \ref{subsec:GMM} reviews the DK-HAC estimation in GMM models
while Section \ref{subsec:IV} considers IV models.

\subsection{\label{subsec:GMM}GMM }

We begin with the GMM setup {[}cf. \citeReferencesSupptwo{hansen:82}{]}.
For a $k$-vector $\beta_{*}$ of unknown parameters, we have the
moment condition $\mathbb{E}m_{t}\left(\beta_{*}\right)=0$ where
$m_{t}\left(\beta\right)$ is a $p$-vector of functions of the data
and parameters where $p\geq k$. The GMM estimator $\widehat{\beta}$
is defined as the solution to $\min_{\beta}m_{T}\left(\beta\right)'\widehat{W}_{2,T}m_{T}\left(\beta\right)$,
where $m_{T}\left(\beta\right)=T^{-1}\sum_{t=1}^{T}m_{t}\left(\beta\right)$
is the sample average of the vector of sample moments $m_{t}\left(\beta\right)$
and $\widehat{W}_{2,T}$ is a (possibly) random, symmetric weighting
matrix. The asymptotic covariance matrix of $\widehat{\beta}$ is
given by $H=\lim_{T\rightarrow\infty}H_{T}$ where
\begin{align*}
H_{T} & =\left(L'_{T}W_{2,T}L_{T}\right)^{-1}L'_{T}W_{2,T}J_{T}W_{2,T}L_{T}\left(L'_{T}W_{2,T}L_{T}\right)^{-1},
\end{align*}
where $L_{T}=T^{-1}\sum_{t=1}^{T}\mathbb{E}m_{t\beta}\left(\beta_{*}\right)$
and $m_{t\beta}\left(\beta\right)$ is the $p\times k$ matrix of
partial derivatives of $m_{t}\left(\beta\right)$, $W_{2,T}$ is a
nonrandom matrix such that $\widehat{W}_{2,T}-W_{2,T}\overset{\mathbb{P}}{\rightarrow}0$,
and $J_{T}=T^{-1}\sum_{s=1}^{T}\sum_{t=1}^{T}\mathbb{E}(m_{t}\left(\beta_{*}\right)m_{s}\left(\beta_{*}\right))'$.
Let $J=\lim_{T\rightarrow\infty}J_{T}$. The consistent estimation
of $H$ boils down to the consistent estimation of $J$ since the
estimation of $L_{T}$ and $W_{2,T}$ is straightforward. $\widehat{W}_{2,T}$
is a natural estimator of $W_{2,T}$ while under regularity conditions
$L_{T}-T^{-1}\sum_{t=1}^{T}m_{t\beta}(\widehat{\beta})\overset{\mathbb{P}}{\rightarrow}0$.
In place of the classical HAC estimators we now estimate $J$ by 
\begin{align}
\widehat{J}_{T}=\sum_{k=-T+1}^{T-1}K_{1}\left(b_{1,T}k\right)\widehat{\Gamma}\left(k\right),\quad\mathrm{where}\quad & \widehat{\Gamma}\left(k\right)\triangleq\frac{n_{T}}{T-n_{T}}\sum_{r=0}^{\left\lfloor \left(T-n_{T}\right)/n_{T}\right\rfloor }\widehat{c}_{T}\left(rn_{T}/T,\,k\right),\label{eq: JHat Supp}
\end{align}
 where 
\begin{align*}
\widehat{c}_{T}\left(rn_{T}/T,\,k\right) & \triangleq\begin{cases}
\left(Tb_{2,T}\right)^{-1}\sum_{s=k+1}^{T}K_{2}^{*}\left(\frac{\left(\left(r+1\right)n_{T}-\left(s+k/2\right)\right)/T}{b_{2,T}}\right)\widehat{m}_{s}\widehat{m}'_{s-k}, & k\geq0\\
\left(Tb_{2,T}\right)^{-1}\sum_{s=-k+1}^{T}K_{2}^{*}\left(\frac{\left(\left(r+1\right)n_{T}-\left(s-k/2\right)\right)/T}{b_{2,T}}\right)\widehat{m}_{s+k}\widehat{m}'_{s}, & k<0
\end{cases},
\end{align*}
 and $\widehat{m}_{s}=m_{s}(\widehat{\beta})$. We can implement $\widehat{J}_{T}$
with the data-dependent methods for selecting $b_{1,T}$ and $b_{2,T}$,
and choose $K_{1}$ and $K_{2}$ on the basis of the optimality results
of Section \ref{Section Optimal-Kernels-and}. For $K_{1}$ one can
use the QS kernel while for $K_{2}$ one can choose $K_{2}=6x\left(1-x\right)$
for $0\leq x\leq1$ and 0 otherwise as suggested in Section \ref{Section Optimal-Kernels-and}.
From the results in Section \ref{Section Data-Dependent-Bandwidths},
\begin{align*}
\widehat{b}_{1,T} & =0.6828\left(\widehat{\phi}\left(2\right)T\widehat{\overline{b}}_{2,T}\right)^{-1/5}\\
\widehat{b}_{2,T}\left(u_{r}\right) & =1.6786\left(\widehat{D}_{1}\left(u_{r}\right)\right){}^{-1/5}\left(\widehat{D}_{2}\left(u_{r}\right)\right)^{1/5}T^{-1/5},\qquad u_{r}=rn_{T}/T,
\end{align*}
where the expressions for $\widehat{\phi}\left(2\right),\,\widehat{D}_{1}\left(u_{r}\right)$
and $\widehat{D}_{2}\left(u_{r}\right)$ are given in the same section.

\subsection{\label{subsec:IV}IV}

Consider the linear model $y_{t}=x'_{t}\beta_{0}+e_{t}$ $\left(t=1,\ldots,\,T\right)$,
where $\beta_{0}\in\Theta\subset\mathbb{R}^{p}$, $y_{t}$ is an observation
on the dependent variable, $x_{t}$ is a $p$-vector of regressors
and $e_{t}$ is an unobserved disturbance potentially autocorrelated.
Suppose the regressor is endogenous: $\mathbb{E}\left(x_{t}e_{t}\right)\neq0$.
The IV estimator $\widehat{\beta}_{\mathrm{IV}}$ is given by $\widehat{\beta}_{\mathrm{IV}}=\left(Z'X\right)^{-1}Z'Y$,
where $Y=\left(y_{1},\ldots,\,y_{T}\right)'$ , $X=\left(x_{1},\ldots,\,x_{T}\right)'$
and $Z=\left(z_{1},\ldots,\,z_{T}\right)'$ where $z_{t}$ is a $p$-vector
of instruments. The asymptotic variance of the IV estimator is given
by the limit of $\mathrm{Var}(\sqrt{T}(\widehat{\beta}_{\mathrm{IV}}-\beta_{0}))=Q_{ZX}^{-1}J_{T}Q_{ZX}^{-1}$
where $Q_{ZX}=T^{-1}\sum_{t=1}^{T}z_{t}x'_{t}$ and $J_{T}=T^{-1}\sum_{s=1}^{T}\sum_{t=1}^{T}$
$\mathbb{E}(e_{s}z_{s}(e_{t}z{}_{t})')$. A natural estimator of $\lim_{T\rightarrow\infty}Q_{ZX}$
is $T^{-1}\sum_{t=1}^{T}z_{t}x'_{t}.$ Let $J=\lim_{T\rightarrow\infty}J_{T}$.
$J$ can be consistently estimated by $\widehat{J}_{T}$ as given
in \eqref{eq: JHat Supp} where $\widehat{m}_{t}$ is replaced by
$\widehat{e}_{t}z_{t}$ where $\widehat{e}_{t}=y_{t}-x'_{t}\widehat{\beta}_{\mathrm{IV}}$. 

\section{\label{Section: Proofs of the Results of Section 2, 4-5}Appendix:
Proofs of the Results of Section \ref{Section: Statistical Enviromnent}
and \ref{Section Optimal-Kernels-and}-\ref{Section Data-Dependent-Bandwidths}}

\subsection{Proofs of the Results of Section \ref{Subsec Segmented-Locally-Stationary}}

\subsubsection{Proof of Theorem \ref{Theorem 2.2 in Dal}}

For $Tu\notin\mathcal{T}$ we use the arguments in the proof of Theorem
2.2 in \citeReferencesSupptwo{dahlhaus:96}. Without loss of generality,
assume $T_{j-1}^{0}<Tu<T_{j}^{0}$ for some $1\leq j\leq m_{0}+1$.
Then, 
\begin{align*}
f_{j,T}\left(u,\,\omega\right) & =\frac{1}{2\pi}\sum_{s=-\infty}^{\infty}\exp\left(-i\omega s\right)\int_{-\pi}^{\pi}\exp\left(i\eta s\right)A_{j,\left\lfloor Tu-s/2\right\rfloor ,T}^{0}\left(\eta\right)\overline{A_{j,\left\lfloor Tu+s/2\right\rfloor ,T}^{0}\left(\eta\right)}d\eta,
\end{align*}
 and
\begin{align*}
f_{j}\left(u,\,\omega\right) & =\frac{1}{2\pi}\sum_{s=-\infty}^{\infty}\exp\left(-i\omega s\right)\int_{-\pi}^{\pi}\exp\left(i\eta s\right)A_{j}\left(u,\,\eta\right)\overline{A_{j}\left(u,\,\eta\right)}d\mu.
\end{align*}
We have, in virtue of standard orthogonality relations,
\begin{align*}
\int_{-\pi}^{\pi} & \left|f_{j,T}\left(u,\,\omega\right)-f_{j}\left(u,\,\omega\right)\right|^{2}d\omega\\
 & =\int_{-\pi}^{\pi}\left|\frac{1}{2\pi}\sum_{s=-\infty}^{\infty}\exp\left(-i\omega s\right)\right.\\
 & \quad\left.\times\left[\int_{-\pi}^{\pi}\exp\left(i\eta s\right)\left(A_{j,\left\lfloor Tu-s/2\right\rfloor ,T}^{0}\left(\eta\right)\overline{A_{j,\left\lfloor Tu+s/2\right\rfloor ,T}^{0}\left(\eta\right)}-A_{j}\left(u,\,\eta\right)\overline{A_{j}\left(u,\,\eta\right)}\right)d\eta\right]\right|^{2}d\omega\\
 & =\frac{1}{2\pi}\sum_{s=-\infty}^{\infty}\left|c_{s,j}\right|^{2}+o\left(1\right),
\end{align*}
 where $c_{s,j}=\int_{-\pi}^{\pi}\exp\left(i\eta s\right)G_{j}\left(s/2T,\,\eta\right)d\eta$
and 
\begin{align*}
G_{j}\left(\frac{s}{2T},\,\eta\right) & =A_{j}\left(u-\frac{s}{2T},\,\eta\right)A_{j}\left(u+\frac{s}{2T},\,-\eta\right)-A_{j}\left(u,\,\eta\right)A_{j}\left(u,\,-\eta\right).
\end{align*}
By well-known results on Fourier coefficients {[}cf. \citeReferencesSupptwo{bary:64},
Chapter 2.3{]}, $\left|c_{s,j}\right|\leq Cs^{-\vartheta}$ and thus
$\sum_{s=n}^{\infty}\left|c_{s,j}\right|^{2}=O(n^{1-2\vartheta}).$
Let $\Delta_{s}\left(\omega\right)=\sum_{r=0}^{s-1}\exp\left(-i\omega r\right).$
Applying summation by parts yields    
\begin{align*}
\sum_{s=0}^{n-1}\left|c_{s,j}\right|^{2} & =\int_{-\pi}^{\pi}\int_{-\pi}^{\pi}\sum_{s=0}^{n-1}\exp\left(-i\left(\omega-\eta\right)s\right)G_{j}\left(\frac{s}{2T},\,\omega\right)\overline{G_{j}\left(\frac{s}{2T},\,\eta\right)}d\omega d\eta\\
 & \leq\int_{-\pi}^{\pi}\int_{-\pi}^{\pi}\left|-\sum_{s=0}^{n-1}[G_{j}\left(\frac{s}{2T},\,\omega\right)\overline{G_{j}\left(\frac{s}{2T},\,\eta\right)}-G_{j}\left(\frac{s-1}{2T},\,\omega\right)\overline{G_{j}\left(\frac{s-1}{2T},\,\eta\right)}]\Delta_{s}\left(\eta-\omega\right)\right.\\
 & \quad\left.+G_{j}\left(\frac{n-1}{2T},\,\omega\right)\overline{G_{j}\left(\frac{n-1}{2T},\,\eta\right)}\Delta_{n}\left(\eta-\omega\right)\right|d\omega d\eta\\
 & =O\left(\frac{n\ln n}{T^{\vartheta}}\right).
\end{align*}
A similar bound holds for $\sum_{s=n}^{\infty}\left|c_{-s,j}\right|^{2}$.
The result for $Tu\notin\mathcal{T}$ follows by choosing $n$ appropriately.
Next, suppose $Tu\in\mathcal{T}$ and $u=T_{j}^{0}/T$. Then, we have
\begin{align*}
f_{j,T}\left(u,\,\omega\right) & =\frac{1}{2\pi}\sum_{s=-\infty}^{\infty}\exp\left(-i\omega s\right)\int_{-\pi}^{\pi}\exp\left(i\eta s\right)A_{j,\left\lfloor Tu-3\left|s\right|/2\right\rfloor ,T}^{0}\left(\eta\right)\overline{A_{j,\left\lfloor Tu-\left|s\right|/2\right\rfloor ,T}^{0}\left(\eta\right)}d\eta
\end{align*}
 and
\begin{align*}
f_{j}\left(u,\,\omega\right) & =\frac{1}{2\pi}\sum_{s=-\infty}^{\infty}\exp\left(-i\omega s\right)\int_{-\pi}^{\pi}\exp\left(i\eta s\right)A_{j}\left(u,\,\eta\right)\overline{A_{j}\left(u,\,\eta\right)}d\eta.
\end{align*}
Proceeding as above, 
\begin{align*}
\int_{-\pi}^{\pi} & \left|f_{T}\left(u,\,\omega\right)-f\left(u,\,\omega\right)\right|^{2}d\omega\\
 & =\int_{-\pi}^{\pi}\left|\frac{1}{2\pi}\sum_{s=-\infty}^{\infty}\exp\left(-i\omega s\right)\right.\\
 & \quad\left.\left[\int_{-\pi}^{\pi}\exp\left(i\eta s\right)A_{j,\left\lfloor uT-3\left|s\right|/2\right\rfloor ,T}^{0}\left(\eta\right)\overline{A_{j,\left\lfloor uT-\left|s\right|/2\right\rfloor ,T}^{0}\left(\eta\right)}d\eta-\int_{-\pi}^{\pi}\exp\left(i\eta s\right)A_{j}\left(u,\,\eta\right)\overline{A_{j}\left(u,\,\eta\right)}d\eta\right]\right|^{2}d\omega\\
 & =\int_{-\pi}^{\pi}\left|\frac{1}{2\pi}\sum_{s=-\infty}^{\infty}\exp\left(-i\omega s\right)\right.\\
 & \quad\left.\left[\int_{-\pi}^{\pi}\exp\left(i\eta s\right)\left(A_{j,\left\lfloor Tu-3\left|s\right|/2\right\rfloor ,T}^{0}\left(\eta\right)\overline{A_{j,\left\lfloor Tu-\left|s\right|/2\right\rfloor ,T}^{0}\left(\eta\right)}-A_{j}\left(u,\,\eta\right)\overline{A_{j}\left(u,\,\eta\right)}\right)d\eta\right]\right|^{2}d\omega\\
 & =\frac{1}{2\pi}\sum_{s=-\infty}^{\infty}\left|c_{s,j}\right|^{2}+o\left(1\right),
\end{align*}
 with $c_{s,j}=\int_{-\pi}^{\pi}\exp\left(i\eta s\right)G_{j}\left(s/2T,\,\eta\right)d\eta$
and 
\begin{align*}
G_{j}\left(\frac{s}{2T},\,\eta\right) & =A_{j}\left(u-\frac{3\left|s\right|}{2T},\,\eta\right)A_{j}\left(u-\frac{\left|s\right|}{2T},\,-\eta\right)-A_{j}\left(u,\,\eta\right)A_{j}\left(u,\,-\eta\right).
\end{align*}
 Using the definition of $\Delta_{s}\left(\omega\right)$ and the
above-mentioned properties of $c_{s,j}$ which continue to hold, 
summation by parts and the Lipschitz continuity of $A_{j}\left(u,\,\cdot\right)$
then imply $\sum_{s=0}^{n-1}\left|c_{s,j}\right|^{2}=O(n\ln n/T^{\vartheta})$.
Since the same bound applies to $\sum_{s=n}^{\infty}\left|c_{-s,j}\right|^{2}$,
we can choose an appropriate $n$ to yield the result for $Tu\in\mathcal{T}$.
$\square$

\subsection{Proofs of the Results of Section \ref{Section Optimal-Kernels-and}}

\subsubsection{Proof of Proposition \ref{Proposition: Optimal Local Covariance}}

We first need to show that $\sqrt{Tb_{2,T}}\left(\widehat{c}_{T}\left(rn_{T}/T,\,k\right)-\widetilde{c}\left(rn_{T}/T,\,k\right)\right)=o_{\mathbb{P}}\left(1\right).$
Without loss of generality, we can focus on the scalar case. From
\eqref{Eq. A.10 Andrews 91}, $\left\Vert \frac{\partial}{\partial\beta'}\widehat{c}_{T}\left(rn_{T}/T,\,k\right)\right\Vert |_{\beta=\bar{\beta}}=O_{\mathbb{P}}\left(1\right).$
A mean-value Taylor's expansion gives 
\begin{align*}
\sqrt{Tb_{2,T}}\left(\widehat{c}_{T}\left(rn_{T}/T,\,k\right)-\widetilde{c}_{T}\left(rn_{T}/T,\,k\right)\right) & =\sqrt{b_{2,T}}\frac{\partial}{\partial\beta'}\widehat{c}_{T}\left(rn_{T}/T,\,k\right)|_{\beta=\bar{\beta}}\sqrt{T}\left(\widehat{\beta}-\beta_{0}\right)\\
 & \leq\sqrt{b_{2,T}}\sup_{r\geq1}\left\Vert \frac{\partial}{\partial\beta'}\widehat{c}\left(rn_{T}/T,\,k\right)\right\Vert |_{\beta=\bar{\beta}}\sqrt{T}\left(\widehat{\beta}-\beta_{0}\right)\\
 & =\sqrt{b_{2,T}}O_{\mathbb{P}}\left(1\right)=o_{\mathbb{P}}\left(1\right).
\end{align*}
 Thus, 
\begin{align*}
\xi_{T} & =\mathrm{vec}\left(\widehat{c}_{T}\left(rn_{T}/T,\,k\right)-\widetilde{c}\left(rn_{T}/T,\,k\right)\right)'\widetilde{W}_{T}\mathrm{vec}\left(\widehat{c}_{T}\left(rn_{T}/T,\,k\right)-\widetilde{c}\left(rn_{T}/T,\,k\right)\right)\overset{\mathbb{P}}{\rightarrow}0.
\end{align*}
Since $\xi_{T}$ is a bounded sequence, $\mathbb{E}\left(\xi_{T}\right)\overset{\mathbb{P}}{\rightarrow}0$.
Hence, given that $\widetilde{W}_{T}\overset{\mathbb{P}}{\rightarrow}\widetilde{W}$,
we have $\mathrm{MSE}(1,\,\widehat{c}_{T}\left(u_{0},\,k\right),\,\widetilde{W}_{T})=\mathrm{MSE}(1,\,\widetilde{c}_{T}\left(u_{0},\,k\right),\,\widetilde{W})+o_{\mathbb{P}}\left(1\right)$.
By using the results of Lemma \ref{Lemma Bias, MSE of c(u,k) Breaks SLS},
the MSE of $\widehat{c}_{T}\left(u_{0},\,k\right)$ for any $u_{0}\in\left(0,\,1\right)$
and any integer $k$, is given by

\begin{align}
\mathbb{E} & \left[\widehat{c}_{T}\left(u_{0},\,k\right)-c\left(u_{0},\,k\right)\right]^{2}\nonumber \\
 & =\frac{1}{4}b_{2,T}^{4}\left(\int_{0}^{1}x^{2}K_{2}\left(x\right)dx\right)^{2}\left(\frac{\partial^{2}}{\partial^{2}u}c\left(u_{0},\,k\right)\right)^{2}\nonumber \\
 & \quad+\frac{1}{Tb_{2,T}}\int_{0}^{1}K_{2}^{2}\left(x\right)dx\sum_{l=-\infty}^{\infty}c\left(u_{0},\,l\right)\left[c\left(u_{0},\,l\right)+c\left(u_{0},\,l+2k\right)\right]\nonumber \\
 & \quad+\frac{1}{Tb_{2,T}}\int_{0}^{1}K_{2}^{2}\left(x\right)dx\sum_{h_{1}=-\infty}^{\infty}\kappa_{V,\left\lfloor Tu_{0}\right\rfloor }\left(-k,\,h_{1},\,h_{1}-k\right)+o\left(b_{2,T}^{4}\right)+o\left(1/\left(b_{2,T}T\right)\right)\nonumber \\
 & \triangleq g\left(K_{2},\,b_{2,T}\right)+o\left(b_{2,T}^{4}\right)+o\left(1/\left(b_{2,T}T\right)\right).\label{eq. Optimal MSE- Scalar case}
\end{align}
Then $g\left(K_{2},\,b_{2,T}\right)=4^{-1}b_{2,T}^{4}H\left(K_{2}\right)D_{1}\left(u_{0}\right)+\left(Tb_{2,T}\right)^{-1}F\left(K_{2}\right)\left(D_{2}\left(u_{0}\right)+D_{3}\left(u_{0}\right)\right)$.
The minimum of $g\left(K_{2},\,b_{2,T}\right)$ in $b_{2,T}$ is determined
by the equation 
\begin{align*}
\frac{\partial}{\partial b_{2,T}}g\left(K_{2},\,b_{2,T}\right) & =b_{2,T}^{3}H\left(K_{2}\right)D_{1}\left(u_{0}\right)-\frac{1}{Tb_{2,T}^{2}}F\left(K_{2}\right)\left(D_{2}\left(u_{0}\right)+D_{3}\left(u_{0}\right)\right)=0.
\end{align*}
The minimum is achieved at 
\[
b_{2,T}^{\mathrm{opt}}=\left[H\left(K_{2}\right)D_{1}\left(u_{0}\right)\right]^{-1/5}\left(F\left(K_{2}\right)\left(D_{2}\left(u_{0}\right)+D_{3}\left(u_{0}\right)\right)\right)^{1/5}T^{-1/5}.
\]
 If $V_{t,T}$ is Gaussian, then the term involving $\kappa_{V,\left\lfloor Tu_{0}\right\rfloor }$
in \eqref{eq. Optimal MSE- Scalar case} is equal to zero and so $D_{3}\left(u_{0}\right)=0$
in $b_{2,T}^{\mathrm{opt}}.$ Next, we minimize $g(K_{2},\,b_{2,T}^{\mathrm{opt}})$
with respect to the class of kernels $K_{2}:\,\mathbb{R}\rightarrow[0,\,\infty]$
that are centered at $x=1/2$ with 
\begin{align}
\int_{\mathbb{R}}K_{2}\left(x\right)dx & =1,\label{eq: Condition 2.17 on Kernel}\\
K_{2}\left(x\right) & =K_{2}\left(1-x\right).\label{Eq: Condition 2.17 on Kernel b}
\end{align}
We use arguments similar to those in Chapter 7 of \citeReferencesSupptwo{priestley:85}
and in \citeReferencesSupptwo{Dahlhaus/Giraitis:98}. Let 
\begin{align*}
\sqrt{K_{2\sigma}\left(x\right)} & =\frac{1}{\sqrt{\sigma}}\left(K_{2}\left(\frac{x-1/2}{\sigma}+\frac{1}{2}\right)\right)^{1/2},\qquad\mathrm{where}\,\sigma\in\left(0,\,\infty\right).
\end{align*}
We have $F\left(K_{2\sigma}\right)=\left(1/\sigma\right)F\left(K_{2}\right)$
and $H\left(K_{2\sigma}\right)=\sigma^{4}H\left(K_{2}\right)$ (with
the integrals in the definition of $F$ and $H$ extended to $\mathbb{R}$
and with the variable of integration $x$ subtracted by $1/2$).
Then, $b_{2,K_{2\sigma},T}^{\mathrm{opt}}=\sigma^{-1}b_{2,T}^{\mathrm{opt}}$
where $b_{2,K_{2\sigma},T}^{\mathrm{opt}}$ is the optimal bandwidth
associated with the kernel $K_{2\sigma}.$ Also, $g(K_{2\sigma},\,b_{2,K_{2\sigma},T}^{\mathrm{opt}})=g(K_{2},\,b_{2,T}^{\mathrm{opt}})$.
We can thus restrict our attention to $K_{2}$ satisfying   
\begin{align}
\int_{\mathbb{R}}\left(x-\frac{1}{2}\right)^{2}K_{2}\left(x\right)dx & =\int_{\mathbb{R}}\left(x-\frac{1}{2}\right)^{2}K_{2}^{\mathrm{opt}}\left(x\right)dx,\label{Eq: 2.19}
\end{align}
where $K_{2}^{\mathrm{opt}}\left(x\right)=6x\left(1-x\right)$ for
$x\in\left[0,\,1\right]$ and $K_{2}^{\mathrm{opt}}\left(x\right)=0$
for $x\notin\left[0,\,1\right]$. Therefore, we have to show that,
for any $K_{2}$ that satisfies \eqref{eq: Condition 2.17 on Kernel}-\eqref{Eq: Condition 2.17 on Kernel b},
\begin{align*}
\int_{\mathbb{R}/\left[0,\,1\right]}K_{2}^{2}\left(x\right)dx+\int_{0}^{1}K_{2}^{2}\left(x\right)dx=\int_{\mathbb{R}}K_{2}^{2}\left(x\right)dx & \geq\int_{\mathbb{R}}\left(K_{2}^{\mathrm{opt}}\left(x\right)\right)^{2}dx=\int_{0}^{1}\left(K_{2}^{\mathrm{opt}}\left(x\right)\right)^{2}dx.
\end{align*}
This is implied by 
\begin{align*}
\int_{0}^{1}K_{2}^{2}\left(x\right)dx & \geq\int_{0}^{1}\left(K_{2}^{\mathrm{opt}}\left(x\right)\right)^{2}dx.
\end{align*}
Let $K_{2}\left(x\right)=K_{2}^{\mathrm{opt}}\left(x\right)+\varepsilon\left(x\right)$,
$x\in\mathbb{R},$ where $\varepsilon>0.$ Since $\int_{\mathbb{R}}\varepsilon^{2}\left(x\right)dx\geq0$
and $K_{2}^{\mathrm{opt}}$ vanishes outside $\left[0,\,1\right]$,
it is sufficient to prove that $\int_{0}^{1}\left(K_{2}^{\mathrm{opt}}\left(x\right)\varepsilon\left(x\right)\right)dx\geq0$
because 
\begin{align*}
\int_{0}^{1}K_{2}^{2}\left(x\right)dx=\int_{0}^{1}\left(K_{2}^{\mathrm{opt}}\left(x\right)+\varepsilon\left(x\right)\right)^{2}dx & \geq\int_{0}^{1}\left(K_{2}^{\mathrm{opt}}\left(x\right)\right)^{2}+2\int_{0}^{1}\left(K_{2}^{\mathrm{opt}}\left(x\right)\varepsilon\left(x\right)\right)dx.
\end{align*}
By \eqref{eq: Condition 2.17 on Kernel}, we have $\int_{\mathbb{R}}\varepsilon\left(x\right)dx=0$,
while  $\int_{\mathbb{R}}\varepsilon\left(x\right)\left(x^{2}-x\right)dx=0$
in view of
\begin{align*}
0=\int_{\mathbb{R}}\left(K_{2}\left(x\right)-K_{2}^{\mathrm{opt}}\left(x\right)\right)\left(x-\frac{1}{2}\right)^{2}dx & =\int_{\mathbb{R}}\left(K_{2}\left(x\right)-K_{2}^{\mathrm{opt}}\left(x\right)\right)\left(x^{2}-x\right)dx+\frac{1}{4}\int_{\mathbb{R}}\varepsilon\left(x\right)dx\\
 & =\int_{\mathbb{R}}\left(K_{2}\left(x\right)-K_{2}^{\mathrm{opt}}\left(x\right)\right)\left(x^{2}-x\right)dx.
\end{align*}
 Note that  $\left(x^{2}-x\right)=x\left(x-1\right).$ Therefore,
we deduce 
\begin{align*}
6 & \int_{\mathbb{R}/\left[0,\,1\right]}x\left(1-x\right)\varepsilon\left(x\right)dx+6\int_{0}^{1}x\left(1-x\right)\varepsilon\left(x\right)dx=0.
\end{align*}
Rearranging the last expression yields,
\begin{align*}
\int_{0}^{1}K_{2}^{\mathrm{opt}}\left(x\right)\varepsilon\left(x\right)dx & =6\int_{\mathbb{R}/\left[0,\,1\right]}x\left(x-1\right)\varepsilon\left(x\right)dx\geq0,
\end{align*}
 because $\varepsilon\left(x\right)\geq0$ and $x\left(x-1\right)\geq0$
for $x\notin\left[0,\,1\right]$. $\square$

\subsubsection{Proof of Theorem \ref{Theorem Optimal Kernels}}

Without loss of generality, we provide the proof for the scalar case.
By Theorem \ref{Theorem 1 -Consistency and Rate}-(iii), if $Tb_{1,T}^{2q+1}b_{2,T}\rightarrow\gamma_{2}\in\left(0,\,\infty\right)$
for some $q\in[0,\,\infty)$ for which $K_{1,q},\,|\int_{0}^{1}f^{\left(q\right)}\left(u,\,0\right)du|\in[0,\,\infty)$,
then 
\begin{align*}
\lim_{T\rightarrow\infty} & \mathrm{MSE}\left(Tb_{1,T}b_{2,T},\,\widehat{J}_{T}^{\mathrm{}}\left(b_{1,T,K_{1}}\right),\,1\right)\\
 & =4\pi^{2}\left[\gamma_{2}K_{1,q}^{2}\left(\int_{0}^{1}f^{\left(q\right)}\left(u,\,0\right)du\right)^{2}+\int K_{1}^{2}\left(y\right)dy\int_{0}^{1}\left(K_{2}\left(x\right)\right)^{2}dx\,\left(\int_{0}^{1}f\left(u,\,0\right)du\right)^{2}\right].
\end{align*}
 We have $Tb_{1,T}^{5}b_{2,T}\rightarrow\gamma$ by assumption. Thus,
we apply Theorem \ref{Theorem 1 -Consistency and Rate}-(iii) with
$q=2$, $K_{1}$ and $b_{1,T,K_{1}}.$ Then, $Tb_{1,T,K_{1}}^{5}b_{2,T}\rightarrow\gamma/\left(\int K_{1}^{2}\left(x\right)dx\right)^{5}$
and 
\begin{align*}
Tb_{1,T}b_{2,T}=Tb_{1,T,K_{1}}b_{2,T}\int K_{1}^{2}\left(x\right)dx & .
\end{align*}
 Therefore, given $K_{1,2}<\infty$, 

\begin{align*}
\lim_{T\rightarrow\infty} & \left(\mathrm{MSE}\left(Tb_{1,T}b_{2,T},\,\widehat{J}_{T}^{\mathrm{}}\left(b_{1,T,K_{1}}\right),\,1\right)-\mathrm{MSE}\left(Tb_{1,T}b_{2,T},\,\widehat{J}_{T}^{\mathrm{QS}}\left(b_{1,T}\right),\,1\right)\right)\\
 & =4\gamma\pi^{2}\left(\int_{0}^{1}f^{\left(q\right)}\left(u,\,0\right)du\right)^{2}\int_{0}^{1}\left(K_{2}^{\mathrm{}}\left(x\right)\right)^{2}dx\left[K_{1,2}^{2}\left(\int K_{1}^{2}\left(y\right)dy\right)^{4}-\left(K_{1,2}^{\mathrm{QS}}\right)^{2}\right].
\end{align*}
Let $\widetilde{K}_{1}\left(\cdot\right)$ and $\widetilde{K}_{1}^{\mathrm{QS}}\left(\cdot\right)$
denote the spectral window generators of $K_{1}\left(\cdot\right)$
and $K_{1}^{\mathrm{QS}}\left(\cdot\right)$, respectively. They have
the following properties: $K_{1,2}=\int_{-\infty}^{\infty}\omega^{2}\widetilde{K}_{1}\left(\omega\right)d\omega$,
$K_{1}\left(0\right)=\int_{-\infty}^{\infty}\widetilde{K}_{1}\left(\omega\right)d\omega$,
and $\int_{-\infty}^{\infty}K_{1}^{2}\left(x\right)dx=\int_{-\infty}^{\infty}\widetilde{K}_{1}^{2}\left(\omega\right)d\omega$.
As in \citeReferencesSupptwo{andrews:91}, the result of the theorem
follows if we can show the following inequality,
\begin{align}
K_{1,2}^{2}\left(\int K_{1}^{2}\left(x\right)dx\right)^{4} & \geq\left(K_{1,2}^{\mathrm{QS}}\right)^{2}\qquad\qquad\mathrm{for}\,\mathrm{all}\,K_{1}\left(\cdot\right)\in\widetilde{\boldsymbol{K}}_{1}.\label{eq (A.20) in A91}
\end{align}
\citeauthor{priestley:85} (1981, Ch. 7.5) showed that $\widetilde{K}_{1}^{\mathrm{QS}}\left(\cdot\right)$
minimizes 
\begin{align}
\int_{-\infty}^{\infty}\omega^{2}\widetilde{K}_{1}\left(\omega\right)d\omega & \left(\int_{-\infty}^{\infty}\widetilde{K}_{1}^{2}\left(\omega\right)d\omega\right)^{2},\label{Eq (A.21) in Andrews 91}
\end{align}
subject to (a) $\int_{-\infty}^{\infty}\widetilde{K}_{1}\left(\omega\right)d\omega=1$,
(b) $\widetilde{K}_{1}\left(\omega\right)\geq0,\,\forall\,\omega\in\mathbb{R}$,
and (c) $\widetilde{K}_{1}\left(\omega\right)=\widetilde{K}_{1}\left(-\omega\right),\,\forall\,\omega\in\mathbb{R}$,
where $K_{1}^{\mathrm{QS}}\left(\omega\right)=\left(5/8\pi\right)\left(1-\omega^{2}/c^{2}\right)$
for $\left|\omega\right|\leq c$ for $c=6\pi/5$. and $K_{1}^{\mathrm{QS}}\left(\omega\right)=0$
otherwise. Note that the inequality \eqref{eq (A.20) in A91} holds
if and only if $\widetilde{K}_{1}^{\mathrm{QS}}\left(\cdot\right)$
minimizes \eqref{Eq (A.21) in Andrews 91}. This proves the inequality
of the theorem. Strict inequality holds when $K_{1}^{\mathrm{QS}}\left(x\right)\neq K_{1}\left(x\right)$
with positive Lebesgue measure. $\square$

\subsubsection{Proof of Corollary \ref{Corollary 1 -Optimal b1 }}

Note that $T^{\frac{2q}{2q+1}}b_{2,T}^{\frac{2q}{2q+1}}=(Tb_{1,T}^{2q+1}b_{2,T})^{-1/\left(2q+1\right)}Tb_{1,T}b_{2,T}=(\gamma^{-1/\left(2q+1\right)}+o\left(1\right))Tb_{1,T}b_{2,T}$.
Thus,
\begin{align}
\lim_{T\rightarrow\infty} & \mathrm{MSE}\left(T^{\frac{2q}{2q+1}}b_{2,T}^{\frac{2q}{2q+1}},\,\widehat{J}_{T}\left(b_{1,T},\,b_{2,T}\right),\,W_{T}\right)\nonumber \\
 & =\gamma^{-1/\left(2q+1\right)}4\pi^{2}\left[\gamma K_{1,q}^{2}\mathrm{vec}\left(\int_{0}^{1}f^{\left(q\right)}\left(u,\,0\right)du\right)'W\mathrm{vec}\left(\int_{0}^{1}f^{\left(q\right)}\left(u,\,0\right)du\right)\right.\label{Eq (A.22) Andrews 91}\\
 & \quad\left.+\int K_{1}^{2}\left(y\right)dy\int_{0}^{1}K_{2}^{2}\left(x\right)dx\,\mathrm{tr}W\left(I_{p^{2}}-C_{pp}\right)\left(\int_{0}^{1}f\left(u,\,0\right)du\right)\otimes\left(\int_{0}^{1}f\left(v,\,0\right)dv\right)\right].\nonumber 
\end{align}
 Minimizing this with respect to $\gamma$ gives
\begin{align*}
\gamma^{2q/\left(2q+1\right)} & K_{1,q}^{2}\mathrm{vec}\left(\int_{0}^{1}f^{\left(q\right)}\left(u,\,0\right)du\right)'W\mathrm{vec}\left(\int_{0}^{1}f^{\left(q\right)}\left(u,\,0\right)du\right)\\
 & =\gamma^{-1/\left(2q+1\right)}\int K_{1}^{2}\left(y\right)dy\int K_{2}^{2}\left(x\right)dx\,\mathrm{tr}W\left(I_{p^{2}}-C_{pp}\right)\left(\int_{0}^{1}f\left(u,\,0\right)du\right)\otimes\left(\int_{0}^{1}f\left(v,\,0\right)dv\right),
\end{align*}
or 
\begin{align*}
\gamma^{\mathrm{opt}} & =\frac{1}{2q}\frac{\int K_{1}^{2}\left(y\right)dy\int K_{2}^{2}\left(x\right)dx\,\mathrm{tr}W\left(I_{p^{2}}+C_{pp}\right)\left(\int_{0}^{1}f\left(u,\,0\right)du\right)\otimes\left(\int_{0}^{1}f\left(v,\,0\right)dv\right)}{K_{1,q}^{2}\mathrm{vec}\left(\int_{0}^{1}f^{\left(q\right)}\left(u,\,0\right)du\right)'W\mathrm{vec}\left(\int_{0}^{1}f^{\left(q\right)}\left(u,\,0\right)du\right)}\\
 & =\left(2qK_{1,q}^{2}\phi\left(q\right)\right)^{-1}\left(\int K_{1}^{2}\left(y\right)dy\int_{0}^{1}K_{2}^{2}\left(x\right)dx\right).
\end{align*}
Note that $\gamma^{\mathrm{opt}}>0$ provided that $0<\phi\left(q\right)<\infty$
and $W$ is positive definite. Hence, $\{b_{1,T}\}$ is optimal in
the sense that $Tb_{1,T}^{2q+1}b_{2,T}\rightarrow\gamma^{\mathrm{opt}}$
if and only if $b_{1,T}=b_{1,T}^{\mathrm{opt}}+o((Tb_{2,T})^{-1/\left(2q+1\right)})$
where $b_{2,T}=O(b_{2,T}^{\mathrm{opt}})$. $\square$

\subsection{Proofs of the Results of Section \ref{Section Data-Dependent-Bandwidths}}

\subsubsection{Proof of Theorem \ref{Theorem 3 Andrews 91}}

Without loss of generality, we assume that $V_{t}$ is a scalar. The
constant $C<\infty$ may vary from line to line. We begin with the
proof of part (ii) because it becomes then simpler to prove part (i).
By Theorem \ref{Theorem 1 -Consistency and Rate}-(ii), $\sqrt{Tb_{\theta_{1},T}b_{\theta_{2},T}}(\widehat{J}_{T}(b_{\theta_{1},T},\,b_{\theta_{2},T})-J_{T})=O_{\mathbb{P}}\left(1\right)$.
It remains to establish the second result of Theorem \ref{Theorem 3 Andrews 91}-(ii).
Let $S_{T}=\left\lfloor b_{\theta_{1},T}^{-r}\right\rfloor $ where
\begin{align*}
r\in & (\max\{\left(8b-5-2q\right)/8\left(b-1\right),\,1.25,\,\left(b/2-1/4\right)/\left(b-1\right),\,q/\left(l-1\right),\,\left(8b-7-6q\right)/8\left(b-1\right)\\
 & \,(b-3/4-q/2)/\left(b-1\right)\},\,\min\left\{ 13q/24+49/48,\,46/48+20q/48,\,7/8+3q/4,\,\left(6+4q\right)/8,\,2\right\} ),
\end{align*}
with $b>1+1/q$ . We will use the following decomposition 
\begin{align}
\widehat{J}_{T}(\widehat{b}_{1,T},\,\widehat{\overline{b}}_{2,T})-\widehat{J}_{T}(b_{\theta_{1},T},\,b_{\theta_{2},T}) & =(\widehat{J}_{T}(\widehat{b}_{1,T},\,\widehat{\overline{b}}_{2,T})-\widehat{J}_{T}(b_{\theta_{1},T},\,\widehat{\overline{b}}_{2,T}))\label{eq (Decomposition J_T proof of Theorem 3 Andrews91)}\\
 & \quad+(\widehat{J}_{T}(b_{\theta_{1},T},\,\widehat{\overline{b}}_{2,T})-\widehat{J}_{T}(b_{\theta_{1},T},\,b_{\theta_{2},T})).\nonumber 
\end{align}
Let 
\begin{align*}
N_{1} & \triangleq\left\{ -S_{T},\,-S_{T}+1,\ldots,\,-1,\,1,\ldots,\,S_{T}-1,\,S_{T}\right\} ,\\
N_{2} & \triangleq\left\{ -T+1,\ldots,\,-S_{T}-1,\,S_{T}+1,\ldots,\,T-1\right\} .
\end{align*}
Let us consider the first term of \eqref{eq (Decomposition J_T proof of Theorem 3 Andrews91)}.
We have 
\begin{align}
T^{8q/10\left(2q+1\right)} & (\widehat{J}_{T}(\widehat{b}_{1,T},\,\widehat{\overline{b}}_{2,T})-\widehat{J}_{T}(b_{\theta_{1},T},\,\widehat{\overline{b}}_{2,T}))\label{Eq. 23}\\
 & =T^{8q/10\left(2q+1\right)}\sum_{k\in N_{1}}(K_{1}(\widehat{b}_{1,T}k)-K_{1}(b_{\theta_{1},T}k))\widehat{\Gamma}\left(k\right)\nonumber \\
 & \quad+T^{8q/10\left(2q+1\right)}\sum_{k\in N_{2}}K_{1}(\widehat{b}_{1,T}k)\widehat{\Gamma}\left(k\right)\nonumber \\
 & \quad-T^{8q/10\left(2q+1\right)}\sum_{k\in N_{2}}K_{1}(b_{\theta_{1},T}k)\widehat{\Gamma}\left(k\right)\nonumber \\
 & \triangleq A_{1,T}+A_{2,T}-A_{3,T}.\nonumber 
\end{align}
 We first show that $A_{1,T}\overset{\mathbb{P}}{\rightarrow}0$.
Let $A_{1,1,T}$ denote $A_{1,T}$ with the summation restricted over
positive integers $k$. Let $\widetilde{n}_{T}=\inf\{T/n_{3,T},\,\sqrt{n_{2,T}}\}$.
We can use the Liptchitz condition on $K_{1}\left(\cdot\right)\in\boldsymbol{K}_{3}$
to yield, 
\begin{align}
\left|A_{1,1,T}\right| & \leq T^{8q/10\left(2q+1\right)}\sum_{k=1}^{S_{T}}C_{2}\left|\widehat{b}_{1,T}-b_{\theta_{1},T}\right|k\left|\widehat{\Gamma}\left(k\right)\right|\label{Eq. 24}\\
 & \leq C\widetilde{n}_{T}\left|\widehat{\phi}\left(q\right)^{1/\left(2q+1\right)}-\phi_{\theta^{*}}^{1/\left(2q+1\right)}\right|\left(\widehat{\phi}\left(q\right)\phi_{\theta^{*}}\right)^{-1/\left(2q+1\right)}\widehat{\overline{b}}_{2,T}^{-1/\left(2q+1\right)}T^{\left(8q-10\right)/10\left(2q+1\right)}\widetilde{n}_{T}^{-1}\sum_{k=1}^{S_{T}}k\left|\widehat{\Gamma}\left(k\right)\right|,\nonumber 
\end{align}
for some $C<\infty$. By Assumption \ref{Assumption E-F-G}-(ii)
($\widetilde{n}_{T}\left|\widehat{\phi}\left(q\right)-\phi_{\theta^{*}}\right|=O_{\mathbb{P}}\left(1\right)$)
and, using the delta method, it suffices to show that $B_{1,T}+B_{2,T}+B_{3,T}\overset{\mathbb{P}}{\rightarrow}0,$
where 
\begin{align}
B_{1,T} & =\widehat{\overline{b}}_{2,T}^{-1/\left(2q+1\right)}T^{\left(8q-10\right)/10\left(2q+1\right)}\widetilde{n}_{T}^{-1}\sum_{k=1}^{S_{T}}k\left|\widehat{\Gamma}\left(k\right)-\widetilde{\Gamma}\left(k\right)\right|,\label{Eq. A.25 Andrews 91}\\
B_{2,T} & =\widehat{\overline{b}}_{2,T}^{-1/\left(2q+1\right)}T^{\left(8q-10\right)/10\left(2q+1\right)}\widetilde{n}_{T}^{-1}\sum_{k=1}^{S_{T}}k\left|\widetilde{\Gamma}\left(k\right)-\Gamma_{T}\left(k\right)\right|,\nonumber \\
B_{3,T} & =\widehat{\overline{b}}_{2,T}^{-1/\left(2q+1\right)}T^{\left(8q-10\right)/10\left(2q+1\right)}\widetilde{n}_{T}^{-1}\sum_{k=1}^{S_{T}}k\left|\Gamma_{T}\left(k\right)\right|,\nonumber 
\end{align}
with $\Gamma_{T}\left(k\right)\triangleq\left(n_{T}/T\right)\sum_{r=0}^{\left\lfloor T/n_{T}\right\rfloor }c\left(rn_{T}/T,\,k\right).$
By a mean-value expansion, we have 
\begin{align}
B_{1,T} & \leq\widehat{\overline{b}}_{2,T}^{-1/\left(2q+1\right)}T^{\left(8q-10\right)/10\left(2q+1\right)}\widetilde{n}_{T}^{-1}T^{-1/2}\sum_{k=1}^{S_{T}}k\left|\left(\frac{\partial}{\partial\beta'}\widehat{\Gamma}\left(k\right)|_{\beta=\overline{\beta}}\right)\sqrt{T}\left(\widehat{\beta}-\beta_{0}\right)\right|\label{Eq. A.26}\\
 & \leq C\widehat{\overline{b}}_{2,T}^{-1/\left(2q+1\right)}T^{\left(8q-10\right)/10\left(2q+1\right)-1/2}\left(Tb_{\theta_{2,T}}\right)^{2r/\left(2q+1\right)}\widetilde{n}_{T}^{-1}\sup_{k\geq1}\left\Vert \frac{\partial}{\partial\beta}\widehat{\Gamma}\left(k\right)|_{\beta=\overline{\beta}}\right\Vert \sqrt{T}\left\Vert \widehat{\beta}-\beta_{0}\right\Vert \nonumber \\
 & \leq C\widehat{\overline{b}}_{2,T}^{\left(-1+2r\right)/\left(2q+1\right)}T^{\left(8q-10\right)/10\left(2q+1\right)-1/2+2r/\left(2q+1\right)}\widetilde{n}_{T}^{-1}\sup_{k\geq1}\left\Vert \frac{\partial}{\partial\beta}\widehat{\Gamma}\left(k\right)|_{\beta=\overline{\beta}}\right\Vert \sqrt{T}\left\Vert \widehat{\beta}-\beta_{0}\right\Vert \overset{\mathbb{P}}{\rightarrow}0,\nonumber 
\end{align}
since $\widetilde{n}_{T}/T^{1/3}\rightarrow\infty$, $r<13q/24+49/48$,
$\sqrt{T}||\widehat{\beta}-\beta_{0}||=O_{\mathbb{P}}\left(1\right)$,
and $\sup_{k\geq1}||\left(\partial/\partial\beta\right)\widehat{\Gamma}\left(k\right)|_{\beta=\overline{\beta}}||=O_{\mathbb{P}}\left(1\right)$
using \eqref{Eq. A.10 Andrews 91} and Assumption \ref{Assumption B}-(ii,iii).
In addition, 
\begin{align}
\mathbb{E}\left(B_{2,T}^{2}\right) & \leq\mathbb{E}\left(\widehat{\overline{b}}_{2,T}^{-2/\left(2q+1\right)}T^{\left(8q-10\right)/5\left(2q+1\right)}\widetilde{n}_{T}^{-2}\sum_{k=1}^{S_{T}}\sum_{j=1}^{S_{T}}kj\left|\widetilde{\Gamma}\left(k\right)-\Gamma_{T}\left(k\right)\right|\left|\widetilde{\Gamma}\left(j\right)-\Gamma_{T}\left(j\right)\right|\right)\label{Eq. A.27}\\
 & \leq\widehat{\overline{b}}_{2,T}^{-2/\left(2q+1\right)-1}T^{\left(8q-10\right)/5\left(2q+1\right)-2/3-1}S_{T}^{4}\sup_{k\geq1}T\widehat{\overline{b}}_{2,T}\mathrm{Var}\left(\widetilde{\Gamma}\left(k\right)\right)\nonumber \\
 & \leq\widehat{\overline{b}}_{2,T}^{-2/\left(2q+1\right)-1}T^{\left(8q-10\right)/5\left(2q+1\right)-2/3-1}\left(Tb_{\theta_{2},T}\right)^{4r/\left(2q+1\right)}\sup_{k\geq1}T\widehat{\overline{b}}_{2,T}\mathrm{Var}\left(\widetilde{\Gamma}\left(k\right)\right)\nonumber \\
 & \leq T^{1/5}T^{2/5\left(2q+1\right)}T^{\left(8q-10\right)/5\left(2q+1\right)-2/3-1}T^{4r/\left(2q+1\right)}T^{-4r/5\left(2q+1\right)}\sup_{k\geq1}T\widehat{\overline{b}}_{2,T}\mathrm{Var}\left(\widetilde{\Gamma}\left(k\right)\right)\rightarrow0,\nonumber 
\end{align}
  given that $\sup_{k\geq1}T\widehat{\overline{b}}_{2,T}\mathrm{Var}(\widetilde{\Gamma}(k))=O\left(1\right)$
using Lemma \ref{Lemma: Tb2*Var(Gamma_k)} and $r<46/48+20q/48$.
Assumption \ref{Assumption E-F-G}-(iii) and $\sum_{k=1}^{\infty}k^{1-l}<\infty$
for $l>2$ yield 
\begin{align}
B_{3,T} & \leq\widehat{\overline{b}}_{2,T}^{-1/\left(2q+1\right)}T^{\left(8q-10\right)/10\left(2q+1\right)}\widetilde{n}_{T}^{-1}C_{3}\sum_{k=1}^{\infty}k^{1-l}\rightarrow0,\label{Eq. A.28}
\end{align}
 where we have used $\widetilde{n}_{T}/T^{3/10}\rightarrow\infty$
and $q<34/4.$  Combining \eqref{Eq. 24}-\eqref{Eq. A.28}, we deduce
that $A_{1,1,T}\overset{\mathbb{P}}{\rightarrow}0$. The same argument
applied to $A_{1,T}$, where the summation now extends over negative
integers $k$, gives $A_{1,T}\overset{\mathbb{P}}{\rightarrow}0$.
Next, we show that $A_{2,T}\overset{\mathbb{P}}{\rightarrow}0$. Again,
we use the notation $A_{2,1,T}$ (resp., $A_{2,2,T}$) to denote $A_{2,T}$
with the summation over positive (resp., negative) integers. Let $A_{2,1,T}=L_{1,T}+L_{2,T}+L_{3,T}$,
where 
\begin{align}
L_{1,T} & =T^{8q/10\left(2q+1\right)}\sum_{k=S_{T}+1}^{T-1}K_{1}\left(\widehat{b}_{1,T}k\right)\left(\widehat{\Gamma}\left(k\right)-\widetilde{\Gamma}\left(k\right)\right),\label{Eq. A.29}\\
L_{2,T} & =L_{2,T}^{A}+L_{2,T}^{B}=T^{8q/10\left(2q+1\right)}\left(\sum_{k=S_{T}+1}^{\left\lfloor D_{T}T^{8/5\left(2q+1\right)}\right\rfloor }+\sum_{k=\left\lfloor D_{T}T^{8/5\left(2q+1\right)}\right\rfloor +1}^{T-1}\right)K_{1}\left(\widehat{b}_{1,T}k\right)\left(\widetilde{\Gamma}\left(k\right)-\Gamma_{T}\left(k\right)\right),\nonumber \\
L_{3,T} & =T^{8q/10\left(2q+1\right)}\sum_{k=S_{T}+1}^{T-1}K_{1}\left(\widehat{b}_{1,T}k\right)\Gamma_{T}\left(k\right).\nonumber 
\end{align}
We apply a mean-value expansion and use $\sqrt{T}(\widehat{\beta}-\beta_{0})=O_{\mathbb{P}}\left(1\right)$
as well as \eqref{Eq. A.10 Andrews 91} to obtain
\begin{align}
\left|L_{1,T}\right| & =T^{8q/10\left(2q+1\right)-1/2}\sum_{k=S_{T}+1}^{T-1}C_{1}\left(\widehat{b}_{1,T}k\right)^{-b}\left|\left(\frac{\partial}{\partial\beta'}\widehat{\Gamma}\left(k\right)\right)|_{\beta=\overline{\beta}}\sqrt{T}\left(\widehat{\beta}-\beta_{0}\right)\right|\label{Eq. (30)}\\
 & =T^{8q/10\left(2q+1\right)-1/2+4b/5\left(2q+1\right)}\sum_{k=S_{T}+1}^{T-1}C_{1}k^{-b}\left|\left(\frac{\partial}{\partial\beta'}\widehat{\Gamma}\left(k\right)\right)|_{\beta=\overline{\beta}}\sqrt{T}\left(\widehat{\beta}-\beta_{0}\right)\right|\nonumber \\
 & =T^{8q/10\left(2q+1\right)-1/2+4b/5\left(2q+1\right)+4r\left(1-b\right)/5\left(2q+1\right)}\left|\left(\frac{\partial}{\partial\beta'}\widehat{\Gamma}\left(k\right)\right)|_{\beta=\overline{\beta}}\sqrt{T}\left(\widehat{\beta}-\beta_{0}\right)\right|\nonumber \\
 & =T^{8q/10\left(2q+1\right)-1/2+4b/5\left(2q+1\right)+4r\left(1-b\right)/5\left(2q+1\right)}O_{\mathbb{P}}\left(1\right)O_{\mathbb{P}}\left(1\right),\nonumber 
\end{align}
 which goes to zero since $r>\left(8b-5-2q\right)/8\left(b-1\right)$.
Let us now consider $L_{2,T}$. We have 
\begin{align}
\left|L_{2,T}^{A}\right| & =T^{\left(8q-1\right)/10\left(2q+1\right)}\sum_{k=S_{T}+1}^{\left\lfloor D_{T}T^{8/5\left(2q+1\right)}\right\rfloor }C_{1}\left(\widehat{b}_{1,T}k\right)^{-b}\left|\widetilde{\Gamma}\left(k\right)-\Gamma_{T}\left(k\right)\right|\label{Eq. 31}\\
 & =C_{1}\left(2qK_{1,q}^{2}\widehat{\phi}\left(q\right)\right)^{b/\left(2q+1\right)}T^{8q/10\left(2q+1\right)+b/\left(2q+1\right)-1/2}\widehat{\overline{b}}_{2,T}^{b/\left(2q+1\right)-1/2}\left(\sum_{k=S_{T}+1}^{\left\lfloor D_{T}T^{8/5\left(2q+1\right)}\right\rfloor }k^{-b}\right)\nonumber \\
 & \quad\times\sqrt{T\widehat{\overline{b}}_{2,T}}\left|\widetilde{\Gamma}\left(k\right)-\Gamma_{T}\left(k\right)\right|.\nonumber 
\end{align}
Note that 
\begin{align}
\mathbb{E} & \left(T^{8q/10\left(2q+1\right)+b/\left(2q+1\right)-1/2}\widehat{\overline{b}}_{2,T}^{b/\left(2q+1\right)-1/2}\sum_{k=S_{T}+1}^{\left\lfloor D_{T}T^{8/5\left(2q+1\right)}\right\rfloor }k^{-b}\sqrt{T\widehat{\overline{b}}_{2,T}}\left|\widetilde{\Gamma}\left(k\right)-\Gamma_{T}\left(k\right)\right|\right)^{2}\label{Eq. 32 Andrews 91}\\
 & \leq T^{8q/5\left(2q+1\right)+2b/\left(2q+1\right)-1}\widehat{\overline{b}}_{2,T}^{2b/\left(2q+1\right)-1}\left(\sum_{k=S_{T}+1}^{\left\lfloor D_{T}T^{8/5\left(2q+1\right)}\right\rfloor }k^{-b}\sqrt{T\widehat{\overline{b}}_{2,T}}\left(\mathrm{Var}\left(\widetilde{\Gamma}\left(k\right)\right)\right)^{1/2}\right)^{2}\nonumber \\
 & =T^{8q/5\left(2q+1\right)+2b/\left(2q+1\right)-1}\widehat{\overline{b}}_{2,T}^{2b/\left(2q+1\right)-1}\left(\sum_{k=S_{T}+1}^{\left\lfloor D_{T}T^{8/5\left(2q+1\right)}\right\rfloor }k^{-b}\right)^{2}O\left(1\right)\nonumber \\
 & =T^{8q/5\left(2q+1\right)+2b/\left(2q+1\right)-1}\widehat{\overline{b}}_{2,T}^{2b/\left(2q+1\right)-1}S_{T}^{2\left(1-b\right)}O\left(1\right)\rightarrow0,\nonumber 
\end{align}
   since $r>1.25$  and $T\widehat{\overline{b}}_{2,T}\mathrm{Var}(\widetilde{\Gamma}\left(k\right))=O\left(1\right)$
as above. Next, 
\begin{align}
\left|L_{2,T}^{B}\right| & =T^{\left(8q-1\right)/10\left(2q+1\right)}\sum_{k=\left\lfloor D_{T}T^{8/5\left(2q+1\right)}\right\rfloor +1}^{T-1}C_{1}\left(\widehat{b}_{1,T}k\right)^{-b}\left|\widetilde{\Gamma}\left(k\right)-\Gamma_{T}\left(k\right)\right|\label{Eq. 31- Lb}\\
 & =C_{1}\left(2qK_{1,q}^{2}\widehat{\phi}\left(q\right)\right)^{b/\left(2q+1\right)}T^{8q/10\left(2q+1\right)+b/\left(2q+1\right)-1/2}\widehat{\overline{b}}_{2,T}^{b/\left(2q+1\right)-1/2}\left(\sum_{k=\left\lfloor D_{T}T^{8/5\left(2q+1\right)}\right\rfloor +1}^{T-1}k^{-b}\right)\nonumber \\
 & \quad\times\sqrt{T\widehat{\overline{b}}_{2,T}}\left|\widetilde{\Gamma}\left(k\right)-\Gamma_{T}\left(k\right)\right|.\nonumber 
\end{align}
Note that 
\begin{align}
\mathbb{E} & \left(T^{8q/10\left(2q+1\right)+b/\left(2q+1\right)-1/2}\widehat{b}_{2,T}^{b/\left(2q+1\right)-1/2}\sum_{k=\left\lfloor D_{T}T^{8/5\left(2q+1\right)}\right\rfloor +1}^{T-1}k^{-b}\sqrt{T\widehat{\overline{b}}_{2,T}}\left|\widetilde{\Gamma}\left(k\right)-\Gamma_{T}\left(k\right)\right|\right)^{2}\label{Eq. 32 Andrews 91 Lb}\\
 & \leq T^{8q/5\left(2q+1\right)+2b/\left(2q+1\right)-1}\widehat{b}_{2,T}^{2b/\left(2q+1\right)-1}\left(\sum_{k=\left\lfloor D_{T}T^{8/5\left(2q+1\right)}\right\rfloor +1}^{T-1}k^{-b}\sqrt{T\widehat{\overline{b}}_{2,T}}\left(\mathrm{Var}\left(\widetilde{\Gamma}\left(k\right)\right)\right)^{1/2}\right)^{2}\nonumber \\
 & =T^{8q/5\left(2q+1\right)+2b/\left(2q+1\right)-1}\widehat{b}_{2,T}^{2b/\left(2q+1\right)-1}\left(\sum_{k=\left\lfloor D_{T}T^{8/5\left(2q+1\right)}\right\rfloor +1}^{T-1}k^{-b}\right)^{2}O\left(1\right)\nonumber \\
 & =T^{8q/5\left(2q+1\right)+2b/\left(2q+1\right)-1}\widehat{b}_{2,T}^{2b/\left(2q+1\right)-1}S_{T}^{2\left(1-b\right)}T^{16\left(1-b\right)/5\left(2q+1\right)}D_{T}^{2}O\left(1\right)\rightarrow0,\nonumber 
\end{align}
 since $r>\left(b/2-1/4\right)/\left(b-1\right)$. Combining \eqref{Eq. 31}
and \eqref{Eq. 32 Andrews 91 Lb} yields $L_{2,T}\overset{\mathbb{P}}{\rightarrow}0$,
since $\widehat{\phi}\left(q\right)=O_{\mathbb{P}}\left(1\right)$.
Let us turn to $L_{3,T}$. By Assumption \ref{Assumption E-F-G}-(iii)
and $\left|K_{1}\left(\cdot\right)\right|\leq1$, we have, 
\begin{align}
\left|L_{3,T}\right| & \leq T^{8q/10\left(2q+1\right)}\sum_{k=S_{T}}^{T-1}C_{3}k^{-l}\leq T^{8q/10\left(2q+1\right)}C_{3}S_{T}^{1-l}\label{Eq. (33)}\\
 & \leq C_{3}T^{8q/10\left(2q+1\right)}T^{-4r\left(l-1\right)/5\left(2q+1\right)}\rightarrow0,\nonumber 
\end{align}
since $r>q/\left(l-1\right)$.  In view of \eqref{Eq. A.29}-\eqref{Eq. (33)},
we deduce that $A_{2,1,T}\overset{\mathbb{P}}{\rightarrow}0$. Applying
the same argument to $A_{2,2,T}$, we have $A_{2,T}\overset{\mathbb{P}}{\rightarrow}0$.
Using similar arguments, one has $A_{3,T}\overset{\mathbb{P}}{\rightarrow}0$.
It remains to show that $T^{8q/10\left(2q+1\right)}(\widehat{J}_{T}(b_{\theta_{1},T},\,\widehat{\overline{b}}_{2,T})-\widehat{J}_{T}(b_{\theta_{1},T},\,b_{\theta_{2},T}))\overset{\mathbb{P}}{\rightarrow}0$.
Let $\widehat{c}_{\theta_{2},T}\left(rn_{T}/T,\,k\right)$ denote
the estimator that uses $b_{\theta_{2},T}$ in place of $\widehat{b}_{2,T}.$
We have for $k\geq0,$ 
\begin{align}
\widehat{c}_{T} & \left(rn_{T}/T,\,k\right)-\widehat{c}_{\theta_{2},T}\left(rn_{T}/T,\,k\right)\nonumber \\
 & =\left(T\overline{b}_{\theta_{2},T}\right)^{-1}\sum_{s=k+1}^{T}\left(K_{2}\left(\frac{\left(\left(r+1\right)n_{T}-\left(s-k/2\right)\right)/T}{\widehat{b}_{2,T}\left(\left(r+1\right)n_{T}/T\right)}\right)-K_{2}\left(\frac{\left(\left(r+1\right)n_{T}-\left(s-k/2\right)\right)/T}{b_{\theta_{2},T}\left(\left(r+1\right)n_{T}/T\right)}\right)\right)\widehat{V}_{s}\widehat{V}{}_{s-k}\nonumber \\
 & \quad+O_{\mathbb{P}}\left(1/T\overline{b}_{\theta_{2},T}\right).\label{eq (c_hat - c_theta2)}
\end{align}
Given Assumption \ref{Assumption E-F-G}-(v,vi,vii) and using the
delta method, we have for $s\in\{Tu-\bigl\lfloor Tb_{\theta_{2},T}\bigr\rfloor,\ldots,\,Tu+\bigl\lfloor Tb_{\theta_{2},T}\bigr\rfloor\}$:
\begin{align}
K_{2} & \left(\frac{\left(Tu-\left(s-k/2\right)\right)/T}{\widehat{b}_{2,T}\left(u\right)}\right)-K_{2}\left(\frac{\left(Tu-\left(s-k/2\right)\right)/T}{b_{\theta_{2},T}\left(u\right)}\right)\label{Eq. K2-K2 fort part (ii)}\\
 & \leq C_{4}\left|\frac{Tu-\left(s-k/2\right)}{T\widehat{b}_{2,T}\left(u\right)}-\frac{Tu-\left(s-k/2\right)}{Tb_{\theta_{2},T}\left(u\right)}\right|\nonumber \\
 & \leq CT^{-4/5-2/5}T^{2/5}\left|\left(\frac{\widehat{D}_{2}\left(u\right)}{\widehat{D}_{1}\left(u\right)}\right)^{-1/5}-\left(\frac{D_{2}\left(u\right)}{D_{1,\theta}\left(u\right)}\right)^{-1/5}\right|\left|Tu-\left(s-k/2\right)\right|\nonumber \\
 & \leq CT^{-4/5-2/5}O_{\mathbb{P}}\left(1\right)\left|Tu-\left(s-k/2\right)\right|.\nonumber 
\end{align}
  Therefore, 
\begin{align}
T^{8q/10\left(2q+1\right)} & \left(\widehat{J}_{T}\left(b_{\theta_{1},T},\,\widehat{\overline{b}}_{2,T}\right)\right.-\left.\widehat{J}_{T}\left(b_{\theta_{1},T},\,b_{\theta_{2},T}\right)\right)\label{Eq. (H1+H2+H3)}\\
 & =T^{8q/10\left(2q+1\right)}\sum_{k=-T+1}^{T-1}K_{1}\left(b_{\theta_{1},T}k\right)\frac{n_{T}}{T}\sum_{r=0}^{\left\lfloor T/n_{T}\right\rfloor }\left(\widehat{c}\left(rn_{T}/T,\,k\right)-\widehat{c}_{\theta_{2},T}\left(rn_{T}/T,\,k\right)\right)\nonumber \\
 & \leq T^{8q/10\left(2q+1\right)}C\sum_{k=-T+1}^{T-1}\bigl|K_{1}\left(b_{\theta_{1},T}k\right)\bigr|\frac{n_{T}}{T}\sum_{r=0}^{\left\lfloor T/n_{T}\right\rfloor }\frac{1}{T\overline{b}_{\theta_{2},T}}\nonumber \\
 & \quad\times\sum_{s=k+1}^{T}\left|K_{2}\left(\frac{\left(\left(r+1\right)n_{T}-\left(s-k/2\right)\right)/T}{\widehat{b}_{2,T}\left(\left(r+1\right)n_{T}/T\right)}\right)-K_{2}\left(\frac{\left(\left(r+1\right)n_{T}-\left(s-k/2\right)\right)/T}{b_{\theta_{2},T}\left(\left(r+1\right)n_{T}/T\right)}\right)\right|\nonumber \\
 & \quad\times\left|\left(\widehat{V}_{s}\widehat{V}_{s-k}-V_{s}V_{s-k}\right)+\left(V_{s}V_{s-k}-\mathbb{E}\left(V_{s}V_{s-k}\right)\right)+\mathbb{E}\left(V_{s}V_{s-k}\right)\right|\nonumber \\
 & \triangleq H_{1,T}+H_{2,T}+H_{3,T}.\nonumber 
\end{align}
We have to show that $H_{1,T}+H_{2,T}+H_{3,T}\overset{\mathbb{P}}{\rightarrow}0$.
Let $H_{1,1,T}$ (resp., $H_{1,2,T}$) be defined as $H_{1,T}$ but
with the sum over $k$ restricted to $k=1,\ldots,\,S_{T}$ (resp.,
$k=S_{T}+1,\ldots,\,T$). By a mean-value expansion,  using \eqref{Eq. K2-K2 fort part (ii)},
\begin{align*}
\left|H_{1,1,T}\right| & \leq CT^{8q/10\left(2q+1\right)}T^{-1/2}\sum_{k=1}^{S_{T}}\left|K_{1}\left(b_{\theta_{1},T}k\right)\right|\frac{n_{T}}{T}\sum_{r=0}^{\left\lfloor T/n_{T}\right\rfloor }\frac{1}{T\overline{b}_{\theta_{2},T}}\\
 & \quad\sum_{s=k+1}^{T}\left|K_{2}\left(\frac{\left(\left(r+1\right)n_{T}-\left(s-k/2\right)\right)/T}{\widehat{\overline{b}}_{2,T}\left(\left(r+1\right)n_{T}/T\right)}\right)-K_{2}\left(\frac{\left(\left(r+1\right)n_{T}-\left(s-k/2\right)\right)/T}{b_{\theta_{2},T}\left(\left(r+1\right)n_{T}/T\right)}\right)\right|\\
 & \quad\times\left\Vert V_{s}\left(\overline{\beta}\right)\frac{\partial}{\partial\beta}V_{s-k}\left(\overline{\beta}\right)+V_{s-k}\left(\overline{\beta}\right)\frac{\partial}{\partial\beta}V_{s}\left(\overline{\beta}\right)\right\Vert \sqrt{T}\left\Vert \widehat{\beta}-\beta_{0}\right\Vert \\
 & \leq CT^{8q/10\left(2q+1\right)}\overline{b}_{\theta_{2},T}^{-1}T^{-1/2-2/5}S_{T}\frac{n_{T}}{T}\sum_{r=0}^{\left\lfloor T/n_{T}\right\rfloor }O_{\mathbb{P}}\left(1\right)\\
 & \quad\times\left(\left(T^{-1}\sum_{s=1}^{T}\sup_{\beta\in\Theta}V_{s}^{2}\left(\beta\right)\right)^{2}\left(T^{-1}\sum_{s=1}^{T}\sup_{\beta\in\Theta}\left\Vert \frac{\partial}{\partial\beta}V_{s}\left(\beta\right)\right\Vert ^{2}\right)^{1/2}\right)\sqrt{T}\left\Vert \widehat{\beta}-\beta_{0}\right\Vert .
\end{align*}
 Using Assumption \ref{Assumption B} the right-hand side above is
such that
\begin{align*}
CT^{8q/10\left(2q+1\right)} & T^{-1/2-2/5}b_{\theta_{2},T}^{-1}S_{T}\frac{n_{T}}{T}\sum_{r=0}^{\left\lfloor T/n_{T}\right\rfloor }O_{\mathbb{P}}\left(1\right)\overset{\mathbb{P}}{\rightarrow}0,
\end{align*}
since $r<7/8+3q/4$. Next, 
\begin{align*}
\left|H_{1,2,T}\right| & \leq CT^{8q/10\left(2q+1\right)}T^{-1/2}\sum_{k=S_{T}+1}^{T-1}\left(b_{\theta_{1},T}k\right)^{-b}\frac{n_{T}}{T}\sum_{r=0}^{\left\lfloor T/n_{T}\right\rfloor }\frac{1}{Tb_{\theta_{2},T}}\\
 & \quad\times\sum_{s=k+1}^{T}\left|K_{2}\left(\frac{\left(\left(r+1\right)n_{T}-\left(s-k/2\right)\right)/T}{\widehat{b}_{2,T}\left(\left(r+1\right)n_{T}/T\right)}\right)-K_{2}\left(\frac{\left(\left(r+1\right)n_{T}-\left(s-k/2\right)\right)/T}{b_{\theta_{2},T}\left(\left(r+1\right)n_{T}/T\right)}\right)\right|\\
 & \quad\times\left\Vert V_{s}\left(\overline{\beta}\right)\frac{\partial}{\partial\beta}V_{s-k}\left(\overline{\beta}\right)+V_{s-k}\left(\overline{\beta}\right)\frac{\partial}{\partial\beta}V_{s}\left(\overline{\beta}\right)\right\Vert \sqrt{T}\left\Vert \widehat{\beta}-\beta_{0}\right\Vert \\
 & \leq CT^{8q/10\left(2q+1\right)}b_{\theta_{2},T}^{-1}T^{-1/2-2/5}b_{\theta_{1},T}^{-b}\sum_{k=S_{T}+1}^{T-1}k^{-b}\frac{n_{T}}{T}\sum_{r=0}^{\left\lfloor T/n_{T}\right\rfloor }O_{\mathbb{P}}\left(1\right)\\
 & \quad\times\left(\left(T^{-1}\sum_{s=1}^{T}\sup_{\beta\in\Theta}V_{s}^{2}\left(\beta\right)\right)^{2}\left(T^{-1}\sum_{s=1}^{T}\sup_{\beta\in\Theta}\left\Vert \frac{\partial}{\partial\beta}V_{s}\left(\beta\right)\right\Vert ^{2}\right)^{1/2}\right)\sqrt{T}\left\Vert \widehat{\beta}-\beta_{0}\right\Vert \\
 & \leq CT^{8q/10\left(2q+1\right)}b_{\theta_{2},T}^{-1}T^{-1/2-2/5}b_{\theta_{1},T}^{-b}\sum_{k=S_{T}+1}^{T-1}k^{-b}O_{\mathbb{P}}\left(1\right)\\
 & \leq CT^{8q/10\left(2q+1\right)}b_{\theta_{2},T}^{-1}T^{-1/2-2/5}b_{\theta_{1},T}^{-b}S_{T}^{1-b}O_{\mathbb{P}}\left(1\right)\\
 & \leq CT^{8q/10\left(2q+1\right)}b_{\theta_{2},T}^{-1}T^{-1/2-2/5}b_{\theta_{1},T}^{-b}b_{\theta_{1},T}^{-r\left(1-b\right)}O_{\mathbb{P}}\left(1\right)\\
 & \leq CT^{8q/10\left(2q+1\right)}b_{\theta_{2},T}^{-1}T^{-1/2-2/5}b_{\theta_{1},T}^{-b}T^{4r\left(1-b\right)/5\left(2q+1\right)}O_{\mathbb{P}}\left(1\right)\rightarrow0,
\end{align*}
 since $r>\left(8b-7-6q\right)/8\left(b-1\right)$. This shows
$H_{1,T}\overset{\mathbb{P}}{\rightarrow}0$. Let $H_{2,1,T}$ (resp.,
$H_{2,2,T}$) be defined as $H_{2,T}$ but with the sum over $k$
restricted to $k=1,\ldots,\,S_{T}$ (resp., $k=S_{T}+1,\ldots,\,T$).
We have 
\begin{align}
\mathbb{E}\left(H_{2,1,T}^{2}\right) & \leq T^{8q/5\left(2q+1\right)}\sum_{k=1}^{S_{T}}\sum_{j=1}^{S_{T}}K_{1}\left(b_{\theta_{1},T}k\right)K_{1}\left(b_{\theta_{1},T}j\right)\left(\frac{n_{T}}{T}\right)^{2}\sum_{r_{1}=0}^{\left\lfloor T/n_{T}\right\rfloor }\sum_{r_{2}=0}^{\left\lfloor T/n_{T}\right\rfloor }\frac{1}{\left(Tb_{\theta_{2},T}\right)^{2}}\label{Eq. (H21) =00003D 0 part (ii)}\\
 & \quad\times\sum_{s=k+1}^{T}\sum_{t=j+1}^{T}\left|K_{2}\left(\frac{\left(\left(r_{1}+1\right)n_{T}-\left(s-k/2\right)\right)/T}{\widehat{b}_{2,T}\left(\left(r_{1}+1\right)n_{T}/T\right)}\right)-K_{2}\left(\frac{\left(\left(r_{1}+1\right)n_{T}-\left(s-k/2\right)\right)/T}{b_{\theta_{2},T}\left(\left(r_{1}+1\right)n_{T}/T\right)}\right)\right|\nonumber \\
 & \quad\times\left|K_{2}\left(\frac{\left(\left(r_{2}+1\right)n_{T}-\left(t-j/2\right)\right)/T}{\widehat{b}_{2,T}\left(\left(r_{2}+1\right)n_{T}/T\right)}\right)-K_{2}\left(\frac{\left(\left(r_{2}+1\right)n_{T}-\left(t-j/2\right)\right)/T}{b_{\theta_{2},T}\left(\left(r_{2}+1\right)n_{T}/T\right)}\right)\right|\nonumber \\
 & \quad\times\left|\mathbb{E}\left(V_{s}V_{s-k}-\mathbb{E}\left(V_{s}V_{s-k}\right)\right)\left(V_{t}V_{t-k}-\mathbb{E}\left(V_{t}V_{t-k}\right)\right)\right|\nonumber \\
 & \leq CT^{8q/5\left(2q+1\right)}S_{T}^{2}T^{-2/5}\left(Tb_{\theta_{2},T}\right)^{-1}\sup_{k\geq1}Tb_{\theta_{2},T}\mathrm{Var}\left(\widetilde{\Gamma}\left(k\right)\right)O_{\mathbb{P}}\left(1\right)\nonumber \\
 & \leq CT^{\left(8q+8r\right)/5\left(2q+1\right)-2/5-1}O_{\mathbb{P}}\left(b_{\theta_{2},T}^{-1}\right)\rightarrow0,\nonumber 
\end{align}
where we have used Lemma \ref{Lemma: Tb2*Var(Gamma_k)} and $r<\left(6+4q\right)/8$.
Turning to $H_{2,2,T},$ 
\begin{align}
\mathbb{E}\left(H_{2,2,T}^{2}\right) & \leq T^{8q/5\left(2q+1\right)-2/5}\left(Tb_{\theta_{2},T}\right)^{-1}b_{\theta_{1},T}^{-2b}\left(\sum_{k=S_{T}+1}^{T-1}k^{-b}\sqrt{Tb_{\theta_{2},T}}\left(\mathrm{Var}\left(\widetilde{\Gamma}\left(k\right)\right)\right)^{1/2}O\left(1\right)\right)^{2}\label{Eq. (H22) =00003D0 part (ii)}\\
 & \leq T^{8q/5\left(2q+1\right)}T^{-2/5-1}b_{\theta_{2},T}^{-1}b_{\theta_{1},T}^{-2b}\left(\sum_{k=S_{T}+1}^{T-1}k^{-b}\sqrt{Tb_{\theta_{2},T}}\left(\mathrm{Var}\left(\widetilde{\Gamma}\left(k\right)\right)\right)^{1/2}\right)^{2}\nonumber \\
 & \leq T^{8q/5\left(2q+1\right)}T^{-2/5-1}b_{\theta_{2},T}^{-1}b_{\theta_{1},T}^{-2b}\left(\sum_{k=S_{T}+1}^{T-1}k^{-b}O\left(1\right)\right)^{2}\nonumber \\
 & \leq T^{8q/5\left(2q+1\right)}T^{-2/5-1}b_{\theta_{2},T}^{-1}b_{\theta_{1},T}^{-2b}S_{T}^{2\left(1-b\right)}\rightarrow0,\nonumber 
\end{align}
since $r>(b-3/4-q/2)/\left(b-1\right)$. Combining \eqref{Eq. (H21) =00003D 0 part (ii)}-\eqref{Eq. (H22) =00003D0 part (ii)}
yields $H_{2,T}\overset{\mathbb{P}}{\rightarrow}0.$  Let $H_{3,1,T}$
(resp., $H_{3,2,T}$) be defined as $H_{3,T}$ but with the sum over
$k$ restricted to $k=1,\ldots,\,S_{T}$ (resp., $k=S_{T}+1,\ldots,\,T$).
Given $\left|K_{1}\left(\cdot\right)\right|\leq1$ and \eqref{Eq. K2-K2 fort part (ii)},
we have 
\begin{align*}
\left|H_{3,1,T}\right| & \leq CT^{8q/10\left(2q+1\right)}T^{-2/5}\sum_{k=1}^{S_{T}}\left|\Gamma_{T}\left(k\right)\right|\\
 & \leq CT^{8q/10\left(2q+1\right)}T^{-2/5}\sum_{k=1}^{\infty}k^{-l}\rightarrow0,
\end{align*}
since $\sum_{k=1}^{\infty}k^{-l}<\infty$ for $l>1$ and $T^{8q/10\left(2q+1\right)}T^{-2/5}\rightarrow0.$
Finally, 
\begin{align*}
\left|H_{3,2,T}\right| & \leq CT^{8q/10\left(2q+1\right)}T^{-2/5}\sum_{k=S_{T}+1}^{T-1}\left|\Gamma_{T}\left(k\right)\right|\\
 & \leq CT^{8q/10\left(2q+1\right)}T^{-2/5}\sum_{k=S_{T}+1}^{T-1}k^{-l}\\
 & \leq CT^{8q/10\left(2q+1\right)}T^{-2/5}S_{T}^{1-l}\\
 & \leq CT^{8q/10\left(2q+1\right)}T^{-2/5}T^{4r\left(1-l\right)/5\left(2q+1\right)}\rightarrow0.
\end{align*}
This completes the proof of part (ii).

We now move to part (i). For some arbitrary $\phi_{\theta^{*}}\in\left(0,\,\infty\right)$,
$\widehat{J}_{T}(b_{\theta_{1},T},\,b_{\theta_{2},T})-J_{T}=o_{\mathbb{P}}\left(1\right)$
by Theorem \ref{Theorem 1 -Consistency and Rate}-(i) since $b_{\theta_{2},T}=O(T^{-1/5})$
and $q>1/2$ imply that $\sqrt{T}b_{1,T}\rightarrow\infty$ holds.
Hence, it remains to show that $\widehat{J}_{T}(b_{\theta_{1},T},\,b_{\theta_{2},T})-\widehat{J}_{T}(\widehat{b}_{1,T},\,\widehat{b}_{2,T})=o_{\mathbb{P}}\left(1\right)$.
Note that this result differs from that of part (ii) only because
the scale factor $T^{8q/10\left(2q+1\right)}$ does not appear, Assumption
\ref{Assumption E-F-G}-(ii) is replaced by part (i) of the same assumption,
Assumption \ref{Assumption E-F-G}-(iii, v, vi) is not imposed, and
$q>1/2$. Let $S_{T}$ be defined as in part (ii) and
\begin{align*}
r\in & (\max\left\{ \left(8b-10q-5\right)/8\left(b-1\right),\,\left(b-1/2-q\right)/\left(b-1\right)\right\} ,\\
 & \min\left\{ 13/16+5q/8,\,\left(3+2q\right)/4,\,1\right\} ),
\end{align*}
with $b>1+1/q$. We will use the decomposition in \eqref{eq (Decomposition J_T proof of Theorem 3 Andrews91)},
and $N_{1}$ and $N_{2}$ as defined after \eqref{eq (Decomposition J_T proof of Theorem 3 Andrews91)}.
Let $A_{1,T},\,A_{2,T}$ and $A_{3,T}$ be as in \eqref{Eq. 23}
without the scale factor $T^{8q/10\left(2q+1\right)}$. Proceeding
as in \eqref{Eq. 24}, we have
\begin{align}
\left|A_{1,1,T}\right| & \leq\sum_{k=1}^{S_{T}}C_{2}\left|\widehat{b}_{1,T}-b_{\theta_{1},T}\right|k\left|\widehat{\Gamma}\left(k\right)\right|\label{Eq. 24-1}\\
 & \leq C\left|\widehat{\phi}\left(q\right)^{1/\left(2q+1\right)}-\phi_{\theta^{*}}^{1/\left(2q+1\right)}\right|\left(\widehat{\phi}\left(q\right)\phi_{\theta^{*}}\right)^{-1/\left(2q+1\right)}\left(T\widehat{\overline{b}}_{2,T}\right)^{-1/\left(2q+1\right)}\sum_{k=1}^{S_{T}}k\left|\widehat{\Gamma}\left(k\right)\right|,\nonumber 
\end{align}
for some $C<\infty$. By Assumption \ref{Assumption E-F-G}-(i),
\begin{align*}
\left|\widehat{\phi}\left(q\right)^{1/\left(2q+1\right)}-\phi_{\theta^{*}}^{1/\left(2q+1\right)}\right|\left(\widehat{\phi}\left(q\right)\phi_{\theta^{*}}\right)^{-1/\left(2q+1\right)} & =O_{\mathbb{P}}\left(1\right).
\end{align*}
Then, it suffices to show that $B_{1,T}+B_{2,T}+B_{3,T}\overset{\mathbb{P}}{\rightarrow}0$,
where 
\begin{align}
B_{1,T} & =\left(T\widehat{\overline{b}}_{2,T}\right)^{-1/\left(2q+1\right)}\sum_{k=1}^{S_{T}}k\left|\widehat{\Gamma}\left(k\right)-\widetilde{\Gamma}\left(k\right)\right|,\label{Eq. A.25 Andrews 91-1}\\
B_{2,T} & =\left(T\widehat{\overline{b}}_{2,T}\right)^{-1/\left(2q+1\right)}\sum_{k=1}^{S_{T}}k\left|\widetilde{\Gamma}\left(k\right)-\Gamma_{T}\left(k\right)\right|,\nonumber \\
B_{3,T} & =\left(T\widehat{\overline{b}}_{2,T}\right)^{-1/\left(2q+1\right)}\sum_{k=1}^{S_{T}}k\left|\Gamma_{T}\left(k\right)\right|.\nonumber 
\end{align}
By a mean-value expansion, we have 
\begin{align}
B_{1,T} & \leq\left(T\widehat{\overline{b}}_{2,T}\right)^{-1/\left(2q+1\right)}T^{-1/2}\sum_{k=1}^{S_{T}}k\left|\left(\frac{\partial}{\partial\beta'}\widehat{\Gamma}\left(k\right)|_{\beta=\overline{\beta}}\right)\sqrt{T}\left(\widehat{\beta}-\beta_{0}\right)\right|\label{Eq. A.26-1}\\
 & \leq C\left(T\widehat{\overline{b}}_{2,T}\right)^{-1/\left(2q+1\right)}\left(Tb_{\theta_{2,T}}\right)^{2r/\left(2q+1\right)}T^{-1/2}\sup_{k\geq1}\left\Vert \frac{\partial}{\partial\beta}\widehat{\Gamma}\left(k\right)|_{\beta=\overline{\beta}}\right\Vert \sqrt{T}\left\Vert \widehat{\beta}-\beta_{0}\right\Vert ,\nonumber 
\end{align}
since $r<13/16+5q/8$, and $\sup_{k\geq1}||\left(\partial/\partial\beta\right)\widehat{\Gamma}\left(k\right)|_{\beta=\overline{\beta}}||=O_{\mathbb{P}}\left(1\right)$
using \eqref{Eq. A.10 Andrews 91} and Assumption \ref{Assumption B}-(ii,iii).
In addition, 
\begin{align}
\mathbb{E}\left(B_{2,T}^{2}\right) & \leq\mathbb{E}\left(\left(T\widehat{\overline{b}}_{2,T}\right)^{-2/\left(2q+1\right)}\sum_{k=1}^{S_{T}}\sum_{j=1}^{S_{T}}kj\left|\widetilde{\Gamma}\left(k\right)-\Gamma_{T}\left(k\right)\right|\left|\widetilde{\Gamma}\left(j\right)-\Gamma_{T}\left(j\right)\right|\right)\label{Eq. A.27-1}\\
 & \leq\mathbb{E}\left(\left(T\widehat{\overline{b}}_{2,T}\right)^{-2/\left(2q+1\right)}\sum_{k=1}^{S_{T}}\sum_{j=1}^{S_{T}}kj\left|\widetilde{\Gamma}\left(k\right)-\Gamma_{T}\left(k\right)\right|\left|\widetilde{\Gamma}\left(j\right)-\Gamma_{T}\left(j\right)\right|\right)\nonumber \\
 & \leq\left(T\widehat{\overline{b}}_{2,T}\right)^{-2/\left(2q+1\right)-1}S_{T}^{4}\sup_{k\geq1}T\widehat{\overline{b}}_{2,T}\mathrm{Var}\left(\widetilde{\Gamma}\left(k\right)\right)\nonumber \\
 & \leq\left(T\widehat{\overline{b}}_{2,T}\right)^{-2/\left(2q+1\right)-1}\left(Tb_{2,T}\right)^{4r/\left(2q+1\right)}\sup_{k\geq1}T\widehat{\overline{b}}_{2,T}\mathrm{Var}\left(\widetilde{\Gamma}\left(k\right)\right)\nonumber \\
 & \leq\widehat{\overline{b}}_{2,T}^{-2/\left(2q+1\right)-1}T^{-1-2/\left(2q+1\right)}T^{16r/5\left(2q+1\right)}\sup_{k\geq1}T\widehat{\overline{b}}_{2,T}\mathrm{Var}\left(\widetilde{\Gamma}\left(k\right)\right)\rightarrow0,\nonumber 
\end{align}
given that $\sup_{k\geq1}T\widehat{\overline{b}}_{2,T}\mathrm{Var}(\widetilde{\Gamma}\left(k\right))=O\left(1\right)$
by Lemma \ref{Lemma: Tb2*Var(Gamma_k)} and $r<\left(3+2q\right)/4$.
The bound in \eqref{Eq. A.28} is replaced by
\begin{align}
B_{3,T} & \leq\left(T\widehat{\overline{b}}_{2,T}\right)^{-1/\left(2q+1\right)}S_{T}\sum_{k=1}^{\infty}\left|\Gamma_{T}\left(k\right)\right|\label{Eq. A.28-1}\\
 & \leq\left(T\widehat{\overline{b}}_{2,T}\right)^{\left(r-1\right)/\left(2q+1\right)}O_{\mathbb{P}}\left(1\right)\rightarrow0,\nonumber 
\end{align}
using Assumption \ref{Assumption A - Dependence}-(i) since $r<1$.
 This gives $A_{1,T}\overset{\mathbb{P}}{\rightarrow}0$. Next,
we show that $A_{2,T}\overset{\mathbb{P}}{\rightarrow}0$. As above,
let $A_{2,1,T}=L_{1,T}+L_{2,T}+L_{3,T}$ where each summand is defined
as in \eqref{Eq. A.29} without the factor $T^{8q/10\left(2q+1\right)}$.
Equation \eqref{Eq. (30)} is then replaced by 
\begin{align}
\left|L_{1,T}\right| & =T^{-1/2}\sum_{k=S_{T}+1}^{T-1}C_{1}\left(\widehat{b}_{1,T}k\right)^{-b}\left|\left(\frac{\partial}{\partial\beta'}\widehat{\Gamma}\left(k\right)\right)|_{\beta=\overline{\beta}}\sqrt{T}\left(\widehat{\beta}-\beta_{0}\right)\right|\label{Eq. (30)-2}\\
 & =T^{-1/2+4b/5\left(2q+1\right)}\sum_{k=S_{T}+1}^{T-1}C_{1}k^{-b}\left|\left(\frac{\partial}{\partial\beta'}\widehat{\Gamma}\left(k\right)\right)|_{\beta=\overline{\beta}}\sqrt{T}\left(\widehat{\beta}-\beta_{0}\right)\right|\nonumber \\
 & =T^{-1/2+4b/5\left(2q+1\right)+4r\left(1-b\right)/5\left(2q+1\right)}\left|\left(\frac{\partial}{\partial\beta'}\widehat{\Gamma}\left(k\right)\right)|_{\beta=\overline{\beta}}\sqrt{T}\left(\widehat{\beta}-\beta_{0}\right)\right|\nonumber \\
 & =T^{-1/2+4b/5\left(2q+1\right)+4r\left(1-b\right)/5\left(2q+1\right)}O\left(1\right)O_{\mathbb{P}}\left(1\right),\nonumber 
\end{align}
 which converges to zero since $r>\left(8b-10q-5\right)/8\left(b-1\right)$.
Also, \eqref{Eq. 31} is replaced by 
\begin{align}
\left|L_{2,T}\right| & =\sum_{k=S_{T}+1}^{T-1}C_{1}\left(\widehat{b}_{1,T}k\right)^{-b}\left|\widetilde{\Gamma}\left(k\right)-\Gamma_{T}\left(k\right)\right|\label{Eq. 31-1}\\
 & =C_{1}\left(qK_{1,q}^{2}\widehat{\phi}\left(q\right)\right)^{b/\left(2q+1\right)}T^{b/\left(2q+1\right)-1/2}\widehat{b}_{2,T}^{b/\left(2q+1\right)-1/2}\left(\sum_{k=S_{T}+1}^{T-1}k^{-b}\right)\sqrt{T\widehat{b}_{2,T}}\left|\widetilde{\Gamma}\left(k\right)-\Gamma_{T}\left(k\right)\right|,\nonumber 
\end{align}
and the bound in \eqref{Eq. 32 Andrews 91} is replaced by, 
\begin{align}
\mathbb{E} & \left(T^{b/\left(2q+1\right)-1/2}\widehat{\overline{b}}_{2,T}^{b/\left(2q+1\right)-1/2}\sum_{k=S_{T}}^{T-1}k^{-b}\sqrt{T\widehat{\overline{b}}_{2,T}}\left|\widetilde{\Gamma}\left(k\right)-\Gamma_{T}\left(k\right)\right|\right)^{2}\label{Eq. 32 Andrews 91-1}\\
 & \leq T^{2b/\left(2q+1\right)-1}\widehat{\overline{b}}_{2,T}^{2b/\left(2q+1\right)-1}\left(\sum_{k=S_{T}}^{T-1}k^{-b}\sqrt{T\widehat{\overline{b}}_{2,T}}\left(\mathrm{Var}\left(\widetilde{\Gamma}\left(k\right)\right)\right)^{1/2}\right)^{2}\nonumber \\
 & =T^{2b/\left(2q+1\right)-1}\widehat{\overline{b}}_{2,T}^{2b/\left(2q+1\right)-1}\left(\sum_{k=S_{T}}^{T-1}k^{-b}\right)^{2}O\left(1\right)\nonumber \\
 & =T^{2b/\left(2q+1\right)-1}\widehat{\overline{b}}_{2,T}^{2b/\left(2q+1\right)-1}S_{T}^{2\left(1-b\right)}O\left(1\right)\rightarrow0,\nonumber 
\end{align}
 since $r>\left(b-1/2-q\right)/\left(b-1\right)$ and $Tb_{2,T}\mathrm{Var}(\widetilde{\Gamma}\left(k\right))=O\left(1\right)$,
as above. Combining \eqref{Eq. 31-1}-\eqref{Eq. 32 Andrews 91-1}
yields $L_{2,T}\overset{\mathbb{P}}{\rightarrow}0$ since $\widehat{\phi}\left(q\right)=O_{\mathbb{P}}\left(1\right)$.
Let us turn to $L_{3,T}$. We have \eqref{Eq. (33)} replaced by,
\begin{align}
\left|\sum_{k=S_{T}+1}^{T-1}K_{1}\left(\widehat{b}_{1,T}k\right)\Gamma_{T}\left(k\right)\right| & \leq\sum_{k=S_{T}+1}^{T-1}\frac{n_{T}}{T}\sum_{r=0}^{\left\lfloor T/n_{T}\right\rfloor }\left|c\left(rn_{T}/T,\,k\right)\right|\label{Eq. (33)-1}\\
 & \leq\sum_{k=S_{T}+1}^{T-1}\sup_{u\in\left[0,\,1\right]}\left|c\left(u,\,k\right)\right|\rightarrow0.\nonumber 
\end{align}
Equations \eqref{Eq. (30)-2}-\eqref{Eq. (33)-1} imply $A_{2,1,T}\overset{\mathbb{P}}{\rightarrow}0$.
Thus, as in the proof of part (ii), we have $A_{2,T}\overset{\mathbb{P}}{\rightarrow}0$
and $A_{3,T}\overset{\mathbb{P}}{\rightarrow}0$. It remains to show
that $(\widehat{J}_{T}(b_{\theta_{1},T},\,\widehat{\overline{b}}_{2,T})-\widehat{J}_{T}(b_{\theta_{1},T},\,\overline{b}_{\theta_{2},T}))\overset{\mathbb{P}}{\rightarrow}0$.
Let $\widehat{c}_{\theta_{2},T}\left(rn_{T}/T,\,k\right)$ be defined
as in part (ii). We have \eqref{eq (c_hat - c_theta2)}, and \eqref{Eq. K2-K2 fort part (ii)}
is replaced by 
\begin{align}
K_{2} & \left(\frac{\left(Tu-\left(s-k/2\right)/T\right)}{\widehat{b}_{2,T}\left(u\right)}\right)-K_{2}\left(\frac{\left(Tu-\left(s-k/2\right)/T\right)}{b_{\theta_{2},T}\left(u\right)}\right)\label{Eq. (K2 - K2) for part (i)}\\
 & \leq C_{4}\left|\frac{Tu-\left(s-k/2\right)}{T\widehat{b}_{2,T}\left(u\right)}-\frac{Tu-\left(s-k/2\right)}{Tb_{\theta_{2},T}\left(u\right)}\right|\nonumber \\
 & \leq C_{4}T^{-1}\left|\frac{Tu-\left(s-k/2\right)\left(\widehat{b}_{2,T}\left(u\right)-b_{\theta_{2},T}\left(u\right)\right)}{\widehat{b}_{2,T}\left(u\right)b_{\theta_{2},T}\left(u\right)}\right|\nonumber \\
 & =C_{4}T^{-4/5}\left(\left(\frac{\widehat{D}_{1}\left(u\right)}{\widehat{D}_{2}\left(u\right)}\right)\left(\frac{D_{1,\theta}\left(u\right)}{D_{2}\left(u\right)}\right)\right)^{1/5}\left|\left(\frac{\widehat{D}_{2}\left(u\right)}{\widehat{D}_{1}\left(u\right)}\right)^{1/5}-\left(\frac{D_{2}\left(u\right)}{D_{1,\theta}\left(u\right)}\right)^{1/5}\right|\left|Tu-\left(s-k/2\right)\right|\nonumber \\
 & =CT^{-4/5}\left|Tu-\left(s-k/2\right)\right|,\nonumber 
\end{align}
for $s\in\left\{ Tu-\left\lfloor Tb_{\theta_{2},T}\left(u\right)\right\rfloor ,\ldots,\,Tu+\left\lfloor Tb_{\theta_{2},T}\left(u\right)\right\rfloor \right\} $,
where $u=\left(r+1\right)n_{T}/T$.  Therefore, 
\begin{align*}
\widehat{J}_{T} & \left(b_{\theta_{1},T},\,\widehat{\overline{b}}_{2,T}\right)-\widehat{J}_{T}\left(b_{\theta_{1},T},\,\overline{b}_{\theta_{2},T}\right)\\
 & =\sum_{k=-T+1}^{T-1}K_{1}\left(b_{\theta_{1},T}k\right)\frac{n_{T}}{T}\sum_{r=0}^{\left\lfloor T/n_{T}\right\rfloor }\left(\widehat{c}\left(rn_{T}/T,\,k\right)-\widehat{c}_{\theta_{2},T}\left(rn_{T}/T,\,k\right)\right)\\
 & \leq C\sum_{k=-T+1}^{T-1}K_{1}\left(b_{\theta_{1},T}k\right)\\
 & \quad\times\frac{n_{T}}{T}\sum_{r=0}^{\left\lfloor T/n_{T}\right\rfloor }\frac{1}{T\overline{b}_{\theta_{2},T}}\sum_{s=k+1}^{T}\left|K_{2}\left(\frac{\left(\left(r+1\right)n_{T}-\left(s-k/2\right)\right)/T}{\widehat{b}_{2,T}\left(\left(r+1\right)n_{T}/T\right)}\right)-K_{2}\left(\frac{\left(\left(r+1\right)n_{T}-\left(s-k/2\right)\right)/T}{b_{\theta_{2},T}\left(\left(r+1\right)n_{T}/T\right)}\right)\right|\\
 & \quad\times\left|\left(\widehat{V}_{s}\widehat{V}_{s-k}-V_{s}V_{s-k}\right)+\left(V_{s}V_{s-k}-\mathbb{E}\left(V_{s}V_{s-k}\right)\right)+\mathbb{E}\left(V_{s}V_{s-k}\right)\right|\\
 & \triangleq H_{1,T}+H_{2,T}+H_{3,T}.
\end{align*}
We have to show that $H_{1,T}+H_{2,T}+H_{3,T}\overset{\mathbb{P}}{\rightarrow}0$.
By a mean-value expansion,  using \eqref{Eq. (K2 - K2) for part (i)},
\begin{align*}
\left|H_{1,T}\right| & \leq CT^{-1/2}\sum_{k=-T+1}^{T-1}\left|K_{1}\left(b_{\theta_{1},T}k\right)\right|\\
 & \quad\times\frac{n_{T}}{T}\sum_{r=0}^{\left\lfloor T/n_{T}\right\rfloor }\frac{1}{Tb_{\theta_{2},T}}\sum_{s=k+1}^{T}\left|K_{2}\left(\frac{\left(\left(r+1\right)n_{T}-\left(s-k/2\right)\right)/T}{\widehat{b}_{2,T}\left(\left(r+1\right)n_{T}/T\right)}\right)-K_{2}\left(\frac{\left(\left(r+1\right)n_{T}-\left(s-k/2\right)\right)/T}{b_{\theta_{2},T}\left(\left(r+1\right)n_{T}/T\right)}\right)\right|\\
 & \quad\times\left\Vert V_{s}\left(\overline{\beta}\right)\frac{\partial}{\partial\beta}V_{s-k}\left(\overline{\beta}\right)+V_{s-k}\left(\overline{\beta}\right)\frac{\partial}{\partial\beta}V_{s}\left(\overline{\beta}\right)\right\Vert \sqrt{T}\left\Vert \widehat{\beta}-\beta_{0}\right\Vert \\
 & \leq Cb_{\theta_{2},T}^{-1}T^{-1/2}\sum_{k=-T+1}^{T-1}\left|K_{1}\left(b_{\theta_{1},T}k\right)\right|\\
 & \quad\times\frac{n_{T}}{T}\sum_{r=0}^{\left\lfloor T/n_{T}\right\rfloor }CO_{\mathbb{P}}\left(1\right)\\
 & \quad\times\left(\left(T^{-1}\sum_{s=1}^{T}\sup_{\beta\in\Theta}V_{s}^{2}\left(\beta\right)\right)^{2}\left(T^{-1}\sum_{s=1}^{T}\sup_{\beta\in\Theta}\left\Vert \frac{\partial}{\partial\beta}V_{s}\left(\beta\right)\right\Vert ^{2}\right)^{1/2}\right)\sqrt{T}\left\Vert \widehat{\beta}-\beta_{0}\right\Vert .
\end{align*}
Using Assumption \ref{Assumption B} and \eqref{Eq. (K2 - K2) for part (i)},
the right-hand side above is such that
\begin{align*}
C & T^{-1/2}b_{\theta_{1},T}^{-1}b_{\theta_{2},T}^{-1}b_{\theta_{1},T}\sum_{k=-T+1}^{T-1}\left|K_{1}\left(b_{\theta_{1},T}k\right)\right|\frac{n_{T}}{T}\sum_{r=0}^{\left\lfloor T/n_{T}\right\rfloor }CO_{\mathbb{P}}\left(1\right)\overset{\mathbb{P}}{\rightarrow}0,
\end{align*}
 since $T^{-1/2}b_{\theta_{1},T}^{-1}b_{\theta_{2},T}^{-1}\rightarrow0$.
This shows $H_{1,T}\overset{\mathbb{P}}{\rightarrow}0$. Let $H_{2,1,T}$
(resp. $H_{2,2,T}$) be defined as $H_{2,T}$ but with the sum over
$k$ restricted to $k=1,\ldots,\,S_{T}$ (resp., $k=S_{T}+1,\ldots,\,T$).
We have 
\begin{align}
\mathbb{E}\left(H_{2,1,T}^{2}\right) & \leq\sum_{k=1}^{S_{T}}\sum_{j=1}^{S_{T}}K_{1}\left(b_{\theta_{1},T}k\right)K_{1}\left(b_{\theta_{1},T}j\right)\label{Eq. (H21) =00003D 0}\\
 & \quad\times\left(\frac{n_{T}}{T}\right)^{2}\sum_{r_{1}=0}^{\left\lfloor T/n_{T}\right\rfloor }\sum_{r_{2}=0}^{\left\lfloor T/n_{T}\right\rfloor }\frac{1}{\left(Tb_{\theta_{2},T}\right)^{2}}\sum_{s=k+1}^{T}\sum_{t=j+1}^{T}\nonumber \\
 & \quad\times\left|K_{2}\left(\frac{\left(\left(r_{1}+1\right)n_{T}-\left(s-k/2\right)\right)/T}{\widehat{b}_{2,T}\left(\left(r_{1}+1\right)n_{T}/T\right)}\right)-K_{2}\left(\frac{\left(\left(r_{1}+1\right)n_{T}-\left(s-k/2\right)\right)/T}{b_{\theta_{2},T}\left(\left(r_{1}+1\right)n_{T}/T\right)}\right)\right|\nonumber \\
 & \quad\times\left|K_{2}\left(\frac{\left(\left(r_{2}+1\right)n_{T}-\left(t-j/2\right)\right)/T}{\widehat{b}_{2,T}\left(\left(r_{2}+1\right)n_{T}/T\right)}\right)-K_{2}\left(\frac{\left(\left(r_{2}+1\right)n_{T}-\left(t-j/2\right)\right)/T}{b_{\theta_{2},T}\left(\left(r_{2}+1\right)n_{T}/T\right)}\right)\right|\nonumber \\
 & \quad\times\left|V_{s}V_{s-k}-\mathbb{E}\left(V_{s}V_{s-k}\right)\left(V_{t}V_{t-k}-\mathbb{E}\left(V_{t}V_{t-k}\right)\right)\right|.\nonumber \\
 & \leq CS_{T}^{2}\left(Tb_{\theta_{2},T}\right)^{-1}\sup_{k\geq1}Tb_{\theta_{2},T}\mathrm{Var}\left(\widetilde{\Gamma}\left(k\right)\right)O_{\mathbb{P}}\left(1\right)\nonumber \\
 & \leq CT^{8r/5\left(2q+1\right)}O_{\mathbb{P}}\left(T^{-1}b_{\theta_{2},T}^{-1}\right)\rightarrow0,\nonumber 
\end{align}
where we have used Lemma \ref{Lemma: Tb2*Var(Gamma_k)}, \eqref{Eq. (K2 - K2) for part (i)}
and $r<3/2$. Turning to $H_{2,2,T},$ 
\begin{align}
\mathbb{E}\left(H_{2,2,T}^{2}\right) & \leq\left(Tb_{\theta_{2},T}\right)^{-1}b_{\theta_{1},T}^{-2b}\left(\sum_{k=S_{T}+1}^{T-1}k^{-b}\sqrt{Tb_{\theta_{2},T}}\left(\mathrm{Var}\left(\widetilde{\Gamma}\left(k\right)\right)\right)^{1/2}O\left(1\right)\right)^{2}\label{Eq. (H22) =00003D0}\\
 & \leq T^{-1}b_{\theta_{2},T}^{-1}b_{\theta_{1},T}^{-2b}\left(\sum_{k=S_{T}+1}^{T-1}k^{-b}\sqrt{Tb_{\theta_{2},T}}\left(\mathrm{Var}\left(\widetilde{\Gamma}\left(k\right)\right)\right)^{1/2}\right)^{2}\nonumber \\
 & \leq T^{-1}b_{\theta_{2},T}^{-1}b_{\theta_{1},T}^{-2b}\left(\sum_{k=S_{T}+1}^{T-1}k^{-b}O\left(1\right)\right)^{2}\nonumber \\
 & \leq T^{-1}b_{\theta_{2},T}^{-1}b_{\theta_{1},T}^{-2b}S_{T}^{2\left(1-b\right)}\rightarrow0,\nonumber 
\end{align}
since $r>\left(b-q-1/2\right)/\left(b-1\right).$ Combining \eqref{Eq. (H21) =00003D 0}-\eqref{Eq. (H22) =00003D0}
yields $H_{2,T}\overset{\mathbb{P}}{\rightarrow}0.$  Given $\left|K_{1}\left(\cdot\right)\right|\leq1$
and \eqref{Eq. (K2 - K2) for part (i)}, we have 
\begin{align*}
\left|H_{3,T}\right| & \leq C\sum_{k=-\infty}^{\infty}\left|\Gamma_{T}\left(k\right)\right|o_{\mathbb{P}}\left(1\right)\rightarrow0.
\end{align*}
This concludes the proof of part (i).

The result of part (iii) follows from the same argument as in Theorem
\ref{Theorem 1 -Consistency and Rate}-(iii) with references to Theorem
\ref{Theorem 1 -Consistency and Rate}-(i,ii) changed to Theorem\textit{
}\ref{Theorem 3 Andrews 91}-(i,ii). $\square$

\bibliographystyleReferencesSupptwo{elsarticle-harv}  
\bibliographyReferencesSupptwo{References_Supp_two}

\clearpage{}

\end{singlespace}
\end{document}